%% file: main.tex
\documentclass[aps, longbibliography, superscriptaddress, floatfix, prx, 10pt, nofootinbib]{revtex4-2}

\usepackage{graphicx}
\usepackage[percent]{overpic}
\newcommand{\steplabel}[3]{\begin{overpic}[width=#1]{#3}\put(2,92){\small\textbf{#2)}}\end{overpic}}
\usepackage[dvipsnames]{xcolor}
\usepackage{dcolumn}
\usepackage{bm}
\usepackage{appendix}
\usepackage{mathrsfs}
\usepackage{tikz}
\usetikzlibrary{quantikz2, backgrounds, decorations, calc}
\definecolor{cpurple}{HTML}{D8D5FA}
\definecolor{cpurplebdr}{HTML}{534AB7}
\definecolor{cteal}{HTML}{C4EDDD}
\definecolor{ctealbdr}{HTML}{0F6E56}
\definecolor{ccoral}{HTML}{F8D7CA}
\definecolor{ccoralbdr}{HTML}{993C1D}
\definecolor{cgray}{HTML}{E2E0D8}
\definecolor{cgraybdr}{HTML}{5F5E5A}
\definecolor{camber}{HTML}{FCE4B0}
\definecolor{camberbdr}{HTML}{854F0B}
\usepackage{amsfonts,amsmath,physics}
\usepackage[hypertexnames=false]{hyperref}
\usepackage{tabularray}
\usepackage{placeins}
\usepackage{booktabs}
\usepackage{cleveref}
\usepackage{booktabs}
\usepackage{array}
\usepackage{comment}
\usepackage{ragged2e}
\usepackage[caption=false]{subfig} 

\makeatletter
\long\def\@caption#1[#2]#3{%
  \par\vskip\abovecaptionskip   
  \addcontentsline{\csname ext@#1\endcsname}{#1}{%
    \protect\numberline{\csname the#1\endcsname}{#2}}%
  \begingroup
    \small
    \@parboxrestore
    \parindent=0pt
    \noindent\parbox{\hsize}{%
      \parindent=0pt
      \leftskip=0pt
      \rightskip=0pt
      \justifying
      \noindent\textbf{\csname fnum@#1\endcsname}\enspace\ignorespaces#3\par
    }%
  \endgroup
  \par\vskip\belowcaptionskip   
}
\makeatother

\input{commands.tex}

\begin{document}

\title{Two Layers, No Swaps: Biplanar SPOQC Architecture Improves Runtime of Fermi-Hubbard Simulation}

\author{Boris Bourdoncle}
\thanks{Alphabetical order (see Section~\ref{sec:contributions} for author contributions). \\ Corresponding authors: boris.bourdoncle@quandela.com, johannes@walruscomputing.com}
\affiliation{Quandela, 7 rue Leonard de Vinci, 91300}
\author{Peter-Jan Derks}
\thanks{Alphabetical order (see Section~\ref{sec:contributions} for author contributions). \\ Corresponding authors: boris.bourdoncle@quandela.com, johannes@walruscomputing.com}
\affiliation{Walrus Computing}
\author{Théo Dessertaine}
\thanks{Alphabetical order (see Section~\ref{sec:contributions} for author contributions). \\ Corresponding authors: boris.bourdoncle@quandela.com, johannes@walruscomputing.com}\affiliation{Quandela, 7 rue Leonard de Vinci, 91300}
\author{Johannes Frank}
\thanks{Alphabetical order (see Section~\ref{sec:contributions} for author contributions). \\ Corresponding authors: boris.bourdoncle@quandela.com, johannes@walruscomputing.com}
\affiliation{Walrus Computing}

\date{\today}

\begin{abstract}
We estimate the cost of simulating the two-dimensional Fermi-Hubbard model on a biplanar spin-optical quantum computing (SPOQC) architecture. 
Qubits are encoded in the honeycomb Floquet code, and we use a circuit-level noise model with explicit timings for each native physical operation.
We benchmark lattice surgery and magic state preparation within each plane, and transversal CNOT gates between corresponding logical qubits across planes.
We compile a plaquette-based Trotterization of the time evolution operator, mapping the two spin sectors of the Fermi-Hubbard model onto two physical planes.
This architectural co-design eliminates fermionic swap operations and reduces the depth of each Trotter step to $4t_\mathrm{synth} + 90$ logical timesteps, where $t_\mathrm{synth}$ is the logical timestep cost of arbitrary-angle rotations, compared to $6t_\mathrm{synth} + 354$ in prior single-plane compilations. 
All error sources --- algorithmic (Trotter), logical noise, magic state infidelity, and rotation synthesis --- are treated jointly within a single 1\% diamond norm budget. 
For an $L\times L$ lattice with hopping amplitude $t$ and on-site interaction strength $U$, setting $L=8$ and $U/t=8$, we estimate a total runtime of approximately $2$ hours using $1.35\times 10^6$ physical qubits. 
We find that fallback-based rotation synthesis methods become a scalability bottleneck: the probability that all $L^2$ parallel rotations succeed on the first attempt vanishes exponentially with system size, causing the failure branch to dominate the expected runtime already at moderate $L$.
\end{abstract}

\maketitle

\tableofcontents

\newpage

\section{Introduction}

The Fermi–Hubbard model~\cite{hubbard1963electron} is a simple yet powerful model of interacting fermions and serves as a central framework for studying strongly correlated systems. Understanding the physics of these systems remains one of the grand challenges in condensed matter physics. Simulating the dynamics of the Fermi–Hubbard model could play a key role in explaining phenomena such as high-temperature superconductivity, which has applications in energy infrastructure through improved grid efficiency, as well as in medical technology, batteries, and energy storage. Even before the discovery of new materials, there are already significant practical stakes. The U.S. Department of Energy is estimated to spend \$117M annually on Fermi–Hubbard-type simulations on high-performance computing systems~\cite{agrawal2024quantifying}. Yet, such simulations become classically intractable at moderate system sizes and were among the first problems proposed for quantum advantage~\cite{abrams1997simulation}.

Encoding fermions and simulating their evolution on a quantum computer requires precise control over many qubits and the execution of deep circuits. Recent NISQ experiments have demonstrated small-scale Fermi--Hubbard dynamics on trapped-ion and superconducting platforms~\cite{alam2025fermionicdynamicstrappedionquantum, alam2025programmabledigitalquantumsimulation, granet2026superconductingpairingcorrelationstrappedion}, but remain limited to short evolution times and modest system sizes. Reaching classically intractable regimes will require fault-tolerant quantum computers (FTQC), which can execute deeper circuits on larger numbers of qubits by suppressing errors using quantum error correction (QEC). However, fault tolerance comes with substantial overheads. Encoding logical qubits and implementing fault-tolerant operations require more physical qubits (space overhead) and more clock cycles (time overhead).

For this reason, end-to-end resource estimation is an essential tool to assess the usefulness of a quantum algorithm, benchmark a qubit platform, and guide both hardware and software development. Its inputs include a QEC code, a fault-tolerant instruction set, a physical error model, a hardware layout, and an algorithm compiled to that instruction set. Its outputs are estimates of the physical qubit count and runtime required to reach a target accuracy. Such studies highlight the strong interdependencies between algorithms, compilation strategies, quantum error-correcting schemes, and hardware architectures, and the need to co-design them. 

The spin-optical quantum computing (SPOQC) architecture~\cite{Gliniasty2024Spin} is a quantum computing platform that combines the spins of semiconductor quantum dots, used as qubits, and the photons that they emit, to perform multi-qubit operations. It offers three key advantages: native parity measurements, connectivity beyond two-dimensional nearest-neighbor (2DNN), and fast gate execution. The ability to directly perform parity measurements makes it well-suited to dynamical QEC codes, such as the honeycomb Floquet code~\cite{HH2021Dynamically, KMT2024Anyon, GNFB2021Fault}. Memory simulations of the honeycomb code with boundaries~\cite{HH2022Boundaries, gidney2022benchmarking} on a planar layout of the SPOQC architecture demonstrated competitive threshold values for the dominant sources of error in quantum-emitter-based platforms, namely photon loss, photon distinguishability, and spin decoherence~\cite{dessertaine2026enhancedfaulttolerancephotonicquantum}.  

\subsection{Our contribution}

In this work, we go beyond the 2DNN layout and the fault-tolerant memory setting. We introduce a biplanar version of the SPOQC architecture together with a full circuit-level noise model, and estimate the resources required to simulate the Fermi-Hubbard model on this platform. We use the planar honeycomb code to protect the information, and logical operations are implemented via lattice surgery within each plane and via transversal $\CNOT$ gates between the planes. 

The biplanar layout can be used for generic compilation onto workspace and memory modules~\cite{litinski2022active, gidney2025factor}, or for problem-specific compilation that leverages a natural mapping between the two qubit sectors and the structure of the target algorithm. Here, we follow the second approach. Building on the compact fermion-to-qubit mapping~\cite{Derby_2021}, we associate one layer of the biplanar layout with the spin-down sector and the other with the spin-up sector. The transversal connections between the two layers enable efficient implementation of onsite interaction terms between spin-up and spin-down modes. 

We implement the plaquette-based Trotterization algorithm (PLAQ)~\cite{Campbell_2021} using a scheme introduced in~\cite{mcardle2025fastcuriousacceleratefaulttolerant}, where all non-overlapping plaquette operations are executed in parallel. This scheme reduces circuit depth at the cost of an increased space overhead. Because we expect that time, rather than space, will be the primary bottleneck for large-scale quantum computation, our approach prioritizes minimizing logical depth.

Using Clifford circuit simulations~\cite{gidney2021stim}, we estimate the resources required to perform lattice-surgery-based Pauli product measurements, transversal $\CNOT$ gates, and magic state preparation. We then show that, on our biplanar layout, each Trotter step requires only $4 t_\mathrm{synth} + 90$ logical timesteps, where $t_\mathrm{synth}$ is the number of logical timesteps needed to synthesize arbitrary-angle rotations, compared to $6t_\mathrm{synth} + 354$ in previous estimates without transversal $\CNOT$ gates~\cite{mcardle2025fastcuriousacceleratefaulttolerant}. We further compute the cost of arbitrary-angle rotation synthesis and of each individual Trotter step. We use the mixed-fallback Clifford+$T$ synthesis method of~\cite{Kliuchnikov2023shorterquantum}. We account for both the expected $T$-count and the additional logical timestep overhead induced by classical feedforward. These synthesis costs are combined with the lattice-surgery overhead of diagonalization circuits and transversal gates to obtain a complete accounting of magic state consumption, logical timesteps and active patches per Trotter step. We find that, while mixed-fallback protocols minimize the expected $T$-count per rotation, they can become a bottleneck in parallelized compilation. When many $Z$-rotations run in parallel, the all-success probability vanishes exponentially with system size, which can inflate the effective synthesis depth beyond that of auxiliary-free strategies.

With that, we can estimate the physical qubit count and runtime required to implement the PLAQ algorithm under a realistic noise model on the biplanar SPOQC architecture. The Fermi-Hubbard model is characterized by lattice size $L$, hopping amplitude $t$, and on-site interaction strength $U$. We set $L=8$, which has previously been identified as the utility-threshold problem size for Fermi-Hubbard simulation~\cite{agrawal2024quantifying}.
This choice already exceeds recent NISQ demonstrations of 2D Fermi--Hubbard dynamics~\cite{alam2025fermionicdynamicstrappedionquantum, alam2025programmabledigitalquantumsimulation, granet2026superconductingpairingcorrelationstrappedion} and lies beyond the reach of classical methods for 2D periodic geometries, while representing a more near-term target than the $30\times 30$ lattices of~\cite{mcardle2025fastcuriousacceleratefaulttolerant}. We set $U/t=8$, which lies in the intermediate coupling regime where state-of-the-art classical methods exhibit uncontrolled systematic errors~\cite{Campbell_2021, zheng2017stripe}. We set the target accuracy to 1\% in diamond norm and partition this error budget into Trotter approximation error, synthesis error, logical error, and magic state infidelity. Putting together our error-correction simulations, our Trotter step compilation for the biplanar layout, our full circuit-level noise model, and the characteristic times of the SPOQC platform, we estimate that the simulations would take approximately $2$ hours and $1.35\times 10^6$ physical qubits.

\subsection{Outline}

The remainder of the paper is organized as follows. In Section~\ref{sec:bilayerSPOQC}, we introduce the biplanar honeycomb SPOQC architecture and describe the associated error model. The logical instruction set we use and the corresponding resource costs are presented in Section~\ref{sec:logical_instructions_for_the_honeycomb_code}. In Section~\ref{sec:simulating_fermi_hubbard}, we describe the compilation of the Fermi–Hubbard simulation algorithm onto the biplanar layout using this instruction set. The results of the end-to-end resource estimation are presented in Section~\ref{sec:resource_estimation}. We conclude with directions for future work in Section~\ref{sec:conclusion}. Additional details on the gate set and noise model, the numerical simulations, the rotation synthesis and the compilation are provided in the Appendices.

\section{A biplanar honeycomb SPOQC architecture}
\label{sec:bilayerSPOQC}

This section is dedicated to the description of the Spin-Optical Quantum Computing (SPOQC) architecture introduced in \cite{Gliniasty2024Spin, dessertaine2026enhancedfaulttolerancephotonicquantum}. In \Cref{sec:elem_modules}, we describe the elementary modules of the SPOQC architecture, as well as the gates we can implement in the SPOQC architecture. In \Cref{sec:physical_noise}, we go through the different sources of noise of the SPOQC architecture simulated in this study. The physical noisy gate set used for simulations is described in Appendix~ \ref{ap:physical_gate_set}. In \Cref{sec:operation_timings}, we discuss the different relevant timescales of the SPOQC architecture and in \Cref{sec:error_model}, we present the physical error model simulated in this study. Finally, in \Cref{sec:bilayer}, we introduce the biplanar layout that will make up the floorplan for the fault-tolerant execution of quantum algorithms. 

\subsection{Elementary modules}
\label{sec:elem_modules}

\subsubsection{Quantum emitters}
\label{sec:quantum-emitters}

The elementary modules of the SPOQC architecture are charged quantum dots placed in a weak static magnetic field, which act as spin-photon interfaces~\cite{lee2019quantum,  appel2022entangling, coste2023high,  cogan2023deterministic, meng2023photonic, su2024continuous}\footnote{Note that the error model we use in this work is specific to this kind of quantum dot, but that the native gate set is not: it can be implemented with any quantum emitters with an embedded degree of freedom, such as~\cite{blinov2004observation, thomas2022efficient, OSullivan2024}.}. The spin degree of freedom of the quantum emitters constitutes the physical qubits of the SPOQC architecture, referred to as spin qubits, or simply spins, in the following. Quantum emitters can emit spin-entangled photons via the emission process~\cite{lee2019quantum, thomas2022efficient, appel2022entangling, coste2023high,  cogan2023deterministic, meng2023photonic, su2024continuous, OSullivan2024}:
\begin{equation}
    \E=\ketbra{0_{\rm sp},0_{\rm ph}}{0_{\rm sp}}+\ketbra{1_{\rm sp},1_{\rm ph}}{1_{\rm sp}},
    \label{eq:em_process}
\end{equation}
where $\ket{0_{\rm sp}}$ and $\ket{1_{\rm sp}}$ correspond to the spin being in the up or down state, respectively, which defines the spin qubit computational basis, while $\ket{0_{\rm ph}}$ and $\ket{1_{\rm ph}}$ correspond to the emitted photon having a right or left circular polarization, respectively. The idealized emission process is equivalent to a $\mathsf{CNOT}$ gate applied between a spin qubit and a photonic qubit in the $\ket{0_{\rm ph}}$ state. 
Here, we neglect physical noise processes affecting emission, as they are not a dominant contribution to the overall error budget.

\subsubsection{Spin control}
\label{sec:spin-control}

Single-qubit gates are implemented by combining two types of spin rotations, enabling arbitrary qubit control. Rotations around the $z$-axis are performed with optical spin rotation pulses (OSRP)~\cite{press2008complete, berezovsky2008picosecond, greilich2009ultrafast, stockill2016quantum}.
An OSRP is a circularly-polarized pulse used to drive one of the two optical transitions of the spin degree of freedom. By controlling the frequency detuning and power of the pulse, the OSRP generates any phase difference between the two spin states without optically exciting the quantum dot. These rotations along the $z$-axis allow for high-fidelity $\Z$, $\Sgate$ and $\T$ gates. 
Rotations around the $y$-axis are performed passively using the Larmor precession of the ground state spin around the magnetic field. The speed of the rotation is determined by the strength of the magnetic field, which acts continuously. A default angle of rotation can be set by synchronizing the delay between periodic optical excitation pulses with the Larmor precession. Furthermore, the angle of this passive rotation can be actively modified by applying OSRPs at appropriate times between the optical excitation pulses to induce a spin echo effect~\cite{hahn1950spin, press2010ultrafast}.

\subsubsection{Spin initialization and measurement modules}
\label{sec:init}

Computational basis initialization \init and measurements $\MZ$ rely on a simple photon detection after emission. Indeed, given the emission process described by Eq.~\eqref{eq:em_process}, measuring the photon in the computational basis amounts to performing the same measurement on the spin with which it is entangled. Therefore, the $\MZ$ operation simply consists in channeling the emitted photon to detectors that measure the polarization of the photon. For a spin initialization in state $\ket{i}$, one simply needs to add to the $\MZ$ gadget a classically controlled gate $\X^{i\oplus m}$, where $m$ is the measurement result of the $\MZ$ gadget. The initialization and measurement gadgets are represented by
\begin{subequations}
   \begin{align}
    \text{\init}&\equiv\tikz[anchor properly]{
    \node at (0,0) {
        \begin{quantikz}
             \qw & \ctrl{1} & \gate[1]{\X^{i\oplus m}} & \qw\rstick{$\ket{i_{\rm sp}}$}\\
               \lstick{$\ket{0_{\rm ph}}$}   & \targ{} & \meter{}\wire[u]{c}\rstick{$m$}
        \end{quantikz}
        };
    },\label{eq:rusinit}\\
    \text{$\MZ$}&\equiv\tikz[anchor properly]{
    \node at (0,0) {
        \begin{quantikz}
             \qw & \ctrl{1} & \\
               \lstick{$\ket{0_{\rm ph}}$}   & \targ{} & \meter{}\rstick{$m$}
        \end{quantikz}
        };
    },\label{eq:rusmz}
\end{align}
\end{subequations}
where the top wire represents the spin qubit, the bottom wire the photonic qubit, and the $\mathsf{CNOT}$ gate the emission process. Through spin control, one can initialize spin qubits in any state and measure in any basis by applying the proper rotation after computational basis initialization and before computational basis measurement. Finally, in the absence of losses, these initialization and measurement schemes succeed with unit probability. However, when losses are present, one might need to repeat this scheme until a photon is actually detected.

\subsubsection{Repeat-Until-Success (RUS) modules}

We complete our physical gate set with $\CZ$ and $\Z\Z$ parity measurement implemented through RUS operations. The core ideas of RUS operations is (1) to perform operations on the spin through the manipulation and detection of the emitted entangled photons, (2) to not alter the information of the spin qubits upon unwanted outcomes of the probabilistic photonic interference, thus allowing the process to be repeated. RUS operations allow near-deterministic operations on spin qubits, as one can repeat until a successful interference is heralded via the detection patterns.

Precisely, the scheme relies on three sub-steps: (1) photonic emission by both emitters and collection of each photon into two waveguides or modes, one for each polarization, (2) channeling of the photons towards a phase-tunable four-mode linear optical interferometer where photons interfere, (3) detection of the photons with Photon Number Resolving (PNR) detectors at the output of the interferometer. In the absence of noise, there are two types of detection patterns: repeat and success patterns. The repeat patterns correspond to two photons being detected in a single output mode. In these cases, no entanglement was generated between the two emitters, but their states are left untouched (up to known single-qubit corrections) and the scheme can therefore be repeated. Success patterns are characterized by one photon detected in one of the two upper modes and the other photon in one of the two lower modes. These patterns herald a successful entangling operation between the two emitters. A tunable phase $\varphi$ in the four-mode linear interferometer of step (2) allows switching between implementing a $\CZ$ gate (up to known single-qubit corrections) for $\varphi=\pi/2$, and a two-qubit $\Z\Z$ parity measurement for $\varphi=0$, upon detecting a success pattern. For each run of the protocol, success and repeat outcomes happen with overall $1/2$ probability each. Calling $N_{RUS}$ the maximum number of allowed cycles, the success probability after $N_{RUS}$ attemps is $1-1/2^{N_{RUS}}$. Throughout the paper, we will choose $N_{RUS}=10$ as it gives over $99.9\%$ probability of successfully implementing the desired operation. In the case where the scheme did not succeed within these $N_{RUS}$ attempts, an abort outcome is recorded \cite{dessertaine2026enhancedfaulttolerancephotonicquantum}. We treat this as an error, see Appendix \ref{ap:physical_gate_set}. In the following, we denote by \ruscz and \rusmzz, $\CZ$ and $\Z\Z$ parity measurement implemented with the above RUS operations. For a more complete description of \ruscz and \rusmzz gates, see \cite{Gliniasty2024Spin, dessertaine2026enhancedfaulttolerancephotonicquantum}.
 
\subsubsection{Photon routing}

One key feature of the SPOQC platform is that it allows for long-range interactions between quantum emitters since entanglement is mediated by the emitted photons. As we will see in the next sections, this allows to go beyond purely nearest-neighbour connectivity, but requires careful routing of the photons such that pairs can meet at the appropriate RUS gate. To that end, the SPOQC architecture uses photon routers: active components in an optical setup that will route light from a specific input to a designated output \cite{Lenzini2017, Hansen2023}. In our setup, we will use $1\to 5$ photon routers meaning that one mode is routed to five outputs. Four of these output modes will be used to implement either \rusmzz or \ruscz gates (see \Cref{sec:bilayer}) and the remaining one will be used for spin initialization and measurements. Routers additionally take classical information as inputs to specify the routing strategy.

\subsection{Physical noise}
\label{sec:physical_noise}

We now describe the sources of noise that affect qubits and operations in the SPOQC architecture. A summary of the most relevant notations, their definitions and simulated values is provided in \Cref{tab:simulated_parameters}. The complete noisy gate set used to carry out the simulations can be found in Appendix \ref{ap:physical_gate_set}. The completely positive trace-preserving (CPTP) maps describing the error channels are represented through their Kraus operators: a channel $\mathcal{E}$ with Kraus operators $\{\mathsf{A}_i\}$ acts on a density matrix $\rho$ as $\mathcal{E}[\rho]=\sum_{i}\mathsf{A}_i\rho\mathsf{A}_i$, and $\mathcal{E}=\sum_{i}[\mathsf{A}_i]$ where $[\mathsf{A}_i]$ is the super-operator associated with $\mathsf{A}_i$.

\begin{table}[h]
    \centering
    \begin{tblr}{
    hline{1, 2}={1.5pt},
    row{1}={font=\bf},
    cells={c, m}, 
    column{3}={3cm},
    }
        Quantity & Definition &Simulated value\\
        $p$ & Overall intensity of physical noise & $p=10^{-2}$\\
        $\varepsilon$ & End-to-end probability to lose a photon &$\varepsilon=0.9\%$\\
        $D$ & Distinguishability between photons from different
        emitters &$D=0.085\%$\\
        $t_c/T_2$ & Ratio between RUS cycle time $t_c$ and decoherence time $T_2$ & $t_c/T_2=0.01\%$\\
        $p_s$ & Single qubit gate infidelity & $p_s=0.005\%$\\
        $N_{RUS}$ & Number of attempts for \ruscz and \rusmzz &$N_{RUS}=10$\\
        $N_i$ & Number of attempts for \init & $N_i=5$\\
        $N_m$ & Number of attempts for $\MZ$ & $N_m=5$
    \end{tblr}
    \caption{Summary of simulated physical parameters and hyper-parameters along with their definitions and the value we used in simulations.}
    \label{tab:simulated_parameters}
\end{table}

\subsubsection{Photon loss} In photonic systems, the most detrimental source of noise is photon loss: photons propagate through a series of components $i$ (e.g., demultiplexer, photonic integrated circuits, detectors), each characterized by an intrinsic transmission $\eta_i$, defined as the average fraction of input light detected at the output. When all optical components are linear, i.e. the transformation between input and output modes is a linear transformation, the overall end-to-end transmission of the setup $\eta$ is the product of $\eta_i$. Photon loss $\varepsilon=1-\eta$ then denotes the fraction of light lost within the setup. Here, we assume that loss is uniform across the setup and characterized by a single value $\varepsilon$.

The noise induced by photon loss is a phase erasure $\frac{1}{2}([\Id]+[\Z])$ acting on the pre-emission density matrix of the corresponding emitter \cite{dessertaine2026enhancedfaulttolerancephotonicquantum}. However, when a photon is lost, no light is detected at the output of the system, so we know that the error channel was applied on the emitter. As a result, this channel, and any channels arising from photon losses are heralded, thereby providing precise information about the affected qubits.

\subsubsection{Emitter distinguishability} Another source of noise in photonic systems is photon distinguishability. In order to properly interfere, photons must be perfectly indistinguishable in all internal degrees of freedom, i.e. degrees of freedom not used to encode the qubit. Denoting by $\ket{\xi}$ the state vector modeling all internal degrees of freedom, the distinguishability $D_{ij}$ between photons $i$ and $j$ can be defined as the deviation from a perfect overlap of the internal degrees of freedom : $D_{ij}=1-\abs{\braket{\xi_i}{\xi_j}}^2$. In the SPOQC architecture, photon distinguishability impacts the entangling RUS operations, which rely on the interference of two photons. We denote by $D_{ab}$ the average distinguishability between photons emitted by a pair of emitters $a$ and $b$, and we assume a uniform value $D$ across all pairs.

\subsubsection{Spin noise} 

Due to interaction between the spin of the quantum dot and the surrounding nuclear bath, idling spin qubits undergo quantum decoherence. In the limit of a large surrounding bath and of a slow relaxation time (which is the case for the quantum emitters we consider), the error due to decoherence after idling time $t$ is given by
\begin{equation}
    (1-p_d(t/T_2))[\Id]+p_d(t/T_2)[\Z],
    \label{eq:deco_channel}
\end{equation}
with $p_d(t/T_2)=(1-\exp(-t/T_2))/2$ and $T_2$ the decoherence time of the quantum emitter (see Appendix C in \cite{Gliniasty2024Spin} for a full derivation). This simple channel is derived from a pure dephasing model which is quite pessimistic in terms of the actual effect of decoherence on the system. Furthermore, it does not capture the effect of decoherence when implementing spin rotations. Indeed, implementing single-qubit gates in the SPOQC architecture is inherently a time-dependent process as it is a combination of both OSRPs and magnetic field precession. Therefore, decoherence will also play a role, along with inherent infidelities coming from the OSRP pulses. For this reason, we model single-qubit gate infidelities with an abstract depolarizing noise channel 
\begin{equation}
    (1-p_s)[\Id]+\frac{p_s}{3}\left([\X]+[\Y]+[\Z]\right),
    \label{eq:single-qb-error-channel}
\end{equation}
with a single error parameter $p_s$, which we assume captures all the physics behind implementing spin rotations. Note that a more accurate description would be a general single-qubit Pauli channel, but it would introduce a complexity beyond the scope of this paper.

\subsection{Operation timings}
\label{sec:operation_timings}

Since they rely on optical pulses or photon-mediated interactions, operations in the SPOQC architecture are fast, which in turns makes QEC cycles fast and offers a reduced total runtime compared to slower platforms. The fastest operations are single-qubit gates. As OSRPs ~\cite{press2008complete, berezovsky2008picosecond, greilich2009ultrafast, stockill2016quantum} rely on picoseconds laser pulses, rotations along the $z$-axis can be accessed on the same timescale. Combined with magnetic field precession, arbitrary angle rotations can be realized in 1 to 10 ns, depending on the Larmor precession time. In the following, we will choose a single-qubit gate time of 5 ns, even for $z$-axis rotations. 

Spin initialization and measurement as well as \ruscz and \rusmzz gates rely on photonic emissions  whose rate depends on the rate of the laser exciting the quantum dot. In \cite{maring2024versatile}, an 80 MHz laser is used which translates into an excitation every 12.5 ns. As photonic emission occurs on a 100 ps timescale, the repetition time can be reduced to 1 ns. We expect the repetition time to be limited by the \textit{feedforward time}, i.e. classical processing time of photonic detector outcomes. Current feedforward times for photonic platforms lie around 23 ns \cite{thiele2025feedforward}, and reducing it needs further development of dedicated integrated electronics. All in all, we will assume a cycle time (comprising excitation and feedforward time) of $t_c=30\,\textrm{ns}$ for all the operations listed above. For spin initialization and measurement, we allow for up to $N_i=N_m=5$ cycles to be performed, translating into 150 ns overall for these operations. For \ruscz and \rusmzz, we allow for up to $N_{RUS}=10$ cycles, such that these gates take 300 ns overall. \Cref{tab:operation_times} summarizes the different timings of the operation in the SPOQC architecture.

Finally, we assume that the reaction time of our architecture, defined as the classical processing time of syndrome decoding, is $10$ \textmu{}s, as in \cite{gidney2025factor}. This will correspond to 33 syndrome extraction cycles of the quantum error-correcting code we will consider, see \Cref{sec:logical_instructions_for_the_honeycomb_code}.

\begin{table}[h]
\centering
    \begin{tblr}{
        hline{1,2}={1.5pt},
        cells={c, m},
        cell{2}{2}={black!5},
        cell{4}{2}={black!5},
        cell{5}{2}={black!5},
        cell{2}{4}={black!5},
        column{4}={6cm}
        }
         \bf Operation & \bf Notation & \bf Timing & \bf Observation\\
         Single-qubit gate time & & $5$ ns & \\
         RUS cycle time & $t_c$ & $30$ ns & Takes classical feedforward into account\\
         \init time & & $150$ ns& Assumes a maximum number of $N_i=5$ attempts\\
         $\MZ$ time & & $150$ ns& Assumes a maximum number of $N_m=5$ attempts\\
         \ruscz time & $t_{RUS}$& $300$ ns& Assumes a maximum number of $N_{RUS}=10$ attempts\\
         \rusmzz time & $t_{RUS}$& $300$ ns& Assumes a maximum number of $N_{RUS}=10$ attempts\\
         Reaction time & $\tau_r$ & $10$ \textmu{}s & Same as in \cite{gidney2025factor}, equivalent to 33 syndrome extraction cycles
    \end{tblr}
    \caption{Summary of physical operation times of the SPOQC architecture considered in this work. For simplicity, we consider that $z$-axis single-qubit gates take the same time as non-diagonal single qubit gates.
    Values for the number of attempts for \init, $\MZ$, \ruscz and \rusmzz are chosen to provide a large success probability of the protocols in a reasonably low amount of time.}
    \label{tab:operation_times}
\end{table}


\subsection{Error model}
\label{sec:error_model}

For different choices of QEC codes and different implementations, Figures 5 in \cite{Gliniasty2024Spin} and 7 in \cite{dessertaine2026enhancedfaulttolerancephotonicquantum} represent fault-tolerant surfaces, which are generalized fault-tolerant thresholds that take into account multiple error mechanisms. In the present study, we fix the relative weights of the different error mechanisms listed in \Cref{sec:physical_noise} such that
\begin{equation}
    \varepsilon=0.9p,\,D=0.085p,\,t_c/T_2=0.01p,\,p_s=0.005p.
\end{equation}
Here $p$ defines the overall intensity of the physical noise in the system, and scalars $0.9,\,0.085,\,0.01,\,0.005$ (that we call biases) represent the relative weights of the error mechanisms. In our simulations, we choose $p=10^{-2}$, setting reasonable long-term parameter values for the hardware used in the SPOQC architecture detailed in Table \ref{tab:simulated_parameters}. 
In the following, we give a brief overview of experimental progress towards reaching the error rates of our noise model.

In all photonic platforms, loss remains the biggest challenge. Reaching an end-to-end transmission above $99\%$ requires optimizing all components that the photons travel through. Sources need to be efficient: over the past decade, this efficiency improved significantly reaching values above $50\% $ \cite{PhysRevLett.126.233601, maring2024versatile} for micropillar cavities, while on-chip sources have demonstrated efficiencies up to $84\%$ \cite{uppu2020scalable}. Ultra-low loss components also need to be designed. SiN integrated beam splitters and mode crossing have shown transmissions reaching $99.988\%$ and $99.963\%$ respectively \cite{psiquantum2025manufacturable}. Integrated Photon Number Resolving (PNR) detectors, chip-to-fiber interface as well as high-bandwidth phase-shifters still need to be improved, with transmission efficiency of $98.7\%$ and $97.7\%$ for the last two \cite{psiquantum2025manufacturable}. 

Concerning indistinguishability, the state of the art for photons coming from different quantum dots in bulk material is $93\%$~\cite{zhai2022quantum}, whose fabrication method could be combined with cavity-enhanced emission \cite{grange2017reducing} to reach an indistinguishability above $99\%$.

Finally, using dynamical decoupling techniques, state-of-the-art spin coherence time of up to 100 \textmu{}s have already been demonstrated \cite{zaporski2023ideal}. With the current state-of-the-art feedforward time of 23 ns \cite{thiele2025feedforward}, the ratio between RUS cycle time and decoherence time is already close to the value used for simulations in this study. Furthermore, single-qubit gates with $99.3\%$ fidelity have been demonstrated~\cite{zaporski2023ideal}, and this value is expected to improve with extended coherence time and overall better source quality.

\subsection{Biplanar Honeycomb layout}
\label{sec:bilayer}

We encode logical information using planar honeycomb Floquet codes \cite{HH2022Boundaries, gidney2022benchmarking}. Compared to the surface code, it allows the highest individual thresholds as well as the overall highest fault-tolerance for memory experiements in the SPOQC architecture \cite{dessertaine2026enhancedfaulttolerancephotonicquantum}. These codes rely on a periodic sequence of entangling two-qubit measurements only requiring a 2DNN connectivity on a honeycomb lattice. Lattice edges are assigned a color, conventionally red, green and blue, that correspond respectively to performing an $XX$, $YY$ or $ZZ$ parity measurement of the qubits connected by the edge. Going trough a measurement schedule of all red, green and blue edges, a single logical qubit is dynamically generated and protected. 
We define the horizontal logical operator $H$ as spanning from left to right across the patch, and the vertical logical operator $V$ as spanning from top to bottom.
We use the implementation of the planar honeycomb Floquet code presented in \cite{gidney2022benchmarking} that can be represented on a regular grid with rectangular faces, see \Cref{fig:bilayer}-$(a)$ for an example of a code patch. With this representation, we call width the number of qubits per row of the patch and height the number of rows of the patch: \Cref{fig:bilayer}-$(a)$ represents a height-$6$ width-$4$ patch. For this planar implementation of the honeycomb code, we consider measurement schedule $\cdots\to$ blue $\to$ red $\to$ blue $\to$ green $\to$ red $\to$ green $\to\cdots$. At the boundary, qubits have a lower degree of connectivity since some colored edges have to be cut when defining the planar patch. Throughout the measurement schedule, whenever edges of a given color are measured, unpaired boundary qubits are measured individually in the corresponding basis ($X$, $Y$, or $Z$ for red, blue, and green edges, respectively). \Cref{fig:circuit_horizontal_memory} shows a circuit implementation and visualization of this schedule on a code patch.

\begin{figure}[t]
  \centering
  \input{biplanar_layout_2.tikz}
  \caption{$(a)$ An example of a $6\times 4$ (height $\times$ width) patch of honeycomb code along with the measurement schedule compatible with this layout from \cite{gidney2022benchmarking}. $(b)$ Bilayer layout studied in this article. Qubits from $U$ (resp. $D$) are laid out on the upper plane (resp. lower) plane. Colored edges represent intra-plane connectivity to implement parity measurements for honeycomb code syndrome extraction and lattice surgery, while black edges represent inter-plane connectivity to implement transversal $\CNOT$ gates. For clarity, we have represented the inter-plane connections only between the upper and lower top-rightmost patches. Dashed edges represent connections between patches that are only active whenever lattice surgery is performed between these patches. In this picture, the upper two left-most patches are undergoing an $HH$ lattice surgery and their patches are merged along the horizontal direction. Similarly for the lower top bottom patches undergoing $VV$ lattice surgery with patches merged along the vertical direction.}
  \label{fig:bilayer}
\end{figure}

\subsubsection{Beyond 2DNN connectivity}

We go beyond purely 2DNN connectivity and consider a biplanar honeycomb layout. In our framework, qubits are partitioned into two sets, or planes, $U$ and $D$, and, within each set, the qubit connectivity is that of a standard honeycomb lattice. We further partition physical qubits into adjacent planar Honeycomb Floquet code patches. We allow for one additional connection between pairs $(u_i, d_i)\in U\times D$ of corresponding qubits labeled $i$ between the two sets. 

The overall biplanar layout is presented on \Cref{fig:bilayer}-$(b)$. We have represented two planes, each comprising four Honeycomb code patches of size $6 \times 4$. Within each plane, colored edges (solid and dashed) represent a physical parity measurement between the edge's qubits, which translates into the routing of the photons emitted by the spin qubits towards a \rusmzz gate. Dashed edges represent edges that will only be active when performing lattice surgery, i.e. a merging of code patches to perform logical multi-qubit Pauli measurements, see \Cref{sec:logical_instructions_for_the_honeycomb_code}. For instance, in \Cref{fig:bilayer}-$(b)$, the two leftmost patches of the upper plane have been merged, and blue edges between these patches are currently active. 
Black edges represent the additional inter-plane connectivity our biplanar architecture allows. These edges are used to implement physical $\CZ$ gates, and, when active, emitted photons are routed towards a \ruscz gate. Using these edges, we can implement transversal $\mathsf{CNOT}$ or $\mathsf{SWAP}$ gates (with additional conjugation with physical $\mathsf{H}$ gates on the target qubits to convert $\CZ$ gates into $\CNOT$ gates) between corresponding code patches in the two planes.

\subsubsection{Compatibility with work/memory compilation}

Although we use the biplanar layout for a problem-specific compilation, as described in \Cref{sec:simulating_fermi_hubbard}, it easily accommodates generic compilation onto compute and memory regions. For instance, in \cite{gidney2025factor}, the author considers a baseline architecture separated into three regions: (1) cold storage, consisting of patches of yoked surface codes optimized for long-time storage of logical qubits, (2) hot storage, consisting of patches of standard surface codes that need to quickly interact with one another, (3) compute, where surface codes from the previous region interact through lattice surgery and are supplied magic states, while the proposition of~\cite{litinski2022active} is based on a partition into workspace and memory modules. 
In our biplanar architecture, one plane can be used for storage and the other for compute. Loading from one plane to the other is implemented with constant-depth circuits using transversal $\mathsf{SWAP}$ gates accessible thanks to the additional connectivity of the architecture. In the compute plane, compilation can be realized in the same way as \cite{Litinski2019gameofsurfacecodes}, but in the biplanar version, the compute plane might offer more routing space, and therefore shallower overall circuit depth compared to a single plane architecture.

\section{Logical instructions}
\label{sec:logical_instructions_for_the_honeycomb_code}

We described how logical qubits are encoded in \Cref{sec:bilayer}. We now estimate the cost of each logical instruction: lattice surgery between patches on the same plane for implementing Pauli product measurements (\Cref{subsec:lattice_surgery}), magic state preparation using factories within a single plane (\Cref{subsec:preparing_magic_states}), and transversal $\mathsf{CNOT}$ gates between corresponding patches across the two planes of the biplanar layout (\Cref{subsec:transversal_cnot_gates}). All estimates are performed using the noise model presented in \Cref{sec:error_model}, the operation timings in \Cref{sec:operation_timings}, and the noisy gate set described in \Cref{tab:physical_gate_set}. 

The noiseless and noisy circuits generated, statistics collected, and code to produce the plots and tables throughout this paper are available at \url{https://doi.org/10.5281/zenodo.19484262}.

\subsection{Lattice surgery}
\label{subsec:lattice_surgery}

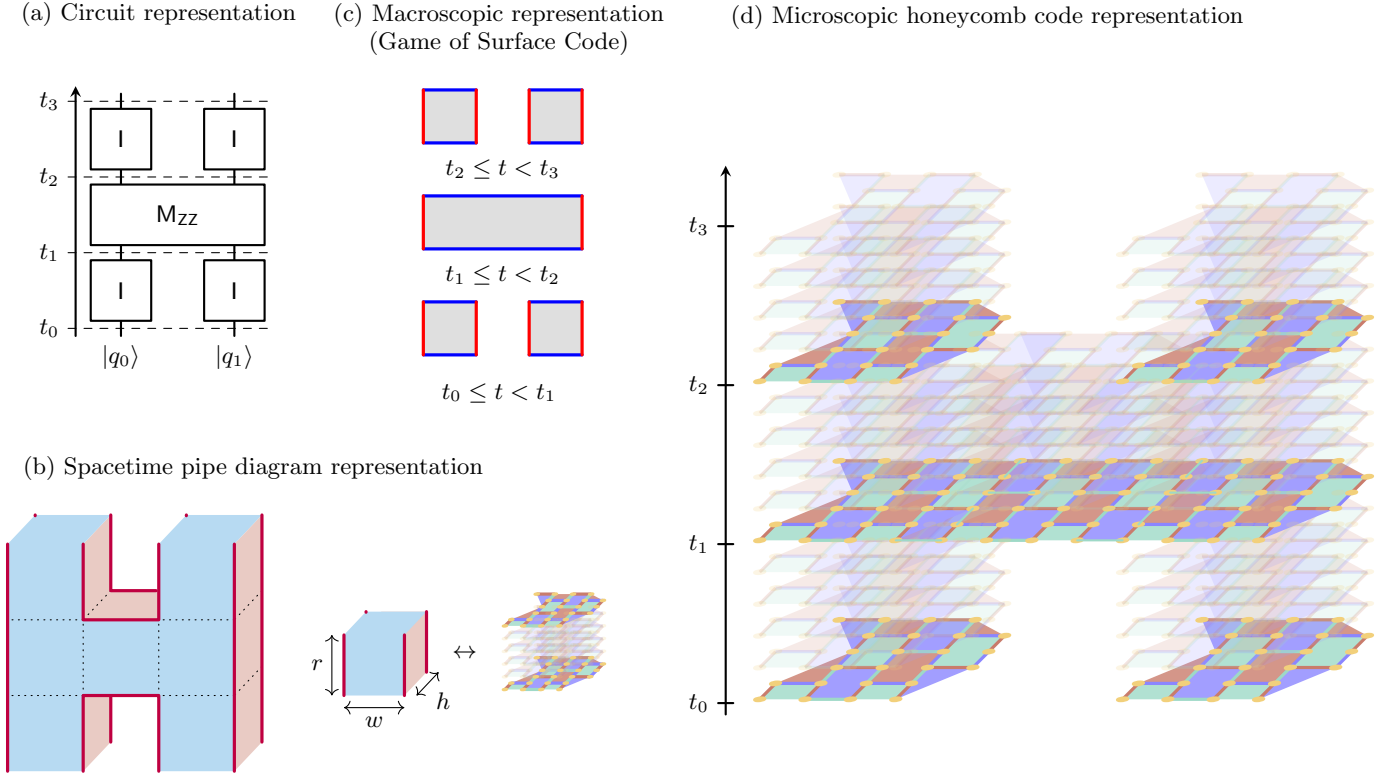
\begin{figure}[h]
    \centering
    \input{lattice_surgery}
    \setcounter{figure}{1}
    \caption{Four representations of a two-qubit $\MZZ$ lattice surgery operation.
(a)~Circuit representation.
(b)~Spacetime pipe diagram, consisting of seven lattice surgery cubes of width $w$, height $h$ and number of rounds $r$.
(c)~Macroscopic representation obtained by slicing the pipe diagram along the time axis. Red and blue boundaries indicate smooth and rough boundaries, respectively.
(d)~Microscopic honeycomb code representation. Here, an auxiliary patch mediates the operation between the two patches; if the patches are adjacent, no auxiliary is needed.}
\label{fig:lattice_surgery_representations}
\end{figure}

\begin{figure}
\includegraphics[width=\textwidth]{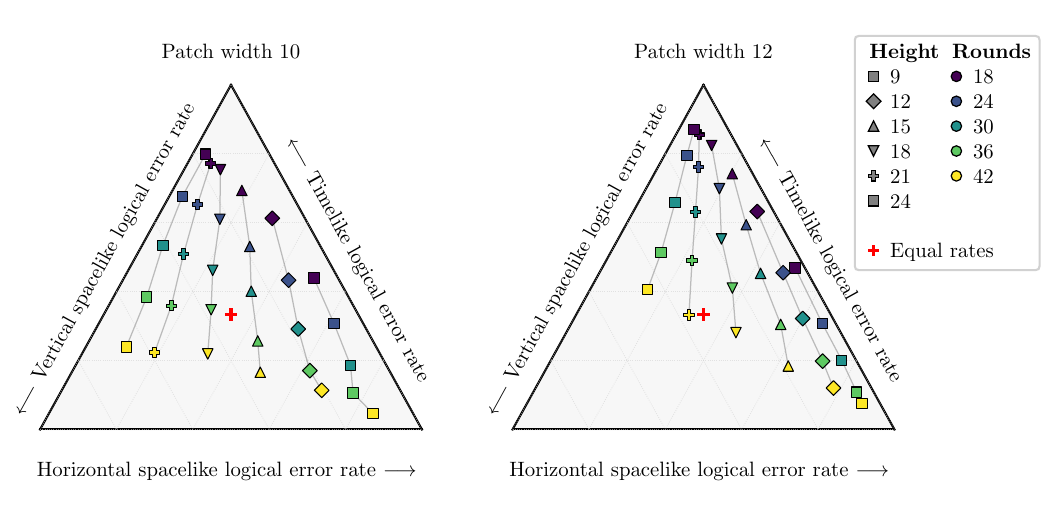}
\caption{Ternary plots of the relative contributions of horizontal spacelike, vertical spacelike, and timelike logical errors to the total logical error rate.
A cube-root transformation improves readability by spreading points away from the corners. Marker shape denotes patch height, colour denotes the number of syndrome measurement rounds, and grey lines connect points of equal height across increasing rounds.
The red cross marks equal error contributions. 
All data were obtained at $p=0.01$ using PyMatching~\cite{higgott2023sparse}. 
For both patch widths, the error rates are closest to equal when the height is $3/5w$ and the number of syndrome rounds is $3/10w$.
The details about the numerical simulations performed to generate these plots are given in \Cref{appendix:numerical_simulations}.}
\label{fig:dimension_ratio_power}
\end{figure}

Lattice surgery~\cite{Horsman_2012, fowler2018low, Litinski2019gameofsurfacecodes} is a technique to perform Pauli product measurements on a 2D lattice of qubits with nearest neighbor connectivity.
\Cref{fig:lattice_surgery_circuit} shows a circuit containing a logical $\MZZ$ gate and four identity gates.
\Cref{fig:lattice_surgery_pipe} shows the corresponding spacetime pipe diagram, the standard representation of lattice surgery operations \cite{TanNiuGidney2024SATScalpel, Gidney2024PipeDiagrams}.
Pipe diagrams are composed of 3D cubes, which we refer to as lattice surgery cubes, or simply cubes.
The pipe diagram in \Cref{fig:lattice_surgery_pipe} contains seven cubes.
The 3D pipe diagram can be converted into a 2D macroscopic representation by taking slices along the time axis, as shown in \Cref{fig:lattice_surgery_macro}.
This macroscopic representation is inspired by the game-of-surface-codes representation \cite{Litinski2019gameofsurfacecodes}.
In the macroscopic representation the colors of each boundary represent whether it is a smooth or rough boundary.
Finally, the microscopic representation in \Cref{fig:lattice_surgery_micro} shows the underlying physical qubit layout.

The physical operations implemented in the bulk of patches throughout a lattice surgery computation are uniform.
Therefore the overhead of a lattice surgery operation can be estimated from the overhead of a single lattice surgery cube.
The spacetime overhead of a lattice surgery cube is defined by its number of qubits in the horizontal and vertical direction (width and height) in a patch, and the number of syndrome extraction rounds.
The aspect ratios of these dimensions are chosen such that the logical error contributions from horizontal spacelike, vertical spacelike, and timelike errors are equal, in order to minimize the total logical error rate for a fixed number of qubits and syndrome extraction rounds.
To find the aspect ratios, we sweep the height and number of syndrome extraction rounds for a cube with fixed width. The results are displayed in \Cref{fig:dimension_ratio_power}.
The optimal width\,:\,height\,:\,rounds ratio is $3:5:10$.
Because our code for constructing circuits requires the height to be a multiple of~$3$ and the number of syndrome extraction rounds to be a multiple of~$6$, we round each up to the nearest such multiple.
The details about the numerical simulations performed to generate \Cref{fig:dimension_ratio_power} are given in Appendix~\ref{appendix:numerical_simulations}.

\begin{figure}[!htbp]                                
\centering                                           
\includegraphics[width=\textwidth]{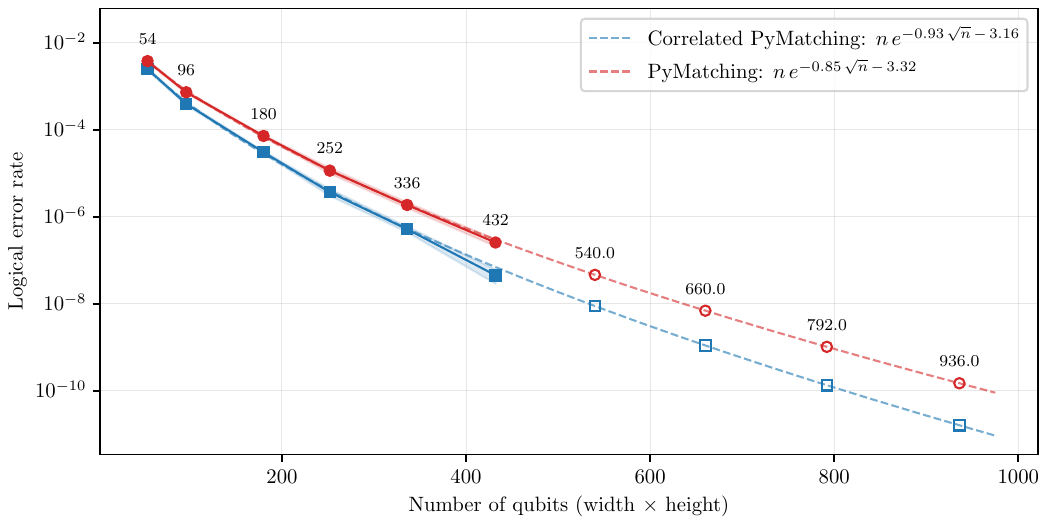}                                               
\caption{Combined spacelike logical error rate $E_{HV} = 1 - (1 - E_H)(1 - E_V)$ as a function of the number of physical qubits (width $\times$ height).
Solid markers show Monte Carlo estimates with shaded 95\% confidence intervals computed using two-level variance estimation over heralding outcomes. 
Dashed lines show exponential fits in log space. 
Open markers indicate extrapolated values at larger block sizes. 
Results are shown for both independent PyMatching~\cite{higgott2023sparse} and correlated matching decoders~\cite{fowler2013optimal}.
All circuits are simulated using Stim~\cite{gidney2021stim}.}
\label{fig:block-size}
\end{figure}

Given the optimal aspect ratio, we characterize the logical error rate as a function of the patch width.
We define the combined spacelike logical error rate $E_{HV} = 1 - (1-E_H)(1-E_V)$, where $E_H$ and $E_V$ are the horizontal and vertical spacelike logical error rates per round, respectively \footnote{Following \cite{gidney2022benchmarking}, we label the logical operators \textit{horizontal} and \textit{vertical} instead of $X$ and $Z$, because the basis of the physical gates implementing the logical operator changes at different timesteps of the circuit implementing the honeycomb code.}.
At any point during a lattice surgery operation, at most two of the three logical error types can occur.
To see this, consider the pipe diagram representation of the operation: each lattice surgery cube has six faces, forming three pairs of opposite faces (i.e., pairs that do not share an edge). 
A logical operator must connect two opposite faces, so there are three possible logical error types, one per pair. 
Since at least one face is connected to an adjacent cube, at least one pair of opposite faces cannot both be boundaries, and hence at most two of the three logical error types can occur simultaneously.
Since our aspect ratio equalizes all three rates, $E_{HV}$ captures the relevant failure probability regardless of which two types are active.
\Cref{fig:block-size} shows $E_{HV}$ as a function of the number of physical qubits per patch at physical error rate $p=0.01$.

We simulate both independent and correlated~\cite{fowler2013optimal} minimum-weight perfect matching decoders, and use the correlated decoder for our resource estimate.
From exponential fits, weighted by the statistical uncertainty of each data point, we can extrapolate to larger patch sizes and derive the number of qubits required to reach any target logical error rate.
We summarize the logical error rate and the corresponding lattice surgery time for a range of patch sizes in \Cref{tab:resource_estimates}.
The details about the numerical simulations performed to generate this plot as well as additional numerical results for $p=0.025$ and $p=0.0075$ are given in Appendix \ref{appendix:numerical_simulations}.

To obtain more accurate estimates and enable explicit circuit-level implementations, one ultimately requires software for generating the corresponding physical circuits. While such tools exist for the surface code~\cite{suau2026tqec}, they are not yet available for the honeycomb code.
Our lattice surgery compilation relies on $Y$-basis measurements of logical patches. For the surface code, such measurements have been realized via twist-based lattice surgery~\cite{chamberland2022circuit} and tangled syndrome extraction~\cite{geher2024tangling}, which provide circuit-level protocols when auxiliary routing space is available on both sides of the patch.
Here, we assume that analogous constructions exist for the honeycomb code and treat their logical cost on equal footing with the other surgery primitives.

\begin{table}[!htbp]
    \centering
    \begin{tblr}{
    hline{1, 2}={1.5pt},
    hline{8}={0.8pt},
    cells={c, m},
    row{1}={font=\bf}
    }
        $\boldsymbol{w}$ & $\boldsymbol{h}$ & Rounds & Qubits & Logical error rate $E_{HV}$ & Lattice surgery time \\
        6 & 9 & 18 & 54 & $(2.48 \pm 0.08) \times 10^{-3}$ & $5.5$ \textmu{}s \\
        8 & 12 & 24 & 96 & $(3.91 \pm 0.14) \times 10^{-4}$ & $7.3$ \textmu{}s \\
        10 & 18 & 36 & 180 & $(3.03 \pm 0.09) \times 10^{-5}$ & $11.0$ \textmu{}s \\
        12 & 21 & 42 & 252 & $(3.65 \pm 0.33) \times 10^{-6}$ & $12.8$ \textmu{}s \\
        14 & 24 & 48 & 336 & $(5.16 \pm 0.38) \times 10^{-7}$ & $14.6$ \textmu{}s \\
        16 & 27 & 54 & 432 & $(4.50 \pm 0.83) \times 10^{-8}$ & $16.5$ \textmu{}s \\
        18$^*$ & 30 & 60 & 540 & $8.82 \times 10^{-9}$ & $18.3$ \textmu{}s \\
        20$^*$ & 33 & 66 & 660 & $1.09 \times 10^{-9}$ & $20.1$ \textmu{}s \\
        22$^*$ & 36 & 72 & 792 & $1.33 \times 10^{-10}$ & $22.0$ \textmu{}s \\
        24$^*$ & 39 & 78 & 936 & $1.60 \times 10^{-11}$ & $23.8$ \textmu{}s \\
        26$^*$ & 42 & 84 & 1092 & $1.90 \times 10^{-12}$ & $25.6$ \textmu{}s \\
        28$^*$ & 48 & 96 & 1344 & $8.02 \times 10^{-14}$ & $29.3$ \textmu{}s \\
        30$^*$ & 51 & 102 & 1530 & $9.25 \times 10^{-15}$ & $31.1$ \textmu{}s
    \end{tblr}
    \caption{Resource estimates for the honeycomb code lattice-surgery patch at physical error rate $p = 0.01$ using the correlated PyMatching decoder. A single round of syndrome extraction takes $305$\,ns.  Rows marked ${}^*$ are extrapolated from a weighted exponential fit $n\,e^{a\sqrt{n}-b}$.}
\label{tab:resource_estimates}
\end{table}

\FloatBarrier

\subsection{Magic state preparation}
\label{subsec:preparing_magic_states}

To the best of our knowledge, no magic state preparation protocols have been designed for the planar honeycomb code. 
We therefore adapt existing surface code protocols, converting between the two codes in both qubit count and syndrome extraction rounds.

\begin{figure*}                                        
\centering                                           
\includegraphics[width=\textwidth]{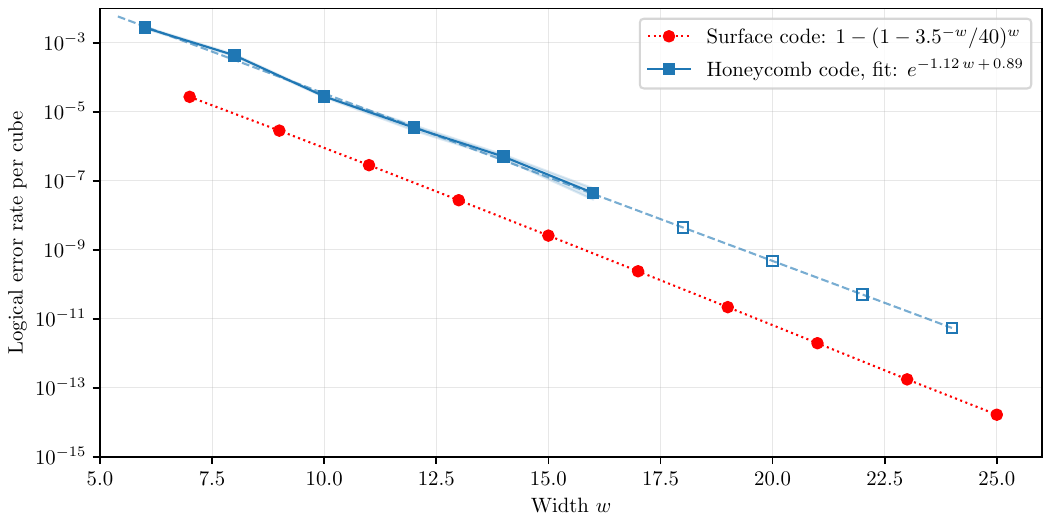}                                               
\caption{To get the same logical error rate per cube, the width of the honeycomb code must be $\approx 1.25\times$ the width of the surface code.
The logical error rate per round is calculated as $p_r = 1 - (1-E)^{1/n_r}$ from the data used to produce \Cref{fig:block-size}.
The data for the surface code is taken from \Ccite[Figure 6]{gidney2025factor}.}
\label{fig:surface_code_comparison}
\end{figure*}

To derive conversion rates, we use the surface code data in \Ccite[Figure 6]{gidney2025factor}.
The error model is the standard depolarizing circuit-level noise model with physical error rate $p=10^{-3}$, which is also commonly used for simulating magic state preparation protocols.
We first find a conversion rate for the patch width. 
\Ccite[Figure 6]{gidney2025factor} contains the logical error rate per round of syndrome extraction as a function of the patch width. 
In a lattice surgery block using the surface code, the number of syndrome extraction rounds is equal to the patch width~\cite{fowler2012surface}.
Here, based on the previous section, the number of syndrome extraction rounds is $10/3$ of the patch width.
Therefore, we compare the logical error rate per $w$ rounds of the surface code to the logical error rate per $10w/3$ rounds of the honeycomb Floquet code.
The results are shown in \Cref{fig:surface_code_comparison}.
We find that the ratio of the patch width of the honeycomb Floquet code to the patch width of the surface code at equal logical error rate per round is $\approx 1.25$.

Now that we have the patch width conversion rate, we can find the qubit number conversion rate.
The number of qubits in the surface code is $2 (w+1)^2$.
The number of qubits in the Hastings-Haah Floquet code is $5w^2/3 $.
Thus the qubit number conversion rate is $5w^2/3  / (2 (w/1.25 + 1)^2) \approx 0.52$.
Because the honeycomb code uses $10w/3 $ rounds, the round conversion rate is $10/3 \times 1.25 \approx 4.2$.

We use the protocols at $p_{\mathrm{phys}}=10^{-3}$ from~\Ccite[Table~1]{Litinski2019magicstate} and, following \cite{mcardle2025fastcuriousacceleratefaulttolerant}, we include prospective $5\times$ reductions in both space and extraction rounds motivated by magic state cultivation~\cite{gidney2024cultivation}.
Then, we apply the qubit and round conversion rates to get the corresponding numbers for the honeycomb code.
The results are shown in \Cref{tab:msf_protocols}.
Note that we are only specifying a number of qubits required here for a magic state factory and do not specify the exact arrangement of the patches in the factory.
We make the simplifying assumption that we can deform the magic state factories into an area that has enough physical qubits.
To account for this assumption and the lack of simulations in this subsection, we eventually assign roughly $3\times$ more qubits for magic state production than our estimate requires in \Cref{sec:t_state_count}.

\begin{table}[h]
\centering
\begin{tblr}{
    width=\linewidth,
    hline{1,3}={1.5pt},
    cells={c,m}
}
\SetCell[r=2]{} Protocol & \SetCell[r=2]{} $p_{\mathrm{out}}$ & \SetCell[c=2]{} Surface code (dist.) &  &\SetCell[c=2]{} Surface code (cult.) & &Hastings-Haah (cult.) \\
\hline{2}{2-3}
         &                    & Qubits & Cycles & Qubits & Cycles & Qubits / Rounds \\
$(15\!\to\!1)_{17,7,7}$                                                      & $4.5\!\times\!10^{-8}$  & 4\,620  & 42.6  & 924    & 8.5   & 480 / 35.7 \\
$(15\!\to\!1)^{6}_{13,5,5}\!\times\!(20\!\to\!4)_{23,11,13}$                & $1.4\!\times\!10^{-10}$ & 43\,300 & 130   & 8\,660 & 26.0  & 4\,500 / 109 \\
$(15\!\to\!1)^{4}_{13,5,5}\!\times\!(20\!\to\!4)_{27,13,15}$                & $2.6\!\times\!10^{-11}$ & 46\,800 & 157   & 9\,360 & 31.4  & 4\,870 / 132 \\
$(15\!\to\!1)^{6}_{11,5,5}\!\times\!(15\!\to\!1)_{25,11,11}$                & $2.7\!\times\!10^{-12}$ & 30\,700 & 82.5  & 6\,140 & 16.5  & 3\,190 / 69.3 \\
$(15\!\to\!1)^{6}_{13,5,5}\!\times\!(15\!\to\!1)_{29,11,13}$                & $3.3\!\times\!10^{-14}$ & 39\,100 & 97.5  & 7\,820 & 19.5  & 4\,070 / 81.9 \\
$(15\!\to\!1)^{6}_{17,7,7}\!\times\!(15\!\to\!1)_{41,17,17}$                & $4.5\!\times\!10^{-20}$ & 73\,400 & 128   & 14\,680& 25.6  & 7\,630 / 108 \\
\end{tblr}
\caption{Magic state factory protocols for $p_{\mathrm{phys}}=10^{-3}$.
Surface code distillation numbers (qubits and cycles) are taken from Table~1 of~\cite{Litinski2019magicstate}.
The cultivation columns apply the assumed $5\times$ reduction in both space and time motivated by~\cite{gidney2024cultivation}, following~\cite{mcardle2025fastcuriousacceleratefaulttolerant}.
The Hastings-Haah columns use the qubit conversion rate of $\approx 0.52$ and the round conversion rate of $\approx 4.2$ explained in the main text.}
\label{tab:msf_protocols}
\end{table}

\FloatBarrier

\subsection{Transversal CNOT gates}
\label{subsec:transversal_cnot_gates}

A transversal $\CNOT$ gate between two code patches is implemented by applying a physical two-qubit gate between each pair of corresponding qubits on the two patches, together with single-qubit basis-change gates. The stabilizer flow generators of a $\CNOT$ gate on a pair of physical qubits are $X_1 \rightarrow X_1X_2$, $X_2 \rightarrow X_2$, $Z_1 \rightarrow Z_1$, and $Z_2 \rightarrow Z_1Z_2$ \cite{mcewen2023relaxing}.
The desired stabilizer flow generators of the logical $\CNOT$ gate on the two horizontal $H_1, H_2$ and two vertical logical operators $V_1, V_2$ are $H_1 \rightarrow H_1H_2$, $H_2 \rightarrow H_2$, $V_1 \rightarrow V_1$, and $V_2 \rightarrow V_1V_2$.

\begin{figure}
  \centering
  \includegraphics[width=\textwidth]{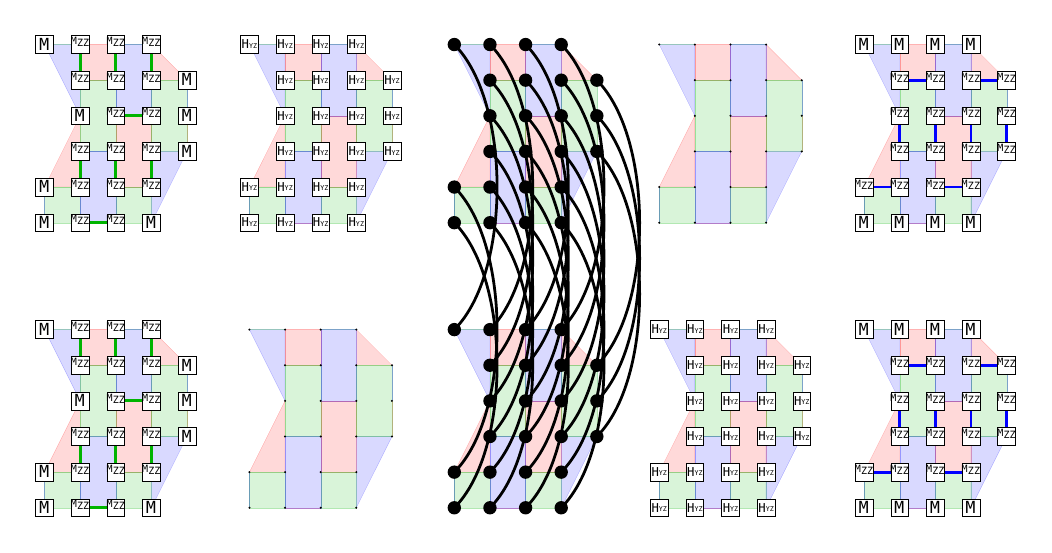}
  \caption{Circuit of the transversal $\CNOT$ gate between two honeycomb code patches (top and bottom), showing two syndrome extraction ticks before the gate, the transversal $\CZ$ tick (center), and two ticks after.
  Colored hexagons indicate the plaquette structure; colored edges show MZZ measurements. 
  An interactive version of the circuit is available in \crumbleTransversalCnot{}. In Crumble, use the E and Q keys to move through the layers of the circuit.}
  \label{fig:transversal_cnot_circuit}
\end{figure}

The transversal $\CNOT$ is applied after measuring $\MYY$ on green edges. 
At this timestep logical operator $H$ consists of single qubit $Y$ operators and logical operator $V$ consists of single qubit $X$ operators.
Thus, to apply a logical $\CNOT$ gate, physical $\CY$ gates are applied. 
In the SPOQC architecture $\CY$ gates can be implemented using the identity $\CY(c, t) = \HXY^{(c)} \cdot \CZ^{(c, t)} \cdot \HXY^{(c)}$. Because $\MYY = \HXY {}_1 \HXY {}_2 \MZZ \HXY {}_1 \HXY {}_2$, the $\HXY$ gates on the control qubits cancel out and the logical $\CNOT$ can be implemented using
\begin{equation}
\label{eq:transversal_cnot}
\bigotimes_{i} \HXY^{(t_i)} \cdot \CZ^{(c_i, t_i)} \cdot \HXY^{(c_i)},
\end{equation}
where $c_i$ and $t_i$ are the $i$-th physical qubits of the control and target patches, respectively, and the tensor product runs over all qubit pairs. 
The full circuit is shown in \Cref{fig:transversal_cnot_circuit}.

Because the gate is transversal, a single physical fault affects at most one qubit in each code block.
This means that the transversal $\CNOT$ does not propagate errors \emph{within} a code block, but it can create correlated errors \emph{between} the two blocks: if a \ruscz gate fails (e.g.\ due to photon loss), both the control and target blocks are affected.
Consequently, the resulting decoding graph cannot be handled directly by PyMatching. We therefore decode the transversal $\CNOT$ circuit using BPOSD~\cite{roffe_decoding_2020, panteleev2021degenerate}. Alternatively, one could adapt minimum-weight perfect matching techniques developed for decoding transversal gates in the surface code~\cite{serra2026decoding, cain2025fast}.

To estimate the logical error rate of a transversal $\CNOT$, we simulate the following circuit.
Two code patches with width $w$ and height $5/3 w$ are initialized, undergo $10/3 w$ rounds of syndrome extraction, then interact via the transversal gate of \Cref{eq:transversal_cnot}, followed by additional syndrome extraction rounds with detectors including measurements on both patches, and are finally measured.
Specifically, the circuit consists of four blocks of six syndrome extraction rounds each (totaling four lattice surgery cubes worth of spacetime volume).
Four types of logical observables can fail in this circuit: the horizontal and vertical logical operators starting at each patch.
The combined logical error rate is
\begin{equation}
E_{\mathrm{CNOT+4}} = 1 - \prod_{k=1}^{4}(1 - e_k),
\end{equation}
where $e_k$ is the per-observable logical error rate.

To isolate the error contribution of the transversal $\CNOT$ from the syndrome extraction overhead, we compare against a four-cube memory baseline obtained from the memory experiments of \Cref{subsec:lattice_surgery} with matching patch sizes and number of syndrome extraction rounds.
The $\CNOT$ error overhead is then defined as
\begin{equation}
\label{eq:cnot_overhead}
p_{\CNOT} = E_{\mathrm{CNOT+4}} - E_{\mathrm{4\text{-}cube}}.
\end{equation}

\input{crumble_url_transversal_cnot}

\begin{figure}[h]
  \centering
  \includegraphics[width=\textwidth]{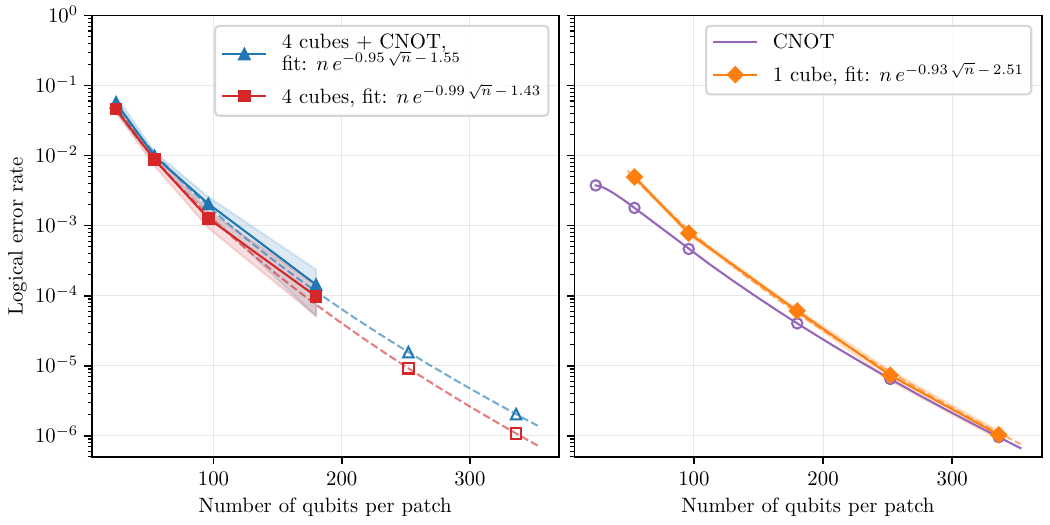}
  \caption{Transversal $\CNOT$ logical error rates at $p = 0.01$ using the BPOSD decoder~\cite{panteleev2021degenerate, roffe_decoding_2020} via the \texttt{stimbposd} package~\cite{higgott_sinter_bposd}.
  Left: combined logical error rate for the four-cube-plus-$\CNOT$ circuit (blue triangles) and the four-cube memory baseline (red squares) as a function of patch size. Solid markers are simulated; open markers are extrapolated. Shaded regions show 95\% confidence intervals.
  Right: $\CNOT$ error overhead $p_{\CNOT}$ (purple circles), defined as the difference between the two curves in the left panel, compared to the single-cube error rate $E_{HV}$ (orange diamonds, reproduced from \Cref{fig:block-size}).
  The $\CNOT$ overhead is comparable to or below the single-cube baseline at all patch sizes.}
  \label{fig:transversal_cnot}
\end{figure}

The results are shown in \Cref{fig:transversal_cnot}.
The left panel shows that the combined logical error rate for the $\CNOT$ circuit (blue) closely tracks the four-cube memory baseline (red), confirming that the transversal gate introduces only a modest overhead.
The right panel isolates the $\CNOT$ contribution $p_{\CNOT}$ (purple) and compares it to the single-cube error rate $E_{HV}$ from \Cref{fig:block-size} (orange).
From exponential fits of the form $n\,e^{a\sqrt{n} + b}$, where $n$ is the number of physical qubits per patch, we can extrapolate $p_{\CNOT}$ to larger patch sizes.
At all patch sizes, the transversal $\CNOT$ overhead is comparable to or smaller than the error of a single lattice surgery cube.
This is expected: the transversal gate is implemented using a depth three circuit that is independent of patch width $w$, so it contributes less spacetime volume susceptible to error than a full cube.
The total duration of the transversal $\CNOT$ gate is $305$ ns.

\FloatBarrier

\section{Simulating the Fermi-Hubbard model on the biplanar SPOQC architecture}
\label{sec:simulating_fermi_hubbard}

This section describes how to fault-tolerantly implement the time-evolution operator of the 2D Fermi--Hubbard Hamiltonian on an $L\times L$ grid with periodic boundary conditions, using the logical instruction set of the biplanar SPOQC architecture introduced in \Cref{sec:logical_instructions_for_the_honeycomb_code}.

\subsection{The 2D Fermi–Hubbard Hamiltonian}
\label{sec:fermi_hubbard_hamiltonian}        
The Hamiltonian of the 2D Fermi-Hubbard model is given by:
    \begin{equation}
    \label{eq:fermi_hubbard_complex_fermionic_operators}
    H = -t \sum_{\langle j,k\rangle}\sum_{\sigma=\uparrow,\downarrow} \left( c^{\dagger}_{j\sigma} c_{k\sigma} + c^{\dagger}_{k\sigma} c_{j\sigma} \right) + U \sum_{j} n_{j\uparrow} n_{j\downarrow} -\mu \sum_{j}\left(n_{j\uparrow}+n_{j\downarrow}\right) = H_h + H_I,
    \end{equation}
where $t$ parametrizes the kinetic energy, $U$ the strength of the on-site repulsion and $\mu$ denotes the chemical potential controlling particle number. The fermionic modes are associated with a site $j$ on a 2D square lattice with periodic boundary conditions, as well as with a spin state $\sigma = \uparrow/\downarrow$.  $\langle j, k\rangle$ denotes that $j$ and $k$ are an adjacent pair of sites. Further, $c^\dagger_{j,\sigma}$ and $ c_{j,\sigma}$ denote the fermionic creation and annihilation operators. The number operator is defined as $n_{j,\sigma} = c^\dagger_{j,\sigma} c_{j,\sigma}$.
The Hamiltonian splits into a hopping part $H_h = H_{h,\uparrow} + H_{h,\downarrow}$, which does not mix the two spin sectors, and an interaction part $H_I$ describing on-site repulsion between electrons of opposite spin on the same site.

Following \cite{Derby_2021, BravyiKitaev2002}, we rewrite \eqref{eq:fermi_hubbard_complex_fermionic_operators} in terms of Majorana \emph{vertex} and \emph{edge} operators $V_{j\sigma}$ and $E_{j\sigma,k\sigma'}$, which generate the full even fermionic operator algebra (see \Cref{appendix:majorana_edge_vertex_operators} for definitions and properties). Under this change of basis the hopping and number operators become:
\begin{equation}
  \label{eq:hopping_term_pair_in_majorana_operators}
    c^{\dagger}_{j\sigma} c_{k\sigma} + c^{\dagger}_{k\sigma} c_{j\sigma} = \frac{1}{2i}\left( E_{j\sigma,k\sigma}V_{k\sigma} + V_{j\sigma} E_{j\sigma,k\sigma} \right), \qquad
    n_{j\sigma} = \frac{1}{2}\left( 1- V_{j\sigma} \right).
\end{equation}
Choosing the chemical potential $\mu = U/2$, as in \cite{Campbell_2021}, cancels the single-body $V_{j\sigma}$ terms in the interaction part and yields,
\begin{equation}
H
= -\frac{t}{2i} \sum_{\langle j,k\rangle}\sum_{\sigma=\uparrow,\downarrow}
\left( E_{j\sigma,k\sigma} V_{k\sigma} + V_{j\sigma} E_{j\sigma,k\sigma} \right)
+ \frac{U}{4} \sum_jV_{j\uparrow}V_{j\downarrow}.
\label{eq:fermi_hubbard_hamiltonian_via_vertex_and_edge_operators}
\end{equation}
Our goal is to approximate the time evolution operator $U(T_\mathrm{sim}) = \exp(-iHT_{\mathrm{sim}})$ for the Hamiltonian \eqref{eq:fermi_hubbard_hamiltonian_via_vertex_and_edge_operators} and simulation time $T_\mathrm{sim}$ on a quantum computer.

\subsection{Fermion-to-Qubit mapping and logical floorplan}
\label{subsec:fermion_to_qubit_mapping}

Simulating fermionic time evolution on a qubit register requires a fermion-to-qubit mapping, i.e. an algebra isomorphism that maps fermionic operators to Pauli operators of the multi-qubit Hilbert space~\cite{BravyiKitaev2002, Derby_2021}. The standard Jordan--Wigner mapping achieves this but, on two-dimensional lattices, maps geometrically local fermionic Hamiltonians to non-local qubit Hamiltonians. When aiming for a maximal degree of parallelization in simulating time evolution, this non-locality becomes a bottleneck, since the Pauli strings of different terms can overlap and prevent simultaneous execution ~\cite{mcardle2025fastcuriousacceleratefaulttolerant}.

\begin{figure}
    \centering
    \raisebox{-0.5\height}{\includegraphics[width=0.35\linewidth]{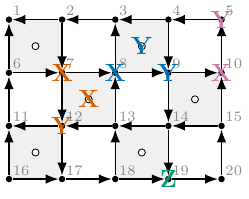}} \quad \raisebox{-0.5\height}{\includegraphics[page=2,width=0.42\linewidth]{graphics/fermion_to_qubit_mapping_compact.pdf}}
\caption{Compact fermion-to-qubit mapping~\cite{Derby_2021} on a square lattice. Fermionic site qubits occupy the vertices and auxiliary qubits sit on every other face in a checkerboard pattern. Vertex operators are mapped to single-qubit $Z$. Edge operators $E_{j,k}$ are mapped to weight-3 Pauli strings. In particular, $X$ on the base $j$, $Y$ on the tip $k$, and $Y$ ($X$) on the auxiliary qubit for horizontal (vertical) edges. The sign is swapped for upward-pointing arrows for consistency with the loop condition \cite{Derby_2021}.}
\label{fig:fermion_to_qubit_mapping_visualization}
\end{figure}

To avoid this, we adopt the compact fermion-to-qubit mapping of~\cite{Derby_2021}, which preserves geometric locality at the cost of a modest qubit overhead. In this construction, fermionic modes are placed on the vertices of a two-dimensional square lattice, with an additional auxiliary qubit on every other lattice face (\Cref{fig:fermion_to_qubit_mapping_visualization}). Vertex operators are mapped to weight-1 Pauli-$Z$ operators, while edge operators become weight-3 Pauli strings of the form $\pm XXY$ and $\pm XYY$, acting on the two endpoint vertices and the shared auxiliary face qubit. The orientation of the lattice edges determines the consistent assignment of signs and Pauli types across the lattice.

\begin{figure}
  \centering
  \includegraphics[width=\linewidth]{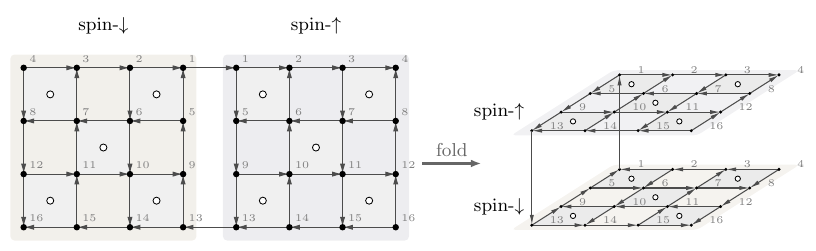}
  \caption{Spinful compact fermion-to-qubit mapping. Left: the two spin sectors are placed on opposite sides of the mapping with mirrored site numbering. Right: folding the layout into two parallel planes. Edge operators connecting different spin sectors become non-local but do not appear in the Fermi--Hubbard Hamiltonian~\eqref{eq:fermi_hubbard_hamiltonian_via_vertex_and_edge_operators}. The biplanar architecture allows the interaction terms $V_{j\uparrow}V_{j\downarrow}$ to be implemented locally via transversal gates.}
  \label{fig:spinful_fermion_to_qubit_mapping}
\end{figure}

For spinful fermions, each site carries two fermionic modes, which must be accomodated for in the qubit lattice. On a single plane this requires interlacing modes of opposite spin. Ref.~\cite{mcardle2025fastcuriousacceleratefaulttolerant} takes this approach and uses fermionic $\mathsf{SWAP}$ gates to dynamically reorder modes as needed. Here, we instead group the sites of opposite spin on opposite sides of the grid, and fold it into parallel planes (\Cref{fig:spinful_fermion_to_qubit_mapping}). 
Edge operators connecting modes of different spin then become non-local, but no such terms appear in the Fermi--Hubbard Hamiltonian in \Cref{eq:fermi_hubbard_hamiltonian_via_vertex_and_edge_operators}.
All terms in our compilation therefore remain local, with the available fault-tolerant gate set of the biplanar SPOQC layout, with the exception of the edge operators that implement the periodic boundary conditions. 

\begin{figure}
\centering
\includegraphics[width=0.8\linewidth]{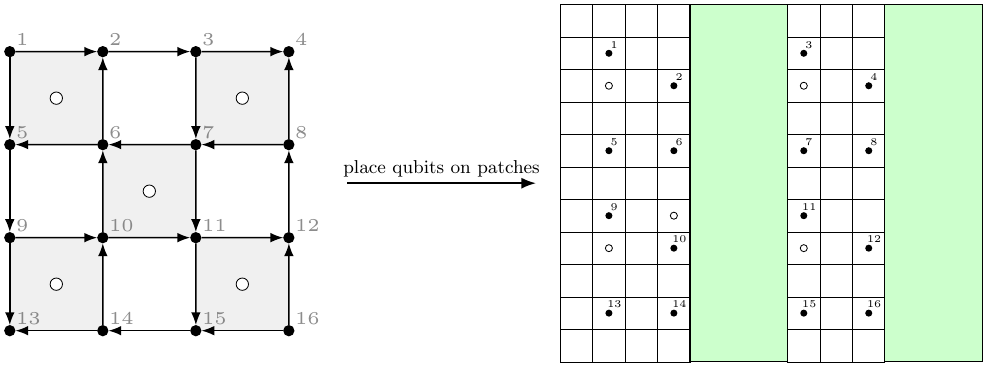}
\caption{Floorplan of code patches for the Fermi-Hubbard problem on a $4\times4$ lattice under the compact fermion-to-qubit mapping. Patches marked with a dot host data qubits that remain initialized throughout the simulation. Unmarked patches provide workspace for lattice-surgery multi-Pauli measurements. Filled dots denote fermionic site qubits (logical $X$ vertical) and empty dots auxiliary qubits (logical $X$ horizontal). Both layers of the biplanar layout share this floorplan. Green aisles are reserved for magic state factories.}
\label{fig:floorplan_four_by_four}
\end{figure}

With this fermion-to-qubit mapping in mind, we can now specify the layout of logical qubits on each plane. The floorplan adopted for both layers, illustrated for a $4\times4$ grid in \Cref{fig:floorplan_four_by_four}, is inspired by~\cite{mcardle2025fastcuriousacceleratefaulttolerant}. Green aisles between data patches are reserved for magic state factories, which are treated as black boxes hosting magic state factories (MSFs) that continuously supply $\ket{T}$ states. Their sizing is fixed by the required $\ket{T}$-state fidelity, and the peak supply rate.  

\subsection{Plaquette trotterization and logical compilation in the biplanar layout}
\label{sec:plaquette_compilation}
We seek a fault-tolerant compilation of the time evolution operator via the plaquette-based Trotterization algorithm (PLAQ) described in \cite{Campbell_2021}. The starting point is the qubit Hamiltonian
\begin{equation}
H
= -\frac{t}{2i} \sum_{\langle j,k\rangle}\sum_{\sigma=\uparrow,\downarrow}
\left( E_{j\sigma,k\sigma} Z_{k\sigma} + Z_{j\sigma} E_{j\sigma,k\sigma} \right)
+ \frac{U}{4} \sum_jZ_{j\uparrow}Z_{j\downarrow}
\label{eq:fermi_hubbard_hamiltonian_via_vertex_and_edge_operators_correct_shift_pauli}
\end{equation}
    where the vertex operators have been mapped to Pauli-$Z$ as determined by the compact fermion-to-qubit mapping of Section~\ref{subsec:fermion_to_qubit_mapping}. We now show how to compile this Hamiltonian into fault-tolerant instructions on the biplanar SPOQC architecture.

\begin{figure}
    \centering
    \includegraphics[width=\linewidth]{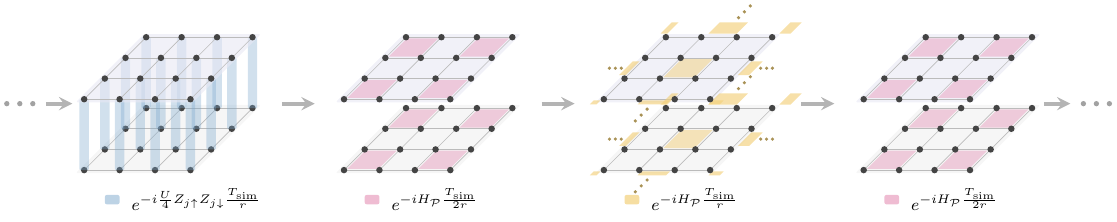}
    \caption{Visualization of a single Trotter step for a $4\times 4$ lattice. The sub-evolutions are shown from left to right in the order they are applied. The interaction sub-evolution $\exp(-iH_I T_{\mathrm{sim}}/r)$ (blue) acts on all $L^2$ sites simultaneously via $Z_\uparrow Z_\downarrow$ terms. The two applications of $\exp(-iH_h^p T_{\mathrm{sim}}/2r)$ (pink) are each supported on $L^2/2$ non-overlapping bulk plaquettes. The sub-evolution $\exp(-iH_h^g T_{\mathrm{sim}}/r)$ (golden) is supported on the complementary plaquette set, including those that implement the periodic boundary conditions. Auxiliary face qubits are omitted for clarity.}
    \label{fig:trotter_step_visualization_linear}
\end{figure}

We decompose the Hamiltonian as $H = H_I + H_h^p + H_h^g$, where $H_h^p$ and $H_h^g$ are two complementary subcollections of hopping terms grouped as plaquettes, indicated by the pink and gold coloring in the middle panels of \Cref{fig:trotter_step_visualization_linear}, satisfying $H_h^p + H_h^g = H_h$ and $[H_h^p, H_h^g] \neq 0$. We approximate the time-evolution operator $U(T_{\mathrm{sim}}) = e^{-iHT_{\mathrm{sim}}}$ using the second-order Trotter product formula in $r$ steps~\cite{Campbell_2021}. The resulting operator-norm error satisfies
\begin{equation}
\epsilon_{\mathrm{alg}} \le \frac{W T_{\mathrm{sim}}^3}{r^2},
\end{equation}
where $W$ is determined by operator-norm bounds on nested commutators of the sub-Hamiltonians~\cite{Campbell_2021}. For a fixed evolution time $T_{\mathrm{sim}}$, the number of Trotter steps $r$ is chosen to achieve a target algorithmic accuracy $\epsilon_{\mathrm{alg}}$, which in turn determines the overall runtime. We return to this in Section~\ref{sec:resource_estimation} with concrete values. Combining the first and last half-steps of adjacent Trotter steps, each step reduces to the four sub-evolutions
\begin{equation}
e^{-iH_I \frac{T_{\mathrm{sim}}}{r}},
\quad
e^{-iH_h^p \frac{T_{\mathrm{sim}}}{2r}},
\quad
e^{-iH_h^g \frac{T_{\mathrm{sim}}}{r}},
\quad
e^{-iH_h^p \frac{T_{\mathrm{sim}}}{2r}}.
\end{equation}
Each of the sub-evolutions follow the same three-step structure. First, a diagonalizing circuit brings the sub-Hamiltonian into diagonal form. Then, $L^2$ single-qubit $Z$-rotations ($L^2/2$ per spin sector) are executed in parallel. The inverse diagonalizing circuit then concludes the sub-evolution. For the $H_h^p$ and $H_h^g$ hopping terms, the two spin sectors are handled independently on their respective planes. A visualization of a single Trotter step is shown in \Cref{fig:trotter_step_visualization_linear}.

We now explain how to compile a full Trotter step into fault-tolerant instructions. We track the number of $\ket{T}$ states consumed, lattice surgery cubes, transversal inter-plane CNOTs, and logical timesteps. Section~\ref{sec:fault_tolerant_synthesis} is dedicated to the fault-tolerant synthesis of arbitrary-angle single-qubit $Z$-rotations shared by all sub-evolutions. Sections~\ref{sec:interaction_term}, \ref{sec:plaquette_evolution}, and \ref{sec:golden_hopping} derive, respectively, the circuits that bring the interaction, pink, and gold sub-Hamiltonians into diagonal form.

\subsubsection{\texorpdfstring{Fault-tolerantly synthesizing $L^2$ single qubit $Z$-rotations}{}}
\label{sec:fault_tolerant_synthesis}

Each of the sub-evolutions requires $L^2$ parallel arbitrary-angle $Z$ rotations.
The synthesis of $\exp(-i\theta Z)$ using only Clifford$+T$ gates admits different strategies \cite{Kliuchnikov2023shorterquantum}, possibly augmented by an auxiliary
qubit, measurements, and classical feed-forward. Each strategy imposes slightly different requirements on the Clifford$+T$ words to be synthesized, and therefore different $T$-count scalings (see details in Appendix~\ref{appendix:synthesis_strategies}). However, every strategy produces diagonal Clifford$+T$ sequences as a subroutine (either as the sole output or as the accept/reject branches of a fallback scheme). The fault-tolerant implementation of these sequences is therefore the common building block, and we describe it here. A diagonal synthesis approximates a $Z$-rotation by a product of $n_T$ Clifford$+T$ pairs,
\begin{equation}
\label{eq:clifford_t_word}
    \exp\!\left(-i\theta Z\right) \overset{\epsilon_{\mathrm{synth}}}{\approx} \prod_{i=1}^{n_T}(\mathsf{C}_iT).
\end{equation}

The Clifford gates $\mathsf{C}_i$ from \Cref{eq:clifford_t_word} can be pushed through the $T$ gates rather than implemented at each step to save on logical timesteps. Since $T = \exp(-\frac{i\pi}{8}Z)$, pushing a Clifford $\mathsf{C}$ through a $T$ gate conjugates the Pauli axis of the $\frac{\pi}{8}$ rotation:
\begin{equation}
\exp\!\left(-i\tfrac{\pi}{8}\,P\right)\mathsf{C} = \mathsf{C}\,\exp\!\left(-i\tfrac{\pi}{8}\,\mathsf{C}^\dagger P \mathsf{C}\right).
\end{equation}
\cite{Litinski_2019}.
Consequently, \Cref{eq:clifford_t_word} can be re-expressed as a sequence of $n_T$ Pauli $\frac{\pi}{8}$ rotations followed by a single-qubit Clifford:

\begin{equation}
\label{eq:z_rotation_synthesis_pauli_pi_over_eight_rotations}
    \exp\!\left(-i\theta Z\right) \overset{\epsilon_{\mathrm{synth}}}{\approx}
    \mathsf{C}\,\exp\!\left(\frac{-i\pi}{8}P_n\right)\cdots \exp\!\left(\frac{-i\pi}{8}P_1\right),
    \quad P_i \in \{\pm X, \pm Y, \pm Z\},\quad \mathsf{C} \in \mathrm{Clifford}.
\end{equation}

Each $\frac{\pi}{8}$-rotation is implemented via a lattice-surgery Pauli-product measurement between the data qubit and an injected magic state (see Fig.~7 of \cite{Litinski2019gameofsurfacecodes}). With probability $50\%$, this requires a Clifford ($\frac{\pi}{4}$-rotation) correction, which can itself be implemented via lattice surgery using an additional qubit initialized in the $\ket{0}$ state (see Fig.~11b of \cite{Litinski2019gameofsurfacecodes}). Conditional Pauli corrections ($\frac{\pi}{2}$-rotations) need not be implemented explicitly but can be handled in software \cite{Litinski2019gameofsurfacecodes}.
In practice, each Clifford correction that arises probabilistically is not implemented immediately but instead pushed through all subsequent $\frac{\pi}{8}$-rotations, dynamically conjugating their Pauli axes. The corrections accumulate in the terminal Clifford $\mathsf{C}$, which can always be implemented with at most $3$ layers of $\frac{\pi}{4}$ rotations.

\begin{figure}
  \centering
\setlength{\tabcolsep}{4pt}
\begin{tabular}{c@{\hspace{0.8cm}}c}
  \begin{tabular}{cc}
    \includegraphics[width=0.22\linewidth]{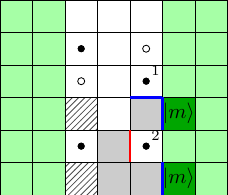} &
    \includegraphics[width=0.22\linewidth]{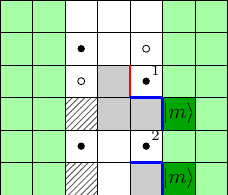} \\
    \multicolumn{2}{c}{\textbf{a)}}
  \end{tabular}
  &
  \begin{tabular}{cc}
    \includegraphics[width=0.22\linewidth]{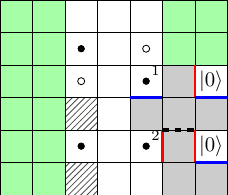} &
    \includegraphics[width=0.22\linewidth]{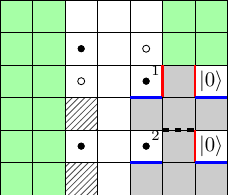} \\
    \multicolumn{2}{c}{\textbf{b)}}
  \end{tabular}
\end{tabular}
\caption{Examples of simultaneous single-qubit Pauli-product measurements within a single plaquette. Gray patches mediate the lattice-surgery measurements. The boundary color specifies the measured Pauli operator (red $=X$, blue $=Z$). Dashed line indicates that mediating patches are not connected. The line-shaded patches reserve auxiliary qubits for fallback gate-synthesis protocols. CNOTs with the rotation qubits would be applied beforehand, so these patches must remain idle during the $\frac{\pi}{8}$ sequence. Note that the qubit labelling is not consistent with \Cref{fig:floorplan_four_by_four} for notational simplicity. \textbf{(a)} shows the measurements implementing the parallel $\exp(-i\frac{\pi}{8}Z_1)\otimes\exp(-i\frac{\pi}{8}X_2)$ and $\exp(-i\frac{\pi}{8}Y_1)\otimes\exp(-i\frac{\pi}{8}Z_2)$ rotations. \textbf{(b)} shows the measurements implementing the corresponding $\frac{\pi}{4}$ rotations, during which no magic states are consumed. The idle footprint of the MSFs can therefore be used to prepare the required $\ket{0}$ auxiliary qubit.}
  \label{fig:single_qubit_pi_over_eight_rotations_lattice_surgery}
\end{figure}

At the level of fault-tolerant instructions, the synthesis reduces to $n_T$ Pauli-product measurements with $\ket{T}$-state injection, followed by up to $3$ measurements implementing $\frac{\pi}{4}$ rotations. Representative patterns of two parallel single-qubit $\frac{\pi}{8}$  and $\frac{\pi}{4}$ rotations implemented via lattice surgery are shown in \Cref{fig:single_qubit_pi_over_eight_rotations_lattice_surgery}. This figure also illustrates that the lattice surgery layout accommodates two simultaneous $Z$-rotations per plaquette at every logical timestep, provided the magic state aisles sustain the required supply rate of two $\ket{T}$ states per plaquette per $\frac{\pi}{8}$ layer. Moreover, the best-scaling synthesis strategies wrap the right-hand side of \Cref{eq:clifford_t_word} in a pair of $\mathsf{CNOT}$s with an auxiliary qubit, followed by a potential fallback synthesized gate conditioned on the auxiliary measurement outcome (see Figure 1 of \cite{Kliuchnikov2023shorterquantum}). The floorplan in \Cref{fig:single_qubit_pi_over_eight_rotations_lattice_surgery} shows that sufficient space is available to allocate this additional auxiliary qubit, so our layout supports both direct and fallback synthesis.

Magic state injection requires classical feed-forward to conjugate the next rotation axis by the appropriate Clifford, introducing a reaction time $\tau_{\mathrm{r}}$ between $\frac{\pi}{8}$ layers. We account for this idle period as $\tau_{\mathrm{r}}/t_{\mathrm{l}}$ logical cubes per initialized logical qubit, where $t_{\mathrm{l}}$ is the logical cycle time, i.e.\ the duration of one lattice surgery cube.

Let now $n_T$ be the (expected) total $T$-count of the synthesis protocol. The fault-tolerant costs for both direct and fallback synthesis are summarized in \Cref{tab:fault_tolerant_cost_synthesis}. Note that different synthesis strategies lead to different expected $T$-count scalings $n_T$. The numbers are given per synthesized rotation, assuming that two rotations are synthesized in parallel within a plaquette. $L^2$ syntheses are performed in parallel within each Trotter sub-evolution, and the next algorithmic step cannot begin until all have completed. Consequently, the logical runtime in the synthesis with fallback case is determined by the worst case (failure branch) among the $L^2$ parallel syntheses. Here $p_{\mathrm{succ}}^{\mathrm{all}}$ denotes the probability that all $L^2$ synthesis attempts succeed simultaneously. The quantities $\tilde{p}_{\mathrm{fail}}$ and $\tilde{p}_{\mathrm{succ}}$ denote the failure and success probabilities conditioned on the event that not all $L^2$ synthesis attempts succeed.

\begin{table}
\centering
\begin{tblr}{
  width=\linewidth,
  colspec={Q[l,wd=0.20\linewidth] Q[l,wd=0.20\linewidth] Q[l,wd=0.52\linewidth]},
  row{1}={font=\bfseries},
  hline{1,2} = {1.5pt},
  hline{3-5} = {0.8pt},
  rows = {valign=t},
}
Resource type & Direct synthesis & Synthesis with fallback \\
Transversal CNOTs
& $0$
& $0$
\\
$\ket{T}$ states consumed
& $n_T$
& $n_T = n_T^{\mathrm{succ}} + p_{\mathrm{fail}}\,n_T^{\mathrm{fb}}$
\\
Logical timesteps
& $n_T\!\left(1+\frac{\tau_{\mathrm{r}}}{t_{\mathrm{l}}}\right)+3$
& $n_T^{\mathrm{succ}}\!\left(1+\frac{\tau_{\mathrm{r}}}{t_{\mathrm{l}}}\right)+7 {}+ \left(1-p_{\mathrm{succ}}^{\mathrm{all}}\right)
\left[n_T^{\mathrm{fb}}\!\left(1+\frac{\tau_{\mathrm{r}}}{t_{\mathrm{l}}}\right)+3\right] $
\\
Active surgery cubes
& $n_T\!\left(5.33+3\frac{\tau_{\mathrm{r}}}{t_{\mathrm{l}}}\right)+23$
& $n_T^{\mathrm{succ}}\!\left(6.33+4\frac{\tau_{\mathrm{r}}}{t_{\mathrm{l}}}\right) + 48 + \left(1-p_{\mathrm{succ}}^{\mathrm{all}}\right)\tilde{p}_{\mathrm{fail}}
\left[n_T^{\mathrm{fb}}\!\left(5.33+3\frac{\tau_{\mathrm{r}}}{t_{\mathrm{l}}}\right)+23\right]+ \left(1-p_{\mathrm{succ}}^{\mathrm{all}}\right)\tilde{p}_{\mathrm{succ}}
\left[n_T^{\mathrm{fb}}\!\left(3+3\frac{\tau_{\mathrm{r}}}{t_{\mathrm{l}}}\right)+9\right] $
\\
\end{tblr}
\caption{Fault-tolerant cost of single-qubit synthesis for direct synthesis (auxiliary-free) and synthesis with fallback. The logical timestep count is obtained by counting the layers of $\frac{\pi}{8}$ rotations required by the protocol, each separated by a reaction time of $\tau_r/t_l$ logical cycles, plus the terminal Clifford. In the fallback case, we additionally account for the $\mathsf{CNOT}$ layers that wrap the rotation sequence, plus the time it takes to implement the fallback if at least one fallback is needed. The active surgery cube count is read off directly from \Cref{fig:single_qubit_pi_over_eight_rotations_lattice_surgery}, averaged over the $X$, $Y$, $Z$ measurement axes (yielding the fractional $0.33$ contribution). The extra unit in the fallback prefactor ($6.33$ vs.\ $5.33$) reflects the control auxiliary qubit that must remain initialized and idle throughout the $\frac{\pi}{8}$ sequence. In the fallback case, the count additionally averages over three per-qubit scenarios conditioned on the global $L^2$-wise synchronization: (i) no qubit triggers the fallback (probability $p_{\mathrm{succ}}^{\mathrm{all}}$), (ii) at least one qubit triggers the fallback but the qubit under consideration succeeded and idles during the fallback round (conditional probability $\tilde{p}_{\mathrm{succ}}$), and (iii) the qubit itself requires the fallback branch (conditional probability $\tilde{p}_{\mathrm{fail}}$).}
\label{tab:fault_tolerant_cost_synthesis}
\end{table}

The maximum $T$-injection rate is $\dfrac{1}{1+\tau_{\mathrm{r}}/t_{\mathrm{l}}}$ per logical timestep, corresponding to at most two rotations synthesized per plaquette. 

Both direct and fallback synthesis employ a channel mixing strategy~\cite{Kliuchnikov2023shorterquantum} that probabilistically selects between an over-rotated and an under-rotated Clifford$+T$ sequence. Since the approximation regions for both branches have equal area, their expected $T$-counts are approximately equal~\cite{Kliuchnikov2023shorterquantum}, so the mixing introduces no relevant additional synchronization overhead beyond that of the fallback mechanism.

\subsubsection{Interaction term \texorpdfstring{$H_I$}{H\_I}}
\label{sec:interaction_term}

The interaction sub-Hamiltonian $H_I = \frac{U}{4}\sum_j Z_{j\uparrow}Z_{j\downarrow}$ couples the spin-up and spin-down layers at each lattice site via a two-qubit $ZZ$ term. A single transversal $\mathsf{CNOT}$ between the two planes at each site serves as the diagonalization circuit, reducing the two-qubit interaction to a single-qubit $Z$ rotation. The same $\mathsf{CNOT}$ also serves as the inverse diagonalization circuit. To balance magic state consumption equally between the two planes, the $\mathsf{CNOT}$ direction is alternated across each plaquette, so that only two single-qubit $Z$ rotations per plaquette are required per plane, as in \Cref{fig:single_qubit_pi_over_eight_rotations_lattice_surgery}. The resulting $L^2/2$ single-qubit $Z$ rotations per plane are fully independent and can be executed in parallel.

The fault-tolerant cost of $\exp(-iH_I\,T_{\mathrm{sim}}/r)$ is then $2L^2$ transversal inter-plane $\mathsf{CNOT}$ gates, which contribute no $|T\rangle$ states, no surgery cubes, and no logical timesteps, plus the synthesis cost of $L^2$ single-qubit $Z$-rotations described in Section~\ref{sec:fault_tolerant_synthesis}.

\subsubsection{Pink hopping term \texorpdfstring{$H_h^p$}{H\_h\^{}p}}
\label{sec:plaquette_evolution}

$H_h^p$ is a sum of mutually commuting plaquette Hamiltonians \cite{Campbell_2021}. We define the \emph{plaquette Hamiltonian} for a plaquette $\mathcal{P}$ as
\begin{equation}
\label{eq:plaquette_hamiltonian}
    H_\mathcal{P} = -t\sum_{\sigma=\uparrow,\downarrow}\sum_{(j,k)\in\partial \mathcal{P}}
    \left(c_{j\sigma}^\dagger c_{k\sigma} + c_{k\sigma}^\dagger c_{j\sigma}\right)
    = H_\mathcal{P}^\uparrow + H_\mathcal{P}^\downarrow.
\end{equation}

\begin{figure}
  \centering
  \includegraphics[width=0.95\linewidth]{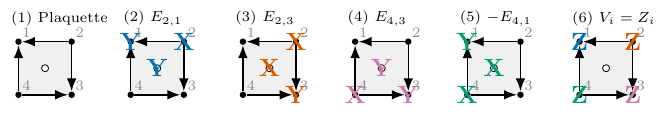}
  \caption{Visualization of the Majorana edge and vertex operators on a single plaquette under the compact fermion-to-qubit mapping. Panels (2)–(5) show the weight-3 Pauli representatives of the four edge operators (each acting on the two endpoint vertices and the auxiliary face qubit), while panel (6) shows the four weight-1 vertex operators $V_i = Z_i$. Note that edge operators which are not explicitly defined can be derived via $E_{ik} = iE_{ij}E_{jk}$~\cite{Derby_2021}.}
  \label{fig:vertex_and_edge_operators_on_single_plaquette}
\end{figure}

Since the pink plaquettes are all structurally identical bulk plaquettes, we derive the time evolution circuit for a single plaquette and apply it in parallel to implement $\exp(-iH_h^p\,T_{\mathrm{sim}}/(2r))$. The vertex and edge operators constituting $H_\mathcal{P}^\sigma$ are shown in \Cref{fig:vertex_and_edge_operators_on_single_plaquette}.

\begin{figure}
\centering
\includegraphics[width=0.7\linewidth]{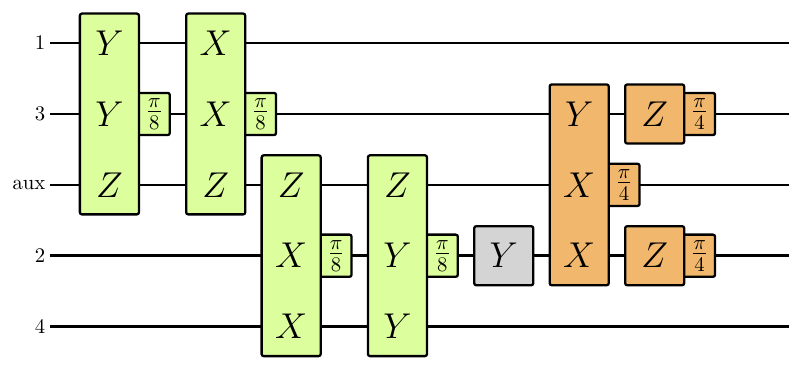}
\caption{Circuit representation of $C^\dagger F_{2,4}^\dagger F_{3,1}^\dagger$. Boxes containing a Pauli string with angle $\varphi$ represent the gate $e^{-iP\varphi}$ following the notation and visuals of~\cite{Litinski_2019}. The single qubit $\frac{\pi}{4}$ rotations on qubit $2$ and $3$ have been pushed through to the end of the circuit, which changes the Pauli axis of the $\frac{\pi}{8}$ rotations w.r.t. \Cref{eq:f_3_1_dagger,eq:f_2_4_dagger} and for $C^\dagger$ from \Cref{eq:clifford_C}. $X_2$ is conjugated to $-Y_2$ and the minus sign can be dropped as a global phase.}
\label{fig:circuit_of_plaquette_time_evolution}
\end{figure}

The plaquette Hamiltonian $H_\mathcal{P}^\sigma$ couples qubits through weight-3 Pauli strings arising from the edge operators. The diagonalizing circuit, derived in Appendix \ref{sec:derivation_of_the_diagonalizing_circuit_for_pink_plaquette}, reduces the time evolution to two independent single-qubit $Z$-rotations, giving the decomposition
\begin{equation}
  \label{eq:time_evolution_unitary_plaquette_hamiltonian}
\exp\!\left(\frac{-iT_{\mathrm{sim}}}{2r}H_\mathcal{P}^\sigma\right)
= F_{3,1}\,F_{2,4}\,C\,e^{i\frac{tT_\mathrm{sim}}{2r}Z_2}\,e^{i\frac{tT_\mathrm{sim}}{2r}Z_3}\,C^\dagger\,F_{2,4}^\dagger\,F_{3,1}^\dagger,
\end{equation}
where we drop $\sigma$ on the right side of \Cref{eq:time_evolution_unitary_plaquette_hamiltonian} for brevity. $F_{3,1}$ and $F_{2,4}$ are fermionic Fourier transforms between the corresponding modes, and $C$ is a Clifford that diagonalizes the inner hopping terms onto separate qubits. Their inverses, which form the first half of the circuit, are given by
\begin{align}
\label{eq:f_3_1_dagger}
F_{3,1}^\dagger &= e^{\frac{-i\pi}{8} X_1Y_3Z_{\mathrm{aux}}}\,
  e^{\frac{i\pi}{8}Y_1X_3Z_{\mathrm{aux}}}\,
  e^{\frac{-i\pi}{4} Z_3},\\
\label{eq:f_2_4_dagger}
F_{2,4}^\dagger &= e^{\frac{i\pi}{8} X_2Y_4Z_{\mathrm{aux}}}\,
  e^{\frac{-i\pi}{8}Y_2X_4Z_{\mathrm{aux}}}\,
  e^{\frac{-i\pi}{4} Z_2},\\
\label{eq:clifford_C}
C^\dagger &= \exp\!\left(\frac{i\pi}{4}Y_2X_3X_{\mathrm{aux}}\right)X_2.
\end{align}
The circuit implementing $C^\dagger F_{2,4}^\dagger F_{3,1}^\dagger$ is shown in \Cref{fig:circuit_of_plaquette_time_evolution}.

\begin{figure}
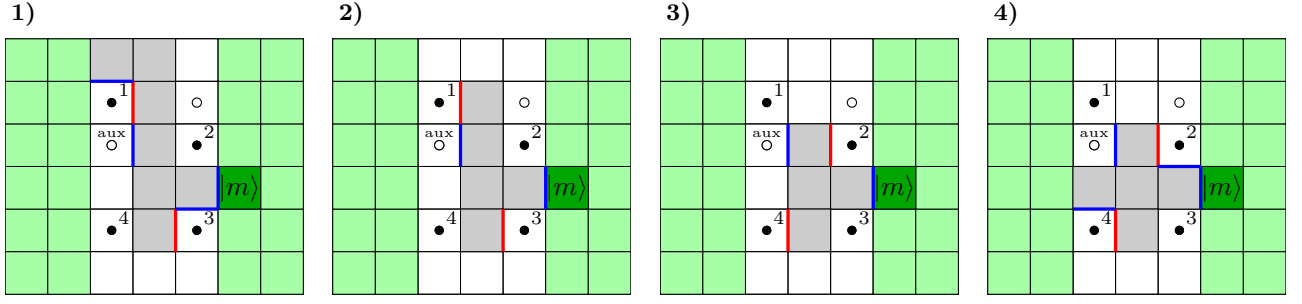

  \centering
  \setlength{\tabcolsep}{4pt} 
  \renewcommand{\arraystretch}{1}
  \begin{tabular}{cccc}
    \steplabel{0.22\linewidth}{1}{graphics/fourier_transform_timesteps_lattice_surgery/step_1} &
    \steplabel{0.22\linewidth}{2}{graphics/fourier_transform_timesteps_lattice_surgery/step_2} &
    \steplabel{0.22\linewidth}{3}{graphics/fourier_transform_timesteps_lattice_surgery/step_3} &
    \steplabel{0.22\linewidth}{4}{graphics/fourier_transform_timesteps_lattice_surgery/step_4}
  \end{tabular}

  \caption{Lattice surgery Multi-Pauli measurement patterns for implementing the $\frac{\pi}{8}$ rotations needed for $C^\dagger F_{2,4}^\dagger F_{3,1}^\dagger$. From left to right: $e^{-i\frac{\pi}{8} Y_1Y_3Z_{\mathrm{aux}}}$, $e^{-i\frac{\pi}{8} X_1X_3Z_{\mathrm{aux}}}$, $e^{-i\frac{\pi}{8} X_2X_4Z_{\mathrm{aux}}}$, $e^{-i\frac{\pi}{8} Y_2Y_4Z_{\mathrm{aux}}}$. Each rotation is realized by measuring the Pauli string appended with a $Z$ on a magic state. Depending on the measurement outcome, $\frac{\pi}{4}$-rotations of the same Pauli string are necessary. Note also that the orientation of logical operators is flipped between auxiliary and magic state qubits w.r.t. fermionic site qubits.}
  \label{fig:fourier_transform_timesteps_lattice_surgery}
\end{figure}

When implementing the circuit of $C^\dagger F_{2,4}^\dagger F_{3,1}^\dagger$ via lattice-surgery Pauli-product measurements and $\ket{T}$-state injections, the bare Pauli $Y_2$ is absorbed into the Pauli frame and tracked in software~\cite{Litinski_2019}.
The two single-qubit $\frac{\pi}{4}$ rotations are merged into the synthesized $Z$-rotations and absorbed into the terminal Clifford $\mathsf{C}$ (see \Cref{eq:z_rotation_synthesis_pauli_pi_over_eight_rotations}), at no additional cost.
The measurement outcomes of the $\frac{\pi}{8}$ rotations determine potential Clifford corrections. Those must in general be known to update the Pauli axis of the subsequent $\frac{\pi}{8}$ rotation. However, all of the $\frac{\pi}{8}$ rotations in \Cref{fig:circuit_of_plaquette_time_evolution} mutually commute. Thus the corrective $\frac{\pi}{4}$ rotations also commute. Successive $\frac{\pi}{8}$-rotation measurements can therefore be performed back-to-back, without waiting for the reaction time. Once all $\frac{\pi}{8}$-rotation measurements are complete, the required $\frac{\pi}{4}$ corrections are applied via further measurements.
The measurement patterns for all four $\frac{\pi}{8}$ rotations are shown in \Cref{fig:fourier_transform_timesteps_lattice_surgery}.

\begin{figure}[h]
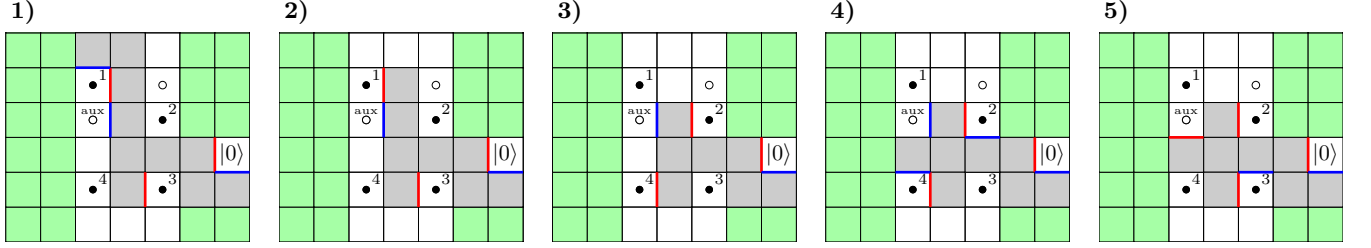

  \centering
  \setlength{\tabcolsep}{4pt} 
  \renewcommand{\arraystretch}{1}
  \begin{tabular}{ccccc}
    \steplabel{0.18\linewidth}{1}{graphics/fourier_transform_timesteps_lattice_surgery_corrections/step_1} &
    \steplabel{0.18\linewidth}{2}{graphics/fourier_transform_timesteps_lattice_surgery_corrections/step_2} &
    \steplabel{0.18\linewidth}{3}{graphics/fourier_transform_timesteps_lattice_surgery_corrections/step_3} &
    \steplabel{0.18\linewidth}{4}{graphics/fourier_transform_timesteps_lattice_surgery_corrections/step_4} &
    \steplabel{0.18\linewidth}{5}{graphics/fourier_transform_timesteps_lattice_surgery_corrections/step_5}
  \end{tabular}

  \caption{Multi-Pauli measurement patterns for the conditional $\frac{\pi}{4}$ Clifford corrections and the mandatory $C^\dagger$ rotation. Steps 1--4 implement the conditional corrections $e^{-i\frac{\pi}{4} Y_1Y_3Z_{\mathrm{aux}}}$, $e^{-i\frac{\pi}{4} X_1X_3Z_{\mathrm{aux}}}$, $e^{-i\frac{\pi}{4} X_2X_4Z_{\mathrm{aux}}}$, $e^{-i\frac{\pi}{4} Y_2Y_4Z_{\mathrm{aux}}}$. Each is realized by measuring the Pauli string with a $\ket{0}$ auxiliary. Step 5 implements the unconditional $e^{i\frac{\pi}{4}\,X_2Y_3X_{\mathrm{aux}}}$ from $C^\dagger$. The corrective Pauli $\frac{\pi}{2}$-rotations need not be implemented explicitly and can be handled in software.}
  \label{fig:fourier_transform_timesteps_lattice_surgery_corrections}
\end{figure}

\Cref{fig:fourier_transform_timesteps_lattice_surgery_corrections} shows the measurement patterns for the potential $\frac{\pi}{4}$ rotation corrections, as well as $e^{-i\frac{\pi}{4}\,X_2Y_3X_{aux}}$ from \Cref{fig:circuit_of_plaquette_time_evolution}. When no correction is needed, the patch sits idle, since all plaquette Hamiltonians are time-evolved in parallel and must remain synchronized.
Note also that, in a parallelized implementation, it is important that these corrective $\frac{\pi}{4}$ rotations are implemented explicitly rather than absorbed into a global deferred Clifford. The support of subsequent $\frac{\pi}{8}$ rotations would otherwise grow with every Trotter step, reducing the available parallelism. In the worst case, the $T$-depth then approaches the total $T$-count.

We derive the fault-tolerant cost of the sub-evolution $\exp(-iH_h^p\,T_{\mathrm{sim}}/(2r))$. It implements $\frac{L^2}{2}$ plaquette time evolutions in parallel ($\frac{L^2}{4}$ per layer), where each plaquette time evolution decomposes into three substeps:
\begin{enumerate}
    \item The diagonalizing circuit $C^\dagger F_{2,4}^\dagger F_{3,1}^\dagger$,
    \item two synthesized single-qubit $Z$-rotations executed in parallel,
    \item the inverse circuit $F_{3,1} F_{2,4} C$.
\end{enumerate}
This totals in $L^2$ single-qubit $Z$-rotations ($\frac{L^2}{2}$ per layer). The fault-tolerant cost of the $Z$-rotations is taken from \Cref{sec:fault_tolerant_synthesis}. The fault-tolerant cost of $C^\dagger F_{2,4}^\dagger F_{3,1}^\dagger$ can be read off from \Cref{fig:fourier_transform_timesteps_lattice_surgery,fig:fourier_transform_timesteps_lattice_surgery_corrections} and is summarized in \Cref{tab:cost_diagonalizing_circuit}. The inverse circuit $F_{3,1} F_{2,4} C$ is structurally identical and incurs the same cost. 
The total cost of $\exp(-iH_h^p\,T_{\mathrm{sim}}/(2r))$ follows from scaling the active surgery cube and $\ket{T}$-state entries of \Cref{tab:cost_diagonalizing_circuit} by $L^2/2$ and adding $L^2$ times the per-rotation cost in \Cref{tab:fault_tolerant_cost_synthesis}. The logical timesteps from both tables are added once rather than multiplied, since the plaquettes and their rotations execute in parallel.

\begin{table}[h]
\centering
\begin{tblr}{
    hline{1,2}={1.5pt},
    row{1}={font=\bf},
    hline{3-6}={0.8pt},
    cells={c, m},
    }
Resource type & Quantity\\
Transversal CNOTs & $0$ \\
$\ket{T}$ states consumed & $8$ \\
Logical timesteps & $18$ \\
Active surgery cubes & $205$ \\
\end{tblr}
\caption{Fault-tolerant cost of the full diagonalization $F_{3,1} F_{2,4} C \cdots C^\dagger F_{2,4}^\dagger F_{3,1}^\dagger$ for a single plaquette, as read off from \Cref{fig:fourier_transform_timesteps_lattice_surgery,fig:fourier_transform_timesteps_lattice_surgery_corrections}. The surgery block count averages over the conditional Clifford corrections, which fire with probability $50\%$ and otherwise leave the patch idle, which costs $6$ active surgery cubes.}
\label{tab:cost_diagonalizing_circuit}
\end{table}

\subsubsection{Golden hopping term \texorpdfstring{$H_h^g$}{H\_h\^{}g}}
\label{sec:golden_hopping}

After the pink sub-evolution, the qubit patches of each golden plaquette are separated by an MSF aisle (see \Cref{fig:floorplan_four_by_four}, plaquette 6-7-11-10). Every second column of patches is shifted through the aisle by growing into the MSF during the first step of the golden diagonalization circuit and shrinking from the trailing side during the next step, incurring no additional logical timesteps. Further, the golden hopping term $H_h^g$ contains both bulk plaquettes and those that realize the periodic boundary conditions. For the bulk plaquettes, the diagonalizing circuit is identical (up to sign flip of the Pauli $\frac{\pi}{8}$-rotations), to the pink case (\Cref{fig:circuit_of_plaquette_time_evolution}). However, for the non-local plaquettes the edge operators that define the Pauli rotation axes of the diagonalization circuit (\Cref{eq:inner_part_of_plaquette_evolution_simplification_through_fermionic_fourier_transform_identities} and \Cref{eq:C1}) have non-local support.
That is because under the compact fermion-to-qubit mapping, an edge operator connecting sites beyond nearest neighbors picks up a $Z$-chain along a connecting path, with $X$ / $Y$ acting on adjacent auxiliary qubit. Edge operators whose $Z$-chains overlap cannot be measured simultaneously with lattice surgery. Diagonalizing the golden plaquettes therefore cannot be fully parallelized. Instead we split them in 3 groups, which are parallelized sequentially:

\begin{enumerate}
    \item \textbf{Bulk plaquettes and the four-corner plaquette.} All golden bulk plaquettes, together with the single plaquette whose four vertices are the four corners of the $L\times L$ square. The edge operators connecting the corners can be chosen such that the $Z$-chains do not overlap with the bulk plaquettes.
    \item \textbf{Vertical-boundary plaquettes.} The edge operators include $Z$-chains which run mutually in parallel and span the full width of the lattice. 
    \item \textbf{Horizontal-boundary plaquettes.} The edge operators include $Z$-chains which run mutually in parallel and span  the full height. 
\end{enumerate}
The logical timestep cost to implement the diagonalizing circuits increases by a factor of $3$ relative to the pink case. The $\ket{T}$-state cost is identical to the pink case. 
The full fault-tolerant cost of the golden diagonalization gadget, i.e. $F_{3,1} F_{2,4} C \cdots C^\dagger F_{2,4}^\dagger F_{3,1}^\dagger$ on every golden plaquette, is derived in \Cref{sec:active_cube_cost_of_golden_diagonalization_circuit} and summarized in \Cref{tab:cost_golden_hopping}.
 \begin{table}
  \centering                                          \begin{tblr}{   
      hline{1,2}={1.5pt},                         row{1}={font=\bf},                          hline{3-6}={0.8pt},
      cells={c, m}, }           
  Resource type & Quantity \\
  Transversal CNOTs & $0$ \\                      
  $\ket{T}$ states consumed & $4L^2$ \\
  Logical timesteps & $ 3\times 9 \times 2 = 54$ \\                             
  Active surgery cubes & $4\left[\frac{210.5\, L^2}{4} + 0.75\, L^2(w_{\mathrm{MSF}}-1) + (104 + 6w_{\mathrm{MSF}})\!\left(\frac{L}{2}-1\right)^{\!2} + (85 + 12w_{\mathrm{MSF}})\!\left(\frac{L}{2}-1\right) +                 
  6w_{\mathrm{MSF}}\right]$ \\   
  \end{tblr}                                           \caption{Fault-tolerant cost of the full golden diagonalization gadget. The $\ket{T}$-state count equals that of the pink case. The factor $4$ accounts for two layers and the forward and inverse circuits. The active surgery block
   count is the base cost plus the four corridor and shuttling overheads enumerated in \Cref{sec:active_cube_cost_of_golden_diagonalization_circuit}.}     
  \label{tab:cost_golden_hopping} 
  \end{table}

Once the diagonalizing circuits have been applied to all golden plaquettes, the synthesized $Z$-rotations are executed in parallel, as in the pink case.

\section{Resource Estimation}
\label{sec:resource_estimation}
We now estimate the fault-tolerant cost of an explicit instance of the
Fermi--Hubbard time-evolution problem. We target an $8 \times 8$ lattice
with periodic boundary conditions at coupling $U/t = 8$, evolved for a
total time $T_{\mathrm{sim}} = 10L = 80$ (in units of inverse hopping
$1/t$), to a target accuracy of $1\%$ in diamond norm. $U/t = 8$ places
us far from the free-fermion limit
$U/t \to 0$ where the dynamics are classically tractable. The system size already exceeds recent
NISQ demonstrations of 2D Fermi--Hubbard
dynamics~\cite{alam2025fermionicdynamicstrappedionquantum, alam2025programmabledigitalquantumsimulation, granet2026superconductingpairingcorrelationstrappedion}, which are moreover restricted to evolution times of at most $\sim 2$,
far below our target. At the same time, $8 \times 8$ dynamics with periodic boundary
conditions and strong coupling lies beyond the reach of classical
methods~\cite{Campbell_2021, zheng2017stripe, oh2026classicalsimulationfreefermionicdynamics}.
Tensor networks in particular suffer from entanglement growth in 2D
periodic geometries, with discrepancies already reported at smaller
sizes~\cite{alam2025fermionicdynamicstrappedionquantum, thompson2025nonzeronoiseextrapolationaccurately}. Compared to the $30 \times 30$ lattices of Ref.~\cite{mcardle2025fastcuriousacceleratefaulttolerant}, an $8 \times 8$ target represents a more near-term benchmark.

Note also, our analysis uses strict worst-case error bounds throughout, so the reported resource estimates are conservative upper bounds. For the time evolution of specific physical observables, recent studies suggest that substantially fewer Trotter steps could suffice in practice~\cite{alam2025fermionicdynamicstrappedionquantum}.

\Cref{fig:resource_pipeline} provides an overview of the 
resource-estimation pipeline. The remainder of this section 
details each step.
\input{resource_pipeline_embedded}

\subsubsection{Error budgeting}
\label{sec:error_budget}
We set the total simulation accuracy target at $1\%$ in diamond norm,
\begin{equation}
\left\|\mathcal{U} - \tilde{\mathcal{E}}\right\|_\diamond \leq 1\%,
\end{equation}
where $\mathcal{U}(\cdot) = U(\cdot)U^\dagger$ is the ideal time-evolution channel and $\tilde{\mathcal{E}}$ is the channel implemented by the noisy fault-tolerant device. Working in diamond norm allows algorithmic approximation errors and hardware noise to be combined in a single budget:
\begin{equation}
\label{eq:error_budget}
\left\|\mathcal{U} - \tilde{\mathcal{E}}\right\|_\diamond
\leq 2\,\epsilon_{\mathrm{alg}}
+ \epsilon_{\mathrm{rot}}
+ \epsilon_{\mathrm{log}}
+ \epsilon_{\mathrm{msf}}.
\end{equation}
The Trotter error $\epsilon_{\mathrm{alg}}$ bounds the operator-norm distance between unitaries. The factor of $2$ converts this to a diamond-norm bound on the corresponding channel~(\cite{Haah_2023}, Proposition 1.6). The rotation synthesis error $\epsilon_{\mathrm{rot}}$, the logical noise $\epsilon_{\mathrm{log}}$, and the magic state infidelity $\epsilon_{\mathrm{msf}}$ are already diamond-norm bounds and enter directly.

We split the total budget equally between algorithmic and hardware contributions:
\begin{equation}
2\,\epsilon_{\mathrm{alg}} + \epsilon_{\mathrm{rot}} = 0.50\%, \qquad
\epsilon_{\mathrm{log}} + \epsilon_{\mathrm{msf}} = 0.50\%.
\end{equation}
Following~\cite{Campbell_2021}, the rotation synthesis budget is set to $1\%$ of the Trotter budget, $\epsilon_{\mathrm{rot}} = 0.01\,\epsilon_{\mathrm{alg}}$.
The hardware budget is split equally between logical noise and magic state infidelity, $\epsilon_{\mathrm{log}} = \epsilon_{\mathrm{msf}} = 0.25\%$. The logical noise budget covers logical errors during both lattice surgery operations and transversal $\CNOT$ operations. \Cref{tab:error_budget} summarizes the budget allocation.

\begin{table}[h]
\centering
\begin{tblr}{
    hline{1,2}={1.5pt},
    row{1}={font=\bf},
    hline{3-6}={0.8pt},
    cells={c, m},
    }
Error source & Budget & Determines \\
Trotter (algorithmic)   & $\epsilon_{\mathrm{alg}} \approx 0.249\%$   & $\#$ Trotter steps $r$ \\
Rotation synthesis      & $\epsilon_{\mathrm{rot}} = 0.01\,\epsilon_{\mathrm{alg}}$ &  $\#T$-gates per rotation \\
Logical noise (cubes + transversal CNOTs) & $\epsilon_{\mathrm{log}} = 0.25\%$ & code distance $d$ \\
Magic state infidelity  & $\epsilon_{\mathrm{msf}} = 0.25\%$   & MSF fidelity target \\
\end{tblr}
\caption{Error budget allocation. The total diamond-norm budget of $1\%$ is split equally between algorithmic errors ($2\,\epsilon_{\mathrm{alg}} + \epsilon_{\mathrm{rot}} = 0.50\%$) and hardware errors ($\epsilon_{\mathrm{log}} + \epsilon_{\mathrm{msf}} = 0.50\%$).}
\label{tab:error_budget}
\end{table}

\subsubsection{Trotter step count}
\label{sec:trotter_steps}

The second-order Trotter error for the PLAQ decomposition satisfies~\cite{Campbell_2021, mcardle2025fastcuriousacceleratefaulttolerant}
\begin{equation}
\|U(T_{\mathrm{sim}}) - S_2(T_{\mathrm{sim}}/r)^r\| \leq W_{\mathrm{PLAQ}}\frac{T_{\mathrm{sim}}^3}{r^2},
\end{equation}
where $T_{\mathrm{sim}}$ is the total simulation time and $W_{\mathrm{PLAQ}} \leq \kappa L^2 t^3$ with
\begin{equation}
\kappa = \frac{1}{24}\left[\frac{3}{2}\left(\frac{U}{t}\right)^2 + 2\left(\frac{U}{t}\right)(2\sqrt{5}+16)+10\right].
\end{equation}
which gives the following lower bound for the number of Trotter steps~\cite{mcardle2025fastcuriousacceleratefaulttolerant}:
\begin{equation}
\label{eq:trotter_steps}
r \geq \frac{\sqrt{\kappa}\,L\,(T_{\mathrm{sim}}t)^{3/2}}{\sqrt{\epsilon_{\mathrm{alg}}}}.
\end{equation}
We set $U/t = 8$ and choose the simulation time $T_{\mathrm{sim}} = 10L$ (in units of $1/t$, i.e.\ $T_{\mathrm{sim}}t = 10L$). With $\epsilon_{\mathrm{alg}} = 0.50\%/2.01$, Eq.~\eqref{eq:trotter_steps} becomes
\begin{equation}
r \geq \frac{\sqrt{\kappa}\,(10)^{3/2}}{\sqrt{\epsilon_{\mathrm{alg}}}}\,L^{5/2} \approx 2695\,L^{5/2}.
\end{equation}
For $L=8$ we get $r\approx4.88 \times 10^5$.

\subsubsection{Synthesis cost per rotation}
\label{sec:synthesis_cost}
We choose the mixed-fallback rotation synthesis method described in \cite{Kliuchnikov2023shorterquantum} and also used in \cite{mcardle2025fastcuriousacceleratefaulttolerant}. The expected $T$-count is given as $n_T = 0.57\log_2(1/\epsilon_\mathrm{synth}) + 8.83$. Here $\epsilon_\mathrm{synth} = \epsilon_\mathrm{rot}/N_\mathrm{rot}$ is the synthesis error per rotation, and $N_\mathrm{rot} = 4 \times L^2 \times r \approx 1.25 \times 10^8$, so $\epsilon_\mathrm{synth} \approx 2\times 10^{-13}$. The expected $T$-count then becomes $n_T = 33$. The success probability for the fallback is at least $0.99$ (see \Ccite[Figure 3]{Kliuchnikov2023shorterquantum}), so we set $p_\mathrm{succ} = 0.99$. The fallback is synthesized according to the mixed-diagonal method described in \cite{Kliuchnikov2023shorterquantum}, with expected $T$-count $n_T^\mathrm{fb} = 1.54\log_2(1/\epsilon_\mathrm{synth}) + 6.85 = 72$. From the relation $n_T = n_T^\mathrm{succ} + p_\mathrm{fail}\,n_T^\mathrm{fb}$, we obtain $n_T^\mathrm{succ} = 33 - 0.01 \times 72 \approx 32$. We use the ratio between reaction time and logical cycle: $\tau_r/t_l = 33/102 \approx 0.324$ , where the logical cycle time depends on the code distance determined in \Cref{sec:trotter_step_cost}. Since the code distance in turn depends on the synthesis cost, we solve this self-referential dependence iteratively and verify that the final values are mutually consistent.
Note that $p_\mathrm{succ}^\mathrm{all} = 0.99^{64}\approx0.526$, where the exponent comes from the $8\times 8$ grid. The conditional failure probability, given that not all $L^2$ syntheses succeeded, is $\tilde{p}_\mathrm{fail} = p_\mathrm{fail}/(1-p_\mathrm{succ}^\mathrm{all}) = 0.01/0.474 \approx 0.0211$, and $\tilde{p}_\mathrm{succ} = 1 - \tilde{p}_\mathrm{fail} \approx 0.979$. Combining all of these values, based on \Cref{tab:fault_tolerant_cost_synthesis}, we obtain an active cube cost of $434$ per rotation. $n_T=33$  $\ket{T}$ states are consumed per rotation.

\subsubsection{Total active cube cost, code distance and physical qubit count}
\label{sec:trotter_step_cost}

We have $4\times L^2$ $Z$-rotations per Trotter step, which corresponds to $434 \times 4L^2 = 1.11 \times 10^5$ active cubes. The interaction subevolution adds $2\times L^2$ transversal $\mathsf{CNOT}$ gates. The two pink time evolutions add $205 \times L^2 = 1.31 \times 10^4$ active cubes. With $w_\mathrm{msf}=2$, the golden time evolution adds another $1.92\times 10^4$ active cubes. This results in $1.43 \times 10^5$ active cubes per Trotter step and a total cube count of $N_L \approx 6.99\times 10^{10}$ active cubes.
We use $2L^2$ transversal $\mathsf{CNOT}$s per layer which is more than three orders of magnitudes below the number of cubes. Therefore we can neglect their contribution.
This puts the target logical error rate per cube at $p_\mathrm{l} = \epsilon_\mathrm{log} / N_L \approx 3.58 \times 10^{-14}$. Meeting this target requires logical patches of width $w=30$ and height $h=51$, operated with $102$ syndrome extraction rounds per logical cycle (\Cref{tab:resource_estimates}). Assuming 2-patch-wide aisles for magic state factories, the floorplan accommodates a total of $882$ logical patches, $336$ of which are allocated for magic state production. This yields a physical qubit count of $882 \times 30 \times 51 \approx 1.35\times 10^6$.

\subsubsection{\texorpdfstring{Total number of $\ket{T}$ states consumed and magic state factory sizing}{}}
\label{sec:t_state_count}
We have $4L^2$ $Z$-rotations per Trotter step, each consuming $33$ $\ket{T}$ states, which adds $33\times 4L^2 = 8448$ $\ket{T}$ states. Each of the three hopping sub-evolutions (two pink and one golden) requires $4L^2$ $\ket{T}$ states for the diagonalization circuits (\Cref{tab:cost_diagonalizing_circuit}), contributing an additional $3 \times 4L^2 = 768$ $\ket{T}$ states. This brings a total of $9216$ $\ket{T}$ states consumed per Trotter step, yielding a total $\ket{T}$-state consumption over the whole algorithm of $N_T\approx4.50\times10^9$. Assuming the error for each $\ket{T}$ state adds up linearly, we must produce them at an error rate of $p_\mathrm{msf} < \epsilon_\mathrm{msf} / N_T \approx 5.56 \times 10^{-13}$, where $\epsilon_\mathrm{msf} = 0.25\%$ comes from our error budget. When $L^2$ rotations are implemented in parallel, the consumption peaks at $L^2$ $\ket{T}$ states every $102 + 33 = 135$ measurement cycles (logical timestep plus reaction time), or $L^2/135 \approx 0.47$ magic states per cycle on average. One factory from \Cref{tab:msf_protocols} at our target fidelity has a rate of $\frac{1}{82}$ magic states per cycle, which requires $39$ of these protocols to run in parallel to meet the peak consumption rate. This amounts to roughly $1.59 \times 10^5$ qubits involved in magic state production. The 2-patch-wide corridors allocate $336$ logical patches for magic state production, which, considering a logical patch size of $30\times 51$, corresponds to $5.14\times 10^5$ qubits, roughly $3\times$ the required amount.

\subsubsection{Time}
\label{sec:time}
The number of logical timesteps needed to implement one Trotter step is $4t_\mathrm{synth} + 90$, where $t_\mathrm{synth}$ is the number of logical timesteps required for gate synthesis and $90$ accounts for the diagonalization circuits of the two pink and the golden sub-evolutions. The transversal $\mathsf{CNOT}$ gates in the interaction sub-evolution add no logical timesteps, as they appear just twice and do not take up full surgery cubes. This compares to $6t_\mathrm{synth} + 354$ from \cite{mcardle2025fastcuriousacceleratefaulttolerant}.

For the mixed-fallback rotation synthesis with $n_T^\mathrm{succ} = 32$, $n_T^\mathrm{fb} = 72$, $\tau_r/t_l = 33/102 \approx 0.324$, and $p_\mathrm{succ}^\mathrm{all} = 0.99^{64} \approx 0.526$, the logical timestep formula from \Cref{tab:fault_tolerant_cost_synthesis} gives $t_\mathrm{synth} \approx 96$. This accounts for the reaction time between $\frac{\pi}{8}$-rotations, the Clifford corrections in the fallback scheme, and the synchronization overhead. Whenever one of the $L^2$ simultaneous $Z$-rotations requires the fallback, all other qubits must idle until it completes. Had we instead chosen the mixed-diagonal direct synthesis, which does not suffer from this synchronization penalty but requires a higher $T$-count ($n_T = 72$ vs.\ an expected $n_T = 33$), the direct synthesis formula gives $t_\mathrm{synth} \approx 98$. The $\ket{T}$ states produced by our chosen factory design are of sufficiently high fidelity to accommodate either strategy.

The total cost per Trotter step is therefore $4 \times 96 + 90 = 474$ logical timesteps, compared to $6 \times 96 + 354 = 930$ for the single-plane compilation of \cite{mcardle2025fastcuriousacceleratefaulttolerant}, a speedup of approximately $1.96\times$ in logical timesteps. Assuming $305$\,ns per syndrome extraction round and $102$ rounds per logical cycle, the logical cycle time is $t_l = 31.1$\,\textmu{}s. The total runtime over $r \approx 4.88 \times 10^5$ Trotter steps is then approximately $2$\,hours, compared to $3$\,hours\,$55$\,minutes for the single-plane compilation of \cite{mcardle2025fastcuriousacceleratefaulttolerant}.

Coincidentally, $t_\mathrm{synth} \approx 96$ for the mixed-fallback and $t_\mathrm{synth} \approx 98$ for the mixed-diagonal direct synthesis place us near the tipping point between the two strategies. Already at $L=8$ the all-success probability is merely half, and it vanishes exponentially with $L^2$. At larger $L$ the failure branch dominates the expected runtime, and the mixed-diagonal direct synthesis would become the faster option despite its higher per-rotation $T$-count.

\section{Conclusion and outlook}
\label{sec:conclusion}

In this work, we carried out a detailed resource estimation for simulating the Fermi–Hubbard model on a biplanar version of the SPOQC architecture. Using the plaquette Trotterization algorithm and a compact fermion-to-qubit mapping, we developed an explicit compilation of the time evolution operator into a set of fault-tolerant logical instructions for the planar honeycomb code together with transversal CNOTS between the two layers. We then derived an estimation of the required resources, including logical timesteps, magic state consumption, runtime and physical qubit counts under a realistic noise model for the SPOQC platform.

A central feature of our approach is the architectural co-design between the algorithm and the hardware. We map the two spin sectors of the Fermi–Hubbard model onto the two planes of the SPOQC architecture, and we exploit the inter-plane connectivity to implement on-site interactions efficiently via transversal operations. While resource estimation work often focuses on minimizing qubit counts, we maximized parallelism in order to reduce logical depth and thus optimize runtime. We estimate that, for a lattice size ($8 \times 8$) and a coupling regime ($U/t=8$) that escape classical simulability, for a target accuracy of 1\%, and for physical noise parameters $\varepsilon=0.9\%$, $D=0.085\%$, $t_c/T_2=0.01\%$ and $p_s=0.005\%$, the Fermi-Hubbard dynamics could be simulated on the biplanar SPOQC architecture in approximately 2h using $1.35 \times 10^6$ physical qubits. In practice the requirements may be lower without further improvements, as, for example, substantially fewer Trotter steps may suffice.

More broadly, this work illustrates the importance of integrated resource estimation that accounts for the full stack, from physical noise models and error correction to compilation and algorithm design. Such end-to-end analyses are essential to identify realistic application targets and guide the development of both quantum hardware and software. In particular, our work highlights that mixed-fallback protocols for rotation synthesis are suboptimal with respect to runtime in large-scale, highly parallelized compilation. 

Several directions remain for future work. 
On the algorithmic side, deriving compilation and resource estimates for the space-optimized version of the Plaquette Trotterization, and comparing it with the time-optimized version analyzed here, could provide valuable insight into hardware-software co-design trade-offs. Extending the analysis to quantum signal processing, instead of Trotterization, would allow us to analyze the performance of our scheme in the regime where higher algorithmic accuracy is required. 
On the error-correction side, an in-depth analysis of magic state cultivation for the SPOQC implementation of the planar honeycomb code is needed to refine our estimates. 
Finally, enriching the SPOQC error model with more detailed noise processes, in particular for single-qubit gate errors, would enable a more advanced end-to-end performance analysis. 

The aim of these directions for future work is to further reduce the resource requirements and improve the accuracy of the estimate.
Such reductions are not unexpected, as over the past 15 years the estimated resource requirements for factoring 2048-bit RSA integers with a superconducting square-grid architecture have dropped from $10^9$ to $10^6$ qubits \cite{gidney2025factor}. 
Achieving reductions of this magnitude will require improvements across the full software stack, including QEC, compilation, and algorithms.

\section*{Author contributions and acknowledgments}

\label{sec:contributions}
Théo Dessertaine worked out the properties of the physical gate set.
Peter-Jan Derks simulated the logical instructions with support from Théo.
Boris Bourdoncle proposed the biplanar architecture and supervised the project.
Johannes Frank designed the layout for the compact fermion to qubit mapping and compiled the algorithm to the logical instructions.
All authors contributed to discussions and writing the manuscript.

We thank Stephen Wein for insightful discussions on the error model, Grégoire de Gliniasty for contributions to the layout proposition, Paul Faehrmann for valuable discussions on the fermion-to-qubit mapping and PLAQ algorithm, Sam McArdle for helpful clarifications on the fallback synthesis method, and Henrik Dreyer for feedback on the Trotter step count. We also thank everyone in Quandela's FTQC team (Pierre Colonna d'Istria, Grégoire de Gliniasty, Aurélie Denys, Katia Hakem, Ewan Murphy) for valuable discussions. This work has been co-funded by the Horizon-CL4 program under the grant agreement 101135288 for the EPIQUE project and by the TUF-TOPIQC program within the French National Quantum Strategy (France 2030). 

\newpage
\appendix

\input{appendix_physical_noisy_gate_set}

\FloatBarrier

\section{Details of the numerical simulations}
\label{appendix:numerical_simulations}

This appendix gives additional details on numerical simulations. \Cref{ap:circuit_presampling} is dedicated to performing logical error rate simulations in the presence of heralded noise. \Cref{ap:simulations_details} provides details on how to simulate memory and stability experiments for the honeycomb Floquet code.\footnote{All simulations were run on an Apple M4 Pro chip with 10 CPU cores.}

\subsection{Circuit presampling}
\label{ap:circuit_presampling}

As explained in \Cref{ap:physical_gate_set}, most of the error channels in the SPOQC error model are heralded. 
Therefore, for most of the basic operation in our architecture, the noise channel that follows will vary depending on some classical signals. 
We use Stim \cite{gidney2021stim} for simulating our quantum circuits, which models noise as fixed channels per instruction. 
To circumvent this, we use a similar simulation pipeline described in \Ccite[Appendix B.2 and Figure 8]{dessertaine2026enhancedfaulttolerancephotonicquantum}. 

In a nutshell, the idea is to perform an \textit{ensemble average} over all possible heralding outcomes. For our simulation, heralded gates are all independent. Therefore, one can independently draw the heralded outcome of each gate according to the probabilities presented in \Cref{tab:physical_gate_set}. These outcomes are recorded in a vector $\vec{h}$, and we then construct a circuit where each heralded gate $i$ is followed by the noise channels corresponding to the drawn outcome $h_i$. We can generate and decode $N_s$ samples from this circuit and decode them to get a logical error rate $\hat{p}(\vec{h})$ conditional on the $\vec{h}$. Repeating this process $N_h$ times, we can average over all conditional logical error rate $\hat{p}(\vec{h}_j)$ computed. 

\subsection{\texorpdfstring{Simulations to generate \Cref{fig:dimension_ratio_power}}{}}
\label{ap:simulations_details}

\begin{figure*}
\centering
\includegraphics[width=\textwidth]{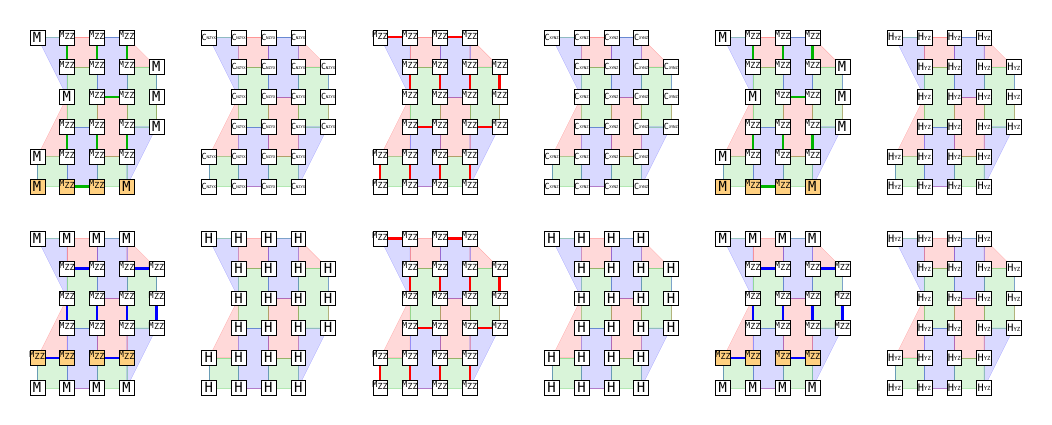}
\caption{Bulk rounds of a horizontal memory experiment on a $6 \times 4$ (height $\times$ width) HCC patch. Each panel shows one timestep of the circuit. Colored edges between $\MZZ$ boxes indicate the lattice edge being measured, with red, green, and blue corresponding to the three edge types of the honeycomb color code. The measurements included in the spacelike horizontal logical observable arehighlighted in orange. An interactive version of the circuit is available in \crumbleHorizontalMemory{}.}
\label{fig:circuit_horizontal_memory}
\end{figure*}

\begin{figure*}
\centering
\includegraphics[width=\textwidth]{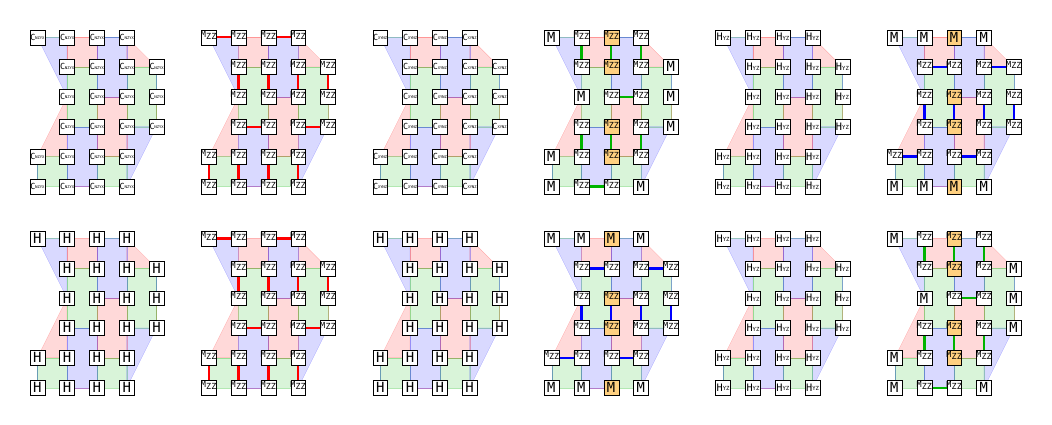}
\caption{Bulk rounds of a vertical memory experiment on a $6 \times 4$ HCC patch. The measurements included in the spacelike vertical logical observable are highlighted in orange. An interactive version of the circuit is available in \crumbleVerticalMemory{}.}
\label{fig:circuit_vertical_memory}
\end{figure*}

\begin{figure*}
\centering
\includegraphics[width=\textwidth]{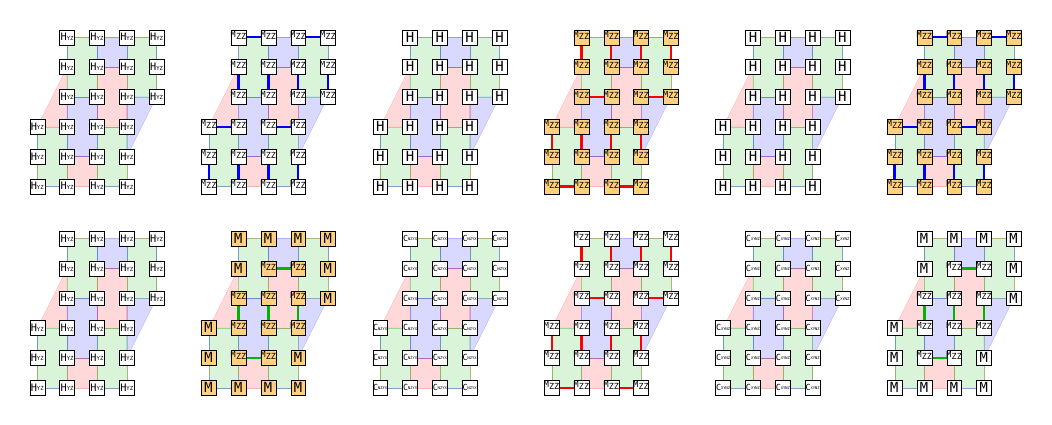}
\caption{Bulk rounds of a horizontal stability experiment on a $6 \times 4$ HCC patch. The measurements included in the  timelike horizontal logical observable are highlighted in orange. An interactive version of the circuit is available in \crumbleHorizontalStability{}.}
\label{fig:circuit_horizontal_stability}
\end{figure*}

\begin{figure*}
\centering
\includegraphics[width=\textwidth]{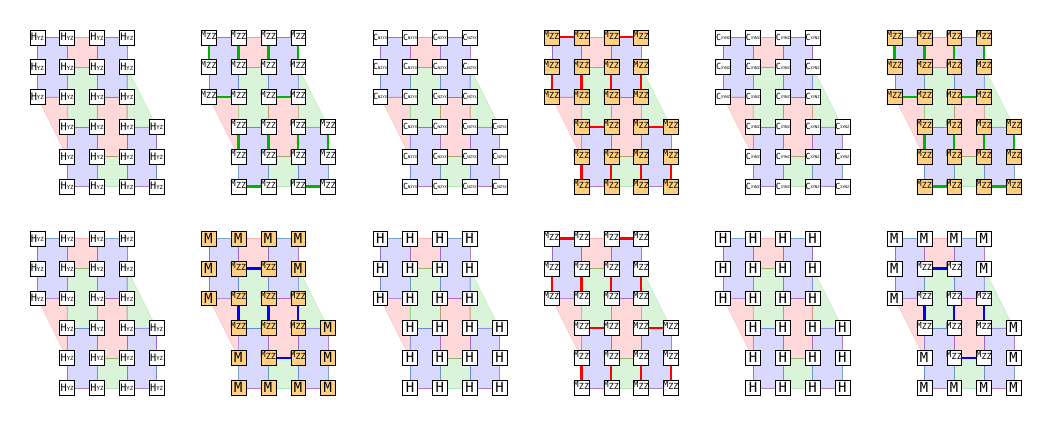}
\caption{Bulk rounds of a vertical stability experiment on a $6 \times 4$ HCC patch. The measurements included in the timelike vertical logical observable are highlighted in orange. An interactive version of the circuit is available in \crumbleVerticalStability{}.}
\label{fig:circuit_vertical_stability}
\end{figure*}

In \Cref{fig:dimension_ratio_power}, the horizontal spacelike, vertical spacelike, and timelike logical error rates are shown for width $w=10$ and $w=12$ for various heights $h$ and number of syndrome extraction rounds $r$.
For a given $(w, h, r)$ we calculate the horizontal spacelike error rates using a horizontal memory experiment.
Six rounds of syndrome extraction of a horizontal memory experiment are shown in \Cref{fig:circuit_horizontal_memory}.
To calculate the vertical spacelike error rates, we use a vertical memory experiment, which is shown in \Cref{fig:circuit_vertical_memory}.
The memory experiments are adaptions of the circuits from \cite{gidney2022benchmarking} to the gateset of the SPOQC architecture given in \Cref{tab:physical_gate_set}.
To calculate the timelike error rates, we use a horizontal stability experiment, which measures the probability of a timelike logical error using the horizontal logical observable.
This circuit is shown in \Cref{fig:circuit_horizontal_stability}.
For completeness, the vertical stability experiment is shown in \Cref{fig:circuit_vertical_stability}.

We simulate the circuits using Stim~\cite{gidney2021stim} and decode using MWPM via the PyMatching package~\cite{higgott2023sparse}.
The number of circuits, shots, and logical errors for each data point in \Cref{fig:dimension_ratio_power} are summarized in \Cref{tab:dimension_ratio_simulation_stats}.

\input{crumble_url_horizontal_memory}

\input{crumble_url_vertical_memory}

\input{crumble_url_horizontal_stability}

\input{crumble_url_vertical_stability}

\begin{table*}[t]
\centering
\begin{tabular}{cc c rrr rrr r @{\hskip 1.5em} c rrr rrr r}
\toprule
 & & & \multicolumn{7}{c}{$w = 10$} & & \multicolumn{7}{c}{$w = 12$} \\
\cmidrule(lr){3-10} \cmidrule(lr){11-18}
 & & & \multicolumn{3}{c}{Shots ($\times 10^6$)} & \multicolumn{3}{c}{Errors} & & & \multicolumn{3}{c}{Shots ($\times 10^6$)} & \multicolumn{3}{c}{Errors} & \\
\cmidrule(lr){4-6} \cmidrule(lr){7-9} \cmidrule(lr){12-14} \cmidrule(lr){15-17}
$h$ & $r$ & Circ. & H & V & S & H & V & S & Time & Circ. & H & V & S & H & V & S & Time \\
\midrule
9 & 18 & 150 & 0.7 & 6.0 & 0.9 & 6\,937 & 124 & 5\,958 & 2\,min & 210 & 1.1 & 11.1 & 0.8 & 12\,630 & 16 & 6\,999 & 4\,min \\
9 & 24 & 50 & 0.6 & 5.0 & 4.7 & 5\,256 & 79 & 4\,869 & 2\,min & 150 & 1.0 & 10.5 & 4.8 & 11\,610 & 19 & 5\,734 & 5\,min \\
9 & 30 & 50 & 0.5 & 5.0 & 5.5 & 5\,313 & 109 & 938 & 3\,min & 150 & 1.0 & 10.5 & 6.1 & 11\,592 & 14 & 1\,243 & 5\,min \\
9 & 36 & 50 & 0.6 & 5.0 & 5.5 & 5\,300 & 207 & 131 & 3\,min & 150 & 1.0 & 10.5 & 6.1 & 11\,651 & 19 & 154 & 6\,min \\
9 & 42 & 50 & 0.6 & 5.0 & 5.5 & 5\,256 & 88 & 8 & 3\,min & 150 & 1.1 & 10.5 & 6.1 & 11\,563 & 17 & 43 & 7\,min \\
\midrule
12 & 18 & 50 & 3.6 & 5.0 & 0.6 & 4\,908 & 157 & 7\,190 & 4\,min & 150 & 5.6 & 10.5 & 0.5 & 10\,645 & 47 & 7\,690 & 9\,min \\
12 & 24 & 200 & 4.3 & 6.5 & 4.4 & 6\,460 & 277 & 6\,478 & 6\,min & 210 & 6.0 & 11.1 & 3.9 & 11\,278 & 49 & 6\,198 & 11\,min \\
12 & 30 & 50 & 3.2 & 5.0 & 6.5 & 4\,982 & 294 & 1\,677 & 6\,min & 150 & 5.4 & 10.5 & 6.1 & 10\,665 & 48 & 1\,864 & 12\,min \\
12 & 36 & 50 & 3.4 & 5.0 & 6.5 & 4\,973 & 277 & 199 & 6\,min & 150 & 5.4 & 10.5 & 6.1 & 10\,645 & 40 & 258 & 12\,min \\
12 & 42 & 50 & 3.3 & 5.0 & 6.5 & 4\,909 & 203 & 44 & 7\,min & 150 & 5.7 & 10.5 & 6.1 & 10\,681 & 41 & 42 & 13\,min \\
\midrule
15 & 18 & 50 & 5.0 & 5.0 & 0.5 & 1\,286 & 419 & 7\,431 & 7\,min & 150 & 10.5 & 10.5 & 0.5 & 3\,775 & 68 & 7\,988 & 18\,min \\
15 & 24 & 50 & 5.0 & 5.0 & 3.6 & 1\,405 & 394 & 6\,701 & 8\,min & 150 & 10.4 & 10.5 & 2.8 & 3\,437 & 84 & 6\,300 & 20\,min \\
15 & 30 & 200 & 6.5 & 6.5 & 6.1 & 1\,535 & 524 & 2\,112 & 12\,min & 210 & 11.1 & 11.1 & 6.1 & 3\,541 & 89 & 2\,386 & 23\,min \\
15 & 36 & 50 & 5.0 & 5.0 & 6.1 & 1\,292 & 411 & 311 & 11\,min & 150 & 10.4 & 10.5 & 6.1 & 3\,976 & 74 & 396 & 23\,min \\
15 & 42 & 50 & 5.0 & 5.0 & 6.1 & 1\,344 & 440 & 62 & 11\,min & 150 & 10.4 & 10.5 & 6.1 & 3\,690 & 96 & 54 & 24\,min \\
\midrule
18 & 18 & 50 & 5.0 & 5.0 & 0.4 & 175 & 693 & 7\,165 & 11\,min & 150 & 10.5 & 10.5 & 0.4 & 548 & 124 & 8\,391 & 28\,min \\
18 & 24 & 50 & 5.0 & 5.0 & 2.7 & 233 & 575 & 6\,278 & 12\,min & 150 & 10.5 & 10.5 & 2.2 & 630 & 115 & 6\,391 & 29\,min \\
18 & 30 & 50 & 5.0 & 5.0 & 6.0 & 219 & 645 & 2\,378 & 14\,min & 150 & 10.5 & 10.5 & 6.0 & 539 & 150 & 2\,651 & 33\,min \\
18 & 36 & 200 & 6.5 & 6.5 & 6.0 & 276 & 725 & 518 & 18\,min & 210 & 11.1 & 11.1 & 6.1 & 656 & 136 & 490 & 35\,min \\
18 & 42 & 50 & 5.0 & 5.0 & 6.0 & 224 & 575 & 79 & 15\,min & 150 & 10.5 & 10.5 & 6.1 & 560 & 132 & 82 & 35\,min \\
\midrule
21 & 18 & 50 & 5.0 & 4.9 & 0.3 & 52 & 1\,067 & 7\,096 & 16\,min & 150 & 10.5 & 10.5 & 0.4 & 92 & 233 & 8\,802 & 46\,min \\
21 & 24 & 50 & 5.0 & 5.0 & 2.3 & 31 & 876 & 6\,262 & 17\,min & 150 & 10.5 & 10.5 & 1.9 & 95 & 183 & 6\,416 & 48\,min \\
21 & 30 & 50 & 5.0 & 5.0 & 6.0 & 32 & 897 & 2\,807 & 20\,min & 150 & 10.5 & 10.5 & 5.9 & 103 & 204 & 3\,355 & 52\,min \\
21 & 36 & 50 & 5.0 & 5.0 & 6.0 & 36 & 881 & 451 & 21\,min & 150 & 10.5 & 10.5 & 6.1 & 103 & 209 & 616 & 52\,min \\
21 & 42 & 200 & 6.5 & 6.5 & 6.0 & 38 & 1\,158 & 59 & 27\,min & 210 & 11.1 & 11.1 & 6.1 & 104 & 207 & 81 & 54\,min \\
\midrule
24 & 18 & 50 & 5.0 & 5.0 & 0.2 & 9 & 1\,098 & 5\,705 & 23\,min & 50 & 5.0 & 5.0 & 0.2 & 14 & 159 & 5\,813 & 30\,min \\
24 & 24 & 50 & 5.0 & 5.0 & 1.7 & 2 & 1\,199 & 5\,114 & 24\,min & 50 & 5.0 & 5.0 & 1.3 & 8 & 122 & 5\,090 & 31\,min \\
24 & 30 & 50 & 5.0 & 5.0 & 5.0 & 3 & 1\,293 & 2\,818 & 26\,min & 50 & 5.0 & 5.0 & 4.8 & 11 & 182 & 3\,406 & 34\,min \\
24 & 36 & 50 & 5.0 & 5.0 & 5.0 & 6 & 1\,340 & 515 & 26\,min & 50 & 5.0 & 5.0 & 5.0 & 9 & 173 & 529 & 35\,min \\
24 & 42 & 50 & 5.0 & 5.0 & 5.0 & 6 & 1\,406 & 69 & 27\,min & 50 & 5.0 & 5.0 & 5.0 & 5 & 128 & 98 & 36\,min \\
\bottomrule
\end{tabular}
\caption{Summary of Monte Carlo sampling runs for the dimension ratio experiment (\Cref{fig:dimension_ratio_power}) at $p = 0.01$ using PyMatching. For each patch height~$h$ and number of syndrome rounds~$r$ we report the number of erasure-pattern circuits, total decoder shots and detected logical errors for horizontal memory~(H), vertical memory~(V), and horizontal stability~(S) experiments, and cumulative CPU time.}
\label{tab:dimension_ratio_simulation_stats}
\end{table*}

\FloatBarrier

\subsection{\texorpdfstring{Simulations to generate \Cref{fig:block-size}}{}}

\begin{figure*}[t]
      \centering
      \includegraphics[width=\textwidth]{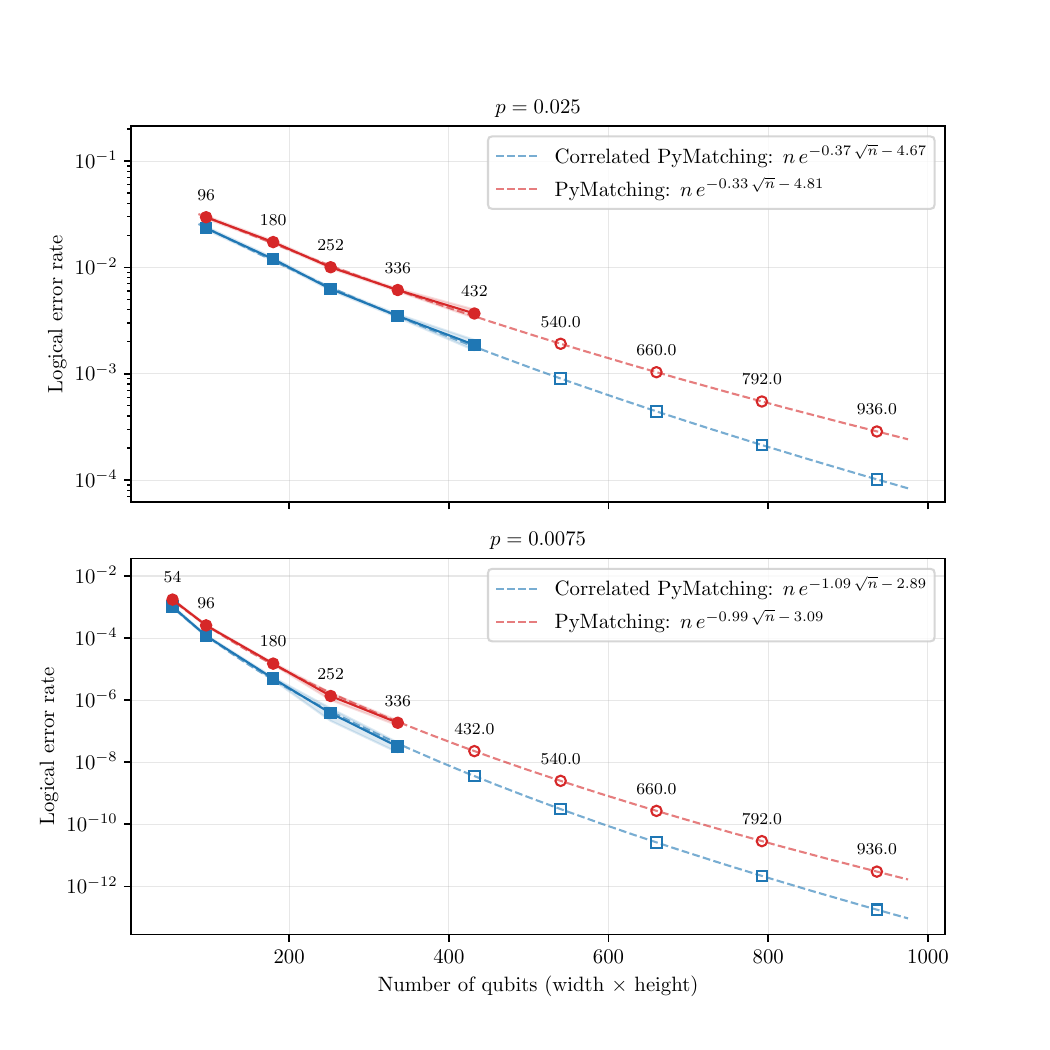}
      \caption{Combined logical error rate $E_{HV} = 1 - (1 - E_H)(1 - E_V)$ as a function of number of qubits (width $\times$ height) for physical error rates $p =
   0.025$ (top) and $p = 0.0075$ (bottom). Solid markers show simulated data with shaded 95\% confidence intervals from two-level Monte Carlo variance
  estimation. Dashed lines are exponential fits in the block size, with hollow markers indicating extrapolated values.}
      \label{fig:block_size_scaling}
\end{figure*}

To generate the data for \Cref{fig:block-size}, we use the same circuits as described in the previous section to perform horizontal and vertical memory experiments.
In addition to the results shown in \Cref{fig:dimension_ratio_power} for $p = 0.01$,
we also perform simulations at physical error rates $p = 0.025$ and $p = 0.0075$.
We decode using both MWPM and correlated MWPM~\cite{fowler2013optimal} via the PyMatching package~\cite{higgott2023sparse}.
For larger patch sizes, the difference in noise between erasure-pattern circuits becomes smaller, but many more shots per circuit are required to observe logical errors; as a result, fewer circuits are simulated at larger sizes.
The results are shown in \Cref{fig:block_size_scaling}.
The number of circuits, shots, and logical errors for each data point in \Cref{fig:block-size} and \Cref{fig:block_size_scaling} are shown in \Cref{tab:simulation_stats}.

\begin{table*}[t]
\centering
\begin{tabular}{cc c rr rr r @{\hskip 1.5em} c rr rr r}
\toprule
 & & & \multicolumn{5}{c}{PyMatching} & & \multicolumn{5}{c}{Correlated PyMatching} \\
\cmidrule(lr){4-8} \cmidrule(lr){10-14}
 & & & \multicolumn{2}{c}{Shots ($\times 10^6$)} & \multicolumn{2}{c}{Errors} & & & \multicolumn{2}{c}{Shots ($\times 10^6$)} & \multicolumn{2}{c}{Errors} & \\
\cmidrule(lr){4-5} \cmidrule(lr){6-7} \cmidrule(lr){11-12} \cmidrule(lr){13-14}
$w$ & $h$ & Circ. & H & V & H & V & Time & Circ. & H & V & H & V & Time \\
\midrule
\multicolumn{14}{c}{$p = 0.025$} \\
\midrule
9 & 15 & 1000 & 5.8 & 3.0 & 110\,474 & 123\,299 & 11\,min & 1000 & 7.5 & 3.9 & 103\,207 & 118\,059 & 28\,min \\
11 & 18 & 1000 & 7.5 & 5.1 & 104\,039 & 112\,463 & 31\,min & 1000 & 9.1 & 7.0 & 86\,613 & 106\,870 & 1\,h \\
13 & 21 & 1000 & 9.1 & 8.0 & 90\,835 & 102\,594 & 1\,h & 1000 & 9.8 & 9.6 & 60\,289 & 78\,038 & 3\,h \\
15 & 24 & 1000 & 9.9 & 9.7 & 67\,550 & 74\,118 & 2\,h & 1000 & 10.0 & 10.0 & 36\,908 & 42\,988 & 7\,h \\
\midrule
\multicolumn{14}{c}{$p = 0.01$} \\
\midrule
9 & 15 & 1050 & 13.2 & 10.7 & 2\,469 & 9\,874 & 14\,min & 1050 & 14.5 & 11.7 & 1\,262 & 5\,143 & 28\,min \\
11 & 18 & 1050 & 14.9 & 14.0 & 771 & 1\,736 & 33\,min & 1050 & 15.0 & 14.8 & 236 & 744 & 1\,h \\
13 & 21 & 1050 & 15.0 & 15.0 & 166 & 303 & 1\,h & 1050 & 15.0 & 15.0 & 44 & 76 & 2\,h \\
15 & 24 & 60 & 59.8 & 59.6 & 106 & 134 & 6\,h & 60 & 60.0 & 60.0 & 29 & 38 & 12\,h \\
17 & 27 & 65 & 58.8 & 68.9 & 18 & 24 & 10\,h & 65 & 54.0 & 54.1 & 2 & 3 & 22\,h \\
\midrule
\multicolumn{14}{c}{$p = 0.0075$} \\
\midrule
9 & 15 & 50 & 49.3 & 22.8 & 2\,259 & 4\,999 & 34\,min & 50 & 50.0 & 43.0 & 799 & 4\,258 & 1\,h \\
11 & 18 & 50 & 50.0 & 50.0 & 348 & 1\,222 & 2\,h & 50 & 50.0 & 50.0 & 116 & 435 & 3\,h \\
13 & 21 & 50 & 50.0 & 50.0 & 54 & 108 & 3\,h & 50 & 50.0 & 50.0 & 20 & 35 & 5\,h \\
\bottomrule
\end{tabular}
\caption{Summary of Monte Carlo sampling runs. For each patch size and physical error rate we report the number of erasure-pattern circuits, total decoder shots and detected logical errors for horizontal~(H) and vertical~(V) memory experiments, and cumulative CPU time for both decoders.}
\label{tab:simulation_stats}
\end{table*}

\FloatBarrier

\subsection{\texorpdfstring{Simulations to generate \Cref{fig:transversal_cnot}}{}}

To generate the data for \Cref{fig:transversal_cnot}, we use two sets of circuits: one where both patches are initialized in the $+1$ eigenstate of the horizontal logical observable, and one where both patches are initialized in the $+1$ eigenstate of the vertical logical observable.
For these circuits, standard MWPM decoding cannot be directly applied. 
We therefore use the BPOSD decoder~\cite{panteleev2021degenerate, roffe_decoding_2020}, which can handle the joint decoding problem across both patches. We note that correlated MWPM-based approaches for decoding across transversal gates have been proposed~\cite{cain2024correlated, serra2026decoding}.
For decoding, the maximum number of belief propagation iterations is set to $30$ and the product sum method is used for belief propagation.
The combination sweep strategy is used for ordered statistics decoding (OSD) with a maximum order of $60$ \cite{roffe_decoding_2020}.
The number of circuits, shots, and logical errors for the transversal $\CNOT$ experiment in \Cref{fig:transversal_cnot} are summarized in \Cref{tab:cnot_simulation_stats}.

\begin{table}[t]
\centering
\begin{tabular}{cc c rr rr r}
\toprule
 & & & \multicolumn{2}{c}{Shots ($\times 10^3$)} & \multicolumn{2}{c}{Errors} & \\
\cmidrule(lr){4-5} \cmidrule(lr){6-7}
$w$ & $h$ & Circ. & H & V & H & V & Time \\
\midrule
4 & 6 & 100 & 77 & 82 & 1\,769 & 1\,520 & 2\,min \\
6 & 9 & 100 & 93 & 99 & 555 & 278 & 31\,min \\
8 & 12 & 242 & 813 & 995 & 876 & 435 & 48\,h \\
10 & 18 & 5 & 95 & 85 & 4 & 8 & 53\,h \\
\bottomrule
\end{tabular}
\caption{Summary of Monte Carlo sampling runs for the transversal $\CNOT$ experiment at $p = 0.01$ using the BPOSD decoder~\cite{panteleev2021degenerate, roffe_decoding_2020} via the \texttt{stimbposd} package~\cite{higgott_sinter_bposd}. For each patch size we report the number of erasure-pattern circuits, total decoder shots and detected logical errors for horizontal~(H) and vertical~(V) directions, and cumulative CPU time.}
\label{tab:cnot_simulation_stats}
\end{table}

\section{Clifford+T Rotation Synthesis Strategies}
\label{appendix:synthesis_strategies}
We summarize here different synthesis strategies introduced in~\cite{Kliuchnikov2023shorterquantum}.
\paragraph{Diagonal approximation.}
The baseline strategy directly approximates $R_Z(\theta)$ by a single Clifford$+T$ word
on the target qubit alone, requiring no auxiliary qubit. The mean $T$-count is
$3.02\log_2(1/\epsilon_{\mathrm{synth}}) + 1.77$, with a worst-case (over target angles) of
$3.02\log_2(1/\epsilon_{\mathrm{synth}}) + 9.19$.
\paragraph{Mixed diagonal approximation.}
Rather than approximating the target \emph{unitary} directly, one targets the approximation
at the level of \emph{quantum channels}. For each shot, one randomly selects between two
Clifford$+T$ words (an under-rotation and an over-rotation) with complementary probabilities,
so that the resulting channel is $\epsilon_{\mathrm{synth}}$-close to the target in diamond norm. Because each individual
word need only hit a less demanding target, the mean $T$-count drops to
$1.52\log_2(1/\epsilon_{\mathrm{synth}}) - 0.01$, with a worst-case (over target angles) of
$1.54\log_2(1/\epsilon_{\mathrm{synth}}) + 6.85$, with no auxiliary qubit required.
\paragraph{Fallback approximation.}
This strategy introduces one auxiliary qubit. A projective rotation is first applied jointly
to the target and auxiliary. With a tunable probability $p_{\mathrm{succ}}$ the target is
left $\epsilon_{\mathrm{synth}}$-close to the desired state (the accept branch), while with probability
$1 - p_{\mathrm{succ}}$ the step fails and the full rotation must be synthesized via
the diagonal strategy (the reject branch), incurring a worst-case $T$-count exceeding
the baseline. In the serial setting the mean $T$-count is $1.03\log_2(1/\epsilon_{\mathrm{synth}}) + 5.75$,
with a worst-case (over target angles) of $1.05\log_2(1/\epsilon_{\mathrm{synth}}) + 11.83$.
The original plaquette Trotterization algorithm of~\cite{Campbell_2021} employs this
strategy. However, in our circuit all $L^2$ rotations within a sub-evolution run
\emph{in parallel}, and the next Trotter step cannot begin until every synthesis is
complete. The probability that at least one of the $L^2$ concurrent syntheses hits
its reject branch is $1 - p_{\mathrm{succ}}^{L^2}$, which approaches one exponentially
fast in system size. This effectively replaces the favorable expected $T$-count with the
worst-case one, potentially eliminating the strategy's advantage over the mixed diagonal
approximation in a highly parallelized compilation.
\paragraph{Mixed fallback approximation.}
The best expected $T$-count is achieved by combining channel mixing with the fallback
structure. For each shot one selects randomly among over- and under-rotated fallback
circuits, giving a mean $T$-count of $0.53\log_2(1/\epsilon_{\mathrm{synth}}) + 4.86$, a roughly six-fold
improvement over the diagonal baseline, with a worst-case (over target angles) of
$0.57\log_2(1/\epsilon_{\mathrm{synth}}) + 8.83$. The strategy requires one auxiliary qubit and inherits
the same worst-case parallelization issue as the plain fallback: the probability that all
$L^2$ concurrent syntheses accept simultaneously vanishes exponentially in $L^2$. This is the synthesis cost assumed
in~\cite{mcardle2025fastcuriousacceleratefaulttolerant}.
The $T$-counts quoted above are furthermore
averages over uniformly random target angles. For the specific angles $\theta = t/(2r)$
arising from a concrete choice of Trotter steps $r$ and simulation time $t$, running the
synthesis algorithm explicitly will in general produce shorter sequences.
\section{Glossary of notation for Section~\ref{sec:simulating_fermi_hubbard}}
\label{appendix:glossary}

\begin{table}[h]
\centering
\label{tab:glossary}
\renewcommand{\arraystretch}{1.25}
\begin{tabular}{ll}
\toprule
\textbf{Symbol} & \textbf{Meaning} \\
\midrule
$L$ & Linear dimension of the $L\times L$ lattice \\
$t$, $U$ & Hopping parameter and on-site repulsion \\
$H_I$ & Interaction Hamiltonian, $\frac{U}{4}\sum_j Z_{j\uparrow}Z_{j\downarrow}$ \\
$H_h^p,\,H_h^g$ & Pink and golden hopping sub-Hamiltonians \\
$V_{j\sigma}$ & Vertex operator; maps to $Z_j$ under compact mapping \\
$E_{j\sigma,k\sigma'}$ & Edge operator; maps to weight-3 Pauli \\
$T_{\mathrm{sim}}$ & Total simulation time in $e^{-iHT_{\mathrm{sim}}}$ \\
$r$ & Number of Trotter steps \\
$\epsilon_{\mathrm{alg}}$ & Trotter (algorithmic) error budget \\
$\epsilon_{\mathrm{synth}}$ & Rotation synthesis accuracy target \\
$n_T$ & $T$-count per synthesized rotation \\
$\tau_{\mathrm{r}}$ & Classical reaction time (feed-forward latency) \\
$t_{\mathrm{l}}$ & Logical cycle time (lattice surgery block time) \\
$\mathcal{P}$, $H_\mathcal{P}$ & Plaquette and its Hamiltonian \\
$F_{j,k}$ & Fermionic Fourier transform between modes $j$ and $k$ \\
$C$ & Diagonalizing Clifford for plaquette time evolution \\
\bottomrule
\end{tabular}
\caption{Notation introduced in Section~4.}
\end{table}

\section{Derivation of the diagonalizing circuit for the pink plaquettes}
\label{sec:derivation_of_the_diagonalizing_circuit_for_pink_plaquette}
In order to implement $\exp(-iH_\mathcal{P}^\sigma\,T_{\mathrm{sim}}/(2r))$ exactly, we follow the fermionic decomposition of~\Ccite[Eq. E10]{Campbell_2021}:
\begin{equation}
\label{eq:plaquette_hamiltonian_time_evolution}
\exp\!\left(\frac{-iT_{\mathrm{sim}}}{2r}H_\mathcal{P}^\sigma\right)
= F_{3\sigma,1\sigma} F_{2\sigma,4\sigma} F_{2\sigma,3\sigma}
\, e^{i 2t \frac{T_{\mathrm{sim}}}{2r} c_{2\sigma}^\dagger c_{2\sigma}}
\, e^{-i 2t \frac{T_{\mathrm{sim}}}{2r} c_{3\sigma}^\dagger c_{3\sigma}}
\, F_{2\sigma,3\sigma}^\dagger F_{2\sigma,4\sigma}^\dagger F_{3\sigma,1\sigma}^\dagger,
\end{equation}
where sites 1, 2, 3, 4 label the plaquette vertices in clockwise order and $F_{j\sigma,k\sigma}$ denotes the fermionic Fourier transform between modes $j\sigma$ and $k\sigma$. The derivation holds for both spin sectors. We drop $\sigma$ for clarity.
To reduce the central number-operator exponentials to single-qubit $Z$-rotations, we use the fact that:
\begin{equation}
\label{eq:inner_part_of_plaquette_evolution_simplification_through_fermionic_fourier_transform_identities}
    F_{j,k}\,e^{i\theta n_j}\,e^{-i\theta n_k}\,F_{j,k}^\dagger = e^{i\theta(c_j^\dagger c_k + c_k^\dagger c_j)} = e^{\frac{\theta}{2}(E_{j,k}V_k + V_jE_{j,k})},
\end{equation}
where the first equality follows from the properties of the fermionic Fourier transform \footnote{Using the action of the two-mode fermionic Fourier transform on the modes
(see Eqs.~(C2)--(C3) of \cite{mcardle2025fastcuriousacceleratefaulttolerant}),
$F_{jk} c_j F_{jk}^\dagger=\tfrac{1}{\sqrt2}(c_j+c_k)$ and
$F_{jk} c_k F_{jk}^\dagger=\tfrac{-i}{\sqrt2}(c_j-c_k)$,
one finds $F_{jk}(n_j-n_k)F_{jk}^\dagger=c_j^\dagger c_k+c_k^\dagger c_j$,
which implies $F_{j,k}e^{i\theta n_j}e^{-i\theta n_k}F_{j,k}^\dagger = e^{i\theta(c_j^\dagger c_k+c_k^\dagger c_j)}$.} and the second from \Cref{eq:hopping_term_pair_in_majorana_operators}. Applying this to the innermost pair $F_{2,3}$ and translating via the fermion-to-qubit mapping (\Cref{fig:vertex_and_edge_operators_on_single_plaquette}) yields
\begin{equation}
    F_{2,3}
\, e^{i 2t \frac{T_{\mathrm{sim}}}{2r} c_{2}^\dagger c_{2}}
\, e^{-i 2t \frac{T_{\mathrm{sim}}}{2r} c_{3}^\dagger c_{3}}
\, F_{2,3}^\dagger = e^{i\frac{tT_{\mathrm{sim}}}{2r}(X_2X_3X_{aux}+ Y_2Y_3X_{aux})} = Ce^{i\frac{tT_{\mathrm{sim}}}{2r}Z_2}e^{i\frac{tT_{\mathrm{sim}}}{2r}Z_3} C^\dagger,
\end{equation}
where the Clifford $C$ is introduced to isolate the two rotations onto separate qubits, enabling them to be synthesized to single qubit gates in parallel. It must satisfy $X_2X_3X_{aux} = CZ_2C^\dagger, \quad
    Y_2Y_3X_{aux} = CZ_3C^\dagger$. One particular choice is:
\begin{equation}
  \label{eq:clifford_C_appendix}
  C^\dagger = \exp\!\left(\frac{i\pi}{4}Y_2X_3X_{aux}\right)X_2
\end{equation}

It remains to translate the outer fermionic Fourier transforms $F_{3,1}$ and $F_{2,4}$ into qubit gates. Using the definition of \Ccite[Equation C1]{mcardle2025fastcuriousacceleratefaulttolerant} and rewriting in terms of edge and vertex operators yields
\begin{equation}
\label{eq:C1}
F_{jk}
= \exp\!\left(-\frac{i\pi}{4} n_j\right)
  \exp\!\left(\frac{i\pi}{4} n_k\right)
  \exp\!\left(\frac{i\pi}{4} f_s\right)
=
    \exp\!\left(\frac{i\pi}{4} V_j\right)
  \exp\!\left(\frac{\pi}{8} V_jE_{jk}\right)
  \exp\!\left(\frac{\pi}{8} E_{jk}V_k\right).
\end{equation}

Where $f_s = I + c_j^\dagger c_k + c_k^\dagger c_j - n_j - n_k$ is the fermionic swapgate.
Plugging in the explicit operators for the sample plaquette, and using the edge-concatenation identity $E_{jl} = iE_{jk}E_{kl}$, which gives $E_{3,1} = iE_{3,2}E_{2,1} = Y_1Y_3Z_{aux}$ and $E_{2,4} = X_2X_4Z_{aux}$, we obtain
\begin{equation}
\label{eq:f_3_1_dagger_and_f_2_4_dagger}
    F_{3,1}^\dagger = e^{\frac{-i\pi}{8} X_1Y_3Z_{aux}}\,
  e^{\frac{i\pi}{8}Y_1X_3Z_{aux}}\,
  e^{\frac{-i\pi}{4} Z_3},
\qquad
  F_{2,4}^\dagger = e^{\frac{i\pi}{8} X_2Y_4Z_{aux}}\,
  e^{\frac{-i\pi}{8}Y_2X_4Z_{aux}}\,
  e^{\frac{-i\pi}{4} Z_2}.
\end{equation}
With that \Cref{eq:plaquette_hamiltonian_time_evolution} is fully described in terms of qubit operations.
The $Z$-rotations on qubit $2$ and $3$ are synthesized fault-tolerantly as described in \Cref{sec:fault_tolerant_synthesis}. 

\section{Active cube cost of golden diagonalization circuit}
\label{sec:active_cube_cost_of_golden_diagonalization_circuit}

We derive the number of active cubes in the diagonalization circuit $C^\dagger F_{2,4}^\dagger F_{3,1}^\dagger$ for one plane. Every golden plaquette, whether bulk or boundary, incurs the same base active cube cost: $102.5$ active cubes for the diagonalization round in which it participates (half the pink cost in \Cref{tab:cost_diagonalizing_circuit}, which counts forward plus inverse), plus $2\times 9 \times 6 = 108$ active cubes for idling during the two rounds in which it does not participate. The total base cost is therefore $210.5 \times L^2/4$ active cubes per layer. On top of this, the shuttling and corridors needed for long-range edge operator measurements add the following overhead (all counts per plane):
\begin{enumerate}
    \item \textbf{Column shift.} Shifting every second column of qubit patches through the MSF aisle once costs $0.5\cdot 1.5L^2\cdot (w_{\mathrm{MSF}}-1)$ active cubes, per plane.
    \item \textbf{Four-corner plaquette corridors.} One corner plaquette per plane. The diagonalization circuit uses three types of corridor. The top-left to bottom-right corridor costs $(11+w_{\mathrm{MSF}})\cdot(L/2-1) + w_{\mathrm{MSF}}$ active cubes per timestep, the bottom-left to top-right corridor costs $(15+w_{\mathrm{MSF}})\cdot(L/2-1) + w_{\mathrm{MSF}}$, and the top-to-bottom corridor costs $7\cdot(L/2-1)$. Of the $8$ timesteps of $F_{2,4}^\dagger F_{3,1}^\dagger$, on average $2$ require no corridor because $2$ of the $4$ potential $\pi/4$ corrections are not needed. The remaining $6$ split evenly: $3$ use the top-left to bottom-right diagonal and $3$ use the bottom-left to top-right diagonal. The $9$th timestep ($C^\dagger$) uses the top-to-bottom corridor. The total corridor overhead is $3\cdot\bigl((11+w_{\mathrm{MSF}})\cdot(L/2-1) + w_{\mathrm{MSF}}\bigr) + 3\cdot\bigl((15+w_{\mathrm{MSF}})\cdot(L/2-1) + w_{\mathrm{MSF}}\bigr) + 7\cdot(L/2-1)$.
    \item \textbf{Top-to-bottom corridors.} $L/2-1$ horizontal-boundary plaquettes per layer. Each requires a top-to-bottom corridor at $8\cdot(L/2-1)$ active cubes per timestep. The corridor is active for $7$ of the $9$ timesteps on average (the remaining $2$ correspond to corrective Cliffords that are not needed). The total corridor overhead per plaquette is $7\cdot 8\cdot(L/2-1)$.
    \item \textbf{Left-to-right corridors.} $L/2-1$ vertical-boundary plaquettes per plane. Each requires a left-to-right corridor at $(8+w_{\mathrm{MSF}})\cdot(L/2-1) + w_{\mathrm{MSF}}$ active cubes per timestep. The corridor is active for $6$ of the $9$ timesteps on average: the $9$th timestep ($C^\dagger$) does not require a left-to-right corridor, and $2$ of the $4$ potential corrective Cliffords are not needed on average. The total corridor overhead per plaquette is $6\cdot\bigl((8+w_{\mathrm{MSF}})\cdot(L/2-1) + w_{\mathrm{MSF}}\bigr)$.
\end{enumerate}
The total active cube cost of implementing $C^\dagger F_{2,4}^\dagger F_{3,1}^\dagger$ on every plaquette of one plane is
\begin{equation}
\frac{210.5\, L^2}{4} + 0.75\, L^2(w_{\mathrm{MSF}}-1) + (104 + 6w_{\mathrm{MSF}})\!\left(\frac{L}{2}-1\right)^{\!2} + (85 + 12w_{\mathrm{MSF}})\!\left(\frac{L}{2}-1\right) + 6w_{\mathrm{MSF}}.
\end{equation}
The inverse circuit, $F_{3,1} F_{2,4} C$, incurs the same cost, and both planes contribute equally. 

Note also that corridors occupy parts of the magic state factory aisles during these steps. However, this does not lead to an undersupply of magic states, since the MSF aisles are sized to produce $L^2$ $\ket{T}$ states every $(1+\tau_r/t_l)$ logical timesteps, whereas we consume only $2L^2$ $\ket{T}$ states over the full $27$ logical timesteps.

\section{Majorana edge and vertex operators}
\label{appendix:majorana_edge_vertex_operators}

Given a set of fermionic modes with creation and annihilation operators $c_j^\dagger$ and $c_j$ satisfying the canonical anti-commutation relations $\{c_i^\dagger, c_j\} = \delta_{ij}$, one defines two Majorana operators per mode:
\begin{equation}
    \gamma_j := c_j + c_j^\dagger, \qquad \bar{\gamma}_j := \frac{c_j - c_j^\dagger}{i}.
\end{equation}
These are Hermitian, traceless, and satisfy $\gamma_j^2 = \bar{\gamma}_j^2 = 1$ and $\{\gamma_j, \bar{\gamma}_j\} = 0$. The Majorana \emph{edge} and \emph{vertex} operators are defined as~\cite{Derby_2021, BravyiKitaev2002}
\begin{equation}
    E_{jk} := -i\gamma_j\gamma_k, \qquad V_j := -i\gamma_j\bar{\gamma}_j.
\end{equation}
Both are Hermitian, traceless, and self-inverse. The edge operators are antisymmetric, $E_{jk} = -E_{kj}$. The vertex operators equal the occupation parity at site $j$, since $V_j = 1 - 2n_j$.

The algebraic structure relevant for fermion-to-qubit mappings is captured by the following (anti-)commutation relations. Operators anti-commute if and only if they share a vertex index:
\begin{equation}
    \{E_{jk}, V_j\} = 0, \qquad \{E_{ij}, E_{jk}\} = 0,
\end{equation}
and for all $i \neq j \neq m \neq n$:
\begin{equation}
    [V_i, V_j] = 0, \qquad [E_{ij}, V_m] = 0, \qquad [E_{ij}, E_{mn}] = 0.
\end{equation}
All even fermionic operators (even products of creation and annihilation operators) can be expressed in terms of edge and vertex operators~\cite{BravyiKitaev2002}. Since all parity-preserving operators are even, these operators suffice to represent every physical fermionic observable. In addition, the product of edge operators around any closed loop of sites $p = \{p_1, p_2, \ldots\}$ satisfies
\begin{equation}
    i^{(|p|-1)} \prod_{i=1}^{|p|-1} E_{p_i p_{i+1}} = 1.
\end{equation}
This loop constraint is enforced in the compact mapping~\cite{Derby_2021} by treating these loop operators as stabilizers and restricting to their joint $+1$ eigenspace.

\bibliographystyle{apsrev4-2}
\bibliography{bibliography}

\end{document}

%% file: commands.tex
\newcommand{\Id}{\mathsf{I}}
\newcommand{\Init}{\mathsf{Init}}
\newcommand{\E}{\mathsf{E}}
\newcommand{\X}{\mathsf{X}}
\newcommand{\Y}{\mathsf{Y}}
\newcommand{\Z}{\mathsf{Z}}
\newcommand{\T}{\mathsf{T}}
\newcommand{\Sgate}{\mathsf{S}}
\newcommand{\CZ}{\mathsf{CZ}}
\newcommand{\CNOT}{\mathsf{CNOT}}
\newcommand{\MZZ}{\mathsf{M}_{\Z\Z}}

\newcommand{\MZ}{\mathsf{M}_{\Z}}
\newcommand{\MYY}{\mathsf{M}_{\Y\Y}}
\renewcommand{\H}{\mathsf{H}}
\newcommand{\HXY}{\mathsf{H}_{\Y\Z}}
\newcommand{\CY}{\mathsf{CY}}

\newcommand{\ruscz}{$\mathsf{RUS}$-$\CZ$~}
\newcommand{\rusmzz}{$\mathsf{RUS}$-$\MZZ$~}
\newcommand{\init}{$\Init$~}

\newcommand{\Ccite}[2][]{%
    \IfSubStr{#2}{,}{Refs.~}{Ref.~}%
    \ifthenelse{\equal{#1}{}}{\cite{#2}}{\cite[#1]{#2}}}

\newcommand{\ccite}[2][]{%
    \IfSubStr{#2}{,}{refs.~}{ref.~}%
    \ifthenelse{\equal{#1}{}}{\cite{#2}}{\cite[#1]{#2}}}
    
\tikzset{anchor properly/.style = {baseline={([yshift=-.5ex]current bounding box.center)}}}

\makeatletter
\def\l@subsubsection#1#2{}
\makeatother

%% file: biplanar_layout_2.tikz
\begin{tikzpicture}[line join=round, line cap=round]
\pgfdeclarelayer{deepbg}
\pgfsetlayers{deepbg,background,main}
\tikzset{
  basis/.style={
    x={(1cm,0cm)},
    y={(0.35cm,0.35cm)}
  }
}

\definecolor{Lavender}{HTML}{8B86FF}
\definecolor{Salmon}{HTML}{C97C6C}
\definecolor{PaleGreen}{HTML}{9ED9C7}
\definecolor{GoldDot}{HTML}{F2D07B}

\def\s{0.6}   
\def\Nx{3}    
\def\Ny{3}    
\pgfmathsetmacro{\rt}{1.7320508075688772} 
\pgfmathsetmacro{\hhex}{0.5*\rt*\s}       

\def\faceop{0.80}
\def\edgelw{0.65pt}
\def\dotr{0.07}
\def\borderlw{0.8pt}

\pgfmathsetmacro{\pad}{0.80*\s}
\def\drop{3} 

\pgfmathsetmacro{\xmin}{-\s}
\pgfmathsetmacro{\xmax}{1.5*\s*\Nx + \s}

\pgfmathsetmacro{\ymin}{-\hhex}
\pgfmathsetmacro{\ymax}{\rt*\s*(\Ny+0.5) + \hhex}

\pgfmathsetmacro{\W}{(\xmax-\xmin) + 2*\pad}
\pgfmathsetmacro{\H}{(\ymax-\ymin) + 2*\pad}

\pgfmathsetmacro{\dx}{\pad - \xmin}
\pgfmathsetmacro{\dy}{\pad - \ymin}

\newcommand{\hexpath}[2]{%
  ({#1+\s/2},{#2})
  \foreach \a in {80,120} { -- ({#1+\s*cos(\a)},{#2+\s*sin(\a)}) } 
  -- ({#1-\s/2},{#2})
  \foreach \a in  {240,300} { -- ({#1+\s*cos(\a)},{#2+\s*sin(\a)}) }
  -- cycle
}

\newcommand{\drawhex}[3]{
  \path[fill=#3, fill fill opacity=\faceop,
        draw=black, draw fill opacity=1, line width=\edgelw]
    \hexpath{#1}{#2};
  \fill[GoldDot] ({#1+\s/2},{#2}) circle (\dotr); 
  \fill[GoldDot] ({#1-\s/2},{#2}) circle (\dotr); 
  \foreach \a in {80,120,240,300}{
    \fill[GoldDot] ({#1+\s*cos(\a)},{#2+\s*sin(\a)}) circle (\dotr);
  }
}

\newcommand{\pickcolor}[3]{
  \pgfmathtruncatemacro{\halfq}{int(#1/2)}
  \pgfmathtruncatemacro{\rax}{#2-\halfq}
  \pgfmathtruncatemacro{\m}{mod(#1+2*\rax,3)}
  \ifnum\m=0 \def#3{Lavender}\fi
  \ifnum\m=1 \def#3{Salmon}\fi
  \ifnum\m=2 \def#3{PaleGreen}\fi
}

\newcommand{\centerxy}[4]{
  \pgfmathsetmacro{#3}{\s*(#1)}
  \pgfmathsetmacro{#4}{\rt*\s*((#2)+mod(#1,2)/2)}
}

\newcommand{\drawplane}[1]{%
  \path[line width=\borderlw, fill fill opacity=#1]
    (0,0) -- (\W,0) -- (\W,\H) -- (0,\H) -- cycle;

  \begin{scope}[reset cm]
    \path[basis] (0,0) coordinate (L0);
    \path[basis] (0,\H) coordinate (L1);

    \path[basis] (\W,0) coordinate (R0);


  \end{scope}
}

\newcommand{\drawpatch}[9]{

\foreach \x in {0,...,5} {
    \pgfmathtruncatemacro{\shift}{(\x==1) || (\x==2) || (\x==3)}
    \foreach \y in {0,...,3} {
        \node (P-#3-X\x-Y\y) at ({#2+\s*(\y+\shift)}, {#1-\s*\x}) {};
        \filldraw[fill=#5, draw=GoldDot, thick] (P-#3-X\x-Y\y) circle (\dotr);
    }

}
\begin{scope}[on background layer]
    \fill[fill=#7, fill opacity=#4] ({P-#3-X0-Y2}.center) -- ({P-#3-X0-Y3}.center) -- ({P-#3-X1-Y2}.center) -- ({P-#3-X2-Y2}.center) -- ({P-#3-X2-Y1}.center) -- ({P-#3-X1-Y1}.center) -- cycle;
    \fill[fill=#7, fill opacity=#4] ({P-#3-X3-Y0}.center) -- ({P-#3-X3-Y1}.center) -- ({P-#3-X5-Y2}.center) -- ({P-#3-X5-Y1}.center) -- cycle;
    \fill[fill=#7, fill opacity=#4] ({P-#3-X3-Y2}.center) -- ({P-#3-X3-Y3}.center) -- ({P-#3-X5-Y3}.center) -- cycle;
    \fill[fill=#7, fill opacity=#4] ({P-#3-X0-Y0}.center) -- ({P-#3-X0-Y1}.center) -- ({P-#3-X2-Y0}.center) -- cycle;
    \fill[fill=#8, fill opacity=#4] ({P-#3-X0-Y1}.center) -- ({P-#3-X0-Y2}.center) -- ({P-#3-X1-Y1}.center) -- ({P-#3-X1-Y0}.center) -- cycle;
    \fill[fill=#8, fill opacity=#4] ({P-#3-X0-Y3}.center) -- ({P-#3-X1-Y3}.center) -- ({P-#3-X1-Y2}.center) -- cycle;
    \fill[fill=#8, fill opacity=#4] ({P-#3-X2-Y1}.center) -- ({P-#3-X2-Y2}.center) -- ({P-#3-X4-Y3}.center) -- ({P-#3-X4-Y2}.center) -- cycle;
    \fill[fill=#8, fill opacity=#4] ({P-#3-X2-Y0}.center) -- ({P-#3-X4-Y1}.center) -- ({P-#3-X4-Y0}.center) -- cycle;
    \fill[fill=#9, fill opacity=#4] ({P-#3-X1-Y0}.center) -- ({P-#3-X1-Y1}.center) -- ({P-#3-X2-Y1}.center) -- ({P-#3-X3-Y1}.center) -- ({P-#3-X3-Y0}.center) -- cycle;
    \fill[fill=#9, fill opacity=#4] ({P-#3-X1-Y2}.center) -- ({P-#3-X1-Y3}.center) -- ({P-#3-X2-Y3}.center) -- ({P-#3-X3-Y3}.center) -- ({P-#3-X3-Y2}.center) -- cycle;
    \fill[fill=#9, fill opacity=#4] ({P-#3-X4-Y2}.center) -- ({P-#3-X4-Y3}.center) -- ({P-#3-X5-Y3}.center) -- ({P-#3-X5-Y2}.center)-- cycle;
    \fill[fill=#9, fill opacity=#4] ({P-#3-X4-Y0}.center) -- ({P-#3-X4-Y1}.center) -- ({P-#3-X5-Y1}.center) -- ({P-#3-X5-Y0}.center)-- cycle;
\end{scope}

\begin{scope}[on background layer]
\draw[very thick, Salmon, #6] ({P-#3-X0-Y0}.center) -- ({P-#3-X0-Y1}.center);
\draw[very thick, Salmon, #6] ({P-#3-X0-Y2}.center) -- ({P-#3-X0-Y3}.center);
\draw[very thick, Salmon, #6] ({P-#3-X1-Y0}.center) -- ({P-#3-X2-Y0}.center);
\draw[very thick, Salmon, #6] ({P-#3-X1-Y1}.center) -- ({P-#3-X2-Y1}.center);
\draw[very thick, Salmon, #6] ({P-#3-X1-Y2}.center) -- ({P-#3-X2-Y2}.center);
\draw[very thick, Salmon, #6] ({P-#3-X1-Y3}.center) -- ({P-#3-X2-Y3}.center);
\draw[very thick, Salmon, #6] ({P-#3-X3-Y0}.center) -- ({P-#3-X3-Y1}.center);
\draw[very thick, Salmon, #6] ({P-#3-X3-Y2}.center) -- ({P-#3-X3-Y3}.center);
\draw[very thick, Salmon, #6] ({P-#3-X4-Y0}.center) -- ({P-#3-X5-Y0}.center);
\draw[very thick, Salmon, #6] ({P-#3-X4-Y1}.center) -- ({P-#3-X5-Y1}.center);
\draw[very thick, Salmon, #6] ({P-#3-X4-Y2}.center) -- ({P-#3-X5-Y2}.center);
\draw[very thick, Salmon, #6] ({P-#3-X4-Y3}.center) -- ({P-#3-X5-Y3}.center);
\draw[very thick, Lavender, #6] ({P-#3-X1-Y0}.center) -- ({P-#3-X1-Y1}.center);
\draw[very thick, Lavender, #6] ({P-#3-X1-Y2}.center) -- ({P-#3-X1-Y3}.center);
\draw[very thick, Lavender, #6] ({P-#3-X2-Y0}.center) -- ({P-#3-X3-Y0}.center);
\draw[very thick, Lavender, #6] ({P-#3-X2-Y1}.center) -- ({P-#3-X3-Y1}.center);
\draw[very thick, Lavender, #6] ({P-#3-X2-Y2}.center) -- ({P-#3-X3-Y2}.center);
\draw[very thick, Lavender, #6] ({P-#3-X2-Y3}.center) -- ({P-#3-X3-Y3}.center);
\draw[very thick, Lavender, #6] ({P-#3-X4-Y0}.center) -- ({P-#3-X4-Y1}.center);
\draw[very thick, Lavender, #6] ({P-#3-X4-Y2}.center) -- ({P-#3-X4-Y3}.center);
\draw[very thick, PaleGreen, #6] ({P-#3-X0-Y1}.center) -- ({P-#3-X1-Y0}.center);
\draw[very thick, PaleGreen, #6] ({P-#3-X0-Y2}.center) -- ({P-#3-X1-Y1}.center);
\draw[very thick, PaleGreen, #6] ({P-#3-X0-Y3}.center) -- ({P-#3-X1-Y2}.center);
\draw[very thick, PaleGreen, #6] ({P-#3-X2-Y1}.center) -- ({P-#3-X2-Y2}.center);
\draw[very thick, PaleGreen, #6] ({P-#3-X3-Y0}.center) -- ({P-#3-X4-Y1}.center);
\draw[very thick, PaleGreen, #6] ({P-#3-X3-Y1}.center) -- ({P-#3-X4-Y2}.center);
\draw[very thick, PaleGreen, #6] ({P-#3-X3-Y2}.center) -- ({P-#3-X4-Y3}.center);
\draw[very thick, PaleGreen, #6] ({P-#3-X5-Y1}.center) -- ({P-#3-X5-Y2}.center);
\end{scope}
}

\begin{scope}[basis, xshift=4cm]
\drawpatch{0}{0}{Ad}{1}{GoldDot}{solid}{Lavender!80}{Salmon!80}{PaleGreen!80}
\drawpatch{0}{4*\s}{Bd}{1}{GoldDot}{solid}{Lavender!80}{Salmon!80}{PaleGreen!80}
\drawpatch{6*\s}{0}{Cd}{1}{GoldDot}{solid}{Lavender!80}{Salmon!80}{PaleGreen!80}
\drawpatch{6*\s}{4*\s}{Dd}{1}{GoldDot}{solid}{Lavender!80}{Salmon!80}{PaleGreen!80}

\drawpatch{15*\s}{-3.5*1.5*\s}{Au}{1}{GoldDot}{solid}{Lavender!80}{Salmon!80}{PaleGreen!80}
\drawpatch{15*\s}{(4-3.5*1.5)*\s}{Bu}{1}{GoldDot}{solid}{Lavender!80}{Salmon!80}{PaleGreen!80}
\drawpatch{21*\s}{-3.5*1.5*\s}{Cu}{1}{GoldDot}{solid}{Lavender!80}{Salmon!80}{PaleGreen!80}
\drawpatch{21*\s}{(4-3.5*1.5)*\s}{Du}{1}{GoldDot}{solid}{Lavender!80}{Salmon!80}{PaleGreen!80}

\begin{pgfonlayer}{deepbg}
    \fill[PaleGreen!80] ({P-Au-X0-Y0}.center) -- ({P-Au-X0-Y1}.center) -- ({P-Cu-X4-Y1}.center) -- ({P-Cu-X4-Y0}.center) --cycle;

    \fill[PaleGreen!80] ({P-Au-X0-Y2}.center) -- ({P-Au-X0-Y3}.center) -- ({P-Cu-X4-Y3}.center) -- ({P-Cu-X4-Y2}.center) --cycle;

    \fill[Salmon!80] ({P-Au-X1-Y0}.center) -- ({P-Au-X1-Y1}.center) -- ({P-Cu-X4-Y2}.center) -- ({P-Cu-X4-Y1}.center) --cycle;

    \fill[Salmon!80] ({P-Au-X1-Y3}.center) -- ({P-Au-X1-Y2}.center) -- ({P-Cu-X5-Y3}.center) --cycle;

    \fill[Salmon!80] ({P-Ad-X2-Y3}.center) -- ({P-Bd-X2-Y0}.center) -- ({P-Bd-X4-Y1}.center)  -- ({P-Bd-X4-Y0}.center) -- cycle;
    
    \fill[Salmon!80] ({P-Ad-X0-Y3}.center) -- ({P-Bd-X0-Y0}.center) -- ({P-Ad-X1-Y3}.center)  -- ({P-Ad-X1-Y2}.center) -- cycle;

    \fill[Lavender!80] ({P-Ad-X3-Y2}.center) -- ({P-Ad-X3-Y3}.center) -- ({P-Bd-X5-Y0}.center) -- ({P-Ad-X5-Y3}.center) --cycle;

    \fill[Lavender!80] ({P-Bd-X0-Y0}.center) -- ({P-Bd-X0-Y1}.center) -- ({P-Bd-X2-Y0}.center) -- ({P-Ad-X2-Y3}.center) --cycle;
\end{pgfonlayer}

\begin{scope}[on background layer]
    \draw[thick, PaleGreen] ({P-Ad-X1-Y3}.center) -- ({P-Bd-X0-Y0}.center);
    \draw[thick, PaleGreen] ({P-Ad-X2-Y3}.center) -- ({P-Bd-X2-Y0}.center);
    \draw[thick, PaleGreen] ({P-Ad-X3-Y3}.center) -- ({P-Bd-X4-Y0}.center);
    \draw[thick, PaleGreen] ({P-Ad-X5-Y3}.center) -- ({P-Bd-X5-Y0}.center);

    \draw[very thick, dotted, PaleGreen] ({P-Cd-X1-Y3}.center) -- ({P-Dd-X0-Y0}.center);
    \draw[very thick, dotted, PaleGreen] ({P-Cd-X2-Y3}.center) -- ({P-Dd-X2-Y0}.center);
    \draw[very thick, dotted, PaleGreen] ({P-Cd-X3-Y3}.center) -- ({P-Dd-X4-Y0}.center);
    \draw[very thick, dotted, PaleGreen] ({P-Cd-X5-Y3}.center) -- ({P-Dd-X5-Y0}.center);

    \draw[very thick, dotted, PaleGreen] ({P-Au-X1-Y3}.center) -- ({P-Bu-X0-Y0}.center);
    \draw[very thick, dotted, PaleGreen] ({P-Au-X2-Y3}.center) -- ({P-Bu-X2-Y0}.center);
    \draw[very thick, dotted, PaleGreen] ({P-Au-X3-Y3}.center) -- ({P-Bu-X4-Y0}.center);
    \draw[very thick, dotted, PaleGreen] ({P-Au-X5-Y3}.center) -- ({P-Bu-X5-Y0}.center);

    \draw[very thick, dotted, PaleGreen] ({P-Cu-X1-Y3}.center) -- ({P-Du-X0-Y0}.center);
    \draw[very thick, dotted, PaleGreen] ({P-Cu-X2-Y3}.center) -- ({P-Du-X2-Y0}.center);
    \draw[very thick, dotted, PaleGreen] ({P-Cu-X3-Y3}.center) -- ({P-Du-X4-Y0}.center);
    \draw[very thick, dotted, PaleGreen] ({P-Cu-X5-Y3}.center) -- ({P-Du-X5-Y0}.center);
\end{scope}

\begin{scope}[on background layer]
    \draw[very thick, dotted, Lavender] ({P-Ad-X0-Y0}.center) -- ({P-Cd-X5-Y0}.center);
    \draw[very thick, dotted, Lavender] ({P-Ad-X0-Y1}.center) -- ({P-Cd-X5-Y1}.center);
    \draw[very thick, dotted, Lavender] ({P-Ad-X0-Y2}.center) -- ({P-Cd-X5-Y2}.center);
    \draw[very thick, dotted, Lavender] ({P-Ad-X0-Y3}.center) -- ({P-Cd-X5-Y3}.center);

    \draw[very thick, dotted, Lavender] ({P-Bd-X0-Y0}.center) -- ({P-Dd-X5-Y0}.center);
    \draw[very thick, dotted, Lavender] ({P-Bd-X0-Y1}.center) -- ({P-Dd-X5-Y1}.center);
    \draw[very thick, dotted, Lavender] ({P-Bd-X0-Y2}.center) -- ({P-Dd-X5-Y2}.center);
    \draw[very thick, dotted, Lavender] ({P-Bd-X0-Y3}.center) -- ({P-Dd-X5-Y3}.center);

    \draw[thick, Lavender] ({P-Au-X0-Y0}.center) -- ({P-Cu-X5-Y0}.center);
    \draw[thick, Lavender] ({P-Au-X0-Y1}.center) -- ({P-Cu-X5-Y1}.center);
    \draw[thick, Lavender] ({P-Au-X0-Y2}.center) -- ({P-Cu-X5-Y2}.center);
    \draw[thick, Lavender] ({P-Au-X0-Y3}.center) -- ({P-Cu-X5-Y3}.center);

    \draw[very thick, dotted, Lavender] ({P-Bu-X0-Y0}.center) -- ({P-Du-X5-Y0}.center);
    \draw[very thick, dotted, Lavender] ({P-Bu-X0-Y1}.center) -- ({P-Du-X5-Y1}.center);
    \draw[very thick, dotted, Lavender] ({P-Bu-X0-Y2}.center) -- ({P-Du-X5-Y2}.center);
    \draw[very thick, dotted, Lavender] ({P-Bu-X0-Y3}.center) -- ({P-Du-X5-Y3}.center);
\end{scope}

    \foreach \x in {0,...,5} {
        \foreach \y in {0,...,3} {
        \begin{pgfonlayer}{deepbg}
            \draw[thin, fill opacity=0.5] ($({P-Du-X\x-Y\y}.center)!0.67!({P-Dd-X\x-Y\y}.center)$) -- ({P-Du-X\x-Y\y}.center);
        \end{pgfonlayer}
        \begin{pgfonlayer}{main}
            \draw[thin, fill opacity=0.5] ($({P-Du-X\x-Y\y}.center)!0.67!({P-Dd-X\x-Y\y}.center)$) -- ({P-Dd-X\x-Y\y}.center) 
            node[pos=1, circle, fill, inner sep=0.8pt] {};
        \end{pgfonlayer}
        \node[circle, fill, inner sep=0.8pt, fill opacity=0.5] at (P-Du-X\x-Y\y) {};
        }
    }

\draw [decorate, decoration={brace,amplitude=3pt,mirror,raise=1ex}]
  ({P-Cu-X0-Y0}.center) -- ({P-Au-X5-Y0}.center) node[midway, basis, rotate=47, yshift=0.5cm]{\scriptsize HH Lattice surgery in progress};

\draw [decorate, decoration={brace,amplitude=3pt,mirror,raise=1ex}]
  ({P-Ad-X5-Y0}.center) -- ($({P-Bd-X5-Y3}.center)+(\s,0)$) node[midway, basis, yshift=-0.5cm]{\scriptsize VV Lattice surgery in progress};

\end{scope}

\begin{scope}[xshift=-3cm, yshift=4.1cm]
    
    \def\hue{100}
    \def\s{0.8}
    \drawpatch{\s/2}{\s/2}{Au}{1}{GoldDot}{solid}{Lavender!\hue}{Salmon!\hue}{PaleGreen!\hue}

\end{scope}

\node[align=center, anchor=west] at (-4,5.5) {$(a)$ Planar honeycomb code patch};
\node[align=center, anchor=west] at (3,5.5) {$(b)$ Biplanar honeycomb layout};

\begin{scope}[xshift=-1.5cm, yshift=-0.75cm]

    \node[align=center] at (\s/2, \s) {Measurement schedule};
    
    \draw[very thick, Lavender] (-2*\s,0) -- (-\s,0) node[pos=0, fill=GoldDot, circle, inner sep=1.8pt] {} node[pos=1, fill=GoldDot, circle, inner sep=1.8pt] {} node[midway, above] {$ZZ$} node[pos=1, right] {\color{black}$\to$};
    
    \draw[very thick, Salmon] (0,0) -- (\s,0) node[pos=0, fill=GoldDot, circle, inner sep=1.8pt] {} node[pos=1, fill=GoldDot, circle, inner sep=1.8pt] {} node[midway, above] {$XX$} node[pos=1, right] {\color{black}$\to$}; 

    \draw[very thick, Lavender] (2*\s,0) -- (3*\s,0) node[pos=0, fill=GoldDot, circle, inner sep=1.8pt] {} node[pos=1, fill=GoldDot, circle, inner sep=1.8pt] {} node[midway, above] {$ZZ$} node[midway, below, yshift=-1pt] {\color{black}$\downarrow$};

    \draw[very thick, PaleGreen] (2*\s,-\s) -- (3*\s,-\s) node[pos=0, fill=GoldDot, circle, inner sep=1.8pt] {} node[pos=1, fill=GoldDot, circle, inner sep=1.8pt] {} node[midway, below] {$YY$} node[pos=0, left] {\color{black}$\leftarrow$};

    \draw[very thick, Salmon] (0,-\s) -- (\s,-\s) node[pos=0, fill=GoldDot, circle, inner sep=1.8pt] {} node[pos=1, fill=GoldDot, circle, inner sep=1.8pt] {} node[midway, below] {$XX$} node[pos=0, left] {\color{black}$\leftarrow$}; 

    \draw[very thick, PaleGreen] (-2*\s,-\s) -- (-\s,-\s) node[pos=0, fill=GoldDot, circle, inner sep=1.8pt] {} node[pos=1, fill=GoldDot, circle, inner sep=1.8pt] {} node[midway, below] {$YY$} node[midway, above, yshift=1pt] {\color{black}$\uparrow$};
\end{scope}

\end{tikzpicture}

%% file: lattice_surgery.tex
\begin{tikzpicture}[line join=round, line cap=round]
\setcounter{figure}{2}
\pgfdeclarelayer{deepbg}
\pgfsetlayers{deepbg,background,main}
\tikzset{
  basis/.style={
    x={(1cm,0cm)},
    y={(0.35cm,0.35cm)}
  }
}

\definecolor{Lavender}{HTML}{8B86FF}
\definecolor{Salmon}{HTML}{C97C6C}
\definecolor{PaleGreen}{HTML}{9ED9C7}
\definecolor{GoldDot}{HTML}{F2D07B}

\def\s{0.6}   
\def\Nx{3}    
\def\Ny{3}    
\pgfmathsetmacro{\rt}{1.7320508075688772} 
\pgfmathsetmacro{\hhex}{0.5*\rt*\s}       

\def\faceop{0.80}
\def\edgelw{0.65pt}
\def\dotr{0.07}
\def\borderlw{0.8pt}

\pgfmathsetmacro{\pad}{0.80*\s}
\def\drop{3} 

\pgfmathsetmacro{\xmin}{-\s}
\pgfmathsetmacro{\xmax}{1.5*\s*\Nx + \s}

\pgfmathsetmacro{\ymin}{-\hhex}
\pgfmathsetmacro{\ymax}{\rt*\s*(\Ny+0.5) + \hhex}

\pgfmathsetmacro{\W}{(\xmax-\xmin) + 2*\pad}
\pgfmathsetmacro{\H}{(\ymax-\ymin) + 2*\pad}

\pgfmathsetmacro{\dx}{\pad - \xmin}
\pgfmathsetmacro{\dy}{\pad - \ymin}

\newcommand{\hexpath}[2]{%
  ({#1+\s/2},{#2})
  \foreach \a in {80,120} { -- ({#1+\s*cos(\a)},{#2+\s*sin(\a)}) } 
  -- ({#1-\s/2},{#2})
  \foreach \a in  {240,300} { -- ({#1+\s*cos(\a)},{#2+\s*sin(\a)}) }
  -- cycle
}

\newcommand{\drawhex}[3]{
  \path[fill=#3, fill fill opacity=\faceop,
        draw=black, draw fill opacity=1, line width=\edgelw]
    \hexpath{#1}{#2};
  \fill[GoldDot] ({#1+\s/2},{#2}) circle (\dotr); 
  \fill[GoldDot] ({#1-\s/2},{#2}) circle (\dotr); 
  \foreach \a in {80,120,240,300}{
    \fill[GoldDot] ({#1+\s*cos(\a)},{#2+\s*sin(\a)}) circle (\dotr);
  }
}

\newcommand{\pickcolor}[3]{
  \pgfmathtruncatemacro{\halfq}{int(#1/2)}
  \pgfmathtruncatemacro{\rax}{#2-\halfq}
  \pgfmathtruncatemacro{\m}{mod(#1+2*\rax,3)}
  \ifnum\m=0 \def#3{Lavender}\fi
  \ifnum\m=1 \def#3{Salmon}\fi
  \ifnum\m=2 \def#3{PaleGreen}\fi
}

\newcommand{\centerxy}[4]{
  \pgfmathsetmacro{#3}{\s*(#1)}
  \pgfmathsetmacro{#4}{\rt*\s*((#2)+mod(#1,2)/2)}
}

\newcommand{\drawplane}[1]{%
  \path[line width=\borderlw, fill fill opacity=#1]
    (0,0) -- (\W,0) -- (\W,\H) -- (0,\H) -- cycle;

  \begin{scope}[reset cm]
    \path[basis] (0,0) coordinate (L0);
    \path[basis] (0,\H) coordinate (L1);

    \path[basis] (\W,0) coordinate (R0);


  \end{scope}
}

\newcommand{\drawpatch}[9]{

\foreach \x in {0,...,5} {
    \pgfmathtruncatemacro{\shift}{(\x==1) || (\x==2) || (\x==3)}
    \foreach \y in {0,...,3} {
        \node (P-#3-X\x-Y\y) at ({#2+\s*(\y+\shift)}, {#1-\s*\x}) {};
        \filldraw[fill=#5, draw=GoldDot, thick, opacity=#4] (P-#3-X\x-Y\y) circle (\dotr);
    }

}

\begin{scope}[on background layer]
    \fill[fill=#7, fill opacity=#4] ({P-#3-X0-Y2}.center) -- ({P-#3-X0-Y3}.center) -- ({P-#3-X1-Y2}.center) -- ({P-#3-X2-Y2}.center) -- ({P-#3-X2-Y1}.center) -- ({P-#3-X1-Y1}.center) -- cycle;
    \fill[fill=#7, fill opacity=#4] ({P-#3-X3-Y0}.center) -- ({P-#3-X3-Y1}.center) -- ({P-#3-X5-Y2}.center) -- ({P-#3-X5-Y1}.center) -- cycle;
    \fill[fill=#7, fill opacity=#4] ({P-#3-X3-Y2}.center) -- ({P-#3-X3-Y3}.center) -- ({P-#3-X5-Y3}.center) -- cycle;
    \fill[fill=#7, fill opacity=#4] ({P-#3-X0-Y0}.center) -- ({P-#3-X0-Y1}.center) -- ({P-#3-X2-Y0}.center) -- cycle;
    \fill[fill=#8, fill opacity=#4] ({P-#3-X0-Y1}.center) -- ({P-#3-X0-Y2}.center) -- ({P-#3-X1-Y1}.center) -- ({P-#3-X1-Y0}.center) -- cycle;
    \fill[fill=#8, fill opacity=#4] ({P-#3-X0-Y3}.center) -- ({P-#3-X1-Y3}.center) -- ({P-#3-X1-Y2}.center) -- cycle;
    \fill[fill=#8, fill opacity=#4] ({P-#3-X2-Y1}.center) -- ({P-#3-X2-Y2}.center) -- ({P-#3-X4-Y3}.center) -- ({P-#3-X4-Y2}.center) -- cycle;
    \fill[fill=#8, fill opacity=#4] ({P-#3-X2-Y0}.center) -- ({P-#3-X4-Y1}.center) -- ({P-#3-X4-Y0}.center) -- cycle;
    \fill[fill=#9, fill opacity=#4] ({P-#3-X1-Y0}.center) -- ({P-#3-X1-Y1}.center) -- ({P-#3-X2-Y1}.center) -- ({P-#3-X3-Y1}.center) -- ({P-#3-X3-Y0}.center) -- cycle;
    \fill[fill=#9, fill opacity=#4] ({P-#3-X1-Y2}.center) -- ({P-#3-X1-Y3}.center) -- ({P-#3-X2-Y3}.center) -- ({P-#3-X3-Y3}.center) -- ({P-#3-X3-Y2}.center) -- cycle;
    \fill[fill=#9, fill opacity=#4] ({P-#3-X4-Y2}.center) -- ({P-#3-X4-Y3}.center) -- ({P-#3-X5-Y3}.center) -- ({P-#3-X5-Y2}.center)-- cycle;
    \fill[fill=#9, fill opacity=#4] ({P-#3-X4-Y0}.center) -- ({P-#3-X4-Y1}.center) -- ({P-#3-X5-Y1}.center) -- ({P-#3-X5-Y0}.center)-- cycle;
\end{scope}

\begin{scope}[on background layer]
\draw[very thick, Salmon, #6, draw opacity=#4] ({P-#3-X0-Y0}.center) -- ({P-#3-X0-Y1}.center);
\draw[very thick, Salmon, #6, draw opacity=#4] ({P-#3-X0-Y2}.center) -- ({P-#3-X0-Y3}.center);
\draw[very thick, Salmon, #6, draw opacity=#4] ({P-#3-X1-Y0}.center) -- ({P-#3-X2-Y0}.center);
\draw[very thick, Salmon, #6, draw opacity=#4] ({P-#3-X1-Y1}.center) -- ({P-#3-X2-Y1}.center);
\draw[very thick, Salmon, #6, draw opacity=#4] ({P-#3-X1-Y2}.center) -- ({P-#3-X2-Y2}.center);
\draw[very thick, Salmon, #6, draw opacity=#4] ({P-#3-X1-Y3}.center) -- ({P-#3-X2-Y3}.center);
\draw[very thick, Salmon, #6, draw opacity=#4] ({P-#3-X3-Y0}.center) -- ({P-#3-X3-Y1}.center);
\draw[very thick, Salmon, #6, draw opacity=#4] ({P-#3-X3-Y2}.center) -- ({P-#3-X3-Y3}.center);
\draw[very thick, Salmon, #6, draw opacity=#4] ({P-#3-X4-Y0}.center) -- ({P-#3-X5-Y0}.center);
\draw[very thick, Salmon, #6, draw opacity=#4] ({P-#3-X4-Y1}.center) -- ({P-#3-X5-Y1}.center);
\draw[very thick, Salmon, #6, draw opacity=#4] ({P-#3-X4-Y2}.center) -- ({P-#3-X5-Y2}.center);
\draw[very thick, Salmon, #6, draw opacity=#4] ({P-#3-X4-Y3}.center) -- ({P-#3-X5-Y3}.center);
\draw[very thick, Lavender, #6, draw opacity=#4] ({P-#3-X1-Y0}.center) -- ({P-#3-X1-Y1}.center);
\draw[very thick, Lavender, #6, draw opacity=#4] ({P-#3-X1-Y2}.center) -- ({P-#3-X1-Y3}.center);
\draw[very thick, Lavender, #6, draw opacity=#4] ({P-#3-X2-Y0}.center) -- ({P-#3-X3-Y0}.center);
\draw[very thick, Lavender, #6, draw opacity=#4] ({P-#3-X2-Y1}.center) -- ({P-#3-X3-Y1}.center);
\draw[very thick, Lavender, #6, draw opacity=#4] ({P-#3-X2-Y2}.center) -- ({P-#3-X3-Y2}.center);
\draw[very thick, Lavender, #6, draw opacity=#4] ({P-#3-X2-Y3}.center) -- ({P-#3-X3-Y3}.center);
\draw[very thick, Lavender, #6, draw opacity=#4] ({P-#3-X4-Y0}.center) -- ({P-#3-X4-Y1}.center);
\draw[very thick, Lavender, #6, draw opacity=#4] ({P-#3-X4-Y2}.center) -- ({P-#3-X4-Y3}.center);
\draw[very thick, PaleGreen, #6, draw opacity=#4] ({P-#3-X0-Y1}.center) -- ({P-#3-X1-Y0}.center);
\draw[very thick, PaleGreen, #6, draw opacity=#4] ({P-#3-X0-Y2}.center) -- ({P-#3-X1-Y1}.center);
\draw[very thick, PaleGreen, #6, draw opacity=#4] ({P-#3-X0-Y3}.center) -- ({P-#3-X1-Y2}.center);
\draw[very thick, PaleGreen, #6, draw opacity=#4] ({P-#3-X2-Y1}.center) -- ({P-#3-X2-Y2}.center);
\draw[very thick, PaleGreen, #6, draw opacity=#4] ({P-#3-X3-Y0}.center) -- ({P-#3-X4-Y1}.center);
\draw[very thick, PaleGreen, #6, draw opacity=#4] ({P-#3-X3-Y1}.center) -- ({P-#3-X4-Y2}.center);
\draw[very thick, PaleGreen, #6, draw opacity=#4] ({P-#3-X3-Y2}.center) -- ({P-#3-X4-Y3}.center);
\draw[very thick, PaleGreen, #6, draw opacity=#4] ({P-#3-X5-Y1}.center) -- ({P-#3-X5-Y2}.center);
\end{scope}
}

\newcommand{\drawlspatch}[2]{
    \drawpatch{#1}{(-3.5*#1/(10*\s))*\s}{Ad}{#2}{GoldDot}{solid}{Lavender!80}{Salmon!80}{PaleGreen!80}
    \drawpatch{#1}{(4-3.5*#1/(10*\s))*\s}{Bd}{#2}{GoldDot}{solid}{Lavender!80}{Salmon!80}{PaleGreen!80}
    \drawpatch{#1}{(8-3.5*#1/(10*\s))*\s}{Cd}{#2}{GoldDot}{solid}{Lavender!80}{Salmon!80}{PaleGreen!80}

\begin{pgfonlayer}{deepbg}

    \fill[Salmon!80, fill opacity=#2] ({P-Ad-X2-Y3}.center) -- ({P-Bd-X2-Y0}.center) -- ({P-Bd-X4-Y1}.center)  -- ({P-Bd-X4-Y0}.center) -- cycle;
    
    \fill[Salmon!80, fill opacity=#2] ({P-Ad-X0-Y3}.center) -- ({P-Bd-X0-Y0}.center) -- ({P-Ad-X1-Y3}.center)  -- ({P-Ad-X1-Y2}.center) -- cycle;

    \fill[Lavender!80, fill opacity=#2] ({P-Ad-X3-Y2}.center) -- ({P-Ad-X3-Y3}.center) -- ({P-Bd-X5-Y0}.center) -- ({P-Ad-X5-Y3}.center) --cycle;

    \fill[Lavender!80, fill opacity=#2] ({P-Bd-X0-Y0}.center) -- ({P-Bd-X0-Y1}.center) -- ({P-Bd-X2-Y0}.center) -- ({P-Ad-X2-Y3}.center) --cycle;

    \fill[Salmon!80, fill opacity=#2] ({P-Bd-X2-Y3}.center) -- ({P-Cd-X2-Y0}.center) -- ({P-Cd-X4-Y1}.center)  -- ({P-Cd-X4-Y0}.center) -- cycle;
    
    \fill[Salmon!80, fill opacity=#2] ({P-Bd-X0-Y3}.center) -- ({P-Cd-X0-Y0}.center) -- ({P-Bd-X1-Y3}.center)  -- ({P-Bd-X1-Y2}.center) -- cycle;

    \fill[Lavender!80, fill opacity=#2] ({P-Bd-X3-Y2}.center) -- ({P-Bd-X3-Y3}.center) -- ({P-Cd-X5-Y0}.center) -- ({P-Bd-X5-Y3}.center) --cycle;

    \fill[Lavender!80, fill opacity=#2] ({P-Cd-X0-Y0}.center) -- ({P-Cd-X0-Y1}.center) -- ({P-Cd-X2-Y0}.center) -- ({P-Bd-X2-Y3}.center) --cycle;
\end{pgfonlayer}
    
    \draw[thick, PaleGreen, opacity=#2] ({P-Ad-X1-Y3}.center) -- ({P-Bd-X0-Y0}.center);
    \draw[thick, PaleGreen, opacity=#2] ({P-Ad-X2-Y3}.center) -- ({P-Bd-X2-Y0}.center);
    \draw[thick, PaleGreen, opacity=#2] ({P-Ad-X3-Y3}.center) -- ({P-Bd-X4-Y0}.center);
    \draw[thick, PaleGreen, opacity=#2] ({P-Ad-X5-Y3}.center) -- ({P-Bd-X5-Y0}.center);
    
    \draw[thick, PaleGreen, opacity=#2] ({P-Bd-X1-Y3}.center) -- ({P-Cd-X0-Y0}.center);
    \draw[thick, PaleGreen, opacity=#2] ({P-Bd-X2-Y3}.center) -- ({P-Cd-X2-Y0}.center);
    \draw[thick, PaleGreen, opacity=#2] ({P-Bd-X3-Y3}.center) -- ({P-Cd-X4-Y0}.center);
    \draw[thick, PaleGreen, opacity=#2] ({P-Bd-X5-Y3}.center) -- ({P-Cd-X5-Y0}.center);
}


\begin{scope}[xshift=-5cm, yshift=2.5cm]

    \node at (0, 4.5) {\refstepcounter{subfigure}\label{fig:lattice_surgery_circuit}(\alph{subfigure})  Circuit representation};
    \begin{scope}[xshift=-0.5cm, yshift=0.25cm]
        \draw[-stealth, thick] (-0.6, 0) -- (-0.6, 3.3);
    \draw[dashed] (-0.7, 0.1) -- (2, 0.1) node[pos=0, left] {$t_0$};
    \draw[dashed] (-0.7, 1.1) -- (2, 1.1) node[pos=0, left] {$t_1$};
    \draw[dashed] (-0.7, 2.1) -- (2, 2.1) node[pos=0, left] {$t_2$};
    \draw[dashed] (-0.7, 3.1) -- (2, 3.1) node[pos=0, left] {$t_3$};
    \draw[thick] (0,0) -- (0,3.2) node[pos=0, below] {$\ket{q_0}$};
    \draw[thick] (1.5,0) -- (1.5,3.2) node[pos=0, below] {$\ket{q_1}$};

    \draw[thick, fill=white] (-0.4, 0.2) rectangle (0.4, 1); 
    \draw[thick, fill=white] (-0.4, 1.2) rectangle (1.9, 2); 
    
    \begin{scope}[xshift=1.5cm]
        \draw[thick, fill=white] (-0.4, 0.2) rectangle (0.4, 1); 
    \end{scope}

    \begin{scope}[xshift=1.5cm, yshift=2cm]
        \draw[thick, fill=white] (-0.4, 0.2) rectangle (0.4, 1); 
    \end{scope}
    \begin{scope}[yshift=2cm]
        \draw[thick, fill=white] (-0.4, 0.2) rectangle (0.4, 1); 
    \end{scope}

    \node at (0, 0.6) {$\Id$};
    \node at (1.5, 0.6) {$\Id$};
    \node at (0, 2.6) {$\Id$};
    \node at (1.5, 2.6) {$\Id$};
    \node at (0.75, 1.6) {$\MZZ$};
    \end{scope}
    
\end{scope}

\begin{scope}[xshift=-4cm, yshift=-3cm]
    \node[align=center] at (0.25,4) {\refstepcounter{subfigure}\label{fig:lattice_surgery_pipe}(\alph{subfigure}) Spacetime pipe diagram representation};
    \begin{scope}[xscale=-1]
         
        \fill[Salmon!40] (1,2,0) -- (2,2,0) -- (2,3,0) -- (1.25, 3, -1) -- (1.25, 2, -1) -- (0.25, 2, -1) -- cycle;
        \fill[Salmon!40] (2,0,0) -- (2, 1, 0) -- (1.25, 1, -1) -- (1.25, 0, -1) -- cycle;
        
        \fill[CornflowerBlue!40] (0,0,0) -- (0, 3, 0) -- (-0.75, 3, -1) -- (0.25, 3, -1) -- (1, 3,0) -- (1, 2, 0) -- (2,2,0) -- (2,3,0) -- (1.25, 3, -1) -- (2.25, 3, -1) -- (3,3,0) -- (3, 0,0) --(2,0,0) -- (2,1,0) -- (1,1,0) -- (1,0,0) -- cycle; 
    
        \fill[Salmon!40] (0,0,0) -- (-0.75, 0, -1)  -- (-0.75, 3, -1) -- (0, 3, 0) -- cycle;
        \draw[dotted] (0,1) -- (1,1);
        \draw[dotted] (0,2) -- (1,2);
        \draw[dotted] (2,1) -- (3,1);
        \draw[dotted] (2,2) -- (3,2);
        \draw[dotted] (1,1) -- (1,2);
        \draw[dotted] (2,1) -- (2,2);
        \draw[dotted] (-0.75, 1,-1) -- (0, 1,0);
        \draw[dotted] (-0.75, 2,-1) -- (0, 2,0);
        \draw[dotted] (1.25, 2,-1) -- (2, 2,0);
    
        \draw[purple, very thick] (0,0) -- (0,3);
        \draw[purple, very thick] (1.25,0, -1) -- (1.25,0.6,-1);
        \draw[purple, very thick] (-0.75, 0, -1) -- (-0.75, 3, -1);
        \draw[purple, very thick] (1,0) -- (1,1) -- (2,1) -- (2,0);
        \draw[purple, very thick] (1,3) -- (1,2) -- (2,2) -- (2,3);
        \draw[purple, very thick] (3,0) -- (3,3);
        \draw[purple, very thick] (0.65,2,-1) -- (1.25,2,-1) -- (1.25,3,-1);
        \draw[purple, very thick] (0.25,3,-1) -- (0.25,2.99,-1);
        \draw[purple, very thick] (2.25,3,-1) -- (2.25,2.99,-1);
    \end{scope}

    \begin{scope}[basis, yshift=1.5cm, xshift=4cm, scale=0.4]
        \drawpatch{0}{0}{Ad}{1}{GoldDot}{solid}{Lavender!80}{Salmon!80}{PaleGreen!80}
        
        \drawpatch{2*\s}{-3.5*0.2*\s}{Au}{0.2}{GoldDot}{solid}{Lavender!80}{Salmon!80}{PaleGreen!80}
        
        \drawpatch{4*\s}{-3.5*0.4*\s}{Au}{0.2}{GoldDot}{solid}{Lavender!80}{Salmon!80}{PaleGreen!80}
        \drawpatch{6*\s}{-3.5*0.6*\s}{Au}{0.2}{GoldDot}{solid}{Lavender!80}{Salmon!80}{PaleGreen!80}

        \drawpatch{8*\s}{-3.5*0.8*\s}{Au}{0.2}{GoldDot}{solid}{Lavender!80}{Salmon!80}{PaleGreen!80}

        \drawpatch{10*\s}{-3.5*1*\s}{Au}{1}{GoldDot}{solid}{Lavender!80}{Salmon!80}{PaleGreen!80}

    \end{scope}

    \begin{scope}[xshift=2.25cm, yshift=1cm, xscale=-0.8, yscale=0.8]
         \fill[Salmon!40] (0,0,0) -- (-0.75, 0, -1) -- (-0.75, 1, -1) -- (0, 1, -1) -- cycle; 
         \fill[CornflowerBlue!40] (0,0,0) -- (0, 1, 0) -- (-0.75, 1, -1) -- (0.25, 1, -1) -- (1, 1,0) -- (1,0,0) -- cycle; 

        \draw[purple, very thick] (0,0) -- (0,1);
        \draw[purple, very thick] (-0.75,0,-1) -- (-0.75,1,-1);
        \draw[purple, very thick] (1,0,0) -- (1,1,0);
        \draw[purple, very thick] (0.25,1,-1) -- (0.25,0.99,-1);
         \node at (-1,0.7) {$\leftrightarrow$}; 
         \draw[<->] (0,-0.2) -- (1,-0.2) node[midway, below] {$w$};
         \draw[<->] (1.2, 0) -- (1.2,1) node[midway, left] {$r$};
         \draw[<->] (-0.2, 0, 0) -- (-0.95, 0, -1) node[midway, below right] {$h$};
    \end{scope}

\end{scope}

\tikzset{cedge/.style={draw=#1}
}
\begin{scope}[xshift=-1.5cm, yshift=2.5cm]
    \node[align=center] at (1, 4.3) {\refstepcounter{subfigure}\label{fig:lattice_surgery_macro}(\alph{subfigure}) Macroscopic representation\\(Game of Surface Code)};
    \begin{scope}[scale=0.7]
        \fill[lightgray!50] (0,0) -- (1,0) -- (1,1) -- (0,1) -- cycle;
        \draw[very thick, blue] (0,0) -- (1,0);
        \draw[very thick, blue] (1,1) -- (0,1);
        \draw[very thick, red] (1,0) -- (1,1);
        \draw[very thick, red] (0,1) -- (0,0);
    \end{scope}
    \begin{scope}[scale=0.7, xshift=2cm]
        \fill[lightgray!50] (0,0) -- (1,0) -- (1,1) -- (0,1) -- cycle;
        \draw[very thick, blue] (0,0) -- (1,0);
        \draw[very thick, blue] (1,1) -- (0,1);
        \draw[very thick, red] (1,0) -- (1,1);
        \draw[very thick, red] (0,1) -- (0,0);
    \end{scope}

    \node at (1,-0.5) {$t_0\leq t< t_1$};
    
    \begin{scope}[scale=0.7, yshift=2cm]
        \fill[lightgray!50] (0,0) -- (3,0) -- (3,1) -- (0,1) -- cycle;
        \draw[very thick, blue] (0,0) -- (3,0);
        \draw[very thick, blue] (3,1) -- (0,1);
        \draw[very thick, red] (3,0) -- (3,1);
        \draw[very thick, red] (0,1) -- (0,0);
        \node at (1.5,-0.5) {$t_1\leq t< t_2$};
    \end{scope}

    \begin{scope}[scale=0.7, yshift=4cm]
        \fill[lightgray!50] (0,0) -- (1,0) -- (1,1) -- (0,1) -- cycle;
        \draw[very thick, blue] (0,0) -- (1,0);
        \draw[very thick, blue] (1,1) -- (0,1);
        \draw[very thick, red] (1,0) -- (1,1);
        \draw[very thick, red] (0,1) -- (0,0);
        \node at (1.5,-0.5) {$t_2\leq t< t_3$};
    \end{scope}
    \begin{scope}[scale=0.7, xshift=2cm, yshift=4cm]
        \fill[lightgray!50] (0,0) -- (1,0) -- (1,1) -- (0,1) -- cycle;
        \draw[very thick, blue] (0,0) -- (1,0);
        \draw[very thick, blue] (1,1) -- (0,1);
        \draw[very thick, red] (1,0) -- (1,1);
        \draw[very thick, red] (0,1) -- (0,0);
    \end{scope}
    
\end{scope}

\begin{scope}[xshift=4cm, yshift=-1cm]
\draw[-stealth, thick] (-1.5,-1.25) -- (-1.5,6);
\draw[thick] (-1.6, -1.1) -- (-1.4,-1.1) node[pos=0, left] {$t_0$};
\draw[thick] (-1.6, 1) -- (-1.4, 1) node[pos=0, left] {$t_1$};
\draw[thick] (-1.6, 3.1) -- (-1.4, 3.1) node[pos=0, left] {$t_2$};
\draw[thick] (-1.6, 5.2) -- (-1.4, 5.2) node[pos=0, left] {$t_3$};
\end{scope}

\begin{scope}[basis, xshift=4cm, yshift=-1cm]

\node at (-6, 22.75) {\refstepcounter{subfigure}\label{fig:lattice_surgery_micro}(\alph{subfigure}) Microscopic honeycomb code representation};

\drawpatch{0}{0}{Ad}{1}{GoldDot}{solid}{Lavender!80}{Salmon!80}{PaleGreen!80}

\drawpatch{0}{8*\s}{Bd}{1}{GoldDot}{solid}{Lavender!80}{Salmon!80}{PaleGreen!80}

\drawpatch{2*\s}{-3.5*0.2*\s}{Au}{0.2}{GoldDot}{solid}{Lavender!80}{Salmon!80}{PaleGreen!80}
\drawpatch{2*\s}{(8-3.5*0.2)*\s}{Bu}{0.2}{GoldDot}{solid}{Lavender!80}{Salmon!80}{PaleGreen!80}
\drawpatch{4*\s}{-3.5*0.4*\s}{Au}{0.2}{GoldDot}{solid}{Lavender!80}{Salmon!80}{PaleGreen!80}
\drawpatch{4*\s}{(8-3.5*0.4)*\s}{Bu}{0.2}{GoldDot}{solid}{Lavender!80}{Salmon!80}{PaleGreen!80}
\drawpatch{6*\s}{-3.5*0.6*\s}{Au}{0.2}{GoldDot}{solid}{Lavender!80}{Salmon!80}{PaleGreen!80}
\drawpatch{6*\s}{(8-3.5*0.6)*\s}{Bu}{0.2}{GoldDot}{solid}{Lavender!80}{Salmon!80}{PaleGreen!80}
\drawpatch{8*\s}{-3.5*0.8*\s}{Au}{0.2}{GoldDot}{solid}{Lavender!80}{Salmon!80}{PaleGreen!80}
\drawpatch{8*\s}{(8-3.5*0.8)*\s}{Bu}{0.2}{GoldDot}{solid}{Lavender!80}{Salmon!80}{PaleGreen!80}

\drawlspatch{10*\s}{1}
\drawlspatch{12*\s}{0.2}
\drawlspatch{14*\s}{0.2}
\drawlspatch{16*\s}{0.2}
\drawlspatch{18*\s}{0.2}

\drawpatch{20*\s}{-3.5*2*\s}{Au}{1}{GoldDot}{solid}{Lavender!80}{Salmon!80}{PaleGreen!80}
\drawpatch{20*\s}{(8-3.5*2)*\s}{Bu}{1}{GoldDot}{solid}{Lavender!80}{Salmon!80}{PaleGreen!80}
\drawpatch{22*\s}{-3.5*2.2*\s}{Au}{0.2}{GoldDot}{solid}{Lavender!80}{Salmon!80}{PaleGreen!80}
\drawpatch{22*\s}{(8-3.5*2.2)*\s}{Bu}{0.2}{GoldDot}{solid}{Lavender!80}{Salmon!80}{PaleGreen!80}
\drawpatch{24*\s}{-3.5*2.4*\s}{Au}{0.2}{GoldDot}{solid}{Lavender!80}{Salmon!80}{PaleGreen!80}
\drawpatch{24*\s}{(8-3.5*2.4)*\s}{Bu}{0.2}{GoldDot}{solid}{Lavender!80}{Salmon!80}{PaleGreen!80}
\drawpatch{26*\s}{-3.5*2.6*\s}{Au}{0.2}{GoldDot}{solid}{Lavender!80}{Salmon!80}{PaleGreen!80}
\drawpatch{26*\s}{(8-3.5*2.6)*\s}{Bu}{0.2}{GoldDot}{solid}{Lavender!80}{Salmon!80}{PaleGreen!80}
\drawpatch{28*\s}{-3.5*2.8*\s}{Au}{0.2}{GoldDot}{solid}{Lavender!80}{Salmon!80}{PaleGreen!80}
\drawpatch{28*\s}{(8-3.5*2.8)*\s}{Bu}{0.2}{GoldDot}{solid}{Lavender!80}{Salmon!80}{PaleGreen!80}

\end{scope}

\end{tikzpicture}

%% file: crumble_url_transversal_cnot.tex
\newcommand{\crumbleTransversalCnot}{\href{http://algassert.com/crumble\#circuit=Q(0.0,0.0)0;Q(0.0,4.0)1;Q(0.0,5.0)2;Q(1.0,0.0)3;Q(1.0,1.0)4;Q(1.0,2.0)5;Q(1.0,3.0)6;Q(1.0,4.0)7;Q(1.0,5.0)8;Q(2.0,0.0)9;Q(2.0,1.0)10;Q(2.0,2.0)11;Q(2.0,3.0)12;Q(2.0,4.0)13;Q(2.0,5.0)14;Q(3.0,0.0)15;Q(3.0,1.0)16;Q(3.0,2.0)17;Q(3.0,3.0)18;Q(3.0,4.0)19;Q(3.0,5.0)20;Q(4.0,1.0)21;Q(4.0,2.0)22;Q(4.0,3.0)23;Q(7.0,0.0)24;Q(7.0,4.0)25;Q(7.0,5.0)26;Q(8.0,0.0)27;Q(8.0,1.0)28;Q(8.0,2.0)29;Q(8.0,3.0)30;Q(8.0,4.0)31;Q(8.0,5.0)32;Q(9.0,0.0)33;Q(9.0,1.0)34;Q(9.0,2.0)35;Q(9.0,3.0)36;Q(9.0,4.0)37;Q(9.0,5.0)38;Q(10.0,0.0)39;Q(10.0,1.0)40;Q(10.0,2.0)41;Q(10.0,3.0)42;Q(10.0,4.0)43;Q(10.0,5.0)44;Q(11.0,1.0)45;Q(11.0,2.0)46;Q(11.0,3.0)47;POLYGON(1,0,0,0.25)_0;POLYGON(1,0,0,0.25)_2;POLYGON(1,0,0,0.25)_3_4_10_9;POLYGON(1,0,0,0.25)_8_14;POLYGON(1,0,0,0.25)_15_16_21;POLYGON(1,0,0,0.25)_20;POLYGON(1,0,0,0.25)_1_7_6_5;POLYGON(1,0,0,0.25)_11_12_13_19_18_17;POLYGON(1,0,0,0.25)_22_23;POLYGON(1,0,0,0.25)_24;POLYGON(1,0,0,0.25)_26;POLYGON(1,0,0,0.25)_27_28_34_33;POLYGON(1,0,0,0.25)_32_38;POLYGON(1,0,0,0.25)_39_40_45;POLYGON(1,0,0,0.25)_44;POLYGON(1,0,0,0.25)_25_31_30_29;POLYGON(1,0,0,0.25)_35_36_37_43_42_41;POLYGON(1,0,0,0.25)_46_47;POLYGON(0,1,0,0.25)_0_3;POLYGON(0,1,0,0.25)_1_2_8_7;POLYGON(0,1,0,0.25)_9_15;POLYGON(0,1,0,0.25)_4_5_6_12_11_10;POLYGON(0,1,0,0.25)_13_14_20_19;POLYGON(0,1,0,0.25)_16_17_18_23_22_21;POLYGON(0,1,0,0.25)_24_27;POLYGON(0,1,0,0.25)_25_26_32_31;POLYGON(0,1,0,0.25)_33_39;POLYGON(0,1,0,0.25)_28_29_30_36_35_34;POLYGON(0,1,0,0.25)_37_38_44_43;POLYGON(0,1,0,0.25)_40_41_42_47_46_45;POLYGON(0,0,1,0.25)_0_5_4_3;POLYGON(0,0,1,0.25)_9_10_11_17_16_15;POLYGON(0,0,1,0.25)_21_22;POLYGON(0,0,1,0.25)_2_1;POLYGON(0,0,1,0.25)_6_7_8_14_13_12;POLYGON(0,0,1,0.25)_18_19_20_23;POLYGON(0,0,1,0.25)_24_29_28_27;POLYGON(0,0,1,0.25)_33_34_35_41_40_39;POLYGON(0,0,1,0.25)_45_46;POLYGON(0,0,1,0.25)_26_25;POLYGON(0,0,1,0.25)_30_31_32_38_37_36;POLYGON(0,0,1,0.25)_42_43_44_47;R_0_1_2_3_4_5_6_7_8_9_10_11_12_13_14_15_16_17_18_19_20_21_22_23_24_25_26_27_28_29_30_31_32_33_34_35_36_37_38_39_40_41_42_43_44_45_46_47;TICK;H_YZ_0_1_2_3_4_5_6_7_8_9_10_11_12_13_14_15_16_17_18_19_20_21_22_23_24_25_26_27_28_29_30_31_32_33_34_35_36_37_38_39_40_41_42_43_44_45_46_47;SHIFT_COORDS(0,0,1);SHIFT_COORDS(0,0,1);TICK;MZZ_1_7_4_10_5_6_11_12_13_19_16_21_17_18_22_23_25_31_28_34_29_30_35_36_37_43_40_45_41_42_46_47;M_0_2_3_8_9_14_15_20_24_26_27_32_33_38_39_44;OBSERVABLE_INCLUDE(0)_rec[-32]_rec[-28];OBSERVABLE_INCLUDE(1)_rec[-24]_rec[-20];TICK;H_0_1_2_3_4_5_6_7_8_9_10_11_12_13_14_15_16_17_18_19_20_21_22_23_24_25_26_27_28_29_30_31_32_33_34_35_36_37_38_39_40_41_42_43_44_45_46_47;TICK;MZZ_0_3_1_2_4_5_6_12_7_8_9_15_10_11_13_14_16_17_18_23_19_20_21_22_24_27_25_26_28_29_30_36_31_32_33_39_34_35_37_38_40_41_42_47_43_44_45_46;DETECTOR(1.5,2,0)_rec[-55]_rec[-54]_rec[-53]_rec[-22]_rec[-21]_rec[-18];DETECTOR(3.5,2,0)_rec[-51]_rec[-50]_rec[-49]_rec[-16]_rec[-15]_rec[-13];DETECTOR(8.5,2,0)_rec[-47]_rec[-46]_rec[-45]_rec[-10]_rec[-9]_rec[-6];DETECTOR(10.5,2,0)_rec[-43]_rec[-42]_rec[-41]_rec[-4]_rec[-3]_rec[-1];TICK;H_0_1_2_3_4_5_6_7_8_9_10_11_12_13_14_15_16_17_18_19_20_21_22_23_24_25_26_27_28_29_30_31_32_33_34_35_36_37_38_39_40_41_42_43_44_45_46_47;SHIFT_COORDS(0,0,1);TICK;MZZ_1_7_4_10_5_6_11_12_13_19_16_21_17_18_22_23_25_31_28_34_29_30_35_36_37_43_40_45_41_42_46_47;M_0_2_3_8_9_14_15_20_24_26_27_32_33_38_39_44;OBSERVABLE_INCLUDE(0)_rec[-32]_rec[-28];OBSERVABLE_INCLUDE(1)_rec[-24]_rec[-20];DETECTOR(1.5,2,0)_rec[-87]_rec[-86]_rec[-85]_rec[-31]_rec[-30]_rec[-29];DETECTOR(3.5,2,0)_rec[-83]_rec[-82]_rec[-81]_rec[-27]_rec[-26]_rec[-25];DETECTOR(0.5,-1,0)_rec[-72]_rec[-70]_rec[-16]_rec[-14];DETECTOR(0.5,5,0)_rec[-88]_rec[-71]_rec[-69]_rec[-32]_rec[-15]_rec[-13];DETECTOR(2.5,-1,0)_rec[-68]_rec[-66]_rec[-12]_rec[-10];DETECTOR(2.5,5,0)_rec[-84]_rec[-67]_rec[-65]_rec[-28]_rec[-11]_rec[-9];DETECTOR(8.5,2,0)_rec[-79]_rec[-78]_rec[-77]_rec[-23]_rec[-22]_rec[-21];DETECTOR(10.5,2,0)_rec[-75]_rec[-74]_rec[-73]_rec[-19]_rec[-18]_rec[-17];DETECTOR(7.5,-1,0)_rec[-64]_rec[-62]_rec[-8]_rec[-6];DETECTOR(7.5,5,0)_rec[-80]_rec[-63]_rec[-61]_rec[-24]_rec[-7]_rec[-5];DETECTOR(9.5,-1,0)_rec[-60]_rec[-58]_rec[-4]_rec[-2];DETECTOR(9.5,5,0)_rec[-76]_rec[-59]_rec[-57]_rec[-20]_rec[-3]_rec[-1];SHIFT_COORDS(0,0,1);TICK;H_YZ_0_1_2_3_4_5_6_7_8_9_10_11_12_13_14_15_16_17_18_19_20_21_22_23_24_25_26_27_28_29_30_31_32_33_34_35_36_37_38_39_40_41_42_43_44_45_46_47;TICK;MZZ_3_4_6_7_8_14_9_10_11_17_12_13_15_16_18_19_27_28_30_31_32_38_33_34_35_41_36_37_39_40_42_43;M_0_1_2_5_20_21_22_23_24_25_26_29_44_45_46_47;OBSERVABLE_INCLUDE(0)_rec[-30]_rec[-14]_rec[-12];OBSERVABLE_INCLUDE(1)_rec[-22]_rec[-6]_rec[-4];DETECTOR(2.5,3,0)_rec[-117]_rec[-116]_rec[-114]_rec[-61]_rec[-60]_rec[-58]_rec[-28]_rec[-27]_rec[-25];DETECTOR(0.515625,3.01562,0)_rec[-120]_rec[-118]_rec[-64]_rec[-62]_rec[-31]_rec[-15]_rec[-13];DETECTOR(4.46875,2.96875,0)_rec[-113]_rec[-57]_rec[-10]_rec[-9];DETECTOR(9.5,3,0)_rec[-109]_rec[-108]_rec[-106]_rec[-53]_rec[-52]_rec[-50]_rec[-20]_rec[-19]_rec[-17];DETECTOR(7.51562,3.01562,0)_rec[-112]_rec[-110]_rec[-56]_rec[-54]_rec[-23]_rec[-7]_rec[-5];DETECTOR(11.4688,2.96875,0)_rec[-105]_rec[-49]_rec[-2]_rec[-1];TICK;C_NZYX_0_1_2_3_4_5_6_7_8_9_10_11_12_13_14_15_16_17_18_19_20_21_22_23_24_25_26_27_28_29_30_31_32_33_34_35_36_37_38_39_40_41_42_43_44_45_46_47;SHIFT_COORDS(0,0,1);TICK;MZZ_0_3_1_2_4_5_6_12_7_8_9_15_10_11_13_14_16_17_18_23_19_20_21_22_24_27_25_26_28_29_30_36_31_32_33_39_34_35_37_38_40_41_42_47_43_44_45_46;TICK;C_XYNZ_0_1_2_3_4_5_6_7_8_9_10_11_12_13_14_15_16_17_18_19_20_21_22_23_24_25_26_27_28_29_30_31_32_33_34_35_36_37_38_39_40_41_42_43_44_45_46_47;TICK;MZZ_3_4_6_7_8_14_9_10_11_17_12_13_15_16_18_19_27_28_30_31_32_38_33_34_35_41_36_37_39_40_42_43;M_0_1_2_5_20_21_22_23_24_25_26_29_44_45_46_47;OBSERVABLE_INCLUDE(0)_rec[-30]_rec[-14]_rec[-12];OBSERVABLE_INCLUDE(1)_rec[-22]_rec[-6]_rec[-4];DETECTOR(2.5,1,0)_rec[-85]_rec[-84]_rec[-82]_rec[-29]_rec[-28]_rec[-26];DETECTOR(1.5,4,0)_rec[-87]_rec[-86]_rec[-83]_rec[-31]_rec[-30]_rec[-27];DETECTOR(0.5,1,0)_rec[-88]_rec[-72]_rec[-69]_rec[-32]_rec[-16]_rec[-13];DETECTOR(4.5,1,0)_rec[-67]_rec[-66]_rec[-11]_rec[-10];DETECTOR(-0.5,4,0)_rec[-71]_rec[-70]_rec[-15]_rec[-14];DETECTOR(3.5,4,0)_rec[-81]_rec[-68]_rec[-65]_rec[-25]_rec[-12]_rec[-9];DETECTOR(9.5,1,0)_rec[-77]_rec[-76]_rec[-74]_rec[-21]_rec[-20]_rec[-18];DETECTOR(8.5,4,0)_rec[-79]_rec[-78]_rec[-75]_rec[-23]_rec[-22]_rec[-19];DETECTOR(7.5,1,0)_rec[-80]_rec[-64]_rec[-61]_rec[-24]_rec[-8]_rec[-5];DETECTOR(11.5,1,0)_rec[-59]_rec[-58]_rec[-3]_rec[-2];DETECTOR(6.5,4,0)_rec[-63]_rec[-62]_rec[-7]_rec[-6];DETECTOR(10.5,4,0)_rec[-73]_rec[-60]_rec[-57]_rec[-17]_rec[-4]_rec[-1];TICK;H_YZ_0_1_2_3_4_5_6_7_8_9_10_11_12_13_14_15_16_17_18_19_20_21_22_23_24_25_26_27_28_29_30_31_32_33_34_35_36_37_38_39_40_41_42_43_44_45_46_47;SHIFT_COORDS(0,0,1);TICK;MZZ_1_7_4_10_5_6_11_12_13_19_16_21_17_18_22_23_25_31_28_34_29_30_35_36_37_43_40_45_41_42_46_47;M_0_2_3_8_9_14_15_20_24_26_27_32_33_38_39_44;OBSERVABLE_INCLUDE(0)_rec[-32]_rec[-28];OBSERVABLE_INCLUDE(1)_rec[-24]_rec[-20];DETECTOR(2.5,3,0)_rec[-149]_rec[-148]_rec[-146]_rec[-116]_rec[-115]_rec[-113]_rec[-60]_rec[-59]_rec[-57]_rec[-29]_rec[-28]_rec[-26];DETECTOR(1.5,0,0)_rec[-151]_rec[-134]_rec[-132]_rec[-120]_rec[-117]_rec[-64]_rec[-61]_rec[-31]_rec[-14]_rec[-12];DETECTOR(1.5,6,0)_rec[-133]_rec[-131]_rec[-118]_rec[-62]_rec[-13]_rec[-11];DETECTOR(9.5,3,0)_rec[-141]_rec[-140]_rec[-138]_rec[-108]_rec[-107]_rec[-105]_rec[-52]_rec[-51]_rec[-49]_rec[-21]_rec[-20]_rec[-18];DETECTOR(8.5,0,0)_rec[-143]_rec[-126]_rec[-124]_rec[-112]_rec[-109]_rec[-56]_rec[-53]_rec[-23]_rec[-6]_rec[-4];DETECTOR(8.5,6,0)_rec[-125]_rec[-123]_rec[-110]_rec[-54]_rec[-5]_rec[-3];SHIFT_COORDS(0,0,1);TICK;H_0_1_2_3_4_5_6_7_8_9_10_11_12_13_14_15_16_17_18_19_20_21_22_23_24_25_26_27_28_29_30_31_32_33_34_35_36_37_38_39_40_41_42_43_44_45_46_47;TICK;MZZ_0_3_1_2_4_5_6_12_7_8_9_15_10_11_13_14_16_17_18_23_19_20_21_22_24_27_25_26_28_29_30_36_31_32_33_39_34_35_37_38_40_41_42_47_43_44_45_46;DETECTOR(1.5,2,0)_rec[-198]_rec[-197]_rec[-194]_rec[-175]_rec[-174]_rec[-173]_rec[-55]_rec[-54]_rec[-53]_rec[-22]_rec[-21]_rec[-18];DETECTOR(3.5,2,0)_rec[-192]_rec[-191]_rec[-189]_rec[-171]_rec[-170]_rec[-169]_rec[-51]_rec[-50]_rec[-49]_rec[-16]_rec[-15]_rec[-13];DETECTOR(8.5,2,0)_rec[-186]_rec[-185]_rec[-182]_rec[-167]_rec[-166]_rec[-165]_rec[-47]_rec[-46]_rec[-45]_rec[-10]_rec[-9]_rec[-6];DETECTOR(10.5,2,0)_rec[-180]_rec[-179]_rec[-177]_rec[-163]_rec[-162]_rec[-161]_rec[-43]_rec[-42]_rec[-41]_rec[-4]_rec[-3]_rec[-1];TICK;H_0_1_2_3_4_5_6_7_8_9_10_11_12_13_14_15_16_17_18_19_20_21_22_23_24_25_26_27_28_29_30_31_32_33_34_35_36_37_38_39_40_41_42_43_44_45_46_47;SHIFT_COORDS(0,0,1);TICK;MZZ_1_7_4_10_5_6_11_12_13_19_16_21_17_18_22_23_25_31_28_34_29_30_35_36_37_43_40_45_41_42_46_47;M_0_2_3_8_9_14_15_20_24_26_27_32_33_38_39_44;OBSERVABLE_INCLUDE(0)_rec[-32]_rec[-28];OBSERVABLE_INCLUDE(1)_rec[-24]_rec[-20];DETECTOR(1.5,2,0)_rec[-87]_rec[-86]_rec[-85]_rec[-31]_rec[-30]_rec[-29];DETECTOR(3.5,2,0)_rec[-83]_rec[-82]_rec[-81]_rec[-27]_rec[-26]_rec[-25];DETECTOR(0.5,-1,0)_rec[-72]_rec[-70]_rec[-16]_rec[-14];DETECTOR(0.5,5,0)_rec[-88]_rec[-71]_rec[-69]_rec[-32]_rec[-15]_rec[-13];DETECTOR(2.5,-1,0)_rec[-68]_rec[-66]_rec[-12]_rec[-10];DETECTOR(2.5,5,0)_rec[-84]_rec[-67]_rec[-65]_rec[-28]_rec[-11]_rec[-9];DETECTOR(8.5,2,0)_rec[-79]_rec[-78]_rec[-77]_rec[-23]_rec[-22]_rec[-21];DETECTOR(10.5,2,0)_rec[-75]_rec[-74]_rec[-73]_rec[-19]_rec[-18]_rec[-17];DETECTOR(7.5,-1,0)_rec[-64]_rec[-62]_rec[-8]_rec[-6];DETECTOR(7.5,5,0)_rec[-80]_rec[-63]_rec[-61]_rec[-24]_rec[-7]_rec[-5];DETECTOR(9.5,-1,0)_rec[-60]_rec[-58]_rec[-4]_rec[-2];DETECTOR(9.5,5,0)_rec[-76]_rec[-59]_rec[-57]_rec[-20]_rec[-3]_rec[-1];SHIFT_COORDS(0,0,1);TICK;H_YZ_0_1_2_3_4_5_6_7_8_9_10_11_12_13_14_15_16_17_18_19_20_21_22_23_24_25_26_27_28_29_30_31_32_33_34_35_36_37_38_39_40_41_42_43_44_45_46_47;TICK;MZZ_3_4_6_7_8_14_9_10_11_17_12_13_15_16_18_19_27_28_30_31_32_38_33_34_35_41_36_37_39_40_42_43;M_0_1_2_5_20_21_22_23_24_25_26_29_44_45_46_47;OBSERVABLE_INCLUDE(0)_rec[-30]_rec[-14]_rec[-12];OBSERVABLE_INCLUDE(1)_rec[-22]_rec[-6]_rec[-4];DETECTOR(2.5,3,0)_rec[-148]_rec[-147]_rec[-145]_rec[-117]_rec[-116]_rec[-114]_rec[-61]_rec[-60]_rec[-58]_rec[-28]_rec[-27]_rec[-25];DETECTOR(0.5,3,0)_rec[-151]_rec[-135]_rec[-133]_rec[-120]_rec[-118]_rec[-64]_rec[-62]_rec[-31]_rec[-15]_rec[-13];DETECTOR(4.5,3,0)_rec[-130]_rec[-129]_rec[-113]_rec[-57]_rec[-10]_rec[-9];DETECTOR(9.5,3,0)_rec[-140]_rec[-139]_rec[-137]_rec[-109]_rec[-108]_rec[-106]_rec[-53]_rec[-52]_rec[-50]_rec[-20]_rec[-19]_rec[-17];DETECTOR(7.5,3,0)_rec[-143]_rec[-127]_rec[-125]_rec[-112]_rec[-110]_rec[-56]_rec[-54]_rec[-23]_rec[-7]_rec[-5];DETECTOR(11.5,3,0)_rec[-122]_rec[-121]_rec[-105]_rec[-49]_rec[-2]_rec[-1];TICK;C_NZYX_0_1_2_3_4_5_6_7_8_9_10_11_12_13_14_15_16_17_18_19_20_21_22_23_24_25_26_27_28_29_30_31_32_33_34_35_36_37_38_39_40_41_42_43_44_45_46_47;SHIFT_COORDS(0,0,1);TICK;MZZ_0_3_1_2_4_5_6_12_7_8_9_15_10_11_13_14_16_17_18_23_19_20_21_22_24_27_25_26_28_29_30_36_31_32_33_39_34_35_37_38_40_41_42_47_43_44_45_46;DETECTOR(2.5,1,0)_rec[-195]_rec[-194]_rec[-192]_rec[-173]_rec[-172]_rec[-170]_rec[-53]_rec[-52]_rec[-50]_rec[-19]_rec[-18]_rec[-16];DETECTOR(1.5,4,0)_rec[-197]_rec[-196]_rec[-193]_rec[-175]_rec[-174]_rec[-171]_rec[-55]_rec[-54]_rec[-51]_rec[-21]_rec[-20]_rec[-17];DETECTOR(9.5,1,0)_rec[-183]_rec[-182]_rec[-180]_rec[-165]_rec[-164]_rec[-162]_rec[-45]_rec[-44]_rec[-42]_rec[-7]_rec[-6]_rec[-4];DETECTOR(8.5,4,0)_rec[-185]_rec[-184]_rec[-181]_rec[-167]_rec[-166]_rec[-163]_rec[-47]_rec[-46]_rec[-43]_rec[-9]_rec[-8]_rec[-5];TICK;C_XYNZ_0_1_2_3_4_5_6_7_8_9_10_11_12_13_14_15_16_17_18_19_20_21_22_23_24_25_26_27_28_29_30_31_32_33_34_35_36_37_38_39_40_41_42_43_44_45_46_47;SHIFT_COORDS(0,0,1);TICK;MZZ_3_4_6_7_8_14_9_10_11_17_12_13_15_16_18_19_27_28_30_31_32_38_33_34_35_41_36_37_39_40_42_43;M_0_1_2_5_20_21_22_23_24_25_26_29_44_45_46_47;OBSERVABLE_INCLUDE(0)_rec[-30]_rec[-14]_rec[-12];OBSERVABLE_INCLUDE(1)_rec[-22]_rec[-6]_rec[-4];DETECTOR(2.5,1,0)_rec[-85]_rec[-84]_rec[-82]_rec[-29]_rec[-28]_rec[-26];DETECTOR(1.5,4,0)_rec[-87]_rec[-86]_rec[-83]_rec[-31]_rec[-30]_rec[-27];DETECTOR(0.5,1,0)_rec[-88]_rec[-72]_rec[-69]_rec[-32]_rec[-16]_rec[-13];DETECTOR(4.5,1,0)_rec[-67]_rec[-66]_rec[-11]_rec[-10];DETECTOR(-0.5,4,0)_rec[-71]_rec[-70]_rec[-15]_rec[-14];DETECTOR(3.5,4,0)_rec[-81]_rec[-68]_rec[-65]_rec[-25]_rec[-12]_rec[-9];DETECTOR(9.5,1,0)_rec[-77]_rec[-76]_rec[-74]_rec[-21]_rec[-20]_rec[-18];DETECTOR(8.5,4,0)_rec[-79]_rec[-78]_rec[-75]_rec[-23]_rec[-22]_rec[-19];DETECTOR(7.5,1,0)_rec[-80]_rec[-64]_rec[-61]_rec[-24]_rec[-8]_rec[-5];DETECTOR(11.5,1,0)_rec[-59]_rec[-58]_rec[-3]_rec[-2];DETECTOR(6.5,4,0)_rec[-63]_rec[-62]_rec[-7]_rec[-6];DETECTOR(10.5,4,0)_rec[-73]_rec[-60]_rec[-57]_rec[-17]_rec[-4]_rec[-1];TICK;H_YZ_0_1_2_3_4_5_6_7_8_9_10_11_12_13_14_15_16_17_18_19_20_21_22_23_24_25_26_27_28_29_30_31_32_33_34_35_36_37_38_39_40_41_42_43_44_45_46_47;SHIFT_COORDS(0,0,1);TICK;MZZ_1_7_4_10_5_6_11_12_13_19_16_21_17_18_22_23_25_31_28_34_29_30_35_36_37_43_40_45_41_42_46_47;M_0_2_3_8_9_14_15_20_24_26_27_32_33_38_39_44;OBSERVABLE_INCLUDE(0)_rec[-32]_rec[-28];OBSERVABLE_INCLUDE(1)_rec[-24]_rec[-20];DETECTOR(2.5,3,0)_rec[-149]_rec[-148]_rec[-146]_rec[-116]_rec[-115]_rec[-113]_rec[-60]_rec[-59]_rec[-57]_rec[-29]_rec[-28]_rec[-26];DETECTOR(1.5,0,0)_rec[-151]_rec[-134]_rec[-132]_rec[-120]_rec[-117]_rec[-64]_rec[-61]_rec[-31]_rec[-14]_rec[-12];DETECTOR(1.5,6,0)_rec[-133]_rec[-131]_rec[-118]_rec[-62]_rec[-13]_rec[-11];DETECTOR(9.5,3,0)_rec[-141]_rec[-140]_rec[-138]_rec[-108]_rec[-107]_rec[-105]_rec[-52]_rec[-51]_rec[-49]_rec[-21]_rec[-20]_rec[-18];DETECTOR(8.5,0,0)_rec[-143]_rec[-126]_rec[-124]_rec[-112]_rec[-109]_rec[-56]_rec[-53]_rec[-23]_rec[-6]_rec[-4];DETECTOR(8.5,6,0)_rec[-125]_rec[-123]_rec[-110]_rec[-54]_rec[-5]_rec[-3];SHIFT_COORDS(0,0,1);TICK;H_0_1_2_3_4_5_6_7_8_9_10_11_12_13_14_15_16_17_18_19_20_21_22_23_24_25_26_27_28_29_30_31_32_33_34_35_36_37_38_39_40_41_42_43_44_45_46_47;TICK;MZZ_0_3_1_2_4_5_6_12_7_8_9_15_10_11_13_14_16_17_18_23_19_20_21_22_24_27_25_26_28_29_30_36_31_32_33_39_34_35_37_38_40_41_42_47_43_44_45_46;DETECTOR(1.5,2,0)_rec[-198]_rec[-197]_rec[-194]_rec[-175]_rec[-174]_rec[-173]_rec[-55]_rec[-54]_rec[-53]_rec[-22]_rec[-21]_rec[-18];DETECTOR(3.5,2,0)_rec[-192]_rec[-191]_rec[-189]_rec[-171]_rec[-170]_rec[-169]_rec[-51]_rec[-50]_rec[-49]_rec[-16]_rec[-15]_rec[-13];DETECTOR(8.5,2,0)_rec[-186]_rec[-185]_rec[-182]_rec[-167]_rec[-166]_rec[-165]_rec[-47]_rec[-46]_rec[-45]_rec[-10]_rec[-9]_rec[-6];DETECTOR(10.5,2,0)_rec[-180]_rec[-179]_rec[-177]_rec[-163]_rec[-162]_rec[-161]_rec[-43]_rec[-42]_rec[-41]_rec[-4]_rec[-3]_rec[-1];TICK;H_0_1_2_3_4_5_6_7_8_9_10_11_12_13_14_15_16_17_18_19_20_21_22_23_24_25_26_27_28_29_30_31_32_33_34_35_36_37_38_39_40_41_42_43_44_45_46_47;SHIFT_COORDS(0,0,1);TICK;MZZ_1_7_4_10_5_6_11_12_13_19_16_21_17_18_22_23_25_31_28_34_29_30_35_36_37_43_40_45_41_42_46_47;M_0_2_3_8_9_14_15_20_24_26_27_32_33_38_39_44;OBSERVABLE_INCLUDE(0)_rec[-32]_rec[-28];OBSERVABLE_INCLUDE(1)_rec[-24]_rec[-20];DETECTOR(1.5,2,0)_rec[-87]_rec[-86]_rec[-85]_rec[-31]_rec[-30]_rec[-29];DETECTOR(3.5,2,0)_rec[-83]_rec[-82]_rec[-81]_rec[-27]_rec[-26]_rec[-25];DETECTOR(0.5,-1,0)_rec[-72]_rec[-70]_rec[-16]_rec[-14];DETECTOR(0.5,5,0)_rec[-88]_rec[-71]_rec[-69]_rec[-32]_rec[-15]_rec[-13];DETECTOR(2.5,-1,0)_rec[-68]_rec[-66]_rec[-12]_rec[-10];DETECTOR(2.5,5,0)_rec[-84]_rec[-67]_rec[-65]_rec[-28]_rec[-11]_rec[-9];DETECTOR(8.5,2,0)_rec[-79]_rec[-78]_rec[-77]_rec[-23]_rec[-22]_rec[-21];DETECTOR(10.5,2,0)_rec[-75]_rec[-74]_rec[-73]_rec[-19]_rec[-18]_rec[-17];DETECTOR(7.5,-1,0)_rec[-64]_rec[-62]_rec[-8]_rec[-6];DETECTOR(7.5,5,0)_rec[-80]_rec[-63]_rec[-61]_rec[-24]_rec[-7]_rec[-5];DETECTOR(9.5,-1,0)_rec[-60]_rec[-58]_rec[-4]_rec[-2];DETECTOR(9.5,5,0)_rec[-76]_rec[-59]_rec[-57]_rec[-20]_rec[-3]_rec[-1];SHIFT_COORDS(0,0,1);TICK;H_YZ_0_1_2_3_4_5_6_7_8_9_10_11_12_13_14_15_16_17_18_19_20_21_22_23_24_25_26_27_28_29_30_31_32_33_34_35_36_37_38_39_40_41_42_43_44_45_46_47;TICK;MZZ_3_4_6_7_8_14_9_10_11_17_12_13_15_16_18_19_27_28_30_31_32_38_33_34_35_41_36_37_39_40_42_43;M_0_1_2_5_20_21_22_23_24_25_26_29_44_45_46_47;OBSERVABLE_INCLUDE(0)_rec[-30]_rec[-14]_rec[-12];OBSERVABLE_INCLUDE(1)_rec[-22]_rec[-6]_rec[-4];DETECTOR(2.5,3,0)_rec[-148]_rec[-147]_rec[-145]_rec[-117]_rec[-116]_rec[-114]_rec[-61]_rec[-60]_rec[-58]_rec[-28]_rec[-27]_rec[-25];DETECTOR(0.5,3,0)_rec[-151]_rec[-135]_rec[-133]_rec[-120]_rec[-118]_rec[-64]_rec[-62]_rec[-31]_rec[-15]_rec[-13];DETECTOR(4.5,3,0)_rec[-130]_rec[-129]_rec[-113]_rec[-57]_rec[-10]_rec[-9];DETECTOR(9.5,3,0)_rec[-140]_rec[-139]_rec[-137]_rec[-109]_rec[-108]_rec[-106]_rec[-53]_rec[-52]_rec[-50]_rec[-20]_rec[-19]_rec[-17];DETECTOR(7.5,3,0)_rec[-143]_rec[-127]_rec[-125]_rec[-112]_rec[-110]_rec[-56]_rec[-54]_rec[-23]_rec[-7]_rec[-5];DETECTOR(11.5,3,0)_rec[-122]_rec[-121]_rec[-105]_rec[-49]_rec[-2]_rec[-1];TICK;C_NZYX_0_1_2_3_4_5_6_7_8_9_10_11_12_13_14_15_16_17_18_19_20_21_22_23_24_25_26_27_28_29_30_31_32_33_34_35_36_37_38_39_40_41_42_43_44_45_46_47;SHIFT_COORDS(0,0,1);TICK;MZZ_0_3_1_2_4_5_6_12_7_8_9_15_10_11_13_14_16_17_18_23_19_20_21_22_24_27_25_26_28_29_30_36_31_32_33_39_34_35_37_38_40_41_42_47_43_44_45_46;DETECTOR(2.5,1,0)_rec[-195]_rec[-194]_rec[-192]_rec[-173]_rec[-172]_rec[-170]_rec[-53]_rec[-52]_rec[-50]_rec[-19]_rec[-18]_rec[-16];DETECTOR(1.5,4,0)_rec[-197]_rec[-196]_rec[-193]_rec[-175]_rec[-174]_rec[-171]_rec[-55]_rec[-54]_rec[-51]_rec[-21]_rec[-20]_rec[-17];DETECTOR(9.5,1,0)_rec[-183]_rec[-182]_rec[-180]_rec[-165]_rec[-164]_rec[-162]_rec[-45]_rec[-44]_rec[-42]_rec[-7]_rec[-6]_rec[-4];DETECTOR(8.5,4,0)_rec[-185]_rec[-184]_rec[-181]_rec[-167]_rec[-166]_rec[-163]_rec[-47]_rec[-46]_rec[-43]_rec[-9]_rec[-8]_rec[-5];TICK;C_XYNZ_0_1_2_3_4_5_6_7_8_9_10_11_12_13_14_15_16_17_18_19_20_21_22_23_24_25_26_27_28_29_30_31_32_33_34_35_36_37_38_39_40_41_42_43_44_45_46_47;SHIFT_COORDS(0,0,1);TICK;MZZ_3_4_6_7_8_14_9_10_11_17_12_13_15_16_18_19_27_28_30_31_32_38_33_34_35_41_36_37_39_40_42_43;M_0_1_2_5_20_21_22_23_24_25_26_29_44_45_46_47;OBSERVABLE_INCLUDE(0)_rec[-30]_rec[-14]_rec[-12];OBSERVABLE_INCLUDE(1)_rec[-22]_rec[-6]_rec[-4];DETECTOR(2.5,1,0)_rec[-85]_rec[-84]_rec[-82]_rec[-29]_rec[-28]_rec[-26];DETECTOR(1.5,4,0)_rec[-87]_rec[-86]_rec[-83]_rec[-31]_rec[-30]_rec[-27];DETECTOR(0.5,1,0)_rec[-88]_rec[-72]_rec[-69]_rec[-32]_rec[-16]_rec[-13];DETECTOR(4.5,1,0)_rec[-67]_rec[-66]_rec[-11]_rec[-10];DETECTOR(-0.5,4,0)_rec[-71]_rec[-70]_rec[-15]_rec[-14];DETECTOR(3.5,4,0)_rec[-81]_rec[-68]_rec[-65]_rec[-25]_rec[-12]_rec[-9];DETECTOR(9.5,1,0)_rec[-77]_rec[-76]_rec[-74]_rec[-21]_rec[-20]_rec[-18];DETECTOR(8.5,4,0)_rec[-79]_rec[-78]_rec[-75]_rec[-23]_rec[-22]_rec[-19];DETECTOR(7.5,1,0)_rec[-80]_rec[-64]_rec[-61]_rec[-24]_rec[-8]_rec[-5];DETECTOR(11.5,1,0)_rec[-59]_rec[-58]_rec[-3]_rec[-2];DETECTOR(6.5,4,0)_rec[-63]_rec[-62]_rec[-7]_rec[-6];DETECTOR(10.5,4,0)_rec[-73]_rec[-60]_rec[-57]_rec[-17]_rec[-4]_rec[-1];SHIFT_COORDS(0,0,1);TICK;H_YZ_0_1_2_3_4_5_6_7_8_9_10_11_12_13_14_15_16_17_18_19_20_21_22_23;TICK;CZ_0_24_1_25_2_26_3_27_4_28_5_29_6_30_7_31_8_32_9_33_10_34_11_35_12_36_13_37_14_38_15_39_16_40_17_41_18_42_19_43_20_44_21_45_22_46_23_47;TICK;H_YZ_24_25_26_27_28_29_30_31_32_33_34_35_36_37_38_39_40_41_42_43_44_45_46_47;TICK;MZZ_1_7_4_10_5_6_11_12_13_19_16_21_17_18_22_23_25_31_28_34_29_30_35_36_37_43_40_45_41_42_46_47;M_0_2_3_8_9_14_15_20_24_26_27_32_33_38_39_44;OBSERVABLE_INCLUDE(1)_rec[-24]_rec[-20];OBSERVABLE_INCLUDE(0)_rec[-32]_rec[-28]_rec[-24]_rec[-20];DETECTOR(2.5,3,0)_rec[-149]_rec[-148]_rec[-146]_rec[-116]_rec[-115]_rec[-113]_rec[-60]_rec[-59]_rec[-57]_rec[-29]_rec[-28]_rec[-26];DETECTOR(1.5,0,0)_rec[-151]_rec[-134]_rec[-132]_rec[-120]_rec[-117]_rec[-64]_rec[-61]_rec[-31]_rec[-14]_rec[-12];DETECTOR(1.5,6,0)_rec[-133]_rec[-131]_rec[-118]_rec[-62]_rec[-13]_rec[-11];DETECTOR(7.75,3,0)_rec[-141]_rec[-140]_rec[-138]_rec[-108]_rec[-107]_rec[-105]_rec[-52]_rec[-51]_rec[-49]_rec[-29]_rec[-28]_rec[-26]_rec[-21]_rec[-20]_rec[-18];DETECTOR(6.75,0,0)_rec[-143]_rec[-126]_rec[-124]_rec[-112]_rec[-109]_rec[-56]_rec[-53]_rec[-31]_rec[-23]_rec[-14]_rec[-12]_rec[-6]_rec[-4];DETECTOR(6.75,6,0)_rec[-125]_rec[-123]_rec[-110]_rec[-54]_rec[-13]_rec[-11]_rec[-5]_rec[-3];SHIFT_COORDS(0,0,1);TICK;H_0_1_2_3_4_5_6_7_8_9_10_11_12_13_14_15_16_17_18_19_20_21_22_23_24_25_26_27_28_29_30_31_32_33_34_35_36_37_38_39_40_41_42_43_44_45_46_47;TICK;MZZ_0_3_1_2_4_5_6_12_7_8_9_15_10_11_13_14_16_17_18_23_19_20_21_22_24_27_25_26_28_29_30_36_31_32_33_39_34_35_37_38_40_41_42_47_43_44_45_46;DETECTOR(8.5,2,0)_rec[-186]_rec[-185]_rec[-182]_rec[-167]_rec[-166]_rec[-165]_rec[-47]_rec[-46]_rec[-45]_rec[-10]_rec[-9]_rec[-6];DETECTOR(10.5,2,0)_rec[-180]_rec[-179]_rec[-177]_rec[-163]_rec[-162]_rec[-161]_rec[-43]_rec[-42]_rec[-41]_rec[-4]_rec[-3]_rec[-1];DETECTOR(4.125,2,0)_rec[-198]_rec[-197]_rec[-194]_rec[-175]_rec[-174]_rec[-173]_rec[-55]_rec[-54]_rec[-53]_rec[-47]_rec[-46]_rec[-45]_rec[-22]_rec[-21]_rec[-18]_rec[-10]_rec[-9]_rec[-6];DETECTOR(6.125,2,0)_rec[-192]_rec[-191]_rec[-189]_rec[-171]_rec[-170]_rec[-169]_rec[-51]_rec[-50]_rec[-49]_rec[-43]_rec[-42]_rec[-41]_rec[-16]_rec[-15]_rec[-13]_rec[-4]_rec[-3]_rec[-1];TICK;H_0_1_2_3_4_5_6_7_8_9_10_11_12_13_14_15_16_17_18_19_20_21_22_23_24_25_26_27_28_29_30_31_32_33_34_35_36_37_38_39_40_41_42_43_44_45_46_47;SHIFT_COORDS(0,0,1);TICK;MZZ_1_7_4_10_5_6_11_12_13_19_16_21_17_18_22_23_25_31_28_34_29_30_35_36_37_43_40_45_41_42_46_47;M_0_2_3_8_9_14_15_20_24_26_27_32_33_38_39_44;OBSERVABLE_INCLUDE(1)_rec[-24]_rec[-20];OBSERVABLE_INCLUDE(0)_rec[-32]_rec[-28]_rec[-24]_rec[-20];DETECTOR(1.5,2,0)_rec[-87]_rec[-86]_rec[-85]_rec[-31]_rec[-30]_rec[-29];DETECTOR(3.5,2,0)_rec[-83]_rec[-82]_rec[-81]_rec[-27]_rec[-26]_rec[-25];DETECTOR(0.5,-1,0)_rec[-72]_rec[-70]_rec[-16]_rec[-14];DETECTOR(0.5,5,0)_rec[-88]_rec[-71]_rec[-69]_rec[-32]_rec[-15]_rec[-13];DETECTOR(2.5,-1,0)_rec[-68]_rec[-66]_rec[-12]_rec[-10];DETECTOR(2.5,5,0)_rec[-84]_rec[-67]_rec[-65]_rec[-28]_rec[-11]_rec[-9];DETECTOR(8.5,2,0)_rec[-79]_rec[-78]_rec[-77]_rec[-23]_rec[-22]_rec[-21];DETECTOR(10.5,2,0)_rec[-75]_rec[-74]_rec[-73]_rec[-19]_rec[-18]_rec[-17];DETECTOR(7.5,-1,0)_rec[-64]_rec[-62]_rec[-8]_rec[-6];DETECTOR(7.5,5,0)_rec[-80]_rec[-63]_rec[-61]_rec[-24]_rec[-7]_rec[-5];DETECTOR(9.5,-1,0)_rec[-60]_rec[-58]_rec[-4]_rec[-2];DETECTOR(9.5,5,0)_rec[-76]_rec[-59]_rec[-57]_rec[-20]_rec[-3]_rec[-1];SHIFT_COORDS(0,0,1);TICK;H_YZ_0_1_2_3_4_5_6_7_8_9_10_11_12_13_14_15_16_17_18_19_20_21_22_23_24_25_26_27_28_29_30_31_32_33_34_35_36_37_38_39_40_41_42_43_44_45_46_47;TICK;MZZ_3_4_6_7_8_14_9_10_11_17_12_13_15_16_18_19_27_28_30_31_32_38_33_34_35_41_36_37_39_40_42_43;M_0_1_2_5_20_21_22_23_24_25_26_29_44_45_46_47;OBSERVABLE_INCLUDE(1)_rec[-22]_rec[-6]_rec[-4];OBSERVABLE_INCLUDE(0)_rec[-30]_rec[-22]_rec[-14]_rec[-12]_rec[-6]_rec[-4];DETECTOR(9.5,3,0)_rec[-140]_rec[-139]_rec[-137]_rec[-109]_rec[-108]_rec[-106]_rec[-53]_rec[-52]_rec[-50]_rec[-20]_rec[-19]_rec[-17];DETECTOR(7.5,3,0)_rec[-143]_rec[-127]_rec[-125]_rec[-112]_rec[-110]_rec[-56]_rec[-54]_rec[-23]_rec[-7]_rec[-5];DETECTOR(11.5,3,0)_rec[-122]_rec[-121]_rec[-105]_rec[-49]_rec[-2]_rec[-1];DETECTOR(5.78125,3,0)_rec[-148]_rec[-147]_rec[-145]_rec[-117]_rec[-116]_rec[-114]_rec[-109]_rec[-108]_rec[-106]_rec[-61]_rec[-60]_rec[-58]_rec[-53]_rec[-52]_rec[-50]_rec[-28]_rec[-27]_rec[-25]_rec[-20]_rec[-19]_rec[-17];DETECTOR(3.78125,3,0)_rec[-151]_rec[-135]_rec[-133]_rec[-120]_rec[-118]_rec[-112]_rec[-110]_rec[-64]_rec[-62]_rec[-56]_rec[-54]_rec[-31]_rec[-23]_rec[-15]_rec[-13]_rec[-7]_rec[-5];DETECTOR(7.78125,3,0)_rec[-130]_rec[-129]_rec[-113]_rec[-105]_rec[-57]_rec[-49]_rec[-10]_rec[-9]_rec[-2]_rec[-1];TICK;C_NZYX_0_1_2_3_4_5_6_7_8_9_10_11_12_13_14_15_16_17_18_19_20_21_22_23_24_25_26_27_28_29_30_31_32_33_34_35_36_37_38_39_40_41_42_43_44_45_46_47;SHIFT_COORDS(0,0,1);TICK;MZZ_0_3_1_2_4_5_6_12_7_8_9_15_10_11_13_14_16_17_18_23_19_20_21_22_24_27_25_26_28_29_30_36_31_32_33_39_34_35_37_38_40_41_42_47_43_44_45_46;DETECTOR(2.5,1,0)_rec[-195]_rec[-194]_rec[-192]_rec[-173]_rec[-172]_rec[-170]_rec[-53]_rec[-52]_rec[-50]_rec[-19]_rec[-18]_rec[-16];DETECTOR(1.5,4,0)_rec[-197]_rec[-196]_rec[-193]_rec[-175]_rec[-174]_rec[-171]_rec[-55]_rec[-54]_rec[-51]_rec[-21]_rec[-20]_rec[-17];DETECTOR(6.10938,1,0)_rec[-183]_rec[-182]_rec[-180]_rec[-165]_rec[-164]_rec[-162]_rec[-53]_rec[-52]_rec[-50]_rec[-45]_rec[-44]_rec[-42]_rec[-19]_rec[-18]_rec[-16]_rec[-7]_rec[-6]_rec[-4];DETECTOR(5.10938,4,0)_rec[-185]_rec[-184]_rec[-181]_rec[-167]_rec[-166]_rec[-163]_rec[-55]_rec[-54]_rec[-51]_rec[-47]_rec[-46]_rec[-43]_rec[-21]_rec[-20]_rec[-17]_rec[-9]_rec[-8]_rec[-5];TICK;C_XYNZ_0_1_2_3_4_5_6_7_8_9_10_11_12_13_14_15_16_17_18_19_20_21_22_23_24_25_26_27_28_29_30_31_32_33_34_35_36_37_38_39_40_41_42_43_44_45_46_47;SHIFT_COORDS(0,0,1);TICK;MZZ_3_4_6_7_8_14_9_10_11_17_12_13_15_16_18_19_27_28_30_31_32_38_33_34_35_41_36_37_39_40_42_43;M_0_1_2_5_20_21_22_23_24_25_26_29_44_45_46_47;OBSERVABLE_INCLUDE(1)_rec[-22]_rec[-6]_rec[-4];OBSERVABLE_INCLUDE(0)_rec[-30]_rec[-22]_rec[-14]_rec[-12]_rec[-6]_rec[-4];DETECTOR(2.5,1,0)_rec[-85]_rec[-84]_rec[-82]_rec[-29]_rec[-28]_rec[-26];DETECTOR(1.5,4,0)_rec[-87]_rec[-86]_rec[-83]_rec[-31]_rec[-30]_rec[-27];DETECTOR(0.5,1,0)_rec[-88]_rec[-72]_rec[-69]_rec[-32]_rec[-16]_rec[-13];DETECTOR(4.5,1,0)_rec[-67]_rec[-66]_rec[-11]_rec[-10];DETECTOR(-0.5,4,0)_rec[-71]_rec[-70]_rec[-15]_rec[-14];DETECTOR(3.5,4,0)_rec[-81]_rec[-68]_rec[-65]_rec[-25]_rec[-12]_rec[-9];DETECTOR(9.5,1,0)_rec[-77]_rec[-76]_rec[-74]_rec[-21]_rec[-20]_rec[-18];DETECTOR(8.5,4,0)_rec[-79]_rec[-78]_rec[-75]_rec[-23]_rec[-22]_rec[-19];DETECTOR(7.5,1,0)_rec[-80]_rec[-64]_rec[-61]_rec[-24]_rec[-8]_rec[-5];DETECTOR(11.5,1,0)_rec[-59]_rec[-58]_rec[-3]_rec[-2];DETECTOR(6.5,4,0)_rec[-63]_rec[-62]_rec[-7]_rec[-6];DETECTOR(10.5,4,0)_rec[-73]_rec[-60]_rec[-57]_rec[-17]_rec[-4]_rec[-1];TICK;H_YZ_0_1_2_3_4_5_6_7_8_9_10_11_12_13_14_15_16_17_18_19_20_21_22_23_24_25_26_27_28_29_30_31_32_33_34_35_36_37_38_39_40_41_42_43_44_45_46_47;SHIFT_COORDS(0,0,1);TICK;MZZ_1_7_4_10_5_6_11_12_13_19_16_21_17_18_22_23_25_31_28_34_29_30_35_36_37_43_40_45_41_42_46_47;M_0_2_3_8_9_14_15_20_24_26_27_32_33_38_39_44;OBSERVABLE_INCLUDE(1)_rec[-24]_rec[-20];OBSERVABLE_INCLUDE(0)_rec[-32]_rec[-28]_rec[-24]_rec[-20];DETECTOR(2.5,3,0)_rec[-149]_rec[-148]_rec[-146]_rec[-116]_rec[-115]_rec[-113]_rec[-60]_rec[-59]_rec[-57]_rec[-29]_rec[-28]_rec[-26];DETECTOR(1.5,0,0)_rec[-151]_rec[-134]_rec[-132]_rec[-120]_rec[-117]_rec[-64]_rec[-61]_rec[-31]_rec[-14]_rec[-12];DETECTOR(1.5,6,0)_rec[-133]_rec[-131]_rec[-118]_rec[-62]_rec[-13]_rec[-11];DETECTOR(9.5,3,0)_rec[-141]_rec[-140]_rec[-138]_rec[-108]_rec[-107]_rec[-105]_rec[-52]_rec[-51]_rec[-49]_rec[-21]_rec[-20]_rec[-18];DETECTOR(8.5,0,0)_rec[-143]_rec[-126]_rec[-124]_rec[-112]_rec[-109]_rec[-56]_rec[-53]_rec[-23]_rec[-6]_rec[-4];DETECTOR(8.5,6,0)_rec[-125]_rec[-123]_rec[-110]_rec[-54]_rec[-5]_rec[-3];SHIFT_COORDS(0,0,1);TICK;H_0_1_2_3_4_5_6_7_8_9_10_11_12_13_14_15_16_17_18_19_20_21_22_23_24_25_26_27_28_29_30_31_32_33_34_35_36_37_38_39_40_41_42_43_44_45_46_47;TICK;MZZ_0_3_1_2_4_5_6_12_7_8_9_15_10_11_13_14_16_17_18_23_19_20_21_22_24_27_25_26_28_29_30_36_31_32_33_39_34_35_37_38_40_41_42_47_43_44_45_46;DETECTOR(1.5,2,0)_rec[-198]_rec[-197]_rec[-194]_rec[-175]_rec[-174]_rec[-173]_rec[-55]_rec[-54]_rec[-53]_rec[-22]_rec[-21]_rec[-18];DETECTOR(3.5,2,0)_rec[-192]_rec[-191]_rec[-189]_rec[-171]_rec[-170]_rec[-169]_rec[-51]_rec[-50]_rec[-49]_rec[-16]_rec[-15]_rec[-13];DETECTOR(8.5,2,0)_rec[-186]_rec[-185]_rec[-182]_rec[-167]_rec[-166]_rec[-165]_rec[-47]_rec[-46]_rec[-45]_rec[-10]_rec[-9]_rec[-6];DETECTOR(10.5,2,0)_rec[-180]_rec[-179]_rec[-177]_rec[-163]_rec[-162]_rec[-161]_rec[-43]_rec[-42]_rec[-41]_rec[-4]_rec[-3]_rec[-1];TICK;H_0_1_2_3_4_5_6_7_8_9_10_11_12_13_14_15_16_17_18_19_20_21_22_23_24_25_26_27_28_29_30_31_32_33_34_35_36_37_38_39_40_41_42_43_44_45_46_47;SHIFT_COORDS(0,0,1);TICK;MZZ_1_7_4_10_5_6_11_12_13_19_16_21_17_18_22_23_25_31_28_34_29_30_35_36_37_43_40_45_41_42_46_47;M_0_2_3_8_9_14_15_20_24_26_27_32_33_38_39_44;OBSERVABLE_INCLUDE(1)_rec[-24]_rec[-20];OBSERVABLE_INCLUDE(0)_rec[-32]_rec[-28]_rec[-24]_rec[-20];DETECTOR(1.5,2,0)_rec[-87]_rec[-86]_rec[-85]_rec[-31]_rec[-30]_rec[-29];DETECTOR(3.5,2,0)_rec[-83]_rec[-82]_rec[-81]_rec[-27]_rec[-26]_rec[-25];DETECTOR(0.5,-1,0)_rec[-72]_rec[-70]_rec[-16]_rec[-14];DETECTOR(0.5,5,0)_rec[-88]_rec[-71]_rec[-69]_rec[-32]_rec[-15]_rec[-13];DETECTOR(2.5,-1,0)_rec[-68]_rec[-66]_rec[-12]_rec[-10];DETECTOR(2.5,5,0)_rec[-84]_rec[-67]_rec[-65]_rec[-28]_rec[-11]_rec[-9];DETECTOR(8.5,2,0)_rec[-79]_rec[-78]_rec[-77]_rec[-23]_rec[-22]_rec[-21];DETECTOR(10.5,2,0)_rec[-75]_rec[-74]_rec[-73]_rec[-19]_rec[-18]_rec[-17];DETECTOR(7.5,-1,0)_rec[-64]_rec[-62]_rec[-8]_rec[-6];DETECTOR(7.5,5,0)_rec[-80]_rec[-63]_rec[-61]_rec[-24]_rec[-7]_rec[-5];DETECTOR(9.5,-1,0)_rec[-60]_rec[-58]_rec[-4]_rec[-2];DETECTOR(9.5,5,0)_rec[-76]_rec[-59]_rec[-57]_rec[-20]_rec[-3]_rec[-1];SHIFT_COORDS(0,0,1);TICK;H_YZ_0_1_2_3_4_5_6_7_8_9_10_11_12_13_14_15_16_17_18_19_20_21_22_23_24_25_26_27_28_29_30_31_32_33_34_35_36_37_38_39_40_41_42_43_44_45_46_47;TICK;MZZ_3_4_6_7_8_14_9_10_11_17_12_13_15_16_18_19_27_28_30_31_32_38_33_34_35_41_36_37_39_40_42_43;M_0_1_2_5_20_21_22_23_24_25_26_29_44_45_46_47;OBSERVABLE_INCLUDE(1)_rec[-22]_rec[-6]_rec[-4];OBSERVABLE_INCLUDE(0)_rec[-30]_rec[-22]_rec[-14]_rec[-12]_rec[-6]_rec[-4];DETECTOR(2.5,3,0)_rec[-148]_rec[-147]_rec[-145]_rec[-117]_rec[-116]_rec[-114]_rec[-61]_rec[-60]_rec[-58]_rec[-28]_rec[-27]_rec[-25];DETECTOR(0.5,3,0)_rec[-151]_rec[-135]_rec[-133]_rec[-120]_rec[-118]_rec[-64]_rec[-62]_rec[-31]_rec[-15]_rec[-13];DETECTOR(4.5,3,0)_rec[-130]_rec[-129]_rec[-113]_rec[-57]_rec[-10]_rec[-9];DETECTOR(9.5,3,0)_rec[-140]_rec[-139]_rec[-137]_rec[-109]_rec[-108]_rec[-106]_rec[-53]_rec[-52]_rec[-50]_rec[-20]_rec[-19]_rec[-17];DETECTOR(7.5,3,0)_rec[-143]_rec[-127]_rec[-125]_rec[-112]_rec[-110]_rec[-56]_rec[-54]_rec[-23]_rec[-7]_rec[-5];DETECTOR(11.5,3,0)_rec[-122]_rec[-121]_rec[-105]_rec[-49]_rec[-2]_rec[-1];TICK;C_NZYX_0_1_2_3_4_5_6_7_8_9_10_11_12_13_14_15_16_17_18_19_20_21_22_23_24_25_26_27_28_29_30_31_32_33_34_35_36_37_38_39_40_41_42_43_44_45_46_47;SHIFT_COORDS(0,0,1);TICK;MZZ_0_3_1_2_4_5_6_12_7_8_9_15_10_11_13_14_16_17_18_23_19_20_21_22_24_27_25_26_28_29_30_36_31_32_33_39_34_35_37_38_40_41_42_47_43_44_45_46;DETECTOR(2.5,1,0)_rec[-195]_rec[-194]_rec[-192]_rec[-173]_rec[-172]_rec[-170]_rec[-53]_rec[-52]_rec[-50]_rec[-19]_rec[-18]_rec[-16];DETECTOR(1.5,4,0)_rec[-197]_rec[-196]_rec[-193]_rec[-175]_rec[-174]_rec[-171]_rec[-55]_rec[-54]_rec[-51]_rec[-21]_rec[-20]_rec[-17];DETECTOR(9.5,1,0)_rec[-183]_rec[-182]_rec[-180]_rec[-165]_rec[-164]_rec[-162]_rec[-45]_rec[-44]_rec[-42]_rec[-7]_rec[-6]_rec[-4];DETECTOR(8.5,4,0)_rec[-185]_rec[-184]_rec[-181]_rec[-167]_rec[-166]_rec[-163]_rec[-47]_rec[-46]_rec[-43]_rec[-9]_rec[-8]_rec[-5];TICK;C_XYNZ_0_1_2_3_4_5_6_7_8_9_10_11_12_13_14_15_16_17_18_19_20_21_22_23_24_25_26_27_28_29_30_31_32_33_34_35_36_37_38_39_40_41_42_43_44_45_46_47;SHIFT_COORDS(0,0,1);TICK;MZZ_3_4_6_7_8_14_9_10_11_17_12_13_15_16_18_19_27_28_30_31_32_38_33_34_35_41_36_37_39_40_42_43;M_0_1_2_5_20_21_22_23_24_25_26_29_44_45_46_47;OBSERVABLE_INCLUDE(1)_rec[-22]_rec[-6]_rec[-4];OBSERVABLE_INCLUDE(0)_rec[-30]_rec[-22]_rec[-14]_rec[-12]_rec[-6]_rec[-4];DETECTOR(2.5,1,0)_rec[-85]_rec[-84]_rec[-82]_rec[-29]_rec[-28]_rec[-26];DETECTOR(1.5,4,0)_rec[-87]_rec[-86]_rec[-83]_rec[-31]_rec[-30]_rec[-27];DETECTOR(0.5,1,0)_rec[-88]_rec[-72]_rec[-69]_rec[-32]_rec[-16]_rec[-13];DETECTOR(4.5,1,0)_rec[-67]_rec[-66]_rec[-11]_rec[-10];DETECTOR(-0.5,4,0)_rec[-71]_rec[-70]_rec[-15]_rec[-14];DETECTOR(3.5,4,0)_rec[-81]_rec[-68]_rec[-65]_rec[-25]_rec[-12]_rec[-9];DETECTOR(9.5,1,0)_rec[-77]_rec[-76]_rec[-74]_rec[-21]_rec[-20]_rec[-18];DETECTOR(8.5,4,0)_rec[-79]_rec[-78]_rec[-75]_rec[-23]_rec[-22]_rec[-19];DETECTOR(7.5,1,0)_rec[-80]_rec[-64]_rec[-61]_rec[-24]_rec[-8]_rec[-5];DETECTOR(11.5,1,0)_rec[-59]_rec[-58]_rec[-3]_rec[-2];DETECTOR(6.5,4,0)_rec[-63]_rec[-62]_rec[-7]_rec[-6];DETECTOR(10.5,4,0)_rec[-73]_rec[-60]_rec[-57]_rec[-17]_rec[-4]_rec[-1];TICK;H_YZ_0_1_2_3_4_5_6_7_8_9_10_11_12_13_14_15_16_17_18_19_20_21_22_23_24_25_26_27_28_29_30_31_32_33_34_35_36_37_38_39_40_41_42_43_44_45_46_47;SHIFT_COORDS(0,0,1);TICK;MZZ_1_7_4_10_5_6_11_12_13_19_16_21_17_18_22_23_25_31_28_34_29_30_35_36_37_43_40_45_41_42_46_47;M_0_2_3_8_9_14_15_20_24_26_27_32_33_38_39_44;OBSERVABLE_INCLUDE(1)_rec[-24]_rec[-20];OBSERVABLE_INCLUDE(0)_rec[-32]_rec[-28]_rec[-24]_rec[-20];DETECTOR(2.5,3,0)_rec[-149]_rec[-148]_rec[-146]_rec[-116]_rec[-115]_rec[-113]_rec[-60]_rec[-59]_rec[-57]_rec[-29]_rec[-28]_rec[-26];DETECTOR(1.5,0,0)_rec[-151]_rec[-134]_rec[-132]_rec[-120]_rec[-117]_rec[-64]_rec[-61]_rec[-31]_rec[-14]_rec[-12];DETECTOR(1.5,6,0)_rec[-133]_rec[-131]_rec[-118]_rec[-62]_rec[-13]_rec[-11];DETECTOR(9.5,3,0)_rec[-141]_rec[-140]_rec[-138]_rec[-108]_rec[-107]_rec[-105]_rec[-52]_rec[-51]_rec[-49]_rec[-21]_rec[-20]_rec[-18];DETECTOR(8.5,0,0)_rec[-143]_rec[-126]_rec[-124]_rec[-112]_rec[-109]_rec[-56]_rec[-53]_rec[-23]_rec[-6]_rec[-4];DETECTOR(8.5,6,0)_rec[-125]_rec[-123]_rec[-110]_rec[-54]_rec[-5]_rec[-3];SHIFT_COORDS(0,0,1);TICK;H_0_1_2_3_4_5_6_7_8_9_10_11_12_13_14_15_16_17_18_19_20_21_22_23_24_25_26_27_28_29_30_31_32_33_34_35_36_37_38_39_40_41_42_43_44_45_46_47;TICK;MZZ_0_3_1_2_4_5_6_12_7_8_9_15_10_11_13_14_16_17_18_23_19_20_21_22_24_27_25_26_28_29_30_36_31_32_33_39_34_35_37_38_40_41_42_47_43_44_45_46;DETECTOR(1.5,2,0)_rec[-198]_rec[-197]_rec[-194]_rec[-175]_rec[-174]_rec[-173]_rec[-55]_rec[-54]_rec[-53]_rec[-22]_rec[-21]_rec[-18];DETECTOR(3.5,2,0)_rec[-192]_rec[-191]_rec[-189]_rec[-171]_rec[-170]_rec[-169]_rec[-51]_rec[-50]_rec[-49]_rec[-16]_rec[-15]_rec[-13];DETECTOR(8.5,2,0)_rec[-186]_rec[-185]_rec[-182]_rec[-167]_rec[-166]_rec[-165]_rec[-47]_rec[-46]_rec[-45]_rec[-10]_rec[-9]_rec[-6];DETECTOR(10.5,2,0)_rec[-180]_rec[-179]_rec[-177]_rec[-163]_rec[-162]_rec[-161]_rec[-43]_rec[-42]_rec[-41]_rec[-4]_rec[-3]_rec[-1];TICK;H_0_1_2_3_4_5_6_7_8_9_10_11_12_13_14_15_16_17_18_19_20_21_22_23_24_25_26_27_28_29_30_31_32_33_34_35_36_37_38_39_40_41_42_43_44_45_46_47;SHIFT_COORDS(0,0,1);TICK;MZZ_1_7_4_10_5_6_11_12_13_19_16_21_17_18_22_23_25_31_28_34_29_30_35_36_37_43_40_45_41_42_46_47;M_0_2_3_8_9_14_15_20_24_26_27_32_33_38_39_44;OBSERVABLE_INCLUDE(1)_rec[-24]_rec[-20];OBSERVABLE_INCLUDE(0)_rec[-32]_rec[-28]_rec[-24]_rec[-20];DETECTOR(1.5,2,0)_rec[-87]_rec[-86]_rec[-85]_rec[-31]_rec[-30]_rec[-29];DETECTOR(3.5,2,0)_rec[-83]_rec[-82]_rec[-81]_rec[-27]_rec[-26]_rec[-25];DETECTOR(0.5,-1,0)_rec[-72]_rec[-70]_rec[-16]_rec[-14];DETECTOR(0.5,5,0)_rec[-88]_rec[-71]_rec[-69]_rec[-32]_rec[-15]_rec[-13];DETECTOR(2.5,-1,0)_rec[-68]_rec[-66]_rec[-12]_rec[-10];DETECTOR(2.5,5,0)_rec[-84]_rec[-67]_rec[-65]_rec[-28]_rec[-11]_rec[-9];DETECTOR(8.5,2,0)_rec[-79]_rec[-78]_rec[-77]_rec[-23]_rec[-22]_rec[-21];DETECTOR(10.5,2,0)_rec[-75]_rec[-74]_rec[-73]_rec[-19]_rec[-18]_rec[-17];DETECTOR(7.5,-1,0)_rec[-64]_rec[-62]_rec[-8]_rec[-6];DETECTOR(7.5,5,0)_rec[-80]_rec[-63]_rec[-61]_rec[-24]_rec[-7]_rec[-5];DETECTOR(9.5,-1,0)_rec[-60]_rec[-58]_rec[-4]_rec[-2];DETECTOR(9.5,5,0)_rec[-76]_rec[-59]_rec[-57]_rec[-20]_rec[-3]_rec[-1];SHIFT_COORDS(0,0,1);TICK;H_YZ_0_1_2_3_4_5_6_7_8_9_10_11_12_13_14_15_16_17_18_19_20_21_22_23_24_25_26_27_28_29_30_31_32_33_34_35_36_37_38_39_40_41_42_43_44_45_46_47;TICK;MZZ_3_4_6_7_8_14_9_10_11_17_12_13_15_16_18_19_27_28_30_31_32_38_33_34_35_41_36_37_39_40_42_43;M_0_1_2_5_20_21_22_23_24_25_26_29_44_45_46_47;OBSERVABLE_INCLUDE(1)_rec[-22]_rec[-6]_rec[-4];OBSERVABLE_INCLUDE(0)_rec[-30]_rec[-22]_rec[-14]_rec[-12]_rec[-6]_rec[-4];DETECTOR(2.5,3,0)_rec[-148]_rec[-147]_rec[-145]_rec[-117]_rec[-116]_rec[-114]_rec[-61]_rec[-60]_rec[-58]_rec[-28]_rec[-27]_rec[-25];DETECTOR(0.5,3,0)_rec[-151]_rec[-135]_rec[-133]_rec[-120]_rec[-118]_rec[-64]_rec[-62]_rec[-31]_rec[-15]_rec[-13];DETECTOR(4.5,3,0)_rec[-130]_rec[-129]_rec[-113]_rec[-57]_rec[-10]_rec[-9];DETECTOR(9.5,3,0)_rec[-140]_rec[-139]_rec[-137]_rec[-109]_rec[-108]_rec[-106]_rec[-53]_rec[-52]_rec[-50]_rec[-20]_rec[-19]_rec[-17];DETECTOR(7.5,3,0)_rec[-143]_rec[-127]_rec[-125]_rec[-112]_rec[-110]_rec[-56]_rec[-54]_rec[-23]_rec[-7]_rec[-5];DETECTOR(11.5,3,0)_rec[-122]_rec[-121]_rec[-105]_rec[-49]_rec[-2]_rec[-1];TICK;C_NZYX_0_1_2_3_4_5_6_7_8_9_10_11_12_13_14_15_16_17_18_19_20_21_22_23_24_25_26_27_28_29_30_31_32_33_34_35_36_37_38_39_40_41_42_43_44_45_46_47;SHIFT_COORDS(0,0,1);TICK;MZZ_0_3_1_2_4_5_6_12_7_8_9_15_10_11_13_14_16_17_18_23_19_20_21_22_24_27_25_26_28_29_30_36_31_32_33_39_34_35_37_38_40_41_42_47_43_44_45_46;DETECTOR(2.5,1,0)_rec[-195]_rec[-194]_rec[-192]_rec[-173]_rec[-172]_rec[-170]_rec[-53]_rec[-52]_rec[-50]_rec[-19]_rec[-18]_rec[-16];DETECTOR(1.5,4,0)_rec[-197]_rec[-196]_rec[-193]_rec[-175]_rec[-174]_rec[-171]_rec[-55]_rec[-54]_rec[-51]_rec[-21]_rec[-20]_rec[-17];DETECTOR(9.5,1,0)_rec[-183]_rec[-182]_rec[-180]_rec[-165]_rec[-164]_rec[-162]_rec[-45]_rec[-44]_rec[-42]_rec[-7]_rec[-6]_rec[-4];DETECTOR(8.5,4,0)_rec[-185]_rec[-184]_rec[-181]_rec[-167]_rec[-166]_rec[-163]_rec[-47]_rec[-46]_rec[-43]_rec[-9]_rec[-8]_rec[-5];TICK;C_XYNZ_0_1_2_3_4_5_6_7_8_9_10_11_12_13_14_15_16_17_18_19_20_21_22_23_24_25_26_27_28_29_30_31_32_33_34_35_36_37_38_39_40_41_42_43_44_45_46_47;SHIFT_COORDS(0,0,1);TICK;MZZ_3_4_6_7_8_14_9_10_11_17_12_13_15_16_18_19_27_28_30_31_32_38_33_34_35_41_36_37_39_40_42_43;M_0_1_2_5_20_21_22_23_24_25_26_29_44_45_46_47;OBSERVABLE_INCLUDE(1)_rec[-22]_rec[-6]_rec[-4];OBSERVABLE_INCLUDE(0)_rec[-30]_rec[-22]_rec[-14]_rec[-12]_rec[-6]_rec[-4];DETECTOR(2.5,1,0)_rec[-85]_rec[-84]_rec[-82]_rec[-29]_rec[-28]_rec[-26];DETECTOR(1.5,4,0)_rec[-87]_rec[-86]_rec[-83]_rec[-31]_rec[-30]_rec[-27];DETECTOR(0.5,1,0)_rec[-88]_rec[-72]_rec[-69]_rec[-32]_rec[-16]_rec[-13];DETECTOR(4.5,1,0)_rec[-67]_rec[-66]_rec[-11]_rec[-10];DETECTOR(-0.5,4,0)_rec[-71]_rec[-70]_rec[-15]_rec[-14];DETECTOR(3.5,4,0)_rec[-81]_rec[-68]_rec[-65]_rec[-25]_rec[-12]_rec[-9];DETECTOR(9.5,1,0)_rec[-77]_rec[-76]_rec[-74]_rec[-21]_rec[-20]_rec[-18];DETECTOR(8.5,4,0)_rec[-79]_rec[-78]_rec[-75]_rec[-23]_rec[-22]_rec[-19];DETECTOR(7.5,1,0)_rec[-80]_rec[-64]_rec[-61]_rec[-24]_rec[-8]_rec[-5];DETECTOR(11.5,1,0)_rec[-59]_rec[-58]_rec[-3]_rec[-2];DETECTOR(6.5,4,0)_rec[-63]_rec[-62]_rec[-7]_rec[-6];DETECTOR(10.5,4,0)_rec[-73]_rec[-60]_rec[-57]_rec[-17]_rec[-4]_rec[-1];SHIFT_COORDS(0,0,1);TICK;M_0_1_2_3_4_5_6_7_8_9_10_11_12_13_14_15_16_17_18_19_20_21_22_23_24_25_26_27_28_29_30_31_32_33_34_35_36_37_38_39_40_41_42_43_44_45_46_47;DETECTOR(1,0.5,0)_rec[-80]_rec[-45]_rec[-44];DETECTOR(3,0.5,0)_rec[-74]_rec[-33]_rec[-32];DETECTOR(2.5,2,0)_rec[-76]_rec[-37]_rec[-31];DETECTOR(2,0.5,0)_rec[-77]_rec[-39]_rec[-38];DETECTOR(2,3.5,0)_rec[-75]_rec[-36]_rec[-35];DETECTOR(1.5,5,0)_rec[-78]_rec[-40]_rec[-34];DETECTOR(1,3.5,0)_rec[-79]_rec[-42]_rec[-41];DETECTOR(3,3.5,0)_rec[-73]_rec[-30]_rec[-29];DETECTOR(1,2,0)_rec[-61]_rec[-43];DETECTOR(0,0,0)_rec[-64]_rec[-48];DETECTOR(4,2,0)_rec[-58]_rec[-26];DETECTOR(4,1,0)_rec[-59]_rec[-27];DETECTOR(0,4,0)_rec[-63]_rec[-47];DETECTOR(0,5,0)_rec[-62]_rec[-46];DETECTOR(4,3,0)_rec[-57]_rec[-25];DETECTOR(3,5,0)_rec[-60]_rec[-28];DETECTOR(1.5,2,0)_rec[-190]_rec[-189]_rec[-186]_rec[-167]_rec[-166]_rec[-165]_rec[-44]_rec[-43]_rec[-42]_rec[-38]_rec[-37]_rec[-36];DETECTOR(3.5,2,0)_rec[-184]_rec[-183]_rec[-181]_rec[-163]_rec[-162]_rec[-161]_rec[-32]_rec[-31]_rec[-30]_rec[-27]_rec[-26]_rec[-25];DETECTOR(8,0.5,0)_rec[-72]_rec[-21]_rec[-20];DETECTOR(10,0.5,0)_rec[-66]_rec[-9]_rec[-8];DETECTOR(9.5,2,0)_rec[-68]_rec[-13]_rec[-7];DETECTOR(9,0.5,0)_rec[-69]_rec[-15]_rec[-14];DETECTOR(9,3.5,0)_rec[-67]_rec[-12]_rec[-11];DETECTOR(8.5,5,0)_rec[-70]_rec[-16]_rec[-10];DETECTOR(8,3.5,0)_rec[-71]_rec[-18]_rec[-17];DETECTOR(10,3.5,0)_rec[-65]_rec[-6]_rec[-5];DETECTOR(8,2,0)_rec[-53]_rec[-19];DETECTOR(7,0,0)_rec[-56]_rec[-24];DETECTOR(11,2,0)_rec[-50]_rec[-2];DETECTOR(11,1,0)_rec[-51]_rec[-3];DETECTOR(7,4,0)_rec[-55]_rec[-23];DETECTOR(7,5,0)_rec[-54]_rec[-22];DETECTOR(11,3,0)_rec[-49]_rec[-1];DETECTOR(10,5,0)_rec[-52]_rec[-4];DETECTOR(8.5,2,0)_rec[-178]_rec[-177]_rec[-174]_rec[-159]_rec[-158]_rec[-157]_rec[-20]_rec[-19]_rec[-18]_rec[-14]_rec[-13]_rec[-12];DETECTOR(10.5,2,0)_rec[-172]_rec[-171]_rec[-169]_rec[-155]_rec[-154]_rec[-153]_rec[-8]_rec[-7]_rec[-6]_rec[-3]_rec[-2]_rec[-1];OBSERVABLE_INCLUDE(1)_rec[-23]_rec[-17]_rec[-11]_rec[-5];OBSERVABLE_INCLUDE(0)_rec[-47]_rec[-41]_rec[-35]_rec[-29]_rec[-23]_rec[-17]_rec[-11]_rec[-5];SHIFT_COORDS(0,0,1)}{Crumble}}

%% file: resource_pipeline_embedded.tex
%
%
%
%
%
%

\begin{figure*}[t]
\centering
\begin{tikzpicture}[
  x=1cm, y=1cm,
  base/.style={
    rounded corners=3pt,
    align=center,
    inner xsep=8pt,
    inner ysep=5pt,
    font=\small,
    line width=0.5pt,
  },
  purplebox/.style ={base, draw=cpurplebdr, fill=cpurple},
  tealbox/.style    ={base, draw=ctealbdr,   fill=cteal},
  coralbox/.style   ={base, draw=ccoralbdr,  fill=ccoral},
  graybox/.style    ={base, draw=cgraybdr,   fill=cgray},
  amberbox/.style   ={base, draw=camberbdr,  fill=camber},
  arr/.style={-stealth, line width=0.6pt, color=black!55},
]

\node[purplebox, text width=11cm] (inputs) at (0, 0) {%
  \textbf{Problem inputs}\\[2pt]
  {\footnotesize $L=8$,\; $U/t=8$,\; $T_{\mathrm{sim}}=80/t$\quad (\Cref{sec:fermi_hubbard_hamiltonian})}%
};

\node[graybox, text width=14cm] (budget) at (0, -2.0) {%
  \textbf{Error budget: 1\,\% diamond norm}\quad{\footnotesize (\Cref{eq:error_budget}, \Cref{tab:error_budget})}\\[3pt]
  {\footnotesize
    $\underbrace{2\epsilon_{\mathrm{alg}} + \epsilon_{\mathrm{rot}}}_{0.50\,\%\;\text{algorithmic}}
    \;+\;
    \underbrace{\epsilon_{\mathrm{log}}}_{0.25\,\%}
    \;+\;
    \underbrace{\epsilon_{\mathrm{msf}}}_{0.25\,\%}
    \;\leq\; 1\,\%$
  }%
};

\draw[arr] (inputs.south) -- (budget.north);

\node[tealbox, text width=6.0cm] (ealg) at (-4.2, -4.5) {%
  \textbf{$\boldsymbol{\epsilon_{\mathrm{alg}}\approx 0.249\,\%}$}\\[2pt]
  {\footnotesize $\;\to\; r \approx 4.88\times 10^{5}$ Trotter steps}\\[1pt]
  {\footnotesize \Cref{eq:trotter_steps}}%
};
\node[tealbox, text width=6.8cm] (erot) at (4.2, -4.5) {%
  \textbf{$\boldsymbol{\epsilon_{\mathrm{rot}} = 0.01\,\epsilon_{\mathrm{alg}}}$}\\[2pt]
  {\footnotesize $\epsilon_{\mathrm{synth}} = \epsilon_{\mathrm{rot}}/N_{\mathrm{rot}}$,\;\;
  $N_{\mathrm{rot}}=4L^2 r$}\\[2pt]
  {\footnotesize $\;\to\; n_T = 33$,\;\;
  $t_{\mathrm{synth}}\approx 96$\; l.t.s.}\\[1pt]
  {\footnotesize Mixed-fallback,\; \Cref{sec:synthesis_cost}}%
};

\draw[arr] ([xshift=-2cm]budget.south) -- ++(0,-0.25) -| (ealg.north);
\draw[arr] ([xshift= 2cm]budget.south) -- ++(0,-0.25) -| (erot.north);

\draw[arr, ctealbdr!70]
  (ealg.east) -- node[above, font=\scriptsize, text=ctealbdr!90] {$r$ sets $N_{\mathrm{rot}}$} (erot.west);

\node[graybox, text width=14cm] (perstep) at (0, -7.4) {%
  \textbf{Per Trotter step}\quad{\footnotesize (\Cref{sec:plaquette_compilation}, \Cref{tab:fault_tolerant_cost_synthesis,tab:cost_diagonalizing_circuit,tab:cost_golden_hopping})}\\[3pt]
  {\footnotesize
    $\underbrace{4\times t_{\mathrm{synth}}}_{\text{synthesis}} + 
     \underbrace{\;90\;}_{\text{diag.\ circuits}}
     = 4\times 96 + 90 = \mathbf{474}$ logical timesteps}\\[4pt]
  {\footnotesize
    $1.43\!\times\! 10^{5}$ active cubes\;\;$\cdot$\;\;
    $9\,216\;\lvert T\rangle$ states}%
};

\draw[arr] (erot.south) -- ([xshift=2cm]perstep.north);

\node[graybox, text width=14cm] (totals) at (0, -9.8) {%
  \textbf{$\boldsymbol{\times\; r}$: totals over full algorithm}\\[3pt]
  {\footnotesize
    $N_L = 1.43\!\times\!10^{5}\times r \approx \mathbf{6.99\times 10^{10}}$ active cubes
    \qquad
    $N_T = 9\,216 \times r \approx \mathbf{4.50\times 10^{9}}\;\lvert T\rangle$ states}%
};

\draw[arr] (perstep.south) -- (totals.north);
\draw[arr] (ealg.south) -- ++(0,-0.4) -| (-8.2,-8.5) -- (-8.2,-9.6) -- (totals.west);

\node[coralbox, text width=6.4cm] (plog) at (-4.0, -12.2) {%
  \textbf{Logical error rate}\\[3pt]
  {\footnotesize
    $p_l = \dfrac{\epsilon_{\mathrm{log}}}{N_L}
    \approx \boldsymbol{3.6\!\times\! 10^{-14}}$}\\[5pt]
  {\footnotesize $\to$ $w\!=\!30$, $h\!=\!51$ from \Cref{tab:resource_estimates}}%
};
\node[coralbox, text width=6.4cm] (pmsf) at (4.0, -12.2) {%
  \textbf{Magic state fidelity}\\[3pt]
  {\footnotesize
    $p_{\mathrm{msf}} = \dfrac{\epsilon_{\mathrm{msf}}}{N_T}
    \approx \boldsymbol{5.6\!\times\! 10^{-13}}$}\\[5pt]
  {\footnotesize $\to$ 39 MSF instances,\; \Cref{tab:msf_protocols}}%
};

\draw[arr] ([xshift=-2cm]totals.south) -- ++(0,-0.3) -| (plog.north);
\draw[arr] ([xshift= 2cm]totals.south) -- ++(0,-0.3) -| (pmsf.north);

\node[amberbox, text width=6.4cm] (runtime) at (-4.0, -15.0) {%
  \textbf{Runtime\;$\boldsymbol{\approx}$\;2\,hours}\\[2pt]
  {\footnotesize $474\times r \times t_{l}$\quad (\Cref{sec:time})}%
};
\node[amberbox, text width=6.4cm] (qubits) at (4.0, -15.0) {%
  \textbf{$\boldsymbol{1.35\times 10^{6}}$ physical qubits}\\[2pt]
  {\footnotesize $882$ patches $\times 30 \times 51$\quad (\Cref{sec:trotter_step_cost})}%
};

\draw[arr] (plog.south) -- (runtime.north);
\draw[arr] (pmsf.south) -- (qubits.north);
\draw[arr, color=black!35] (plog.south east) -- (qubits.north west);


\begin{scope}[shift={(0, -16.5)}]
  \fill[cteal,   draw=ctealbdr,  rounded corners=2pt, line width=0.4pt] (-4.5,0) rectangle ++(0.5,0.35);
  \node[anchor=west, font=\footnotesize] at (-3.85, 0.175) {Algorithmic};
 
  \fill[ccoral,  draw=ccoralbdr, rounded corners=2pt, line width=0.4pt] (-1.5,0) rectangle ++(0.5,0.35);
  \node[anchor=west, font=\footnotesize] at (-0.85, 0.175) {Hardware};
 
  \fill[camber,  draw=camberbdr, rounded corners=2pt, line width=0.4pt] (1.5,0)  rectangle ++(0.5,0.35);
  \node[anchor=west, font=\footnotesize] at (2.15, 0.175)  {Final outputs};
\end{scope}

\end{tikzpicture}
\caption{Overview of the resource-estimation pipeline for the Fermi--Hubbard simulation, discussed in
Section~\ref{sec:resource_estimation}.}
\label{fig:resource_pipeline}
\end{figure*}

%% file: appendix_physical_noisy_gate_set.tex
\section{Physical gate set and noise model of the SPOQC architecture}
\label{ap:physical_gate_set}

In this appendix, we present all the relevant quantities to build the noisy get set of the SPOQC architecture. We will not get into too many details, but rather try to give as many intuitions as possible. The interested reader can refer to \cite{dessertaine2026enhancedfaulttolerancephotonicquantum} for more details regarding the computations. Most of this section will be dedicated to describing how the noise affects the RUS gate. As in the main text, CPTP maps describing the error channels are represented through their Kraus operators: a channel $\mathcal{E}$ with Kraus operators $\{\mathsf{A}_i\}$ acts on a density matrix $\rho$ as $\mathcal{E}[\rho]=\sum_{i}\mathsf{A}_i\rho\mathsf{A}_i$, and $\mathcal{E}=\sum_{i}[\mathsf{A}_i]$ where $[\mathsf{A}_i]$ is the super-operator associated with $\mathsf{A}_i$.

\subsection{The RUS gate}

This section is dedicated to describing the different level of noise modelling for the RUS gate. We will go from the physical level to the more abstract level of quantum channels.

\subsubsection{Physical RUS probabilities and error channels for a single RUS cycle}

As explained in the main text, the RUS gate relies on 3 sub-steps: (1) emission of photons by the quantum emitters, (2) linear-optical interference of these photons in a dedicated interferometer \cite{Gliniasty2024Spin, dessertaine2026enhancedfaulttolerancephotonicquantum}, (3) detection of the photons in photon number resolving (PNR) detectors. We will assume that these steps are perfect, aside from photon loss. We leave emission errors (multi-photon emission, non-isometric emission...) as well as imperfections in the linear-optical interferometer for future work. \Cref{tab:physical_error_probabilities}
 and \Cref{tab:physical_channels} summarize the main notations and definitions used in this section.
 
The RUS gate is a quantum instrument where each photonic detection pattern from step (3) corresponds to a certain quantum operation on the associated emitters. A photonic detection pattern is the record of the number of photons in each output mode, where modes are labeled from 0 to 3, e.g. (2,0,0,0) means there are two photons in mode 0. In a perfect setting, the only possible detection patterns are $(2,0,0,0), (0,2,0,0), (0,0,2,0), (0,0,0,2)$ or $(1,0,1,0), (0,1,1,0), (1,0,0,1), (0,1,0,1)$. The first four patterns herald a so-called \textit{repeat}: in these cases a single $Z$ gate is applied on the first or the second emitter depending on the pattern. We can track this Pauli term which essentially means that the information encoded on the emitter has not bee unaltered. Therefore, the process can be repeated. The second four patterns correspond to a so-called \textit{success}: in these cases, a $\CZ$ gate or a two qubit parity measurement $\MZZ$ has been applied, depending on a tunable phase in the interferometer of step (2). In the $\CZ$ case, a single-qubit Clifford correction needs to be applied depending on which exact detection pattern was recorded. In the $\MZZ$ case, each pattern corresponds to a measurement value: $(1,0,1,0)$ and $(0,1,0,1)$ correspond to measuring the $+1$ eigenvalue of $ZZ$, while $(0,1,1,0)$ and $(1,0,0,1)$ to measuring $-1$. With $\varepsilon$ the probability to lose a photon, the probability to get one successful detection pattern is $=(1-\varepsilon)^2/8$, which gives an overall success probability of $p_s(\varepsilon)=(1-\varepsilon)^2/2$. The probability to get one repeat outcome is $(2-D)(1-\varepsilon)^2/16$, with $D$ the distinguishability between emitted photons, giving an overall repeat probability of $p_{r, indist.}(\varepsilon)=(2-D)(1-\varepsilon)^2/4$.

As hinted in the previous formula, photonic distinguishability plays a role in the possible detection patterns. When photons are indistinguishable, two photons arriving at the same time on both arms of a balanced beam-splitter at the same time can only can only bunch at the output, this is the Hong-Ou-Mandel effect. In our case, distinguishable photons can imperfectly interfere to create detections patterns $(1,1,0,0)$ and $(0,0,1,1)$. They occur with overall probability $p_{r,dist.}(\varepsilon)=D(1-\varepsilon)^2/4$. These patterns can be treated in the same way as repeat patterns as they only herald $Z$ gates applied on either emitter, we therefore consider an overall repeat probability $p_r(\varepsilon)=p_{r,indist.}(\varepsilon)+p_{r,dist.}(\varepsilon)=(1-\varepsilon)^2/2$. Distiguishability also affects the fidelity of the operations for successful outcome patterns as photons interfered imperfectly. In the $\CZ$ case, this translates into a two-qubit Pauli channel $\mathcal{C}_{dist.\,\CZ,ab}=(1-D)[\Id]/2+(1-D)[\Z_a\Z_b]/2$. In the $\MZZ$ case, this translates into a classical measurement flip $\mathcal{C}_{dist.\,\MZZ,ab}=[\Id]\otimes\left((1+D)[\Id]/2+(1-D)[\X]/2\right)$, where the left part of the tensor product refers to the quantum state and the right part the classical register and we use $\X$ to denote a measurement flip $+1\leftrightarrow -1$. See \cite{dessertaine2026enhancedfaulttolerancephotonicquantum} for more details.

So far, all the patterns discussed above had 2 photons at the output. Because of photon loss, it is possible to get patterns with only one output photon $(1,0,0,0), (0,1,0,0), (0,0,1,0), (0,0,0,1)$, \textit{one-loss} case, or no photon $(0,0,0,0)$, \textit{two-loss} case. As explained in the main text, whenever a photon is lost, the associated spin $i$ undergoes a full dephasing $\mathcal{C}_i=([\Id]+[\Z_i])/2$. This channel translates into heralded channels for the \textit{one-loss} and \textit{two-loss} cases. In the \textit{one-loss} case, a single photon is detected, we therefore need to apply $\mathcal{C}_i$ to one of the emitter and leave the other one untouched. However, since we cannot know what the original emitter of the lost photon was, we need to dephase either spin with equal probability, resulting in the heralded error channel $\mathcal{C}_{loss,ab}^{(1)}=\left(\mathcal{C}_a+\mathcal{C}_b\right)/2$, with $a,b$ the original quantum emitters. Each of the one-loss detection pattern occurs with probability $\varepsilon(1-\varepsilon)/2$, which gives an overall one-loss probability of $p_1(\varepsilon)=2\varepsilon(1-\varepsilon)$. The two-loss case is simpler since both photons were lost and therefore both emitters are dephased. The associated channel is $\mathcal{C}_{loss,ab}^{(\infty)}=\mathcal{C}_a\mathcal{C}_b$, where the multiplication of the channel stands for composition. The two-loss probability is $p_2(\varepsilon)=\varepsilon^2$.

\begin{table}[h]
    \centering
    \begin{tblr}{
    hline{1,2}={1.5pt},
    hline{3-8}={0.8pt},
    cells={c, m},
    width = \linewidth,
    colspec = {X X X X},
    }
         \bf Notation & \bf Expression & \bf Description & \bf Noiseless value\\
         $p_s(\varepsilon)$ & $(1-\varepsilon)^2/2$ & Success probability of the RUS gate & $1/2$\\
         $p_{r,indist.}(\varepsilon)$ & $(2-D)(1-\varepsilon)^2/4$ & Probability to get a bunching outcome & $1/2$\\
         $p_{r,dist.}(\varepsilon)$ & $D(1-\varepsilon)^2/4$ & Probability to get a photon in each mode of the upper or lower pair of modes & $0$\\
         $p_{r}(\varepsilon)$ & $(1-\varepsilon)^2/2$ & Repeat probability of the RUS gate & $1/2$\\
         $p_1(\varepsilon)$ & $2(1-\varepsilon)\varepsilon$ & Probability to lose a photon & $0$\\
         $p_2(\varepsilon)$ & $\varepsilon^2$ & Probability to lose two photons & $0$\\
    \end{tblr}
    \caption{Table of probabilities associated with the events of the RUS gate. One readily checks that $p_s(\varepsilon)+p_r(\varepsilon)+p_1(\varepsilon)+p_2(\varepsilon)=1$, for any value of $\varepsilon$.}
    \label{tab:physical_error_probabilities}
\end{table}

\begin{table}[h]
    \centering
    \begin{tblr}{
    hline{1,2}={1.5pt},
    hline{3-8}={0.8pt},
    cells={c, m},
    row{1}={font=\bf},
    width = \linewidth,
    colspec = {X X X},
    }
         Notation & Expression & Description\\
         $\mathcal{C}_i$ & $\dfrac{1}{2}\left([\Id]+[\Z_i]\right)$ & Phase erasure of emitter $i$ \\
         $\mathcal{C}_{loss,ab}^{(1)}$ & $\dfrac{1}{2}\left(\mathcal{C}_a+\mathcal{C}_b\right)=\dfrac{1}{2}[\Id]+\dfrac{1}{4}([\Z_a]+[\Z_b])$ & Heralded error channel following the loss of a single photon in a RUS cycle \\
         $\mathcal{C}_{loss,ab}^{(k)}$ & $\left(\mathcal{C}_{loss,ab}^{(1)}\right)^k=\left(\dfrac{1}{4}+\dfrac{1}{2^{k+1}}\right)[\Id]+\dfrac{1}{4}([\Z_a]+[\Z_b])+\left(\dfrac{1}{4}-\dfrac{1}{2^{k+1}}\right)[\Z_a\Z_b]$ & $k$-fold composition of the heralded channel $\mathcal{C}_{loss,ab}^{(1)}$\\
         $\mathcal{C}_{loss,ab}^{(\infty)}$ & $\mathcal{C}_a\mathcal{C}_b=\dfrac{1}{4}\left([\Id]+ [\Z_a]+[\Z_b]+[\Z_a\Z_b]\right)$ & Heralded error channel following the loss of two photons in a RUS cycle, also equivalent to the infinite composition of $\mathcal{C}_{loss,ab}^{(1)}$, see \cite{dessertaine2026enhancedfaulttolerancephotonicquantum}\\
         $\mathcal{C}_{dist.\,\CZ,ab}$ & $\dfrac{1+D}{2}[\Id]+\dfrac{1-D}{2}[\Z_a\Z_b]$ & Distinguishability induced error channel after a successful \ruscz performed with partially distinguishable photons\\
         $\mathcal{C}_{dist.\,\MZZ,ab}$ & $[\Id]\otimes\left(\dfrac{1+D}{2}[\Id]+\dfrac{1-D}{2}[\X]\right)$ & Distinguishability induced classical measurement flip after a successful \rusmzz performed with partially distinguishable photons. The left (resp. right) part of the tensor product models the quantum state (resp. the classical measurement space)\\
    \end{tblr}
    \caption{Table of the notations and definitions used to model the different noise channel of the RUS cycle.}
    \label{tab:physical_channels}
\end{table}

\subsubsection{Noise channels for multiple cycles and heralding probabilities}

So far, we have considered the effects of the noise after a single detection of the photons and therefore over a single cycle of the RUS gate. However, as the name suggests, cycles are repeated until we get a successful outcome, or whenever we reached a maximum number of trials $N_{RUS}$, and we need to consider the overall channels after $N_{RUS}$ cycles. To simplify simulations for the present study, we assume that we wait exactly $N_{RUS}$ attempts for each RUS gate, even if the RUS gate might have stopped before this cap. It allows to consider layers of $N_{RUS}$ cycles over which RUS gates are executed in parallel, independently of one another. This is a slight simplification with regards to \cite{dessertaine2026enhancedfaulttolerancephotonicquantum}, where one starts the next layer of RUS gates as soon as all gates in the current layer have stopped, regardless if the cap is reached. With this assumptio, we overestimate the noise induced by decoherence but we the time to perform the layer from every single RUS gate execution in the layer.

In the same way as in \cite{dessertaine2026enhancedfaulttolerancephotonicquantum}, several detection histories result in the same overall noise channel after $N_{RUS}$ cycles. For instance, it does not matter whether a successful outcome occurred after 3 or 5 repeat outcomes, and we thus treat these two possibilities in the same way. Accounting for all the possibilities, we can construct all of the heralded noise channels to characterize the RUS gates.

For \ruscz gates, there are 4 possible noise channels after $N_{RUS}$ trials each occurring with given probabilities summarized in \Cref{tab:heralding_probabilities_cz}. The possibilities are:
\begin{enumerate}
    \item A pure success, preceded only by repeat outcomes. This happens with a probability $p_0(N_{RUS}, \varepsilon)$ and the associated noise channel is $\mathcal{C}_{dist., \CZ}$ due to distinguishability lowering the fidelity of the successful outcome. 
    \item A success, preceded by exactly $1\leq k<N_{RUS}$ outcomes where one photon was lost and no outcome where two photon were lost, which happens with probability $p_k(N_{RUS}, \varepsilon)$. The associated noise channel is the $k$-fold composition of the one-loss channel, see \Cref{tab:physical_channels}, followed by the distinguishabily induced channel: $\mathcal{C}_{dist., \CZ}\mathcal{C}_{loss}^{(k)}$.
    \item A failure of the \ruscz, characterized by the occurrence of a single event were two photons were lost, which happens with probability $p_f(N_{RUS}, \varepsilon)$.. In this case, the $\CZ$ gate has not been performed\footnote{In simulations, we still assume that the $\CZ$ gate is performed and followed by $\mathcal{C}_{loss}^{(\infty)}$. This does not matter from the point of view of the quantum state since $\mathcal{C}_{loss}^{(\infty)}\CZ=\mathcal{C}_{loss}^{(\infty)}$, but it simplifies simulations since one does not have to remove $\CZ$ gates from the circuit.} and the associated noise channel is a complete dephasing of both spins $\mathcal{C}_{loss}^{(\infty)}$.
    \item An abort outcome when no success nor failure have been recorded within the allocated $N_{RUS}$ attempts, which a probability $p_a(N_{RUS}, \varepsilon)$.. We treat this outcome as a failure, see \cite{dessertaine2026enhancedfaulttolerancephotonicquantum}.
\end{enumerate}

\begin{table}[h]
    \centering
    \begin{tblr}{
    hline{1,2}={1.5pt},
    hline{3-6}={0.8pt},
    cells={c, m}, 
    width=\linewidth,
    colspec={c X c c},
    column{1}={2cm},
    column{3}={6cm},
    column{4}={2.5cm}
    }
         \bf Notation & \bf Expression & \bf Description & \bf Noiseless value\\
         $p_0(N_{RUS},\varepsilon)$ & $p_s(\varepsilon)\dfrac{1-p_r(\varepsilon)^{N_{RUS}}}{1-p_r(\varepsilon)}$ & Probability to perform a successful \ruscz without losses & $1-\dfrac{1}{2^{N_{RUS}}}$
         \\
         $p_k(N_{RUS},\varepsilon)$ & $p_s(\varepsilon)\sum_{t=1}^{N_{RUS}}\binom{t-1}{k}p_1(\varepsilon)^{k}p_r(\varepsilon)^{t-1-k}$ & Probability to perform a successful \ruscz with $k$ one-photon loss events and no two-photon loss events & $0$
         \\
         $p_f(N_{RUS},\varepsilon)$ & $p_2(\varepsilon)\dfrac{1-(p_r(\varepsilon)+p_1(\varepsilon))^{N_{RUS}}}{1-p_r(\varepsilon)-p_1(\varepsilon)}$ & Probability to perform a successful \ruscz with at least one two-photon loss event & $0$
         \\
         $p_a(N_{RUS},\varepsilon)$ & $(p_r(\varepsilon)+p_1(\varepsilon))^{N_{RUS}}$ & Probability to not succeed nor fail the \ruscz in the allocated $N_{RUS}$ RUS cycles & $\dfrac{1}{2^{N_{RUS}}}$\\
    \end{tblr}
    \caption{Table of channel probabilities for \ruscz. On can check that, for any values of $N_{RUS}$ and $\varepsilon$, heralding probabilities sum to 1, i.e. $p_0+\sum_{k}p_k+p_f+p_a=1$.}
    \label{tab:heralding_probabilities_cz}
\end{table}

For \rusmzz, there are 3 possible noise channels after $N_{RUS}$ trials each occurring with given probabilities summarized in \Cref{tab:heralding_probabilities_mzz}. The possibilities are:
\begin{enumerate}
    \item A pure success, preceded only by repeat outcomes. This happens with a probability $q_0(N_{RUS}, \varepsilon)$ and the associated noise channel is $\mathcal{C}_{dist., \MZZ}$ due to distinguishability flipping the classical measurement record. 
    \item A success, preceded by any event where one or two photons were lost which happens with probability $q_e(N_{RUS}, \varepsilon)$\footnote{After a successful $Z_aZ_b$ measurement with outcome $m$, the operator $mZ_aZ_b$ is a stabilizer of the quantum state $\rho$. As a consequence, $\rho=(mZ_aZ_b)\rho(mZ_aZ_b)=Z_aZ_b\rho Z_aZ_b$, since $m^2=1$. From this identity, we can derive $Z_a\rho Z_a=Z_b\rho Z_b$. Applying these to the definitions of the loss channels from \Cref{tab:physical_channels}, we get that $\mathcal{C}_{loss,ab}^{(k)}=\mathcal{C}_a=\mathcal{C}_b$, for $1\leq k\leq \infty$, meaning that any one of these channels is equivalent to dephasing spin $a$ or $b$.}. The associated noise channel is a phase erasure on one of the two spins (we will choose the first one for concreteness) followed by the distinguishability induced classical measurement flip: $\mathcal{C}_a\mathcal{C}_{dist.\,\MZZ}$.
    \item An abort outcome when no success nor failure have been recorded within the allocated $N_{RUS}$ attempts, which a probability $q_a(N_{RUS}, \varepsilon)$. We treat this outcome as a successful measurement followed by both a phase erasure and a measurement erasure $\mathcal{C}_a\otimes\left(\dfrac{1}{2}([\Id]+[\X])\right)$, see \cite{dessertaine2026enhancedfaulttolerancephotonicquantum}.
\end{enumerate}

\begin{table}[h]
    \centering
    \begin{tblr}{
    hline{1,2}={1.5pt},
    hline{3-5}={0.8pt},
    cells={c, m},
    width=\linewidth,
    colspec={c X c c},
    column{1}={2cm},
    column{3}={6cm},
    column{4}={2.5cm}
    }
         \bf Notation & \bf Expression & \bf Description & \bf Noiseless value\\
         $q_0(N_{RUS},\varepsilon)$ & $p_s(\varepsilon)\dfrac{1-p_r(\varepsilon)^{N_{RUS}}}{1-p_r(\varepsilon)}$ & 
             Probability to perform a successful \rusmzz parity measurement without losses
          & $1-\dfrac{1}{2^{N_{RUS}}}$
         \\
         $q_e(N_{RUS},\varepsilon)$ & $1-(1-p_s(\varepsilon))^{N_{RUS}}-p_s(\varepsilon)\dfrac{1-p_r(\varepsilon)^{N_{RUS}}}{1-p_r(\varepsilon)}$ & Probability to perform a successful \rusmzz with at least one loss event & $0$
         \\
         $q_a(N_{RUS},\varepsilon)$ & $(1-p_s(\varepsilon))^{N_{RUS}}$ & Probability to not succeed nor fail  the RUS $\Z\Z$ parity measurement in the allocated $N_{RUS}$ RUS cycles & $\dfrac{1}{2^{N_{RUS}}}$\\
    \end{tblr}
    \caption{Table of channel probabilities for \rusmzz. On can check that, for any values of $N_{RUS}$ and $\varepsilon$, heralding probabilities sum to 1, i.e. $q_0+q_e+q_a=1$.}
    \label{tab:heralding_probabilities_mzz}
\end{table}

\subsection{Spin initialization and measurement}

As explained in the main text, spin initialization and measurement are performed through emission and detection of a photon. In the absence of losses, this process is deterministic as the emitted photon will always be measured. Whenever losses occur, this process may have to be repeated as long as no photons are detected. Each time a photon is lost, the associated spin $a$ undergoes a phase erasure $\mathcal{C}_a$. Since photonic detection performs a projective $Z$ measurement on the spin, the phase erasure will have no effect after the measurement. Consequently, we can ignore the effect of photon loss when implementing spin initialization and measurement. As with the RUS gates, we cap the number of attempts at spin initialization $N_i$ and measurement $N_m$ to $N_i=N_m=5$. With probability $\varepsilon^{N_i}$ (or $\varepsilon^{N_m}$ for measurement), there is a possibility that, for all attempts, the emitted photon was lost. This case is similar to the abort case of the RUS gate. For spin initialization, we will simulate this outcome as an initialization followed by a complete depolarization of the spin $([\Id]+[\X]+[\Y]+[\Z])/4$, while for spin measurement, we will treat this as a measurement loss $[\Id]\otimes\frac{1}{2}\left([\Id]+[\X]\right)$.

\subsection{Idling time}

To simulate the effect of decoherence from \Cref{eq:deco_channel} in the main text, we split our circuits into layers of operations that can be implemented in parallel. Within each layer, we compute the idling time of each qubits assuming that: (a) qubits being part of a RUS gate are idling for a time $N_{RUS}t_c$, (c) qubits being initialized (resp. measured) are idling for a time $N_{i}t_c$ (resp. $N_mt_c$), (b) qubits undergoing single qubit gates are not idling during the implementation of the single qubit gate, as decoherence is taken into account in the single qubit error channel \Cref{eq:single-qb-error-channel}.

\subsection{Physical gate set}

\begin{table}[h]
\centering
    \begin{tblr}{
        hline{1,2}={1.5pt},
        hline{3,4,6,8,12, 15}={1pt},
        hline{5,7, 9-11, 13-14}={0.2pt},
        cells={c, m},
        cell{2}{4}={black!10},
        cell{3}{4}={black!10},
        cell{4}{2}={black!10},
        cell{6}{2}={black!10},
        width=\linewidth,
        colspec={c X X X},
        column{1}={5cm}
        }
     \bf Operation & \bf Error Channel & \bf Heralded & \bf Probability\\
     Idling during time $t$& $(1-p_d(t/T_2))[\Id]+p_d(t/T_2)[\Z]$ & No & \\
     Single-qubit rotation & $(1-p_s)[\Id]+\dfrac{p_s}{3}([\X]+[\Y]+[\Z])$ &  No & \\
    \SetCell[r=2]{} $\Z$-basis initialization \init &  & Yes & $1-\varepsilon^{N_i}$\\
    & $\dfrac{1}{4}\left([\Id]+[\X]+[\Y]+[\Z]\right)$ & Yes & $\varepsilon^{N_i}$ \\
    \SetCell[r=2]{} $\Z$-basis measurement $\MZ$ &  & Yes & $1-\varepsilon^{N_m}$\\
    & $[\Id]\otimes\dfrac{1}{2}\left([\Id]+[\X]\right)$ & Yes & $\varepsilon^{N_m}$\\
    \SetCell[r=4]{} Control-$\Z$ $\CZ_{ab}$ & $\mathcal{C}_{dist.\,\CZ}$ & Yes & $p_0(N_{RUS},\varepsilon)$\\
    & $\mathcal{C}_{loss}^{(k)}\mathcal{C}_{dist.\,\CZ}$ & Yes & $p_k(N_{RUS},\varepsilon)$\\
    & $\mathcal{C}_{loss}^{(\infty)}$ & Yes & $p_f(N_{RUS},\varepsilon)$\\
    & $\mathcal{C}_{loss}^{(\infty)}$ & Yes & $p_a(N_{RUS},\varepsilon)$\\
    \SetCell[r=3]{} $\Z_a\Z_b$ parity measurement $\MZZ$ & $\mathcal{C}_{dist.\,\MZZ}$ & Yes & $q_0(N_{RUS},\varepsilon)$\\
    & $\mathcal{C}_a\mathcal{C}_{dist.\,\MZZ}$ & Yes & $q_e(N_{RUS},\varepsilon)$\\
    & $\mathcal{C}_a\otimes\left(\dfrac{1}{2}([\Id]+[\X])\right)$ & Yes & $q_a(N_{RUS},\varepsilon)$\\
    \end{tblr}
    \caption{Noisy gate set for the SPOQC architecture.}
    \label{tab:physical_gate_set}
\end{table}

Gathering all results from the previous sections, \Cref{tab:physical_gate_set} summarizes the physical noisy gate set that we use in our simulation. For each quantum operation, we indicate all possible heralded channels along with their probabilities. All of these channels are applied the operation they are attached to.

%% file: crumble_url_horizontal_memory.tex
\newcommand{\crumbleHorizontalMemory}{\href{http://algassert.com/crumble\#circuit=Q(0.0,0.0)0;Q(0.0,4.0)1;Q(0.0,5.0)2;Q(1.0,0.0)3;Q(1.0,1.0)4;Q(1.0,2.0)5;Q(1.0,3.0)6;Q(1.0,4.0)7;Q(1.0,5.0)8;Q(2.0,0.0)9;Q(2.0,1.0)10;Q(2.0,2.0)11;Q(2.0,3.0)12;Q(2.0,4.0)13;Q(2.0,5.0)14;Q(3.0,0.0)15;Q(3.0,1.0)16;Q(3.0,2.0)17;Q(3.0,3.0)18;Q(3.0,4.0)19;Q(3.0,5.0)20;Q(4.0,1.0)21;Q(4.0,2.0)22;Q(4.0,3.0)23;POLYGON(1,0,0,0.25)_0;POLYGON(1,0,0,0.25)_2;POLYGON(1,0,0,0.25)_3_4_10_9;POLYGON(1,0,0,0.25)_8_14;POLYGON(1,0,0,0.25)_15_16_21;POLYGON(1,0,0,0.25)_20;POLYGON(1,0,0,0.25)_1_7_6_5;POLYGON(1,0,0,0.25)_11_12_13_19_18_17;POLYGON(1,0,0,0.25)_22_23;POLYGON(0,1,0,0.25)_0_3;POLYGON(0,1,0,0.25)_1_2_8_7;POLYGON(0,1,0,0.25)_9_15;POLYGON(0,1,0,0.25)_4_5_6_12_11_10;POLYGON(0,1,0,0.25)_13_14_20_19;POLYGON(0,1,0,0.25)_16_17_18_23_22_21;POLYGON(0,0,1,0.25)_0_5_4_3;POLYGON(0,0,1,0.25)_9_10_11_17_16_15;POLYGON(0,0,1,0.25)_21_22;POLYGON(0,0,1,0.25)_2_1;POLYGON(0,0,1,0.25)_6_7_8_14_13_12;POLYGON(0,0,1,0.25)_18_19_20_23;R_0_1_2_3_4_5_6_7_8_9_10_11_12_13_14_15_16_17_18_19_20_21_22_23;TICK;H_YZ_0_1_2_3_4_5_6_7_8_9_10_11_12_13_14_15_16_17_18_19_20_21_22_23;SHIFT_COORDS(0,0,1);TICK;MZZ_1_7_4_10_5_6_11_12_13_19_16_21_17_18_22_23;M_0_2_3_8_9_14_15_20;OBSERVABLE_INCLUDE(0)_rec[-16]_rec[-12];TICK;H_0_1_2_3_4_5_6_7_8_9_10_11_12_13_14_15_16_17_18_19_20_21_22_23;TICK;MZZ_0_3_1_2_4_5_6_12_7_8_9_15_10_11_13_14_16_17_18_23_19_20_21_22;DETECTOR(1.5,2,0)_rec[-27]_rec[-26]_rec[-25]_rec[-10]_rec[-9]_rec[-6];DETECTOR(3.5,2,0)_rec[-23]_rec[-22]_rec[-21]_rec[-4]_rec[-3]_rec[-1];TICK;H_0_1_2_3_4_5_6_7_8_9_10_11_12_13_14_15_16_17_18_19_20_21_22_23;SHIFT_COORDS(0,0,1);TICK;MZZ_1_7_4_10_5_6_11_12_13_19_16_21_17_18_22_23;M_0_2_3_8_9_14_15_20;OBSERVABLE_INCLUDE(0)_rec[-16]_rec[-12];DETECTOR(1.5,2,0)_rec[-43]_rec[-42]_rec[-41]_rec[-15]_rec[-14]_rec[-13];DETECTOR(3.5,2,0)_rec[-39]_rec[-38]_rec[-37]_rec[-11]_rec[-10]_rec[-9];DETECTOR(0.5,-1,0)_rec[-36]_rec[-34]_rec[-8]_rec[-6];DETECTOR(0.5,5,0)_rec[-44]_rec[-35]_rec[-33]_rec[-16]_rec[-7]_rec[-5];DETECTOR(2.5,-1,0)_rec[-32]_rec[-30]_rec[-4]_rec[-2];DETECTOR(2.5,5,0)_rec[-40]_rec[-31]_rec[-29]_rec[-12]_rec[-3]_rec[-1];SHIFT_COORDS(0,0,1);TICK;H_YZ_0_1_2_3_4_5_6_7_8_9_10_11_12_13_14_15_16_17_18_19_20_21_22_23;TICK;MZZ_3_4_6_7_8_14_9_10_11_17_12_13_15_16_18_19;M_0_1_2_5_20_21_22_23;OBSERVABLE_INCLUDE(0)_rec[-14]_rec[-6]_rec[-4];DETECTOR(2.5,3,0)_rec[-57]_rec[-56]_rec[-54]_rec[-29]_rec[-28]_rec[-26]_rec[-12]_rec[-11]_rec[-9];DETECTOR(0.515625,3.01562,0)_rec[-60]_rec[-58]_rec[-32]_rec[-30]_rec[-15]_rec[-7]_rec[-5];DETECTOR(4.46875,2.96875,0)_rec[-53]_rec[-25]_rec[-2]_rec[-1];TICK;C_NZYX_0_1_2_3_4_5_6_7_8_9_10_11_12_13_14_15_16_17_18_19_20_21_22_23;SHIFT_COORDS(0,0,1);TICK;MZZ_0_3_1_2_4_5_6_12_7_8_9_15_10_11_13_14_16_17_18_23_19_20_21_22;TICK;C_XYNZ_0_1_2_3_4_5_6_7_8_9_10_11_12_13_14_15_16_17_18_19_20_21_22_23;TICK;MZZ_3_4_6_7_8_14_9_10_11_17_12_13_15_16_18_19;M_0_1_2_5_20_21_22_23;OBSERVABLE_INCLUDE(0)_rec[-14]_rec[-6]_rec[-4];DETECTOR(2.5,1,0)_rec[-41]_rec[-40]_rec[-38]_rec[-13]_rec[-12]_rec[-10];DETECTOR(1.5,4,0)_rec[-43]_rec[-42]_rec[-39]_rec[-15]_rec[-14]_rec[-11];DETECTOR(0.5,1,0)_rec[-44]_rec[-36]_rec[-33]_rec[-16]_rec[-8]_rec[-5];DETECTOR(4.5,1,0)_rec[-31]_rec[-30]_rec[-3]_rec[-2];DETECTOR(-0.5,4,0)_rec[-35]_rec[-34]_rec[-7]_rec[-6];DETECTOR(3.5,4,0)_rec[-37]_rec[-32]_rec[-29]_rec[-9]_rec[-4]_rec[-1];TICK;H_YZ_0_1_2_3_4_5_6_7_8_9_10_11_12_13_14_15_16_17_18_19_20_21_22_23;SHIFT_COORDS(0,0,1);TICK;MZZ_1_7_4_10_5_6_11_12_13_19_16_21_17_18_22_23;M_0_2_3_8_9_14_15_20;OBSERVABLE_INCLUDE(0)_rec[-16]_rec[-12];DETECTOR(2.5,3,0)_rec[-73]_rec[-72]_rec[-70]_rec[-56]_rec[-55]_rec[-53]_rec[-28]_rec[-27]_rec[-25]_rec[-13]_rec[-12]_rec[-10];DETECTOR(1.5,0,0)_rec[-75]_rec[-66]_rec[-64]_rec[-60]_rec[-57]_rec[-32]_rec[-29]_rec[-15]_rec[-6]_rec[-4];DETECTOR(1.5,6,0)_rec[-65]_rec[-63]_rec[-58]_rec[-30]_rec[-5]_rec[-3];SHIFT_COORDS(0,0,1);TICK;H_0_1_2_3_4_5_6_7_8_9_10_11_12_13_14_15_16_17_18_19_20_21_22_23;TICK;MZZ_0_3_1_2_4_5_6_12_7_8_9_15_10_11_13_14_16_17_18_23_19_20_21_22;DETECTOR(1.5,2,0)_rec[-98]_rec[-97]_rec[-94]_rec[-87]_rec[-86]_rec[-85]_rec[-27]_rec[-26]_rec[-25]_rec[-10]_rec[-9]_rec[-6];DETECTOR(3.5,2,0)_rec[-92]_rec[-91]_rec[-89]_rec[-83]_rec[-82]_rec[-81]_rec[-23]_rec[-22]_rec[-21]_rec[-4]_rec[-3]_rec[-1];TICK;H_0_1_2_3_4_5_6_7_8_9_10_11_12_13_14_15_16_17_18_19_20_21_22_23;SHIFT_COORDS(0,0,1);TICK;MZZ_1_7_4_10_5_6_11_12_13_19_16_21_17_18_22_23;M_0_2_3_8_9_14_15_20;OBSERVABLE_INCLUDE(0)_rec[-16]_rec[-12];DETECTOR(1.5,2,0)_rec[-43]_rec[-42]_rec[-41]_rec[-15]_rec[-14]_rec[-13];DETECTOR(3.5,2,0)_rec[-39]_rec[-38]_rec[-37]_rec[-11]_rec[-10]_rec[-9];DETECTOR(0.5,-1,0)_rec[-36]_rec[-34]_rec[-8]_rec[-6];DETECTOR(0.5,5,0)_rec[-44]_rec[-35]_rec[-33]_rec[-16]_rec[-7]_rec[-5];DETECTOR(2.5,-1,0)_rec[-32]_rec[-30]_rec[-4]_rec[-2];DETECTOR(2.5,5,0)_rec[-40]_rec[-31]_rec[-29]_rec[-12]_rec[-3]_rec[-1];SHIFT_COORDS(0,0,1);TICK;H_YZ_0_1_2_3_4_5_6_7_8_9_10_11_12_13_14_15_16_17_18_19_20_21_22_23;TICK;MZZ_3_4_6_7_8_14_9_10_11_17_12_13_15_16_18_19;M_0_1_2_5_20_21_22_23;OBSERVABLE_INCLUDE(0)_rec[-14]_rec[-6]_rec[-4];DETECTOR(2.5,3,0)_rec[-72]_rec[-71]_rec[-69]_rec[-57]_rec[-56]_rec[-54]_rec[-29]_rec[-28]_rec[-26]_rec[-12]_rec[-11]_rec[-9];DETECTOR(0.5,3,0)_rec[-75]_rec[-67]_rec[-65]_rec[-60]_rec[-58]_rec[-32]_rec[-30]_rec[-15]_rec[-7]_rec[-5];DETECTOR(4.5,3,0)_rec[-62]_rec[-61]_rec[-53]_rec[-25]_rec[-2]_rec[-1];TICK;C_NZYX_0_1_2_3_4_5_6_7_8_9_10_11_12_13_14_15_16_17_18_19_20_21_22_23;SHIFT_COORDS(0,0,1);TICK;MZZ_0_3_1_2_4_5_6_12_7_8_9_15_10_11_13_14_16_17_18_23_19_20_21_22;DETECTOR(2.5,1,0)_rec[-95]_rec[-94]_rec[-92]_rec[-85]_rec[-84]_rec[-82]_rec[-25]_rec[-24]_rec[-22]_rec[-7]_rec[-6]_rec[-4];DETECTOR(1.5,4,0)_rec[-97]_rec[-96]_rec[-93]_rec[-87]_rec[-86]_rec[-83]_rec[-27]_rec[-26]_rec[-23]_rec[-9]_rec[-8]_rec[-5];TICK;C_XYNZ_0_1_2_3_4_5_6_7_8_9_10_11_12_13_14_15_16_17_18_19_20_21_22_23;SHIFT_COORDS(0,0,1);TICK;MZZ_3_4_6_7_8_14_9_10_11_17_12_13_15_16_18_19;M_0_1_2_5_20_21_22_23;OBSERVABLE_INCLUDE(0)_rec[-14]_rec[-6]_rec[-4];DETECTOR(2.5,1,0)_rec[-41]_rec[-40]_rec[-38]_rec[-13]_rec[-12]_rec[-10];DETECTOR(1.5,4,0)_rec[-43]_rec[-42]_rec[-39]_rec[-15]_rec[-14]_rec[-11];DETECTOR(0.5,1,0)_rec[-44]_rec[-36]_rec[-33]_rec[-16]_rec[-8]_rec[-5];DETECTOR(4.5,1,0)_rec[-31]_rec[-30]_rec[-3]_rec[-2];DETECTOR(-0.5,4,0)_rec[-35]_rec[-34]_rec[-7]_rec[-6];DETECTOR(3.5,4,0)_rec[-37]_rec[-32]_rec[-29]_rec[-9]_rec[-4]_rec[-1];TICK;H_YZ_0_1_2_3_4_5_6_7_8_9_10_11_12_13_14_15_16_17_18_19_20_21_22_23;SHIFT_COORDS(0,0,1);TICK;MZZ_1_7_4_10_5_6_11_12_13_19_16_21_17_18_22_23;M_0_2_3_8_9_14_15_20;OBSERVABLE_INCLUDE(0)_rec[-16]_rec[-12];DETECTOR(2.5,3,0)_rec[-73]_rec[-72]_rec[-70]_rec[-56]_rec[-55]_rec[-53]_rec[-28]_rec[-27]_rec[-25]_rec[-13]_rec[-12]_rec[-10];DETECTOR(1.5,0,0)_rec[-75]_rec[-66]_rec[-64]_rec[-60]_rec[-57]_rec[-32]_rec[-29]_rec[-15]_rec[-6]_rec[-4];DETECTOR(1.5,6,0)_rec[-65]_rec[-63]_rec[-58]_rec[-30]_rec[-5]_rec[-3];SHIFT_COORDS(0,0,1);TICK;H_0_1_2_3_4_5_6_7_8_9_10_11_12_13_14_15_16_17_18_19_20_21_22_23;TICK;MZZ_0_3_1_2_4_5_6_12_7_8_9_15_10_11_13_14_16_17_18_23_19_20_21_22;DETECTOR(1.5,2,0)_rec[-98]_rec[-97]_rec[-94]_rec[-87]_rec[-86]_rec[-85]_rec[-27]_rec[-26]_rec[-25]_rec[-10]_rec[-9]_rec[-6];DETECTOR(3.5,2,0)_rec[-92]_rec[-91]_rec[-89]_rec[-83]_rec[-82]_rec[-81]_rec[-23]_rec[-22]_rec[-21]_rec[-4]_rec[-3]_rec[-1];TICK;H_0_1_2_3_4_5_6_7_8_9_10_11_12_13_14_15_16_17_18_19_20_21_22_23;SHIFT_COORDS(0,0,1);TICK;MZZ_1_7_4_10_5_6_11_12_13_19_16_21_17_18_22_23;M_0_2_3_8_9_14_15_20;OBSERVABLE_INCLUDE(0)_rec[-16]_rec[-12];DETECTOR(1.5,2,0)_rec[-43]_rec[-42]_rec[-41]_rec[-15]_rec[-14]_rec[-13];DETECTOR(3.5,2,0)_rec[-39]_rec[-38]_rec[-37]_rec[-11]_rec[-10]_rec[-9];DETECTOR(0.5,-1,0)_rec[-36]_rec[-34]_rec[-8]_rec[-6];DETECTOR(0.5,5,0)_rec[-44]_rec[-35]_rec[-33]_rec[-16]_rec[-7]_rec[-5];DETECTOR(2.5,-1,0)_rec[-32]_rec[-30]_rec[-4]_rec[-2];DETECTOR(2.5,5,0)_rec[-40]_rec[-31]_rec[-29]_rec[-12]_rec[-3]_rec[-1];SHIFT_COORDS(0,0,1);TICK;H_YZ_0_1_2_3_4_5_6_7_8_9_10_11_12_13_14_15_16_17_18_19_20_21_22_23;TICK;MZZ_3_4_6_7_8_14_9_10_11_17_12_13_15_16_18_19;M_0_1_2_5_20_21_22_23;OBSERVABLE_INCLUDE(0)_rec[-14]_rec[-6]_rec[-4];DETECTOR(2.5,3,0)_rec[-72]_rec[-71]_rec[-69]_rec[-57]_rec[-56]_rec[-54]_rec[-29]_rec[-28]_rec[-26]_rec[-12]_rec[-11]_rec[-9];DETECTOR(0.5,3,0)_rec[-75]_rec[-67]_rec[-65]_rec[-60]_rec[-58]_rec[-32]_rec[-30]_rec[-15]_rec[-7]_rec[-5];DETECTOR(4.5,3,0)_rec[-62]_rec[-61]_rec[-53]_rec[-25]_rec[-2]_rec[-1];TICK;C_NZYX_0_1_2_3_4_5_6_7_8_9_10_11_12_13_14_15_16_17_18_19_20_21_22_23;SHIFT_COORDS(0,0,1);TICK;MZZ_0_3_1_2_4_5_6_12_7_8_9_15_10_11_13_14_16_17_18_23_19_20_21_22;DETECTOR(2.5,1,0)_rec[-95]_rec[-94]_rec[-92]_rec[-85]_rec[-84]_rec[-82]_rec[-25]_rec[-24]_rec[-22]_rec[-7]_rec[-6]_rec[-4];DETECTOR(1.5,4,0)_rec[-97]_rec[-96]_rec[-93]_rec[-87]_rec[-86]_rec[-83]_rec[-27]_rec[-26]_rec[-23]_rec[-9]_rec[-8]_rec[-5];TICK;C_XYNZ_0_1_2_3_4_5_6_7_8_9_10_11_12_13_14_15_16_17_18_19_20_21_22_23;SHIFT_COORDS(0,0,1);TICK;MZZ_3_4_6_7_8_14_9_10_11_17_12_13_15_16_18_19;M_0_1_2_5_20_21_22_23;OBSERVABLE_INCLUDE(0)_rec[-14]_rec[-6]_rec[-4];DETECTOR(2.5,1,0)_rec[-41]_rec[-40]_rec[-38]_rec[-13]_rec[-12]_rec[-10];DETECTOR(1.5,4,0)_rec[-43]_rec[-42]_rec[-39]_rec[-15]_rec[-14]_rec[-11];DETECTOR(0.5,1,0)_rec[-44]_rec[-36]_rec[-33]_rec[-16]_rec[-8]_rec[-5];DETECTOR(4.5,1,0)_rec[-31]_rec[-30]_rec[-3]_rec[-2];DETECTOR(-0.5,4,0)_rec[-35]_rec[-34]_rec[-7]_rec[-6];DETECTOR(3.5,4,0)_rec[-37]_rec[-32]_rec[-29]_rec[-9]_rec[-4]_rec[-1];TICK;H_YZ_0_1_2_3_4_5_6_7_8_9_10_11_12_13_14_15_16_17_18_19_20_21_22_23;SHIFT_COORDS(0,0,1);TICK;MZZ_1_7_4_10_5_6_11_12_13_19_16_21_17_18_22_23;M_0_2_3_8_9_14_15_20;OBSERVABLE_INCLUDE(0)_rec[-16]_rec[-12];DETECTOR(2.5,3,0)_rec[-73]_rec[-72]_rec[-70]_rec[-56]_rec[-55]_rec[-53]_rec[-28]_rec[-27]_rec[-25]_rec[-13]_rec[-12]_rec[-10];DETECTOR(1.5,0,0)_rec[-75]_rec[-66]_rec[-64]_rec[-60]_rec[-57]_rec[-32]_rec[-29]_rec[-15]_rec[-6]_rec[-4];DETECTOR(1.5,6,0)_rec[-65]_rec[-63]_rec[-58]_rec[-30]_rec[-5]_rec[-3];SHIFT_COORDS(0,0,1);TICK;H_0_1_2_3_4_5_6_7_8_9_10_11_12_13_14_15_16_17_18_19_20_21_22_23;TICK;MZZ_0_3_1_2_4_5_6_12_7_8_9_15_10_11_13_14_16_17_18_23_19_20_21_22;DETECTOR(1.5,2,0)_rec[-98]_rec[-97]_rec[-94]_rec[-87]_rec[-86]_rec[-85]_rec[-27]_rec[-26]_rec[-25]_rec[-10]_rec[-9]_rec[-6];DETECTOR(3.5,2,0)_rec[-92]_rec[-91]_rec[-89]_rec[-83]_rec[-82]_rec[-81]_rec[-23]_rec[-22]_rec[-21]_rec[-4]_rec[-3]_rec[-1];TICK;H_0_1_2_3_4_5_6_7_8_9_10_11_12_13_14_15_16_17_18_19_20_21_22_23;SHIFT_COORDS(0,0,1);TICK;MZZ_1_7_4_10_5_6_11_12_13_19_16_21_17_18_22_23;M_0_2_3_8_9_14_15_20;OBSERVABLE_INCLUDE(0)_rec[-16]_rec[-12];DETECTOR(1.5,2,0)_rec[-43]_rec[-42]_rec[-41]_rec[-15]_rec[-14]_rec[-13];DETECTOR(3.5,2,0)_rec[-39]_rec[-38]_rec[-37]_rec[-11]_rec[-10]_rec[-9];DETECTOR(0.5,-1,0)_rec[-36]_rec[-34]_rec[-8]_rec[-6];DETECTOR(0.5,5,0)_rec[-44]_rec[-35]_rec[-33]_rec[-16]_rec[-7]_rec[-5];DETECTOR(2.5,-1,0)_rec[-32]_rec[-30]_rec[-4]_rec[-2];DETECTOR(2.5,5,0)_rec[-40]_rec[-31]_rec[-29]_rec[-12]_rec[-3]_rec[-1];SHIFT_COORDS(0,0,1);TICK;H_YZ_0_1_2_3_4_5_6_7_8_9_10_11_12_13_14_15_16_17_18_19_20_21_22_23;TICK;MZZ_3_4_6_7_8_14_9_10_11_17_12_13_15_16_18_19;M_0_1_2_5_20_21_22_23;OBSERVABLE_INCLUDE(0)_rec[-14]_rec[-6]_rec[-4];DETECTOR(2.5,3,0)_rec[-72]_rec[-71]_rec[-69]_rec[-57]_rec[-56]_rec[-54]_rec[-29]_rec[-28]_rec[-26]_rec[-12]_rec[-11]_rec[-9];DETECTOR(0.5,3,0)_rec[-75]_rec[-67]_rec[-65]_rec[-60]_rec[-58]_rec[-32]_rec[-30]_rec[-15]_rec[-7]_rec[-5];DETECTOR(4.5,3,0)_rec[-62]_rec[-61]_rec[-53]_rec[-25]_rec[-2]_rec[-1];TICK;C_NZYX_0_1_2_3_4_5_6_7_8_9_10_11_12_13_14_15_16_17_18_19_20_21_22_23;SHIFT_COORDS(0,0,1);TICK;MZZ_0_3_1_2_4_5_6_12_7_8_9_15_10_11_13_14_16_17_18_23_19_20_21_22;DETECTOR(2.5,1,0)_rec[-95]_rec[-94]_rec[-92]_rec[-85]_rec[-84]_rec[-82]_rec[-25]_rec[-24]_rec[-22]_rec[-7]_rec[-6]_rec[-4];DETECTOR(1.5,4,0)_rec[-97]_rec[-96]_rec[-93]_rec[-87]_rec[-86]_rec[-83]_rec[-27]_rec[-26]_rec[-23]_rec[-9]_rec[-8]_rec[-5];TICK;C_XYNZ_0_1_2_3_4_5_6_7_8_9_10_11_12_13_14_15_16_17_18_19_20_21_22_23;SHIFT_COORDS(0,0,1);TICK;MZZ_3_4_6_7_8_14_9_10_11_17_12_13_15_16_18_19;M_0_1_2_5_20_21_22_23;OBSERVABLE_INCLUDE(0)_rec[-14]_rec[-6]_rec[-4];DETECTOR(2.5,1,0)_rec[-41]_rec[-40]_rec[-38]_rec[-13]_rec[-12]_rec[-10];DETECTOR(1.5,4,0)_rec[-43]_rec[-42]_rec[-39]_rec[-15]_rec[-14]_rec[-11];DETECTOR(0.5,1,0)_rec[-44]_rec[-36]_rec[-33]_rec[-16]_rec[-8]_rec[-5];DETECTOR(4.5,1,0)_rec[-31]_rec[-30]_rec[-3]_rec[-2];DETECTOR(-0.5,4,0)_rec[-35]_rec[-34]_rec[-7]_rec[-6];DETECTOR(3.5,4,0)_rec[-37]_rec[-32]_rec[-29]_rec[-9]_rec[-4]_rec[-1];SHIFT_COORDS(0,0,1);TICK;M_0_1_2_3_4_5_6_7_8_9_10_11_12_13_14_15_16_17_18_19_20_21_22_23;DETECTOR(1,0.5,0)_rec[-40]_rec[-21]_rec[-20];DETECTOR(3,0.5,0)_rec[-34]_rec[-9]_rec[-8];DETECTOR(2.5,2,0)_rec[-36]_rec[-13]_rec[-7];DETECTOR(2,0.5,0)_rec[-37]_rec[-15]_rec[-14];DETECTOR(2,3.5,0)_rec[-35]_rec[-12]_rec[-11];DETECTOR(1.5,5,0)_rec[-38]_rec[-16]_rec[-10];DETECTOR(1,3.5,0)_rec[-39]_rec[-18]_rec[-17];DETECTOR(3,3.5,0)_rec[-33]_rec[-6]_rec[-5];DETECTOR(1,2,0)_rec[-29]_rec[-19];DETECTOR(0,0,0)_rec[-32]_rec[-24];DETECTOR(4,2,0)_rec[-26]_rec[-2];DETECTOR(4,1,0)_rec[-27]_rec[-3];DETECTOR(0,4,0)_rec[-31]_rec[-23];DETECTOR(0,5,0)_rec[-30]_rec[-22];DETECTOR(4,3,0)_rec[-25]_rec[-1];DETECTOR(3,5,0)_rec[-28]_rec[-4];DETECTOR(1.5,2,0)_rec[-94]_rec[-93]_rec[-90]_rec[-83]_rec[-82]_rec[-81]_rec[-20]_rec[-19]_rec[-18]_rec[-14]_rec[-13]_rec[-12];DETECTOR(3.5,2,0)_rec[-88]_rec[-87]_rec[-85]_rec[-79]_rec[-78]_rec[-77]_rec[-8]_rec[-7]_rec[-6]_rec[-3]_rec[-2]_rec[-1];OBSERVABLE_INCLUDE(0)_rec[-23]_rec[-17]_rec[-11]_rec[-5]}{Crumble}}

%% file: crumble_url_vertical_memory.tex
\newcommand{\crumbleVerticalMemory}{\href{http://algassert.com/crumble\#circuit=Q(0.0,0.0)0;Q(0.0,4.0)1;Q(0.0,5.0)2;Q(1.0,0.0)3;Q(1.0,1.0)4;Q(1.0,2.0)5;Q(1.0,3.0)6;Q(1.0,4.0)7;Q(1.0,5.0)8;Q(2.0,0.0)9;Q(2.0,1.0)10;Q(2.0,2.0)11;Q(2.0,3.0)12;Q(2.0,4.0)13;Q(2.0,5.0)14;Q(3.0,0.0)15;Q(3.0,1.0)16;Q(3.0,2.0)17;Q(3.0,3.0)18;Q(3.0,4.0)19;Q(3.0,5.0)20;Q(4.0,1.0)21;Q(4.0,2.0)22;Q(4.0,3.0)23;POLYGON(1,0,0,0.25)_0;POLYGON(1,0,0,0.25)_2;POLYGON(1,0,0,0.25)_3_4_10_9;POLYGON(1,0,0,0.25)_8_14;POLYGON(1,0,0,0.25)_15_16_21;POLYGON(1,0,0,0.25)_20;POLYGON(1,0,0,0.25)_1_7_6_5;POLYGON(1,0,0,0.25)_11_12_13_19_18_17;POLYGON(1,0,0,0.25)_22_23;POLYGON(0,1,0,0.25)_0_3;POLYGON(0,1,0,0.25)_1_2_8_7;POLYGON(0,1,0,0.25)_9_15;POLYGON(0,1,0,0.25)_4_5_6_12_11_10;POLYGON(0,1,0,0.25)_13_14_20_19;POLYGON(0,1,0,0.25)_16_17_18_23_22_21;POLYGON(0,0,1,0.25)_0_5_4_3;POLYGON(0,0,1,0.25)_9_10_11_17_16_15;POLYGON(0,0,1,0.25)_21_22;POLYGON(0,0,1,0.25)_2_1;POLYGON(0,0,1,0.25)_6_7_8_14_13_12;POLYGON(0,0,1,0.25)_18_19_20_23;R_0_1_2_3_4_5_6_7_8_9_10_11_12_13_14_15_16_17_18_19_20_21_22_23;SHIFT_COORDS(0,0,1);TICK;MZZ_1_7_4_10_5_6_11_12_13_19_16_21_17_18_22_23;M_0_2_3_8_9_14_15_20;OBSERVABLE_INCLUDE(0)_rec[-13]_rec[-4]_rec[-3];DETECTOR(1.5,1,0)_rec[-15];DETECTOR(3.5,1,0)_rec[-11];DETECTOR(1,2.5,0)_rec[-14];DETECTOR(0.5,4,0)_rec[-16];DETECTOR(3,2.5,0)_rec[-10];DETECTOR(2.5,4,0)_rec[-12];DETECTOR(2,2.5,0)_rec[-13];DETECTOR(4,2.5,0)_rec[-9];DETECTOR(0,0,0)_rec[-8];DETECTOR(0,5,0)_rec[-7];DETECTOR(2,0,0)_rec[-4];DETECTOR(1,0,0)_rec[-6];DETECTOR(2,5,0)_rec[-3];DETECTOR(1,5,0)_rec[-5];DETECTOR(3,0,0)_rec[-2];DETECTOR(3,5,0)_rec[-1];SHIFT_COORDS(0,0,1);TICK;H_0_1_2_3_4_5_6_7_8_9_10_11_12_13_14_15_16_17_18_19_20_21_22_23;TICK;MZZ_0_3_1_2_4_5_6_12_7_8_9_15_10_11_13_14_16_17_18_23_19_20_21_22;TICK;H_0_1_2_3_4_5_6_7_8_9_10_11_12_13_14_15_16_17_18_19_20_21_22_23;TICK;MZZ_1_7_4_10_5_6_11_12_13_19_16_21_17_18_22_23;M_0_2_3_8_9_14_15_20;OBSERVABLE_INCLUDE(0)_rec[-13]_rec[-4]_rec[-3];DETECTOR(1.5,2,0)_rec[-43]_rec[-42]_rec[-41]_rec[-15]_rec[-14]_rec[-13];DETECTOR(3.5,2,0)_rec[-39]_rec[-38]_rec[-37]_rec[-11]_rec[-10]_rec[-9];DETECTOR(0.5,-1,0)_rec[-36]_rec[-34]_rec[-8]_rec[-6];DETECTOR(0.5,5,0)_rec[-44]_rec[-35]_rec[-33]_rec[-16]_rec[-7]_rec[-5];DETECTOR(2.5,-1,0)_rec[-32]_rec[-30]_rec[-4]_rec[-2];DETECTOR(2.5,5,0)_rec[-40]_rec[-31]_rec[-29]_rec[-12]_rec[-3]_rec[-1];SHIFT_COORDS(0,0,1);TICK;H_YZ_0_1_2_3_4_5_6_7_8_9_10_11_12_13_14_15_16_17_18_19_20_21_22_23;TICK;MZZ_3_4_6_7_8_14_9_10_11_17_12_13_15_16_18_19;M_0_1_2_5_20_21_22_23;OBSERVABLE_INCLUDE(0)_rec[-13]_rec[-11];TICK;C_NZYX_0_1_2_3_4_5_6_7_8_9_10_11_12_13_14_15_16_17_18_19_20_21_22_23;TICK;MZZ_0_3_1_2_4_5_6_12_7_8_9_15_10_11_13_14_16_17_18_23_19_20_21_22;DETECTOR(2.5,1,0)_rec[-25]_rec[-24]_rec[-22]_rec[-7]_rec[-6]_rec[-4];DETECTOR(1.5,4,0)_rec[-27]_rec[-26]_rec[-23]_rec[-9]_rec[-8]_rec[-5];TICK;C_XYNZ_0_1_2_3_4_5_6_7_8_9_10_11_12_13_14_15_16_17_18_19_20_21_22_23;SHIFT_COORDS(0,0,1);TICK;MZZ_3_4_6_7_8_14_9_10_11_17_12_13_15_16_18_19;M_0_1_2_5_20_21_22_23;OBSERVABLE_INCLUDE(0)_rec[-13]_rec[-11];DETECTOR(2.5,1,0)_rec[-41]_rec[-40]_rec[-38]_rec[-13]_rec[-12]_rec[-10];DETECTOR(1.5,4,0)_rec[-43]_rec[-42]_rec[-39]_rec[-15]_rec[-14]_rec[-11];DETECTOR(0.5,1,0)_rec[-44]_rec[-36]_rec[-33]_rec[-16]_rec[-8]_rec[-5];DETECTOR(4.5,1,0)_rec[-31]_rec[-30]_rec[-3]_rec[-2];DETECTOR(-0.5,4,0)_rec[-35]_rec[-34]_rec[-7]_rec[-6];DETECTOR(3.5,4,0)_rec[-37]_rec[-32]_rec[-29]_rec[-9]_rec[-4]_rec[-1];TICK;H_YZ_0_1_2_3_4_5_6_7_8_9_10_11_12_13_14_15_16_17_18_19_20_21_22_23;SHIFT_COORDS(0,0,1);TICK;MZZ_1_7_4_10_5_6_11_12_13_19_16_21_17_18_22_23;M_0_2_3_8_9_14_15_20;OBSERVABLE_INCLUDE(0)_rec[-13]_rec[-4]_rec[-3];DETECTOR(2.5,3,0)_rec[-73]_rec[-72]_rec[-70]_rec[-56]_rec[-55]_rec[-53]_rec[-28]_rec[-27]_rec[-25]_rec[-13]_rec[-12]_rec[-10];DETECTOR(1.5,0,0)_rec[-75]_rec[-66]_rec[-64]_rec[-60]_rec[-57]_rec[-32]_rec[-29]_rec[-15]_rec[-6]_rec[-4];DETECTOR(1.5,6,0)_rec[-65]_rec[-63]_rec[-58]_rec[-30]_rec[-5]_rec[-3];SHIFT_COORDS(0,0,1);TICK;H_0_1_2_3_4_5_6_7_8_9_10_11_12_13_14_15_16_17_18_19_20_21_22_23;TICK;MZZ_0_3_1_2_4_5_6_12_7_8_9_15_10_11_13_14_16_17_18_23_19_20_21_22;DETECTOR(1.5,2,0)_rec[-98]_rec[-97]_rec[-94]_rec[-87]_rec[-86]_rec[-85]_rec[-27]_rec[-26]_rec[-25]_rec[-10]_rec[-9]_rec[-6];DETECTOR(3.5,2,0)_rec[-92]_rec[-91]_rec[-89]_rec[-83]_rec[-82]_rec[-81]_rec[-23]_rec[-22]_rec[-21]_rec[-4]_rec[-3]_rec[-1];TICK;H_0_1_2_3_4_5_6_7_8_9_10_11_12_13_14_15_16_17_18_19_20_21_22_23;SHIFT_COORDS(0,0,1);TICK;MZZ_1_7_4_10_5_6_11_12_13_19_16_21_17_18_22_23;M_0_2_3_8_9_14_15_20;OBSERVABLE_INCLUDE(0)_rec[-13]_rec[-4]_rec[-3];DETECTOR(1.5,2,0)_rec[-43]_rec[-42]_rec[-41]_rec[-15]_rec[-14]_rec[-13];DETECTOR(3.5,2,0)_rec[-39]_rec[-38]_rec[-37]_rec[-11]_rec[-10]_rec[-9];DETECTOR(0.5,-1,0)_rec[-36]_rec[-34]_rec[-8]_rec[-6];DETECTOR(0.5,5,0)_rec[-44]_rec[-35]_rec[-33]_rec[-16]_rec[-7]_rec[-5];DETECTOR(2.5,-1,0)_rec[-32]_rec[-30]_rec[-4]_rec[-2];DETECTOR(2.5,5,0)_rec[-40]_rec[-31]_rec[-29]_rec[-12]_rec[-3]_rec[-1];SHIFT_COORDS(0,0,1);TICK;H_YZ_0_1_2_3_4_5_6_7_8_9_10_11_12_13_14_15_16_17_18_19_20_21_22_23;TICK;MZZ_3_4_6_7_8_14_9_10_11_17_12_13_15_16_18_19;M_0_1_2_5_20_21_22_23;OBSERVABLE_INCLUDE(0)_rec[-13]_rec[-11];DETECTOR(2.5,3,0)_rec[-72]_rec[-71]_rec[-69]_rec[-57]_rec[-56]_rec[-54]_rec[-29]_rec[-28]_rec[-26]_rec[-12]_rec[-11]_rec[-9];DETECTOR(0.5,3,0)_rec[-75]_rec[-67]_rec[-65]_rec[-60]_rec[-58]_rec[-32]_rec[-30]_rec[-15]_rec[-7]_rec[-5];DETECTOR(4.5,3,0)_rec[-62]_rec[-61]_rec[-53]_rec[-25]_rec[-2]_rec[-1];TICK;C_NZYX_0_1_2_3_4_5_6_7_8_9_10_11_12_13_14_15_16_17_18_19_20_21_22_23;SHIFT_COORDS(0,0,1);TICK;MZZ_0_3_1_2_4_5_6_12_7_8_9_15_10_11_13_14_16_17_18_23_19_20_21_22;DETECTOR(2.5,1,0)_rec[-95]_rec[-94]_rec[-92]_rec[-85]_rec[-84]_rec[-82]_rec[-25]_rec[-24]_rec[-22]_rec[-7]_rec[-6]_rec[-4];DETECTOR(1.5,4,0)_rec[-97]_rec[-96]_rec[-93]_rec[-87]_rec[-86]_rec[-83]_rec[-27]_rec[-26]_rec[-23]_rec[-9]_rec[-8]_rec[-5];TICK;C_XYNZ_0_1_2_3_4_5_6_7_8_9_10_11_12_13_14_15_16_17_18_19_20_21_22_23;SHIFT_COORDS(0,0,1);TICK;MZZ_3_4_6_7_8_14_9_10_11_17_12_13_15_16_18_19;M_0_1_2_5_20_21_22_23;OBSERVABLE_INCLUDE(0)_rec[-13]_rec[-11];DETECTOR(2.5,1,0)_rec[-41]_rec[-40]_rec[-38]_rec[-13]_rec[-12]_rec[-10];DETECTOR(1.5,4,0)_rec[-43]_rec[-42]_rec[-39]_rec[-15]_rec[-14]_rec[-11];DETECTOR(0.5,1,0)_rec[-44]_rec[-36]_rec[-33]_rec[-16]_rec[-8]_rec[-5];DETECTOR(4.5,1,0)_rec[-31]_rec[-30]_rec[-3]_rec[-2];DETECTOR(-0.5,4,0)_rec[-35]_rec[-34]_rec[-7]_rec[-6];DETECTOR(3.5,4,0)_rec[-37]_rec[-32]_rec[-29]_rec[-9]_rec[-4]_rec[-1];TICK;H_YZ_0_1_2_3_4_5_6_7_8_9_10_11_12_13_14_15_16_17_18_19_20_21_22_23;SHIFT_COORDS(0,0,1);TICK;MZZ_1_7_4_10_5_6_11_12_13_19_16_21_17_18_22_23;M_0_2_3_8_9_14_15_20;OBSERVABLE_INCLUDE(0)_rec[-13]_rec[-4]_rec[-3];DETECTOR(2.5,3,0)_rec[-73]_rec[-72]_rec[-70]_rec[-56]_rec[-55]_rec[-53]_rec[-28]_rec[-27]_rec[-25]_rec[-13]_rec[-12]_rec[-10];DETECTOR(1.5,0,0)_rec[-75]_rec[-66]_rec[-64]_rec[-60]_rec[-57]_rec[-32]_rec[-29]_rec[-15]_rec[-6]_rec[-4];DETECTOR(1.5,6,0)_rec[-65]_rec[-63]_rec[-58]_rec[-30]_rec[-5]_rec[-3];SHIFT_COORDS(0,0,1);TICK;H_0_1_2_3_4_5_6_7_8_9_10_11_12_13_14_15_16_17_18_19_20_21_22_23;TICK;MZZ_0_3_1_2_4_5_6_12_7_8_9_15_10_11_13_14_16_17_18_23_19_20_21_22;DETECTOR(1.5,2,0)_rec[-98]_rec[-97]_rec[-94]_rec[-87]_rec[-86]_rec[-85]_rec[-27]_rec[-26]_rec[-25]_rec[-10]_rec[-9]_rec[-6];DETECTOR(3.5,2,0)_rec[-92]_rec[-91]_rec[-89]_rec[-83]_rec[-82]_rec[-81]_rec[-23]_rec[-22]_rec[-21]_rec[-4]_rec[-3]_rec[-1];TICK;H_0_1_2_3_4_5_6_7_8_9_10_11_12_13_14_15_16_17_18_19_20_21_22_23;SHIFT_COORDS(0,0,1);TICK;MZZ_1_7_4_10_5_6_11_12_13_19_16_21_17_18_22_23;M_0_2_3_8_9_14_15_20;OBSERVABLE_INCLUDE(0)_rec[-13]_rec[-4]_rec[-3];DETECTOR(1.5,2,0)_rec[-43]_rec[-42]_rec[-41]_rec[-15]_rec[-14]_rec[-13];DETECTOR(3.5,2,0)_rec[-39]_rec[-38]_rec[-37]_rec[-11]_rec[-10]_rec[-9];DETECTOR(0.5,-1,0)_rec[-36]_rec[-34]_rec[-8]_rec[-6];DETECTOR(0.5,5,0)_rec[-44]_rec[-35]_rec[-33]_rec[-16]_rec[-7]_rec[-5];DETECTOR(2.5,-1,0)_rec[-32]_rec[-30]_rec[-4]_rec[-2];DETECTOR(2.5,5,0)_rec[-40]_rec[-31]_rec[-29]_rec[-12]_rec[-3]_rec[-1];SHIFT_COORDS(0,0,1);TICK;H_YZ_0_1_2_3_4_5_6_7_8_9_10_11_12_13_14_15_16_17_18_19_20_21_22_23;TICK;MZZ_3_4_6_7_8_14_9_10_11_17_12_13_15_16_18_19;M_0_1_2_5_20_21_22_23;OBSERVABLE_INCLUDE(0)_rec[-13]_rec[-11];DETECTOR(2.5,3,0)_rec[-72]_rec[-71]_rec[-69]_rec[-57]_rec[-56]_rec[-54]_rec[-29]_rec[-28]_rec[-26]_rec[-12]_rec[-11]_rec[-9];DETECTOR(0.5,3,0)_rec[-75]_rec[-67]_rec[-65]_rec[-60]_rec[-58]_rec[-32]_rec[-30]_rec[-15]_rec[-7]_rec[-5];DETECTOR(4.5,3,0)_rec[-62]_rec[-61]_rec[-53]_rec[-25]_rec[-2]_rec[-1];TICK;C_NZYX_0_1_2_3_4_5_6_7_8_9_10_11_12_13_14_15_16_17_18_19_20_21_22_23;SHIFT_COORDS(0,0,1);TICK;MZZ_0_3_1_2_4_5_6_12_7_8_9_15_10_11_13_14_16_17_18_23_19_20_21_22;DETECTOR(2.5,1,0)_rec[-95]_rec[-94]_rec[-92]_rec[-85]_rec[-84]_rec[-82]_rec[-25]_rec[-24]_rec[-22]_rec[-7]_rec[-6]_rec[-4];DETECTOR(1.5,4,0)_rec[-97]_rec[-96]_rec[-93]_rec[-87]_rec[-86]_rec[-83]_rec[-27]_rec[-26]_rec[-23]_rec[-9]_rec[-8]_rec[-5];TICK;C_XYNZ_0_1_2_3_4_5_6_7_8_9_10_11_12_13_14_15_16_17_18_19_20_21_22_23;SHIFT_COORDS(0,0,1);TICK;MZZ_3_4_6_7_8_14_9_10_11_17_12_13_15_16_18_19;M_0_1_2_5_20_21_22_23;OBSERVABLE_INCLUDE(0)_rec[-13]_rec[-11];DETECTOR(2.5,1,0)_rec[-41]_rec[-40]_rec[-38]_rec[-13]_rec[-12]_rec[-10];DETECTOR(1.5,4,0)_rec[-43]_rec[-42]_rec[-39]_rec[-15]_rec[-14]_rec[-11];DETECTOR(0.5,1,0)_rec[-44]_rec[-36]_rec[-33]_rec[-16]_rec[-8]_rec[-5];DETECTOR(4.5,1,0)_rec[-31]_rec[-30]_rec[-3]_rec[-2];DETECTOR(-0.5,4,0)_rec[-35]_rec[-34]_rec[-7]_rec[-6];DETECTOR(3.5,4,0)_rec[-37]_rec[-32]_rec[-29]_rec[-9]_rec[-4]_rec[-1];TICK;H_YZ_0_1_2_3_4_5_6_7_8_9_10_11_12_13_14_15_16_17_18_19_20_21_22_23;SHIFT_COORDS(0,0,1);TICK;MZZ_1_7_4_10_5_6_11_12_13_19_16_21_17_18_22_23;M_0_2_3_8_9_14_15_20;OBSERVABLE_INCLUDE(0)_rec[-13]_rec[-4]_rec[-3];DETECTOR(2.5,3,0)_rec[-73]_rec[-72]_rec[-70]_rec[-56]_rec[-55]_rec[-53]_rec[-28]_rec[-27]_rec[-25]_rec[-13]_rec[-12]_rec[-10];DETECTOR(1.5,0,0)_rec[-75]_rec[-66]_rec[-64]_rec[-60]_rec[-57]_rec[-32]_rec[-29]_rec[-15]_rec[-6]_rec[-4];DETECTOR(1.5,6,0)_rec[-65]_rec[-63]_rec[-58]_rec[-30]_rec[-5]_rec[-3];SHIFT_COORDS(0,0,1);TICK;H_0_1_2_3_4_5_6_7_8_9_10_11_12_13_14_15_16_17_18_19_20_21_22_23;TICK;MZZ_0_3_1_2_4_5_6_12_7_8_9_15_10_11_13_14_16_17_18_23_19_20_21_22;DETECTOR(1.5,2,0)_rec[-98]_rec[-97]_rec[-94]_rec[-87]_rec[-86]_rec[-85]_rec[-27]_rec[-26]_rec[-25]_rec[-10]_rec[-9]_rec[-6];DETECTOR(3.5,2,0)_rec[-92]_rec[-91]_rec[-89]_rec[-83]_rec[-82]_rec[-81]_rec[-23]_rec[-22]_rec[-21]_rec[-4]_rec[-3]_rec[-1];TICK;H_0_1_2_3_4_5_6_7_8_9_10_11_12_13_14_15_16_17_18_19_20_21_22_23;SHIFT_COORDS(0,0,1);TICK;MZZ_1_7_4_10_5_6_11_12_13_19_16_21_17_18_22_23;M_0_2_3_8_9_14_15_20;OBSERVABLE_INCLUDE(0)_rec[-13]_rec[-4]_rec[-3];DETECTOR(1.5,2,0)_rec[-43]_rec[-42]_rec[-41]_rec[-15]_rec[-14]_rec[-13];DETECTOR(3.5,2,0)_rec[-39]_rec[-38]_rec[-37]_rec[-11]_rec[-10]_rec[-9];DETECTOR(0.5,-1,0)_rec[-36]_rec[-34]_rec[-8]_rec[-6];DETECTOR(0.5,5,0)_rec[-44]_rec[-35]_rec[-33]_rec[-16]_rec[-7]_rec[-5];DETECTOR(2.5,-1,0)_rec[-32]_rec[-30]_rec[-4]_rec[-2];DETECTOR(2.5,5,0)_rec[-40]_rec[-31]_rec[-29]_rec[-12]_rec[-3]_rec[-1];SHIFT_COORDS(0,0,1);TICK;H_YZ_0_1_2_3_4_5_6_7_8_9_10_11_12_13_14_15_16_17_18_19_20_21_22_23;TICK;MZZ_3_4_6_7_8_14_9_10_11_17_12_13_15_16_18_19;M_0_1_2_5_20_21_22_23;OBSERVABLE_INCLUDE(0)_rec[-13]_rec[-11];DETECTOR(2.5,3,0)_rec[-72]_rec[-71]_rec[-69]_rec[-57]_rec[-56]_rec[-54]_rec[-29]_rec[-28]_rec[-26]_rec[-12]_rec[-11]_rec[-9];DETECTOR(0.5,3,0)_rec[-75]_rec[-67]_rec[-65]_rec[-60]_rec[-58]_rec[-32]_rec[-30]_rec[-15]_rec[-7]_rec[-5];DETECTOR(4.5,3,0)_rec[-62]_rec[-61]_rec[-53]_rec[-25]_rec[-2]_rec[-1];TICK;C_NZYX_0_1_2_3_4_5_6_7_8_9_10_11_12_13_14_15_16_17_18_19_20_21_22_23;SHIFT_COORDS(0,0,1);TICK;MZZ_0_3_1_2_4_5_6_12_7_8_9_15_10_11_13_14_16_17_18_23_19_20_21_22;DETECTOR(2.5,1,0)_rec[-95]_rec[-94]_rec[-92]_rec[-85]_rec[-84]_rec[-82]_rec[-25]_rec[-24]_rec[-22]_rec[-7]_rec[-6]_rec[-4];DETECTOR(1.5,4,0)_rec[-97]_rec[-96]_rec[-93]_rec[-87]_rec[-86]_rec[-83]_rec[-27]_rec[-26]_rec[-23]_rec[-9]_rec[-8]_rec[-5];TICK;C_XYNZ_0_1_2_3_4_5_6_7_8_9_10_11_12_13_14_15_16_17_18_19_20_21_22_23;SHIFT_COORDS(0,0,1);TICK;MZZ_3_4_6_7_8_14_9_10_11_17_12_13_15_16_18_19;M_0_1_2_5_20_21_22_23;OBSERVABLE_INCLUDE(0)_rec[-13]_rec[-11];DETECTOR(2.5,1,0)_rec[-41]_rec[-40]_rec[-38]_rec[-13]_rec[-12]_rec[-10];DETECTOR(1.5,4,0)_rec[-43]_rec[-42]_rec[-39]_rec[-15]_rec[-14]_rec[-11];DETECTOR(0.5,1,0)_rec[-44]_rec[-36]_rec[-33]_rec[-16]_rec[-8]_rec[-5];DETECTOR(4.5,1,0)_rec[-31]_rec[-30]_rec[-3]_rec[-2];DETECTOR(-0.5,4,0)_rec[-35]_rec[-34]_rec[-7]_rec[-6];DETECTOR(3.5,4,0)_rec[-37]_rec[-32]_rec[-29]_rec[-9]_rec[-4]_rec[-1];TICK;H_YZ_0_1_2_3_4_5_6_7_8_9_10_11_12_13_14_15_16_17_18_19_20_21_22_23;SHIFT_COORDS(0,0,1);TICK;M_0_1_2_3_4_5_6_7_8_9_10_11_12_13_14_15_16_17_18_19_20_21_22_23;DETECTOR(2.5,3,0)_rec[-81]_rec[-80]_rec[-78]_rec[-64]_rec[-63]_rec[-61]_rec[-36]_rec[-35]_rec[-33]_rec[-13]_rec[-12]_rec[-11]_rec[-7]_rec[-6]_rec[-5];DETECTOR(2.5,1,0)_rec[-47]_rec[-46]_rec[-44]_rec[-37]_rec[-36]_rec[-34]_rec[-15]_rec[-14]_rec[-13]_rec[-9]_rec[-8]_rec[-7];DETECTOR(1.5,4,0)_rec[-49]_rec[-48]_rec[-45]_rec[-39]_rec[-38]_rec[-35]_rec[-18]_rec[-17]_rec[-16]_rec[-12]_rec[-11]_rec[-10];DETECTOR(1.5,0.25,0)_rec[-83]_rec[-74]_rec[-72]_rec[-68]_rec[-65]_rec[-40]_rec[-37]_rec[-21]_rec[-20]_rec[-15]_rec[-14];DETECTOR(1.5,5.5,0)_rec[-73]_rec[-71]_rec[-66]_rec[-38]_rec[-16]_rec[-10];OBSERVABLE_INCLUDE(0)_rec[-15]_rec[-14]_rec[-12]_rec[-11]}{Crumble}}

%% file: crumble_url_horizontal_stability.tex
\newcommand{\crumbleHorizontalStability}{\href{http://algassert.com/crumble\#circuit=Q(0.0,4.0)0;Q(0.0,5.0)1;Q(0.0,6.0)2;Q(1.0,1.0)3;Q(1.0,2.0)4;Q(1.0,3.0)5;Q(1.0,4.0)6;Q(1.0,5.0)7;Q(1.0,6.0)8;Q(2.0,1.0)9;Q(2.0,2.0)10;Q(2.0,3.0)11;Q(2.0,4.0)12;Q(2.0,5.0)13;Q(2.0,6.0)14;Q(3.0,1.0)15;Q(3.0,2.0)16;Q(3.0,3.0)17;Q(3.0,4.0)18;Q(3.0,5.0)19;Q(3.0,6.0)20;Q(4.0,1.0)21;Q(4.0,2.0)22;Q(4.0,3.0)23;POLYGON(1,0,0,0.25)_2_1;POLYGON(1,0,0,0.25)_3_9;POLYGON(1,0,0,0.25)_7_8_14_13;POLYGON(1,0,0,0.25)_15_21;POLYGON(1,0,0,0.25)_19_20;POLYGON(1,0,0,0.25)_0_6_5_4;POLYGON(1,0,0,0.25)_10_11_12_18_17_16;POLYGON(1,0,0,0.25)_22_23;POLYGON(0,1,0,0.25)_0_1_2_8_7_6;POLYGON(0,1,0,0.25)_3_4_5_11_10_9;POLYGON(0,1,0,0.25)_12_13_14_20_19_18;POLYGON(0,1,0,0.25)_15_16_17_23_22_21;POLYGON(0,0,1,0.25)_4_3;POLYGON(0,0,1,0.25)_2_8;POLYGON(0,0,1,0.25)_9_10_16_15;POLYGON(0,0,1,0.25)_14_20;POLYGON(0,0,1,0.25)_21_22;POLYGON(0,0,1,0.25)_1_0;POLYGON(0,0,1,0.25)_5_6_7_13_12_11;POLYGON(0,0,1,0.25)_17_18_19_23;R_0_1_2_3_4_5_6_7_8_9_10_11_12_13_14_15_16_17_18_19_20_21_22_23;SHIFT_COORDS(0,0,1);TICK;MZZ_0_6_1_2_3_9_4_5_7_8_10_11_12_18_13_14_15_21_16_17_19_20_22_23;DETECTOR(0,5.5,0)_rec[-11];DETECTOR(1.5,1,0)_rec[-10];DETECTOR(2,5.5,0)_rec[-5];DETECTOR(1,5.5,0)_rec[-8];DETECTOR(3.5,1,0)_rec[-4];DETECTOR(3,5.5,0)_rec[-2];DETECTOR(1,2.5,0)_rec[-9];DETECTOR(0.5,4,0)_rec[-12];DETECTOR(3,2.5,0)_rec[-3];DETECTOR(2.5,4,0)_rec[-6];DETECTOR(2,2.5,0)_rec[-7];DETECTOR(4,2.5,0)_rec[-1];SHIFT_COORDS(0,0,1);TICK;H_0_1_2_3_4_5_6_7_8_9_10_11_12_13_14_15_16_17_18_19_20_21_22_23;TICK;MZZ_0_1_2_8_3_4_5_11_6_7_9_10_12_13_14_20_15_16_17_23_18_19_21_22;TICK;H_0_1_2_3_4_5_6_7_8_9_10_11_12_13_14_15_16_17_18_19_20_21_22_23;TICK;MZZ_0_6_1_2_3_9_4_5_7_8_10_11_12_18_13_14_15_21_16_17_19_20_22_23;DETECTOR(0.5,5,0)_rec[-36]_rec[-35]_rec[-32]_rec[-12]_rec[-11]_rec[-8];DETECTOR(1.5,2,0)_rec[-34]_rec[-33]_rec[-31]_rec[-10]_rec[-9]_rec[-7];DETECTOR(2.5,5,0)_rec[-30]_rec[-29]_rec[-26]_rec[-6]_rec[-5]_rec[-2];DETECTOR(3.5,2,0)_rec[-28]_rec[-27]_rec[-25]_rec[-4]_rec[-3]_rec[-1];SHIFT_COORDS(0,0,1);TICK;H_YZ_0_1_2_3_4_5_6_7_8_9_10_11_12_13_14_15_16_17_18_19_20_21_22_23;TICK;MZZ_5_6_7_13_10_16_11_12_17_18;M_0_1_2_3_4_8_9_14_15_19_20_21_22_23;TICK;C_NZYX_0_1_2_3_4_5_6_7_8_9_10_11_12_13_14_15_16_17_18_19_20_21_22_23;TICK;MZZ_0_1_2_8_3_4_5_11_6_7_9_10_12_13_14_20_15_16_17_23_18_19_21_22;DETECTOR(1.5,4,0)_rec[-31]_rec[-30]_rec[-28]_rec[-9]_rec[-8]_rec[-6];TICK;C_XYNZ_0_1_2_3_4_5_6_7_8_9_10_11_12_13_14_15_16_17_18_19_20_21_22_23;SHIFT_COORDS(0,0,1);TICK;MZZ_5_6_7_13_10_16_11_12_17_18;M_0_1_2_3_4_8_9_14_15_19_20_21_22_23;DETECTOR(1.5,4,0)_rec[-50]_rec[-49]_rec[-47]_rec[-19]_rec[-18]_rec[-16];DETECTOR(0.5,1,0)_rec[-42]_rec[-41]_rec[-11]_rec[-10];DETECTOR(0.5,7,0)_rec[-43]_rec[-40]_rec[-12]_rec[-9];DETECTOR(2.5,1,0)_rec[-48]_rec[-39]_rec[-37]_rec[-17]_rec[-8]_rec[-6];DETECTOR(2.5,7,0)_rec[-38]_rec[-35]_rec[-7]_rec[-4];DETECTOR(4.5,1,0)_rec[-34]_rec[-33]_rec[-3]_rec[-2];DETECTOR(-0.5,4,0)_rec[-45]_rec[-44]_rec[-14]_rec[-13];DETECTOR(3.5,4,0)_rec[-46]_rec[-36]_rec[-32]_rec[-15]_rec[-5]_rec[-1];TICK;H_YZ_0_1_2_3_4_5_6_7_8_9_10_11_12_13_14_15_16_17_18_19_20_21_22_23;SHIFT_COORDS(0,0,1);TICK;MZZ_0_6_1_2_3_9_4_5_7_8_10_11_12_18_13_14_15_21_16_17_19_20_22_23;DETECTOR(2.5,3,0)_rec[-69]_rec[-68]_rec[-65]_rec[-60]_rec[-59]_rec[-58]_rec[-29]_rec[-28]_rec[-27]_rec[-7]_rec[-6]_rec[-3];SHIFT_COORDS(0,0,1);TICK;H_0_1_2_3_4_5_6_7_8_9_10_11_12_13_14_15_16_17_18_19_20_21_22_23;TICK;MZZ_0_1_2_8_3_4_5_11_6_7_9_10_12_13_14_20_15_16_17_23_18_19_21_22;DETECTOR(0.5,5,0)_rec[-98]_rec[-97]_rec[-94]_rec[-86]_rec[-85]_rec[-82]_rec[-24]_rec[-23]_rec[-20]_rec[-12]_rec[-11]_rec[-8];DETECTOR(1.5,2,0)_rec[-96]_rec[-95]_rec[-93]_rec[-84]_rec[-83]_rec[-81]_rec[-22]_rec[-21]_rec[-19]_rec[-10]_rec[-9]_rec[-7];DETECTOR(2.5,5,0)_rec[-92]_rec[-91]_rec[-88]_rec[-80]_rec[-79]_rec[-76]_rec[-18]_rec[-17]_rec[-14]_rec[-6]_rec[-5]_rec[-2];DETECTOR(3.5,2,0)_rec[-90]_rec[-89]_rec[-87]_rec[-78]_rec[-77]_rec[-75]_rec[-16]_rec[-15]_rec[-13]_rec[-4]_rec[-3]_rec[-1];TICK;H_0_1_2_3_4_5_6_7_8_9_10_11_12_13_14_15_16_17_18_19_20_21_22_23;SHIFT_COORDS(0,0,1);TICK;MZZ_0_6_1_2_3_9_4_5_7_8_10_11_12_18_13_14_15_21_16_17_19_20_22_23;DETECTOR(0.5,5,0)_rec[-36]_rec[-35]_rec[-32]_rec[-12]_rec[-11]_rec[-8];DETECTOR(1.5,2,0)_rec[-34]_rec[-33]_rec[-31]_rec[-10]_rec[-9]_rec[-7];DETECTOR(2.5,5,0)_rec[-30]_rec[-29]_rec[-26]_rec[-6]_rec[-5]_rec[-2];DETECTOR(3.5,2,0)_rec[-28]_rec[-27]_rec[-25]_rec[-4]_rec[-3]_rec[-1];SHIFT_COORDS(0,0,1);TICK;H_YZ_0_1_2_3_4_5_6_7_8_9_10_11_12_13_14_15_16_17_18_19_20_21_22_23;TICK;MZZ_5_6_7_13_10_16_11_12_17_18;M_0_1_2_3_4_8_9_14_15_19_20_21_22_23;DETECTOR(2.5,3,0)_rec[-72]_rec[-71]_rec[-70]_rec[-50]_rec[-49]_rec[-46]_rec[-26]_rec[-25]_rec[-22]_rec[-17]_rec[-16]_rec[-15];DETECTOR(-0.5,6,0)_rec[-68]_rec[-67]_rec[-54]_rec[-30]_rec[-13]_rec[-12];DETECTOR(1.5,0,0)_rec[-66]_rec[-63]_rec[-53]_rec[-29]_rec[-11]_rec[-8];DETECTOR(1.5,6,0)_rec[-73]_rec[-64]_rec[-62]_rec[-51]_rec[-48]_rec[-27]_rec[-24]_rec[-18]_rec[-9]_rec[-7];DETECTOR(3.5,0,0)_rec[-61]_rec[-58]_rec[-47]_rec[-23]_rec[-6]_rec[-3];DETECTOR(3.5,6,0)_rec[-60]_rec[-59]_rec[-45]_rec[-21]_rec[-5]_rec[-4];DETECTOR(0.5,3,0)_rec[-74]_rec[-69]_rec[-65]_rec[-55]_rec[-52]_rec[-31]_rec[-28]_rec[-19]_rec[-14]_rec[-10];DETECTOR(4.5,3,0)_rec[-57]_rec[-56]_rec[-44]_rec[-20]_rec[-2]_rec[-1];TICK;C_NZYX_0_1_2_3_4_5_6_7_8_9_10_11_12_13_14_15_16_17_18_19_20_21_22_23;SHIFT_COORDS(0,0,1);TICK;MZZ_0_1_2_8_3_4_5_11_6_7_9_10_12_13_14_20_15_16_17_23_18_19_21_22;DETECTOR(1.5,4,0)_rec[-95]_rec[-94]_rec[-92]_rec[-86]_rec[-85]_rec[-83]_rec[-31]_rec[-30]_rec[-28]_rec[-9]_rec[-8]_rec[-6];TICK;C_XYNZ_0_1_2_3_4_5_6_7_8_9_10_11_12_13_14_15_16_17_18_19_20_21_22_23;SHIFT_COORDS(0,0,1);TICK;MZZ_5_6_7_13_10_16_11_12_17_18;M_0_1_2_3_4_8_9_14_15_19_20_21_22_23;DETECTOR(1.5,4,0)_rec[-50]_rec[-49]_rec[-47]_rec[-19]_rec[-18]_rec[-16];DETECTOR(0.5,1,0)_rec[-42]_rec[-41]_rec[-11]_rec[-10];DETECTOR(0.5,7,0)_rec[-43]_rec[-40]_rec[-12]_rec[-9];DETECTOR(2.5,1,0)_rec[-48]_rec[-39]_rec[-37]_rec[-17]_rec[-8]_rec[-6];DETECTOR(2.5,7,0)_rec[-38]_rec[-35]_rec[-7]_rec[-4];DETECTOR(4.5,1,0)_rec[-34]_rec[-33]_rec[-3]_rec[-2];DETECTOR(-0.5,4,0)_rec[-45]_rec[-44]_rec[-14]_rec[-13];DETECTOR(3.5,4,0)_rec[-46]_rec[-36]_rec[-32]_rec[-15]_rec[-5]_rec[-1];TICK;H_YZ_0_1_2_3_4_5_6_7_8_9_10_11_12_13_14_15_16_17_18_19_20_21_22_23;SHIFT_COORDS(0,0,1);TICK;MZZ_0_6_1_2_3_9_4_5_7_8_10_11_12_18_13_14_15_21_16_17_19_20_22_23;DETECTOR(2.5,3,0)_rec[-69]_rec[-68]_rec[-65]_rec[-60]_rec[-59]_rec[-58]_rec[-29]_rec[-28]_rec[-27]_rec[-7]_rec[-6]_rec[-3];SHIFT_COORDS(0,0,1);TICK;H_0_1_2_3_4_5_6_7_8_9_10_11_12_13_14_15_16_17_18_19_20_21_22_23;TICK;MZZ_0_1_2_8_3_4_5_11_6_7_9_10_12_13_14_20_15_16_17_23_18_19_21_22;DETECTOR(0.5,5,0)_rec[-98]_rec[-97]_rec[-94]_rec[-86]_rec[-85]_rec[-82]_rec[-24]_rec[-23]_rec[-20]_rec[-12]_rec[-11]_rec[-8];DETECTOR(1.5,2,0)_rec[-96]_rec[-95]_rec[-93]_rec[-84]_rec[-83]_rec[-81]_rec[-22]_rec[-21]_rec[-19]_rec[-10]_rec[-9]_rec[-7];DETECTOR(2.5,5,0)_rec[-92]_rec[-91]_rec[-88]_rec[-80]_rec[-79]_rec[-76]_rec[-18]_rec[-17]_rec[-14]_rec[-6]_rec[-5]_rec[-2];DETECTOR(3.5,2,0)_rec[-90]_rec[-89]_rec[-87]_rec[-78]_rec[-77]_rec[-75]_rec[-16]_rec[-15]_rec[-13]_rec[-4]_rec[-3]_rec[-1];TICK;H_0_1_2_3_4_5_6_7_8_9_10_11_12_13_14_15_16_17_18_19_20_21_22_23;SHIFT_COORDS(0,0,1);TICK;MZZ_0_6_1_2_3_9_4_5_7_8_10_11_12_18_13_14_15_21_16_17_19_20_22_23;DETECTOR(0.5,5,0)_rec[-36]_rec[-35]_rec[-32]_rec[-12]_rec[-11]_rec[-8];DETECTOR(1.5,2,0)_rec[-34]_rec[-33]_rec[-31]_rec[-10]_rec[-9]_rec[-7];DETECTOR(2.5,5,0)_rec[-30]_rec[-29]_rec[-26]_rec[-6]_rec[-5]_rec[-2];DETECTOR(3.5,2,0)_rec[-28]_rec[-27]_rec[-25]_rec[-4]_rec[-3]_rec[-1];SHIFT_COORDS(0,0,1);TICK;H_YZ_0_1_2_3_4_5_6_7_8_9_10_11_12_13_14_15_16_17_18_19_20_21_22_23;TICK;MZZ_5_6_7_13_10_16_11_12_17_18;M_0_1_2_3_4_8_9_14_15_19_20_21_22_23;DETECTOR(2.5,3,0)_rec[-72]_rec[-71]_rec[-70]_rec[-50]_rec[-49]_rec[-46]_rec[-26]_rec[-25]_rec[-22]_rec[-17]_rec[-16]_rec[-15];DETECTOR(-0.5,6,0)_rec[-68]_rec[-67]_rec[-54]_rec[-30]_rec[-13]_rec[-12];DETECTOR(1.5,0,0)_rec[-66]_rec[-63]_rec[-53]_rec[-29]_rec[-11]_rec[-8];DETECTOR(1.5,6,0)_rec[-73]_rec[-64]_rec[-62]_rec[-51]_rec[-48]_rec[-27]_rec[-24]_rec[-18]_rec[-9]_rec[-7];DETECTOR(3.5,0,0)_rec[-61]_rec[-58]_rec[-47]_rec[-23]_rec[-6]_rec[-3];DETECTOR(3.5,6,0)_rec[-60]_rec[-59]_rec[-45]_rec[-21]_rec[-5]_rec[-4];DETECTOR(0.5,3,0)_rec[-74]_rec[-69]_rec[-65]_rec[-55]_rec[-52]_rec[-31]_rec[-28]_rec[-19]_rec[-14]_rec[-10];DETECTOR(4.5,3,0)_rec[-57]_rec[-56]_rec[-44]_rec[-20]_rec[-2]_rec[-1];TICK;C_NZYX_0_1_2_3_4_5_6_7_8_9_10_11_12_13_14_15_16_17_18_19_20_21_22_23;SHIFT_COORDS(0,0,1);TICK;MZZ_0_1_2_8_3_4_5_11_6_7_9_10_12_13_14_20_15_16_17_23_18_19_21_22;DETECTOR(1.5,4,0)_rec[-95]_rec[-94]_rec[-92]_rec[-86]_rec[-85]_rec[-83]_rec[-31]_rec[-30]_rec[-28]_rec[-9]_rec[-8]_rec[-6];TICK;C_XYNZ_0_1_2_3_4_5_6_7_8_9_10_11_12_13_14_15_16_17_18_19_20_21_22_23;SHIFT_COORDS(0,0,1);TICK;MZZ_5_6_7_13_10_16_11_12_17_18;M_0_1_2_3_4_8_9_14_15_19_20_21_22_23;DETECTOR(1.5,4,0)_rec[-50]_rec[-49]_rec[-47]_rec[-19]_rec[-18]_rec[-16];DETECTOR(0.5,1,0)_rec[-42]_rec[-41]_rec[-11]_rec[-10];DETECTOR(0.5,7,0)_rec[-43]_rec[-40]_rec[-12]_rec[-9];DETECTOR(2.5,1,0)_rec[-48]_rec[-39]_rec[-37]_rec[-17]_rec[-8]_rec[-6];DETECTOR(2.5,7,0)_rec[-38]_rec[-35]_rec[-7]_rec[-4];DETECTOR(4.5,1,0)_rec[-34]_rec[-33]_rec[-3]_rec[-2];DETECTOR(-0.5,4,0)_rec[-45]_rec[-44]_rec[-14]_rec[-13];DETECTOR(3.5,4,0)_rec[-46]_rec[-36]_rec[-32]_rec[-15]_rec[-5]_rec[-1];TICK;H_YZ_0_1_2_3_4_5_6_7_8_9_10_11_12_13_14_15_16_17_18_19_20_21_22_23;SHIFT_COORDS(0,0,1);TICK;MZZ_0_6_1_2_3_9_4_5_7_8_10_11_12_18_13_14_15_21_16_17_19_20_22_23;DETECTOR(2.5,3,0)_rec[-69]_rec[-68]_rec[-65]_rec[-60]_rec[-59]_rec[-58]_rec[-29]_rec[-28]_rec[-27]_rec[-7]_rec[-6]_rec[-3];SHIFT_COORDS(0,0,1);TICK;H_0_1_2_3_4_5_6_7_8_9_10_11_12_13_14_15_16_17_18_19_20_21_22_23;TICK;MZZ_0_1_2_8_3_4_5_11_6_7_9_10_12_13_14_20_15_16_17_23_18_19_21_22;DETECTOR(0.5,5,0)_rec[-98]_rec[-97]_rec[-94]_rec[-86]_rec[-85]_rec[-82]_rec[-24]_rec[-23]_rec[-20]_rec[-12]_rec[-11]_rec[-8];DETECTOR(1.5,2,0)_rec[-96]_rec[-95]_rec[-93]_rec[-84]_rec[-83]_rec[-81]_rec[-22]_rec[-21]_rec[-19]_rec[-10]_rec[-9]_rec[-7];DETECTOR(2.5,5,0)_rec[-92]_rec[-91]_rec[-88]_rec[-80]_rec[-79]_rec[-76]_rec[-18]_rec[-17]_rec[-14]_rec[-6]_rec[-5]_rec[-2];DETECTOR(3.5,2,0)_rec[-90]_rec[-89]_rec[-87]_rec[-78]_rec[-77]_rec[-75]_rec[-16]_rec[-15]_rec[-13]_rec[-4]_rec[-3]_rec[-1];OBSERVABLE_INCLUDE(0)_rec[-12]_rec[-11]_rec[-10]_rec[-9]_rec[-8]_rec[-7]_rec[-6]_rec[-5]_rec[-4]_rec[-3]_rec[-2]_rec[-1];TICK;H_0_1_2_3_4_5_6_7_8_9_10_11_12_13_14_15_16_17_18_19_20_21_22_23;SHIFT_COORDS(0,0,1);TICK;MZZ_0_6_1_2_3_9_4_5_7_8_10_11_12_18_13_14_15_21_16_17_19_20_22_23;DETECTOR(0.5,5,0)_rec[-36]_rec[-35]_rec[-32]_rec[-12]_rec[-11]_rec[-8];DETECTOR(1.5,2,0)_rec[-34]_rec[-33]_rec[-31]_rec[-10]_rec[-9]_rec[-7];DETECTOR(2.5,5,0)_rec[-30]_rec[-29]_rec[-26]_rec[-6]_rec[-5]_rec[-2];DETECTOR(3.5,2,0)_rec[-28]_rec[-27]_rec[-25]_rec[-4]_rec[-3]_rec[-1];OBSERVABLE_INCLUDE(0)_rec[-12]_rec[-11]_rec[-10]_rec[-9]_rec[-8]_rec[-7]_rec[-6]_rec[-5]_rec[-4]_rec[-3]_rec[-2]_rec[-1];SHIFT_COORDS(0,0,1);TICK;H_YZ_0_1_2_3_4_5_6_7_8_9_10_11_12_13_14_15_16_17_18_19_20_21_22_23;TICK;MZZ_5_6_7_13_10_16_11_12_17_18;M_0_1_2_3_4_8_9_14_15_19_20_21_22_23;DETECTOR(2.5,3,0)_rec[-72]_rec[-71]_rec[-70]_rec[-50]_rec[-49]_rec[-46]_rec[-26]_rec[-25]_rec[-22]_rec[-17]_rec[-16]_rec[-15];DETECTOR(-0.5,6,0)_rec[-68]_rec[-67]_rec[-54]_rec[-30]_rec[-13]_rec[-12];DETECTOR(1.5,0,0)_rec[-66]_rec[-63]_rec[-53]_rec[-29]_rec[-11]_rec[-8];DETECTOR(1.5,6,0)_rec[-73]_rec[-64]_rec[-62]_rec[-51]_rec[-48]_rec[-27]_rec[-24]_rec[-18]_rec[-9]_rec[-7];DETECTOR(3.5,0,0)_rec[-61]_rec[-58]_rec[-47]_rec[-23]_rec[-6]_rec[-3];DETECTOR(3.5,6,0)_rec[-60]_rec[-59]_rec[-45]_rec[-21]_rec[-5]_rec[-4];DETECTOR(0.5,3,0)_rec[-74]_rec[-69]_rec[-65]_rec[-55]_rec[-52]_rec[-31]_rec[-28]_rec[-19]_rec[-14]_rec[-10];DETECTOR(4.5,3,0)_rec[-57]_rec[-56]_rec[-44]_rec[-20]_rec[-2]_rec[-1];OBSERVABLE_INCLUDE(0)_rec[-19]_rec[-18]_rec[-17]_rec[-16]_rec[-15]_rec[-14]_rec[-13]_rec[-12]_rec[-11]_rec[-10]_rec[-9]_rec[-8]_rec[-7]_rec[-6]_rec[-5]_rec[-4]_rec[-3]_rec[-2]_rec[-1];TICK;C_NZYX_0_1_2_3_4_5_6_7_8_9_10_11_12_13_14_15_16_17_18_19_20_21_22_23;SHIFT_COORDS(0,0,1);TICK;MZZ_0_1_2_8_3_4_5_11_6_7_9_10_12_13_14_20_15_16_17_23_18_19_21_22;DETECTOR(1.5,4,0)_rec[-95]_rec[-94]_rec[-92]_rec[-86]_rec[-85]_rec[-83]_rec[-31]_rec[-30]_rec[-28]_rec[-9]_rec[-8]_rec[-6];TICK;C_XYNZ_0_1_2_3_4_5_6_7_8_9_10_11_12_13_14_15_16_17_18_19_20_21_22_23;SHIFT_COORDS(0,0,1);TICK;MZZ_5_6_7_13_10_16_11_12_17_18;M_0_1_2_3_4_8_9_14_15_19_20_21_22_23;DETECTOR(1.5,4,0)_rec[-50]_rec[-49]_rec[-47]_rec[-19]_rec[-18]_rec[-16];DETECTOR(0.5,1,0)_rec[-42]_rec[-41]_rec[-11]_rec[-10];DETECTOR(0.5,7,0)_rec[-43]_rec[-40]_rec[-12]_rec[-9];DETECTOR(2.5,1,0)_rec[-48]_rec[-39]_rec[-37]_rec[-17]_rec[-8]_rec[-6];DETECTOR(2.5,7,0)_rec[-38]_rec[-35]_rec[-7]_rec[-4];DETECTOR(4.5,1,0)_rec[-34]_rec[-33]_rec[-3]_rec[-2];DETECTOR(-0.5,4,0)_rec[-45]_rec[-44]_rec[-14]_rec[-13];DETECTOR(3.5,4,0)_rec[-46]_rec[-36]_rec[-32]_rec[-15]_rec[-5]_rec[-1];TICK;H_YZ_0_1_2_3_4_5_6_7_8_9_10_11_12_13_14_15_16_17_18_19_20_21_22_23;SHIFT_COORDS(0,0,1);TICK;M_0_1_2_3_4_5_6_7_8_9_10_11_12_13_14_15_16_17_18_19_20_21_22_23;DETECTOR(2.5,3,0)_rec[-81]_rec[-80]_rec[-77]_rec[-72]_rec[-71]_rec[-70]_rec[-41]_rec[-40]_rec[-39]_rec[-14]_rec[-13]_rec[-12]_rec[-8]_rec[-7]_rec[-6];DETECTOR(1.5,4,0)_rec[-52]_rec[-51]_rec[-49]_rec[-43]_rec[-42]_rec[-40]_rec[-19]_rec[-18]_rec[-17]_rec[-13]_rec[-12]_rec[-11]}{Crumble}}

%% file: crumble_url_vertical_stability.tex
\newcommand{\crumbleVerticalStability}{\href{http://algassert.com/crumble\#circuit=Q(0.0,0.0)0;Q(0.0,1.0)1;Q(0.0,2.0)2;Q(1.0,0.0)3;Q(1.0,1.0)4;Q(1.0,2.0)5;Q(1.0,3.0)6;Q(1.0,4.0)7;Q(1.0,5.0)8;Q(2.0,0.0)9;Q(2.0,1.0)10;Q(2.0,2.0)11;Q(2.0,3.0)12;Q(2.0,4.0)13;Q(2.0,5.0)14;Q(3.0,0.0)15;Q(3.0,1.0)16;Q(3.0,2.0)17;Q(3.0,3.0)18;Q(3.0,4.0)19;Q(3.0,5.0)20;Q(4.0,3.0)21;Q(4.0,4.0)22;Q(4.0,5.0)23;POLYGON(1,0,0,0.25)_1_0;POLYGON(1,0,0,0.25)_3_4_10_9;POLYGON(1,0,0,0.25)_8_14;POLYGON(1,0,0,0.25)_15_16;POLYGON(1,0,0,0.25)_20_23;POLYGON(1,0,0,0.25)_2_7_6_5;POLYGON(1,0,0,0.25)_11_12_13_19_18_17;POLYGON(1,0,0,0.25)_21_22;POLYGON(0,1,0,0.25)_0_3;POLYGON(0,1,0,0.25)_2_1;POLYGON(0,1,0,0.25)_8_7;POLYGON(0,1,0,0.25)_9_15;POLYGON(0,1,0,0.25)_4_5_6_12_11_10;POLYGON(0,1,0,0.25)_13_14_20_19;POLYGON(0,1,0,0.25)_16_17_18_21;POLYGON(0,1,0,0.25)_22_23;POLYGON(0,0,1,0.25)_0_1_2_5_4_3;POLYGON(0,0,1,0.25)_9_10_11_17_16_15;POLYGON(0,0,1,0.25)_6_7_8_14_13_12;POLYGON(0,0,1,0.25)_18_19_20_23_22_21;R_0_1_2_3_4_5_6_7_8_9_10_11_12_13_14_15_16_17_18_19_20_21_22_23;TICK;H_YZ_0_1_2_3_4_5_6_7_8_9_10_11_12_13_14_15_16_17_18_19_20_21_22_23;SHIFT_COORDS(0,0,1);TICK;MZZ_4_10_5_6_11_12_13_19_17_18;M_0_1_2_3_7_8_9_14_15_16_20_21_22_23;TICK;H_0_1_2_3_4_5_6_7_8_9_10_11_12_13_14_15_16_17_18_19_20_21_22_23;TICK;MZZ_0_3_1_2_4_5_6_12_7_8_9_15_10_11_13_14_16_17_18_21_19_20_22_23;DETECTOR(1.5,2,0)_rec[-31]_rec[-30]_rec[-29]_rec[-10]_rec[-9]_rec[-6];TICK;H_0_1_2_3_4_5_6_7_8_9_10_11_12_13_14_15_16_17_18_19_20_21_22_23;SHIFT_COORDS(0,0,1);TICK;MZZ_4_10_5_6_11_12_13_19_17_18;M_0_1_2_3_7_8_9_14_15_16_20_21_22_23;DETECTOR(1.5,2,0)_rec[-50]_rec[-49]_rec[-48]_rec[-19]_rec[-18]_rec[-17];DETECTOR(0.5,-1,0)_rec[-45]_rec[-42]_rec[-14]_rec[-11];DETECTOR(-0.5,2,0)_rec[-44]_rec[-43]_rec[-13]_rec[-12];DETECTOR(0.5,5,0)_rec[-41]_rec[-40]_rec[-10]_rec[-9];DETECTOR(2.5,-1,0)_rec[-39]_rec[-37]_rec[-8]_rec[-6];DETECTOR(2.5,5,0)_rec[-47]_rec[-38]_rec[-35]_rec[-16]_rec[-7]_rec[-4];DETECTOR(3.5,2,0)_rec[-46]_rec[-36]_rec[-34]_rec[-15]_rec[-5]_rec[-3];DETECTOR(4.5,5,0)_rec[-33]_rec[-32]_rec[-2]_rec[-1];SHIFT_COORDS(0,0,1);TICK;H_YZ_0_1_2_3_4_5_6_7_8_9_10_11_12_13_14_15_16_17_18_19_20_21_22_23;TICK;MZZ_0_1_2_5_3_4_6_7_8_14_9_10_11_17_12_13_15_16_18_19_20_23_21_22;DETECTOR(2.5,3,0)_rec[-60]_rec[-59]_rec[-58]_rec[-29]_rec[-28]_rec[-27]_rec[-6]_rec[-5]_rec[-3];TICK;C_NZYX_0_1_2_3_4_5_6_7_8_9_10_11_12_13_14_15_16_17_18_19_20_21_22_23;SHIFT_COORDS(0,0,1);TICK;MZZ_0_3_1_2_4_5_6_12_7_8_9_15_10_11_13_14_16_17_18_21_19_20_22_23;TICK;C_XYNZ_0_1_2_3_4_5_6_7_8_9_10_11_12_13_14_15_16_17_18_19_20_21_22_23;TICK;MZZ_0_1_2_5_3_4_6_7_8_14_9_10_11_17_12_13_15_16_18_19_20_23_21_22;DETECTOR(0.5,1,0)_rec[-36]_rec[-35]_rec[-34]_rec[-12]_rec[-11]_rec[-10];DETECTOR(2.5,1,0)_rec[-31]_rec[-30]_rec[-28]_rec[-7]_rec[-6]_rec[-4];DETECTOR(1.5,4,0)_rec[-33]_rec[-32]_rec[-29]_rec[-9]_rec[-8]_rec[-5];DETECTOR(3.5,4,0)_rec[-27]_rec[-26]_rec[-25]_rec[-3]_rec[-2]_rec[-1];TICK;H_YZ_0_1_2_3_4_5_6_7_8_9_10_11_12_13_14_15_16_17_18_19_20_21_22_23;SHIFT_COORDS(0,0,1);TICK;MZZ_4_10_5_6_11_12_13_19_17_18;M_0_1_2_3_7_8_9_14_15_16_20_21_22_23;DETECTOR(2.5,3,0)_rec[-72]_rec[-71]_rec[-70]_rec[-49]_rec[-48]_rec[-46]_rec[-25]_rec[-24]_rec[-22]_rec[-17]_rec[-16]_rec[-15];DETECTOR(-0.5,0,0)_rec[-69]_rec[-68]_rec[-55]_rec[-31]_rec[-14]_rec[-13];DETECTOR(1.5,0,0)_rec[-74]_rec[-66]_rec[-63]_rec[-53]_rec[-50]_rec[-29]_rec[-26]_rec[-19]_rec[-11]_rec[-8];DETECTOR(1.5,6,0)_rec[-64]_rec[-62]_rec[-51]_rec[-27]_rec[-9]_rec[-7];DETECTOR(3.5,0,0)_rec[-61]_rec[-60]_rec[-47]_rec[-23]_rec[-6]_rec[-5];DETECTOR(3.5,6,0)_rec[-59]_rec[-56]_rec[-45]_rec[-21]_rec[-4]_rec[-1];DETECTOR(0.5,3,0)_rec[-73]_rec[-67]_rec[-65]_rec[-54]_rec[-52]_rec[-30]_rec[-28]_rec[-18]_rec[-12]_rec[-10];DETECTOR(4.5,3,0)_rec[-58]_rec[-57]_rec[-44]_rec[-20]_rec[-3]_rec[-2];SHIFT_COORDS(0,0,1);TICK;H_0_1_2_3_4_5_6_7_8_9_10_11_12_13_14_15_16_17_18_19_20_21_22_23;TICK;MZZ_0_3_1_2_4_5_6_12_7_8_9_15_10_11_13_14_16_17_18_21_19_20_22_23;DETECTOR(1.5,2,0)_rec[-96]_rec[-95]_rec[-92]_rec[-86]_rec[-85]_rec[-84]_rec[-31]_rec[-30]_rec[-29]_rec[-10]_rec[-9]_rec[-6];TICK;H_0_1_2_3_4_5_6_7_8_9_10_11_12_13_14_15_16_17_18_19_20_21_22_23;SHIFT_COORDS(0,0,1);TICK;MZZ_4_10_5_6_11_12_13_19_17_18;M_0_1_2_3_7_8_9_14_15_16_20_21_22_23;DETECTOR(1.5,2,0)_rec[-50]_rec[-49]_rec[-48]_rec[-19]_rec[-18]_rec[-17];DETECTOR(0.5,-1,0)_rec[-45]_rec[-42]_rec[-14]_rec[-11];DETECTOR(-0.5,2,0)_rec[-44]_rec[-43]_rec[-13]_rec[-12];DETECTOR(0.5,5,0)_rec[-41]_rec[-40]_rec[-10]_rec[-9];DETECTOR(2.5,-1,0)_rec[-39]_rec[-37]_rec[-8]_rec[-6];DETECTOR(2.5,5,0)_rec[-47]_rec[-38]_rec[-35]_rec[-16]_rec[-7]_rec[-4];DETECTOR(3.5,2,0)_rec[-46]_rec[-36]_rec[-34]_rec[-15]_rec[-5]_rec[-3];DETECTOR(4.5,5,0)_rec[-33]_rec[-32]_rec[-2]_rec[-1];SHIFT_COORDS(0,0,1);TICK;H_YZ_0_1_2_3_4_5_6_7_8_9_10_11_12_13_14_15_16_17_18_19_20_21_22_23;TICK;MZZ_0_1_2_5_3_4_6_7_8_14_9_10_11_17_12_13_15_16_18_19_20_23_21_22;DETECTOR(2.5,3,0)_rec[-68]_rec[-67]_rec[-65]_rec[-60]_rec[-59]_rec[-58]_rec[-29]_rec[-28]_rec[-27]_rec[-6]_rec[-5]_rec[-3];TICK;C_NZYX_0_1_2_3_4_5_6_7_8_9_10_11_12_13_14_15_16_17_18_19_20_21_22_23;SHIFT_COORDS(0,0,1);TICK;MZZ_0_3_1_2_4_5_6_12_7_8_9_15_10_11_13_14_16_17_18_21_19_20_22_23;DETECTOR(0.5,1,0)_rec[-98]_rec[-97]_rec[-96]_rec[-86]_rec[-85]_rec[-84]_rec[-24]_rec[-23]_rec[-22]_rec[-12]_rec[-11]_rec[-10];DETECTOR(2.5,1,0)_rec[-93]_rec[-92]_rec[-90]_rec[-81]_rec[-80]_rec[-78]_rec[-19]_rec[-18]_rec[-16]_rec[-7]_rec[-6]_rec[-4];DETECTOR(1.5,4,0)_rec[-95]_rec[-94]_rec[-91]_rec[-83]_rec[-82]_rec[-79]_rec[-21]_rec[-20]_rec[-17]_rec[-9]_rec[-8]_rec[-5];DETECTOR(3.5,4,0)_rec[-89]_rec[-88]_rec[-87]_rec[-77]_rec[-76]_rec[-75]_rec[-15]_rec[-14]_rec[-13]_rec[-3]_rec[-2]_rec[-1];TICK;C_XYNZ_0_1_2_3_4_5_6_7_8_9_10_11_12_13_14_15_16_17_18_19_20_21_22_23;SHIFT_COORDS(0,0,1);TICK;MZZ_0_1_2_5_3_4_6_7_8_14_9_10_11_17_12_13_15_16_18_19_20_23_21_22;DETECTOR(0.5,1,0)_rec[-36]_rec[-35]_rec[-34]_rec[-12]_rec[-11]_rec[-10];DETECTOR(2.5,1,0)_rec[-31]_rec[-30]_rec[-28]_rec[-7]_rec[-6]_rec[-4];DETECTOR(1.5,4,0)_rec[-33]_rec[-32]_rec[-29]_rec[-9]_rec[-8]_rec[-5];DETECTOR(3.5,4,0)_rec[-27]_rec[-26]_rec[-25]_rec[-3]_rec[-2]_rec[-1];TICK;H_YZ_0_1_2_3_4_5_6_7_8_9_10_11_12_13_14_15_16_17_18_19_20_21_22_23;SHIFT_COORDS(0,0,1);TICK;MZZ_4_10_5_6_11_12_13_19_17_18;M_0_1_2_3_7_8_9_14_15_16_20_21_22_23;DETECTOR(2.5,3,0)_rec[-72]_rec[-71]_rec[-70]_rec[-49]_rec[-48]_rec[-46]_rec[-25]_rec[-24]_rec[-22]_rec[-17]_rec[-16]_rec[-15];DETECTOR(-0.5,0,0)_rec[-69]_rec[-68]_rec[-55]_rec[-31]_rec[-14]_rec[-13];DETECTOR(1.5,0,0)_rec[-74]_rec[-66]_rec[-63]_rec[-53]_rec[-50]_rec[-29]_rec[-26]_rec[-19]_rec[-11]_rec[-8];DETECTOR(1.5,6,0)_rec[-64]_rec[-62]_rec[-51]_rec[-27]_rec[-9]_rec[-7];DETECTOR(3.5,0,0)_rec[-61]_rec[-60]_rec[-47]_rec[-23]_rec[-6]_rec[-5];DETECTOR(3.5,6,0)_rec[-59]_rec[-56]_rec[-45]_rec[-21]_rec[-4]_rec[-1];DETECTOR(0.5,3,0)_rec[-73]_rec[-67]_rec[-65]_rec[-54]_rec[-52]_rec[-30]_rec[-28]_rec[-18]_rec[-12]_rec[-10];DETECTOR(4.5,3,0)_rec[-58]_rec[-57]_rec[-44]_rec[-20]_rec[-3]_rec[-2];SHIFT_COORDS(0,0,1);TICK;H_0_1_2_3_4_5_6_7_8_9_10_11_12_13_14_15_16_17_18_19_20_21_22_23;TICK;MZZ_0_3_1_2_4_5_6_12_7_8_9_15_10_11_13_14_16_17_18_21_19_20_22_23;DETECTOR(1.5,2,0)_rec[-96]_rec[-95]_rec[-92]_rec[-86]_rec[-85]_rec[-84]_rec[-31]_rec[-30]_rec[-29]_rec[-10]_rec[-9]_rec[-6];TICK;H_0_1_2_3_4_5_6_7_8_9_10_11_12_13_14_15_16_17_18_19_20_21_22_23;SHIFT_COORDS(0,0,1);TICK;MZZ_4_10_5_6_11_12_13_19_17_18;M_0_1_2_3_7_8_9_14_15_16_20_21_22_23;DETECTOR(1.5,2,0)_rec[-50]_rec[-49]_rec[-48]_rec[-19]_rec[-18]_rec[-17];DETECTOR(0.5,-1,0)_rec[-45]_rec[-42]_rec[-14]_rec[-11];DETECTOR(-0.5,2,0)_rec[-44]_rec[-43]_rec[-13]_rec[-12];DETECTOR(0.5,5,0)_rec[-41]_rec[-40]_rec[-10]_rec[-9];DETECTOR(2.5,-1,0)_rec[-39]_rec[-37]_rec[-8]_rec[-6];DETECTOR(2.5,5,0)_rec[-47]_rec[-38]_rec[-35]_rec[-16]_rec[-7]_rec[-4];DETECTOR(3.5,2,0)_rec[-46]_rec[-36]_rec[-34]_rec[-15]_rec[-5]_rec[-3];DETECTOR(4.5,5,0)_rec[-33]_rec[-32]_rec[-2]_rec[-1];SHIFT_COORDS(0,0,1);TICK;H_YZ_0_1_2_3_4_5_6_7_8_9_10_11_12_13_14_15_16_17_18_19_20_21_22_23;TICK;MZZ_0_1_2_5_3_4_6_7_8_14_9_10_11_17_12_13_15_16_18_19_20_23_21_22;DETECTOR(2.5,3,0)_rec[-68]_rec[-67]_rec[-65]_rec[-60]_rec[-59]_rec[-58]_rec[-29]_rec[-28]_rec[-27]_rec[-6]_rec[-5]_rec[-3];TICK;C_NZYX_0_1_2_3_4_5_6_7_8_9_10_11_12_13_14_15_16_17_18_19_20_21_22_23;SHIFT_COORDS(0,0,1);TICK;MZZ_0_3_1_2_4_5_6_12_7_8_9_15_10_11_13_14_16_17_18_21_19_20_22_23;DETECTOR(0.5,1,0)_rec[-98]_rec[-97]_rec[-96]_rec[-86]_rec[-85]_rec[-84]_rec[-24]_rec[-23]_rec[-22]_rec[-12]_rec[-11]_rec[-10];DETECTOR(2.5,1,0)_rec[-93]_rec[-92]_rec[-90]_rec[-81]_rec[-80]_rec[-78]_rec[-19]_rec[-18]_rec[-16]_rec[-7]_rec[-6]_rec[-4];DETECTOR(1.5,4,0)_rec[-95]_rec[-94]_rec[-91]_rec[-83]_rec[-82]_rec[-79]_rec[-21]_rec[-20]_rec[-17]_rec[-9]_rec[-8]_rec[-5];DETECTOR(3.5,4,0)_rec[-89]_rec[-88]_rec[-87]_rec[-77]_rec[-76]_rec[-75]_rec[-15]_rec[-14]_rec[-13]_rec[-3]_rec[-2]_rec[-1];OBSERVABLE_INCLUDE(0)_rec[-12]_rec[-11]_rec[-10]_rec[-9]_rec[-8]_rec[-7]_rec[-6]_rec[-5]_rec[-4]_rec[-3]_rec[-2]_rec[-1];TICK;C_XYNZ_0_1_2_3_4_5_6_7_8_9_10_11_12_13_14_15_16_17_18_19_20_21_22_23;SHIFT_COORDS(0,0,1);TICK;MZZ_0_1_2_5_3_4_6_7_8_14_9_10_11_17_12_13_15_16_18_19_20_23_21_22;DETECTOR(0.5,1,0)_rec[-36]_rec[-35]_rec[-34]_rec[-12]_rec[-11]_rec[-10];DETECTOR(2.5,1,0)_rec[-31]_rec[-30]_rec[-28]_rec[-7]_rec[-6]_rec[-4];DETECTOR(1.5,4,0)_rec[-33]_rec[-32]_rec[-29]_rec[-9]_rec[-8]_rec[-5];DETECTOR(3.5,4,0)_rec[-27]_rec[-26]_rec[-25]_rec[-3]_rec[-2]_rec[-1];OBSERVABLE_INCLUDE(0)_rec[-12]_rec[-11]_rec[-10]_rec[-9]_rec[-8]_rec[-7]_rec[-6]_rec[-5]_rec[-4]_rec[-3]_rec[-2]_rec[-1];TICK;H_YZ_0_1_2_3_4_5_6_7_8_9_10_11_12_13_14_15_16_17_18_19_20_21_22_23;SHIFT_COORDS(0,0,1);TICK;MZZ_4_10_5_6_11_12_13_19_17_18;M_0_1_2_3_7_8_9_14_15_16_20_21_22_23;DETECTOR(2.5,3,0)_rec[-72]_rec[-71]_rec[-70]_rec[-49]_rec[-48]_rec[-46]_rec[-25]_rec[-24]_rec[-22]_rec[-17]_rec[-16]_rec[-15];DETECTOR(-0.5,0,0)_rec[-69]_rec[-68]_rec[-55]_rec[-31]_rec[-14]_rec[-13];DETECTOR(1.5,0,0)_rec[-74]_rec[-66]_rec[-63]_rec[-53]_rec[-50]_rec[-29]_rec[-26]_rec[-19]_rec[-11]_rec[-8];DETECTOR(1.5,6,0)_rec[-64]_rec[-62]_rec[-51]_rec[-27]_rec[-9]_rec[-7];DETECTOR(3.5,0,0)_rec[-61]_rec[-60]_rec[-47]_rec[-23]_rec[-6]_rec[-5];DETECTOR(3.5,6,0)_rec[-59]_rec[-56]_rec[-45]_rec[-21]_rec[-4]_rec[-1];DETECTOR(0.5,3,0)_rec[-73]_rec[-67]_rec[-65]_rec[-54]_rec[-52]_rec[-30]_rec[-28]_rec[-18]_rec[-12]_rec[-10];DETECTOR(4.5,3,0)_rec[-58]_rec[-57]_rec[-44]_rec[-20]_rec[-3]_rec[-2];OBSERVABLE_INCLUDE(0)_rec[-19]_rec[-18]_rec[-17]_rec[-16]_rec[-15]_rec[-14]_rec[-13]_rec[-12]_rec[-11]_rec[-10]_rec[-9]_rec[-8]_rec[-7]_rec[-6]_rec[-5]_rec[-4]_rec[-3]_rec[-2]_rec[-1];SHIFT_COORDS(0,0,1);TICK;H_0_1_2_3_4_5_6_7_8_9_10_11_12_13_14_15_16_17_18_19_20_21_22_23;TICK;MZZ_0_3_1_2_4_5_6_12_7_8_9_15_10_11_13_14_16_17_18_21_19_20_22_23;DETECTOR(1.5,2,0)_rec[-96]_rec[-95]_rec[-92]_rec[-86]_rec[-85]_rec[-84]_rec[-31]_rec[-30]_rec[-29]_rec[-10]_rec[-9]_rec[-6];TICK;H_0_1_2_3_4_5_6_7_8_9_10_11_12_13_14_15_16_17_18_19_20_21_22_23;SHIFT_COORDS(0,0,1);TICK;MZZ_4_10_5_6_11_12_13_19_17_18;M_0_1_2_3_7_8_9_14_15_16_20_21_22_23;DETECTOR(1.5,2,0)_rec[-50]_rec[-49]_rec[-48]_rec[-19]_rec[-18]_rec[-17];DETECTOR(0.5,-1,0)_rec[-45]_rec[-42]_rec[-14]_rec[-11];DETECTOR(-0.5,2,0)_rec[-44]_rec[-43]_rec[-13]_rec[-12];DETECTOR(0.5,5,0)_rec[-41]_rec[-40]_rec[-10]_rec[-9];DETECTOR(2.5,-1,0)_rec[-39]_rec[-37]_rec[-8]_rec[-6];DETECTOR(2.5,5,0)_rec[-47]_rec[-38]_rec[-35]_rec[-16]_rec[-7]_rec[-4];DETECTOR(3.5,2,0)_rec[-46]_rec[-36]_rec[-34]_rec[-15]_rec[-5]_rec[-3];DETECTOR(4.5,5,0)_rec[-33]_rec[-32]_rec[-2]_rec[-1];SHIFT_COORDS(0,0,1);TICK;H_YZ_0_1_2_3_4_5_6_7_8_9_10_11_12_13_14_15_16_17_18_19_20_21_22_23;TICK;MZZ_0_1_2_5_3_4_6_7_8_14_9_10_11_17_12_13_15_16_18_19_20_23_21_22;DETECTOR(2.5,3,0)_rec[-68]_rec[-67]_rec[-65]_rec[-60]_rec[-59]_rec[-58]_rec[-29]_rec[-28]_rec[-27]_rec[-6]_rec[-5]_rec[-3];TICK;C_NZYX_0_1_2_3_4_5_6_7_8_9_10_11_12_13_14_15_16_17_18_19_20_21_22_23;SHIFT_COORDS(0,0,1);TICK;MZZ_0_3_1_2_4_5_6_12_7_8_9_15_10_11_13_14_16_17_18_21_19_20_22_23;DETECTOR(0.5,1,0)_rec[-98]_rec[-97]_rec[-96]_rec[-86]_rec[-85]_rec[-84]_rec[-24]_rec[-23]_rec[-22]_rec[-12]_rec[-11]_rec[-10];DETECTOR(2.5,1,0)_rec[-93]_rec[-92]_rec[-90]_rec[-81]_rec[-80]_rec[-78]_rec[-19]_rec[-18]_rec[-16]_rec[-7]_rec[-6]_rec[-4];DETECTOR(1.5,4,0)_rec[-95]_rec[-94]_rec[-91]_rec[-83]_rec[-82]_rec[-79]_rec[-21]_rec[-20]_rec[-17]_rec[-9]_rec[-8]_rec[-5];DETECTOR(3.5,4,0)_rec[-89]_rec[-88]_rec[-87]_rec[-77]_rec[-76]_rec[-75]_rec[-15]_rec[-14]_rec[-13]_rec[-3]_rec[-2]_rec[-1];TICK;C_XYNZ_0_1_2_3_4_5_6_7_8_9_10_11_12_13_14_15_16_17_18_19_20_21_22_23;SHIFT_COORDS(0,0,1);TICK;MZZ_0_1_2_5_3_4_6_7_8_14_9_10_11_17_12_13_15_16_18_19_20_23_21_22;DETECTOR(0.5,1,0)_rec[-36]_rec[-35]_rec[-34]_rec[-12]_rec[-11]_rec[-10];DETECTOR(2.5,1,0)_rec[-31]_rec[-30]_rec[-28]_rec[-7]_rec[-6]_rec[-4];DETECTOR(1.5,4,0)_rec[-33]_rec[-32]_rec[-29]_rec[-9]_rec[-8]_rec[-5];DETECTOR(3.5,4,0)_rec[-27]_rec[-26]_rec[-25]_rec[-3]_rec[-2]_rec[-1];SHIFT_COORDS(0,0,1);TICK;M_0_1_2_3_4_5_6_7_8_9_10_11_12_13_14_15_16_17_18_19_20_21_22_23;DETECTOR(1,0.5,0)_rec[-34]_rec[-21]_rec[-20];DETECTOR(0.5,2,0)_rec[-35]_rec[-22]_rec[-19];DETECTOR(0,0.5,0)_rec[-36]_rec[-24]_rec[-23];DETECTOR(3,0.5,0)_rec[-28]_rec[-9]_rec[-8];DETECTOR(2.5,2,0)_rec[-30]_rec[-13]_rec[-7];DETECTOR(2,0.5,0)_rec[-31]_rec[-15]_rec[-14];DETECTOR(2,3.5,0)_rec[-29]_rec[-12]_rec[-11];DETECTOR(1.5,5,0)_rec[-32]_rec[-16]_rec[-10];DETECTOR(1,3.5,0)_rec[-33]_rec[-18]_rec[-17];DETECTOR(4,3.5,0)_rec[-25]_rec[-3]_rec[-2];DETECTOR(3.5,5,0)_rec[-26]_rec[-4]_rec[-1];DETECTOR(3,3.5,0)_rec[-27]_rec[-6]_rec[-5];DETECTOR(1.5,2,0)_rec[-89]_rec[-88]_rec[-85]_rec[-79]_rec[-78]_rec[-77]_rec[-20]_rec[-19]_rec[-18]_rec[-14]_rec[-13]_rec[-12]}{Crumble}}